%% file: These.tex
\documentclass[a4paper,11pt,onecolumn]{mathese}

\usepackage{graphicx,floatflt}
\usepackage{amsmath}
\usepackage{amssymb}
\addto\captionsenglish{}

\newcommand{\dd}{\mathrm{d}} 
\newcommand{\Mpl}{M_\mathrm{p}} 
\newcommand{\mpl}{m_\mathrm{p}} 

\newcommand{\epsone}{\epsilon_1}
\newcommand{\epstwo}{\epsilon_2}
\newcommand{\rr}{\mathrm}
\newcommand{\lik}{\mathcal L}
\newcommand{\vect}[1]{\boldsymbol{#1}}

\newcommand{\vev}{{\textrm{vev}}}

\newcommand{\deriv}[2]{#1_{\negthinspace,#2}}

\newcommand{\calS}{\mathcal{S}}
\newcommand{\calM}{\mathcal{M}}
\newcommand{\calI}{\mathcal{I}}
\newcommand{\calH}{\mathcal{H}}
\newcommand{\calL}{\mathcal{L}}

\newcommand{\uc}{\mathrm{c}}
\newcommand{\ud}{\mathrm{d}}
\newcommand{\uB}{\mathrm{B}}

\newcommand{\uini}{\mathrm{ini}}
\newcommand{\ui}{\mathrm{i}}
\newcommand{\uend}{\mathrm{end}}

\newcommand{\coupling}{\kappa}

\newcommand{\uhit}{\mathrm{hit}}

\def\lsim{\mathrel{\lower4pt\hbox{$\sim$}}  
\hskip-12.5pt\raise1.6pt\hbox{$<$}\;}  
  
\def\gsim{\mathrel{\lower4pt\hbox{$\sim$}}  
\hskip-12.5pt\raise1.6pt\hbox{$>$}\;}

\def\beq{\begin{equation}}
\def\eeq{\end{equation}}

\def\beqa{\begin{eqnarray}}
\def\eeqa{\end{eqnarray}}

\begin{document}

\input{titre}

\newpage 

\frontmatter 
\addcontentsline{toc}{chapter}{Remerciements} 
\input{remerciements.tex}


\tableofcontents 
\addcontentsline{toc}{chapter}{Table of contents}

\input{abreviations.tex}

\addcontentsline{toc}{chapter}{Notations and conventions}

\setlength{\parskip}{1.2em} 
\mainmatter 
 
\input{intro}

\addcontentsline{toc}{chapter}{Introduction} 


\part{General context}

\input{FLRWmodel}

\input{The_inflationary_paradigm}

\part{Multi-field dynamics of hybrid inflationary models}

\input{hybrid}

\input{Slow-roll_violations}

\input{Initial_conditions}

\input{waterfall_phase}

\input{bounce} 

\part{21-cm Forecasts}
\input{21cmb}

\input{conclu}

\addcontentsline{toc}{chapter}{Conclusion and perspectives} 
 
\begin{appendix} 
\input{annexes/FFTT}

\input{annexes/fisher}

\input{annexes/mcmc}

\end{appendix} 
 
\bibliographystyle{hunsrt}
\addcontentsline{toc}{chapter}{Bibliography}
\bibliography{biblio} 
 \backmatter 
\end{document}

%% file: titre.tex
\thispagestyle{empty}
\begin{flushleft}
\begin{tabular}{clcl}
\parbox[c]{3cm}{\includegraphics[width=3cm]{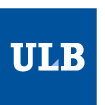}}&\begin{tabular}{l}\Large  Universit\'e Libre de Bruxelles \\
 Facult\'e des Sciences, D\'epartement de Physique\\ \\
										Service de Physique Th\'eorique (ULB)\\
								     Centre de Cosmologie, de Physique des Particules \\
								    et de Ph\'enom\'enologie (UCL)
										                 \end{tabular}
&\begin{tabular}{l}
  \parbox[c]{2.cm}{\includegraphics[width=1.5cm]{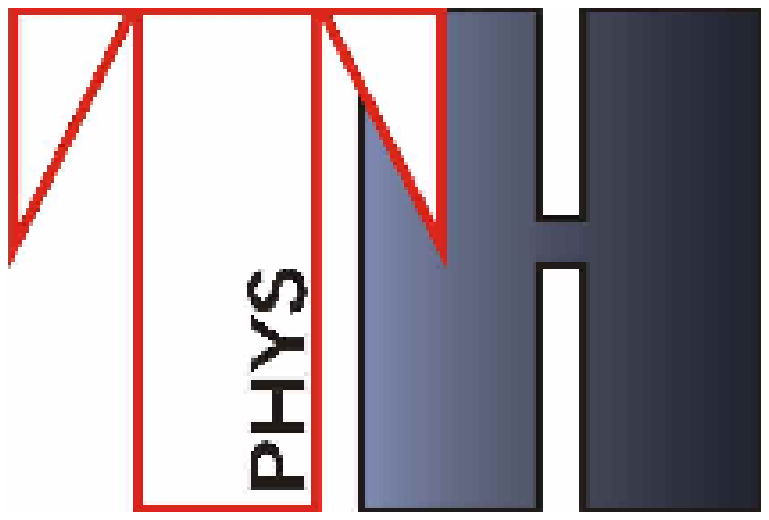}} 
  
    \vspace{1mm} 
    
\\
\vspace{1mm} 

\parbox[c]{2.cm}{\includegraphics[width=1.5cm]{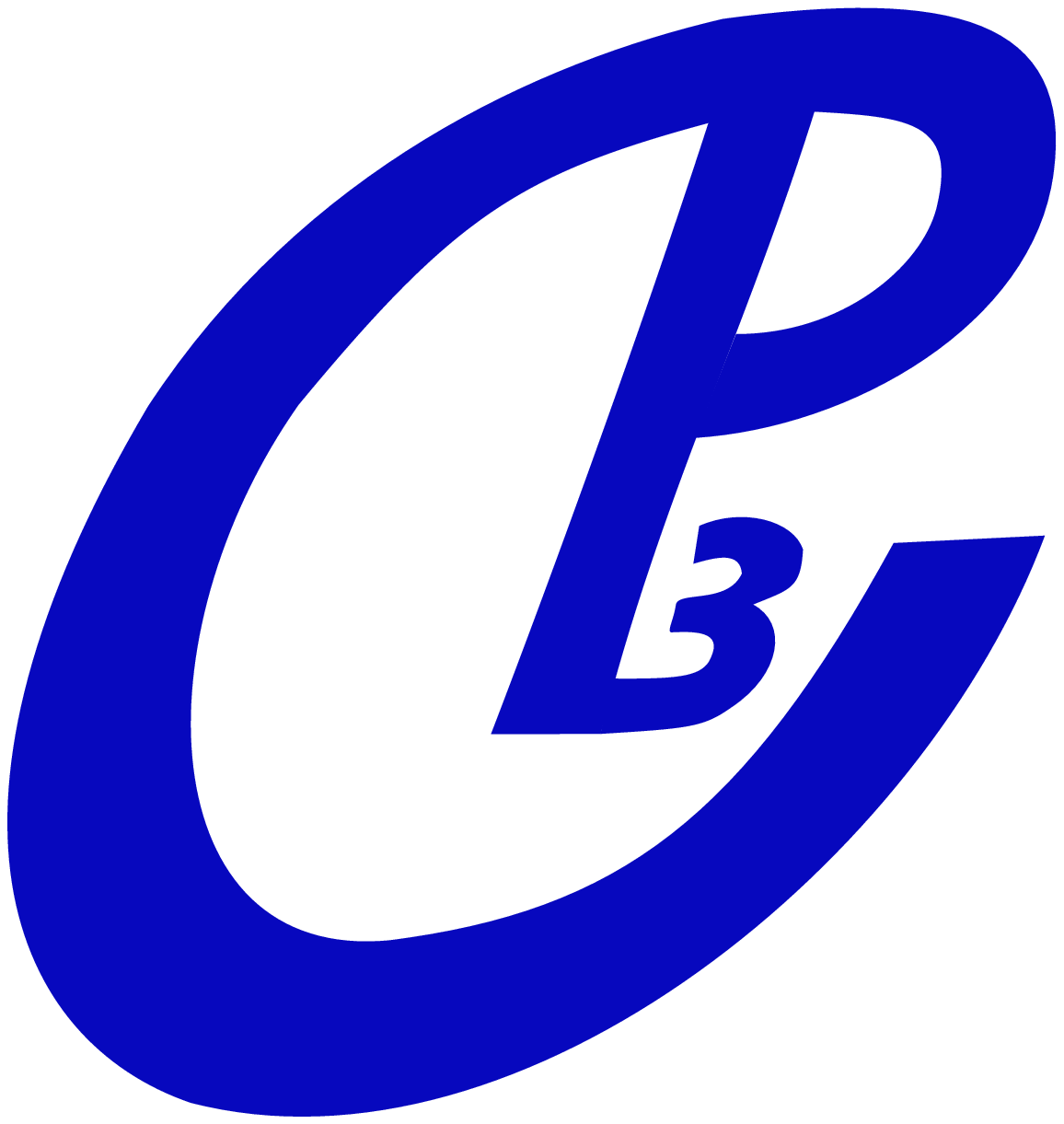}} 
 \end{tabular}
\end{tabular} 
\end{flushleft}

 \vspace*{\stretch{12}}
 \begin{center}
 \begin{tabular}{|c p{.80\textwidth}c|}\hline
 \rule{0ex}{0pt}\rule{0pt}{3ex}
&\begin{center}\huge \textbf{\textit{Hybrid Inflation: \\ Multi-field Dynamics and Cosmological Constraints}}\end{center}
 &\rule{0ex}{0pt} 
\\[6.5ex] \hline
 \end{tabular}
 
 \vspace*{3ex}
 \Large S\'ebastien Clesse
 \end{center}
 \vspace*{\stretch{15}}
 %
 \begin{center}\noindent
 Th\`ese pr\'esent\'ee en vue de l'obtention du titre de 
 
\textbf{Docteur en Sciences}

Juin 2011

\end{center}

\vspace*{\stretch{1.75}} 
\begin{minipage}[b]{.70\linewidth}
   \begin{flushleft}
 \begin{tabular}{ll}  
  \textbf{Promoteurs: } &  Professeur  C. Ringeval (UCL, Louvain)  \\
  & Professeur  M. Tytgat (ULB, Bruxelles)   \\
 \textbf{Jury:} &   Professeur A.C. Davis (Cambridge University) \\
 \phantom{\textbf{Jury: }} &    Professeur T. Hambye (ULB, Bruxelles)\\
\phantom{\textbf{Jury: }} &   Professeur J. Martin (IAP, Paris) \\
\phantom{\textbf{Jury: }} &    Professeur S. Van Eck (ULB, Bruxelles)
\end{tabular}


   \end{flushleft}
  \end{minipage} \hfill
\vspace*{\stretch{1}}

\newpage
\thispagestyle{empty}
\null

\newpage

\thispagestyle{empty}
\begin{flushleft}
\begin{tabular}{clcl}
\parbox[c]{3cm}{\includegraphics[width=3cm]{./figures/ulbquadri_mini.eps}}&\begin{tabular}{l}\Large  Universit\'e Libre de Bruxelles \\
 Facult\'e des Sciences, D\'epartement de Physique\\ \\
										Service de Physique Th\'eorique (ULB)\\
								     Centre de Cosmologie, de Physique des Particules \\
								    et de Ph\'enom\'enologie (UCL)
										                 \end{tabular}
&\begin{tabular}{l}
  \parbox[c]{2.cm}{\includegraphics[width=1.5cm]{./figures/spt.eps}} 
  
    \vspace{1mm} 
    
\\
\vspace{1mm} 

\parbox[c]{2.cm}{\includegraphics[width=1.5cm]{./figures/cp3.eps}} 
 \end{tabular}
\end{tabular} 
\end{flushleft}

 \vspace*{\stretch{12}}
 \begin{center}
 \begin{tabular}{|c p{.80\textwidth}c|}\hline
 \rule{0ex}{0pt}\rule{0pt}{3ex}
&\begin{center}\huge \textbf{\textit{Hybrid Inflation: \\ Multi-field Dynamics and Cosmological Constraints}}\end{center}
 &\rule{0ex}{0pt} 
\\[6.5ex] \hline
 \end{tabular}
 
 \vspace*{3ex}
 \Large S\'ebastien Clesse
 \end{center}
 \vspace*{\stretch{15}}
 %
 \begin{center}\noindent
 Th\`ese pr\'esent\'ee en vue de l'obtention du titre de 
 
\textbf{Docteur en Sciences}

Juin 2011

\end{center}

\vspace*{\stretch{1.75}} 
\begin{minipage}[b]{.70\linewidth}
   \begin{flushleft}
 \begin{tabular}{ll}  
  \textbf{Promoteurs: } &  Professeur  C. Ringeval (UCL, Louvain)  \\
  & Professeur  M. Tytgat (ULB, Bruxelles)   \\
 \textbf{Jury:} &   Professeur A.C. Davis (Cambridge University) \\
 \phantom{\textbf{Jury: }} &    Professeur T. Hambye (ULB, Bruxelles)\\
\phantom{\textbf{Jury: }} &   Professeur J. Martin (IAP, Paris) \\
\phantom{\textbf{Jury: }} &    Professeur S. Van Eck (ULB, Bruxelles)
\end{tabular}


   \end{flushleft}
  \end{minipage} \hfill
\vspace*{\stretch{1}}

 \newpage
 
\thispagestyle{empty} 
\null

\newpage

  \normalsize{ \textbf{Abstract:}  Hybrid models of inflation are particularly interesting and well motivated, since easily embedded in various high energy frameworks like supersymmetry/supergravity, Grand Unified Theories or extra-dimensional theories.  
   If the original hybrid model is often considered as disfavored, because it generically predicts a blue spectrum of scalar perturbations, realistic hybrid models can be in agreement with CMB observations.  
  The dynamics of hybrid models is usually approximated by the evolution of a scalar field slowly rolling along a nearly flat valley.  Inflation ends with a waterfall phase, due to a tachyonic instability.  This final phase is usually assumed to be nearly instantaneous.  
  
In this thesis, we go beyond these approximations and analyze the exact 2-field non-linear dynamics of hybrid models.  Several non trivial effects are put in evidence:  1) the possible violation of the slow-roll conditions along the valley induce the non existence of inflation at small field values.  Provided super-planckian fields, the scalar spectrum of the original model is red, in agreement with CMB observations, independently of the position of the critical instability point. 
 2) Contrary to what was thought, the initial field values leading to inflation are not fine-tuned along the valley but also occupy a considerable part of the field space exterior to it.  They form a complex connected structure with fractal boundaries that is the basin of attraction of the valley.  Using bayesian methods, their distribution in the whole parameter space, including initial velocities, is studied.  Natural bounds on the potential parameters are derived.   3)  For the original model, after the field evolution along the valley, inflation continues for more than 60 e-folds along the waterfall trajectories in some part of the parameter space.  Observable predictions are modified, and the scalar power spectrum of adiabatic perturbations is generically red, possibly in agreement with CMB observations.  Moreover, topological defects are conveniently stretched outside the observable Universe.  
  4)  The analysis of the initial conditions is extended to the case of a closed Universe, in which the initial singularity is replaced by a classical bounce.  Contrary to some other scenarios, due to the attractor nature of the valley, the field values in the contracting phase do not need to be extremely fine-tuned to generate a bounce followed by a phase of hybrid inflation.  
 
 In the third part of the thesis, we study how the present CMB constraints on the cosmological parameters could be ameliorated with the observation of the 21cm cosmic background from the dark ages and the reionization, by the future generation of giant radio-telescope.   Assuming ideal foreground removals, forecasts on the cosmological parameters are determined for a characteristic Fast Fourier Transform Telescope experiment, by using both Fisher matrix and MCMC methods.   }

\newpage
 
\thispagestyle{empty} 
\null

\newpage

%% file: remerciements.tex

 \vspace*{3cm}

{\it


Une page ne serait pas suffisante si je devais nommer tous ceux qui, de pr\`es ou de loin, ont contribu\'e \`a la r\'eussite de ces quatre ann\'ees de doctorat.   Je citerai tout de m\^eme mes collaborateurs directs, avec qui j'ai eu l'honneur et le tr\`es grand plaisir de travailler:  Dr. Jonathan Rocher, Dr. Marc Lilley, Dr. Larissa Lorenz, Dr. Laura H. Lopez et Dr. H. Tashiro.   Mes remerciements les plus sinc\`eres vont aussi naturellement \`a tous les membres du Service de Physique Th\'eorique de l'ULB et du Centre de Cosmologie, de Physique des Particules et de Ph\'enom\'enologie de l'UCL, en particulier les doctorants et post-doctorants, pour leur accueil, pour les \'echanges d'id\'ees, et pour l'ambiance toujours chaleureuse et  amicale.    

D'autre part, j'ai \'et\'e particuli\`erement touch\'e par la grande disponibilit\'e de personnalit\'es de renom pour r\'epondre \`a mes questions parfois na\"\i ves.  Je pense en particulier aux discussions fructueuses entretenues avec Juan Garcia-Bellido, David Lyth, Patrick Peter, Jer\^ome Martin et Andr\'e Fuzfa.  

J'ai l'immense privil\`ege de compter comme membres du jury Anne-Christine Davis, Sophie Van Eck, Jer\^ome martin et Thomas Hambye.    Je les remercie pour leur pr\'esence et pour l'int\'er\^et qu'ils ont port\'e \`a mes travaux.   

Ces quatre ann\'ees de doctorat ont \'et\'e financ\'ees par le Fond pour la Recherche dans l'Industrie et l'Agriculture (FRIA).  Je remercie cet organisme et particuli\`erement les membres des commissions scientifiques, pour la confiance accord\'ee \`a mon projet de recherche en Cosmologie.

Enfin, je suis extr\^emement reconnaissant du soutien et de la confiance que mes promoteurs, Michel Tytgat et Christophe Ringeval, m'ont accord\'es.  Toujours disponibles, ouverts \`a de nouvelles id\'ees,  ils sont sans conteste les \'el\'ements d\'ecisifs qui ont conduit \`a l'accomplissement de mes travaux et \`a la r\'edaction de cette th\`ese.   Plus particuli\`erement, je dois \`a Christophe l'apprentissage de la rigueur scientifique, mes comp\'etences num\'eriques et la d\'ecouverte des indices de style romain.   Je profite aussi de ces quelques lignes pour lui adresser mes plus plates excuses pour avoir \'et\'e si souvent un  ``imbitable goret''.  Michel et sa patience inestimable m'ont appris l'intuition physique et les formalismes li\'es \`a la cosmologie 21cm.   Tous deux ont r\'eussi a tirer le meilleur de moi, tout en me laissant la libert\'e n\'ecessaire \`a mon \'epanouissmeent.  Pour toutes ces raisons, et pour bien plus encore, je vous adresse, Michel et Christophe, un immense et chaleureux Merci!

Mais le plus important \`a mes yeux est le soutien inconditionnel d'Alice et de ma famille.   Je remercie plus particuli\`erement Alice pour avoir support\'e mon stress et mes sautes d'humeur durant la phase de r\'edaction de cette th\`ese.  

\vspace{3mm}
Merci \`a tous ceux que j'oublie, qui se reconna\^itront...
}

\newpage

\null

\newpage

{\flushright

\null 
\vspace{10.cm}
\textit{``L'essentiel est invisible'',}

Antoine de Saint-Exup\'ery

\vspace{2cm}
\textit{``Ce n'est pas parce qu'on fait des choses s\'erieuses \\
qu'il faut le faire en tirant une gueule d'enterrement!''} 

Alain Moussiaux

}

%% file: abreviations.tex
\chapternonum{Abbreviation List} 
\label{ch:abreviations} 
 
\noindent 
 
\noindent 
\begin{tabular}{ll} 
2DF & Two degree field  \\
ACBAR & Arcminute Cosmology Bolometer Array Receiver \\
BAO & Baryonic Acoustic Oscillations \\
BBN      & Big Bang Nucleosynthesis \\ 
CDM      & Cold Dark Matter\\
CDMS &  Cryogenic Dark Matter Search   \\
CMB      & Cosmic Microwave Background \\ 
DM       & Dark Matter\\ 
dof      & degree of freedom\\
e.o.m. & equation of motion \\
FFTT & Fast Fourier Transform Telescope \\
FL & Friedmann-Lema\^itre \\
FLRW     & Friedmann-Lema\^itre-Robertson-Walker\\ 
GR & General Relativity \\
HI & neutral hydrogen \\
i.c. &  Initial Condition \\
IGM & Inter Galactic Medium \\
KG & Klein-Gordon \\
LHC & Large Hadron Collider  \\
LOFAR & LOw Frequency ARray \\
LSS & Large Scale Structures \\
MCMC & Monte-Carlo-Markov-Chain \\
MWA & Murchison Widefield Array \\
PNGB & Pseudo Nambu Goldstone Boson \\
QUaD & Q and U Extragalactic Sub-mm Telescope at DASI \\
SDSS & Sloan Digital Sky Survey  \\
SM      & Standard Model\\ 
SKA  & Square Kilometre Array \\
SUGRA & Supergravity \\
SUSY     & Supersymmetry\\ 
vev      & Vacuum Expectation Value\\
WMAP    & Wilkinson Microwave Anisotropy Probe\\
WIMP     & Weakly Interacting Massive Particle\\
 \end{tabular}

%% file: intro.tex
\chapternonum{Introduction and motivations}
\label{chap:intro}



Since the 1990's and the COBE experiment, the cosmology has entered into an era of high precision. Measurements of the anisotropies in the Cosmic Microwave Background (CMB)  have become increasingly accurate. Combined with the observations of the large scale structures and the type-Ia supernovae, they have permitted to measure and constrain with accuracy the parameters of the standard cosmological model.  The main contributions to the energy density of the Universe today are the dark energy (71\%) and the dark matter (23\%).  But their nature and origin still have to be understood.  

Moreover, for the cosmological model to be in accord with observations, the inhomogeneities  at the origin of the galaxies are required to follow precise statistical properties in the early Universe.  
These can be obtained in a natural way if a phase of quasi-exponentially accelerated expansion is assumed to occur in the early stages of the Universe's evolution.  Such a phase of inflation can be realized if the Universe is filled with one or more scalar fields slowly rolling along their potential.   However, such scalar fields cannot be integrated in the standard model (SM) of particle physics.   A major challenge is thus to identify these fields in a theory beyond the standard model.  

Among the zoo of inflationary models, the hybrid class is particularly interesting and motivated since such models are easily embedded in various high energy frameworks like supersymmetry/supergravity (SUSY/SUGRA)~\cite{Halyo:1996pp,Binetruy:1996xj,Dvali:1994ms,
Kallosh:2003ux}, Grand Unified Theories (GUT)~\cite{Jeannerot:1997is,
Jeannerot:2003qv} or extra-dimensional theories (see e.g. Refs.~\cite{Dvali:1998pa,Brax:2007xq,Koyama:2003yc,Fukuyama:2008dv,Linde:2005dd,ArkaniHamed:2003wu}).   It is therefore important to determine correctly and accurately their dynamics and their observable signatures.  

In hybrid models, inflation takes place in the false vacuum along a nearly flat valley of the scalar field potential.  The accelerated expansion ends due to a Higgs-type tachyonic instability that forces the field trajectories to deviate and to reach one of the global minima of the potential during the so-called waterfall phase.  A phase of tachyonic preheating~\cite{Felder:2000hj,Felder:2001kt} is triggered during the waterfall when the mass of the transverse field becomes larger than the Hubble expansion rate.  Topological defects like cosmic strings can be formed when the initial symmetry is broken.  They can affect the primordial power spectrum of curvature perturbations.

\begin{figure}[h] \label{fig:pot_intro}  \begin{center}
 \includegraphics[width=13.0cm]{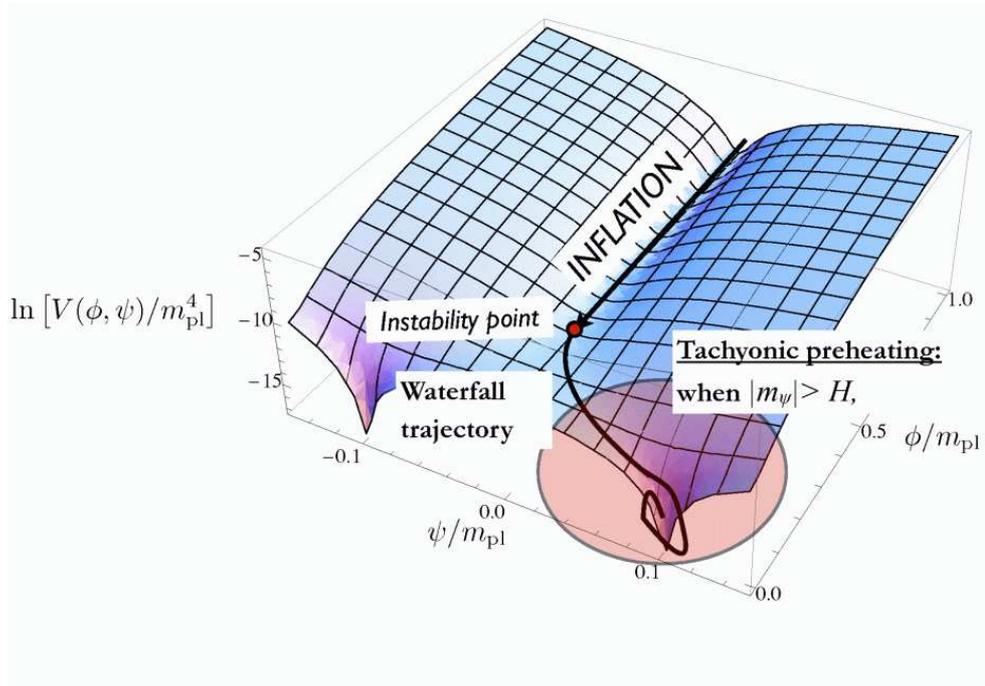}
   \caption{Logarithm of the scalar field potential $V(\phi,\psi)$ for the original model of hybrid inflation, where  $\phi$ is the inflaton field and $\psi$ is an auxiliary field.  Inflation can take place in the false vacuum, along the nearly flat valley, and ends with a waterfall phase due to the tachyonic instability.  A phase of tachyonic preheating is triggered when the mass $m_\psi$ of the auxiliary field becomes larger than $H$, the Hubble expansion rate. $m_{\rr{Pl}} $ is the Planck mass. } \end{center}
\end{figure}

In the usual way to study hybrid models, it is assumed that inflation is realized when the fields are slowly evolving in the valley along an effective 1-field potential.  It is also assumed that inflation ends instantaneously once the trajectories have reached the critical instability point.  In that 1-field slow-roll description, the observable quantities like the amplitude of the primordial power spectrum of curvature perturbations, its spectral index, and the ratio between tensor and scalar metric perturbations can be calculated after a Taylor expansion of the power spectra about a pivot length scale.  

For the original hybrid model~\cite{Linde:1993cn,Copeland:1994vg}, the predictions on the spectral index are such that it is disfavored by CMB experiments.  For its supersymmetric (SUSY) version, the F-term hybrid model~\cite{Dvali:1994ms}, the flat valley along which inflation takes place is lifted up by radiative and supergravity corrections ~\cite{Coleman:1973jx,Senoguz:2003zw,Jeannerot:2005mc,Jeannerot:2006jj}.   The spectral index can be in accord with observations, but the model is in tension with the data~\cite{Battye:2010hg}.   Many other hybrid-type models have been proposed, from various high energy frameworks, and their 1-field slow-roll predictions have been determined.  They are more or less in agreement with observations.

In the first part of this thesis, the standard cosmological model, the inflationary paradigm and the present status of observations are explained.  In the second part, we analyze the exact 2-field classical dynamics of hybrid models and put in evidence  several non trivial effects.  

\newpage
\begin{enumerate}
\item  \textbf{Slow-roll violations (chapter 4):}  For the original hybrid model, the slow-roll conditions can be violated during the field evolution along the valley.  Determining the effects of these slow-roll violations require the integration of the exact field dynamics.  We show that such violations induce the non existence of inflation at small field values.  This modifies the primordial power spectrum of curvature perturbations that can be in agreement with CMB observations, provided super-planckian initial field values.  These results are independent of the position of the critical instability point.   

\item \textbf{Set of initial field values (chapter 5):}   In a flat Universe, for generating more than 60 e-folds of accelerated expansion, the 2-field trajectories were usually required to be initially fine-tuned in a very narrow band along the inflationary valley or in some subdominant isolated points outside it.  From a more precise investigation of the dynamics, we have shown with C. Ringeval and J. Rocher \cite{Clesse:2008pf,Clesse:2009ur,Clesse:2009zd,Clesse:2010ht} that original hybrid inflation does not suffer from any fine-tuning problem, even when the fields are restricted to be sub-planckian.   Because of the attractor nature of the inflationary valley, a non-negligible part of the field trajectories initially exterior to the valley reach the slow-roll regime along it, after some oscillations around the bottom of the potential.    
We show that the set of successful initial field values is connected, of dimension two and possesses a fractal boundary of infinite length exploring the whole field space.

The relative area covered by successful initial field values depends on the potential parameters.  Therefore, a Monte-Carlo-Markov-Chain (MCMC) bayesian analysis is performed on the whole parameter space consisting of the initial field values, field velocities and potential parameters.   For each of these parameters, we give the marginalized posterior probability distributions such that inflation is long enough to solve the standard cosmological problems.   It is found that inflation is realized more probably by field trajectories starting outside the valley.   Natural bounds on potential parameters are also deduced.  

Finally, the genericity of our results are confirmed for 5 other hybrid models from various framework, namely the SUSY/SUGRA F-term, smooth and shifted hybrid models, as well as the radion assisted gauge inflation model.

\item \textbf{Inflation along waterfall trajectories (chapter 6):}  For the original hybrid model, the exact integration of the classical 2-field trajectories reveals that inflation can continue for more than 60 e-folds after crossing the critical instability point\cite{Clesse:2010iz}.   We first check that the classical dynamics is not spoiled by quantum back-reactions of the fields.  Then, by performing a MCMC analysis of the parameter space, we show that inflation along the waterfall trajectories lasts for more than $60$ e-folds in a large part of this space.   When this occurs, the predictions on the spectral index are modified, and the primordial power spectrum of curvature perturbations is possibly in agreement with the CMB constraints.   Moreover, the topological defects formed when the initial symmetry is broken at the critical instability point are diluted by the subsequent phase of inflation along the waterfall trajectories.  They become therefore non-observable.  

\item \textbf{Classical bounce plus hybrid inflation (chapter 7):}  With M. Lilley and L. Lorenz, we have extended the analysis of the chapter 5 to the case of a closed Universe, for which the initial singularity is replaced by a classical bounce.  Contrary to previously proposed scenarios, we show that the initial conditions in the contracting phase do not need to be extremely fine-tuned for hybrid inflation to be triggered after the bounce, provided that spatial curvature was initially sufficiently large.   

\end{enumerate}

Currently the best constraints on inflationary models come from the observations of the CMB temperature anisotropies.  In the future, a major challenge will consist in improving CMB measurements and in observing new cosmological signals.  

In the third part of the thesis, we are interested in one of these promising signals: the 21cm cosmic background.  This could be used to probe the dark ages and the reionization epochs.   The 21cm cosmic background is induced by the transitions between the hyperfine ground states of the neutral hydrogen (HI) atoms.   The signal corresponds to a stimulated emission or an absorption of 21cm CMB photons.  Compared to CMB, the 21cm signal is in principle observable over a wide range redshifts ($7 \lesssim z \lesssim 200$).  The observation of its anisotropies is expected to improve in the future the accuracy of the cosmological parameter measurements.  

With L. H. Lopez, C. Ringeval, H. Tashiro and M. Tytgat, we focus on a concept of 21cm dedicated giant radio-telescope, the Fast Fourier Transform Telescope (FFTT), and analyze its ability to put significant constraints on the cosmological parameters.  Our first motivation was to determine forecasts directly on the parameters of some inflation models, including hybrid ones, as well as on the reheating energy scale.  This objective is on his way and we give here a particular attention to the forecasts on the spectral index of the primordial power spectrum of curvature perturbations.  More specifically, we compare the interest of observing a 21cm signal from the dark ages and from the reionization.   We show that the observation of the 21cm signal from the dark ages should only contribute to put significant constraints on the spectral index for idealistic configurations of the FFTT experiment.  For the signal from the reionization, we obtain forecasts similar to those of Ref.~\cite{Mao:2008ug}.
  

The thesis is organized as follows:  In the first part, the general context is introduced and explained.  In chapter 1, we introduce the standard cosmological model, its observational confirmations and the current bounds on its parameters, as well as several problems and unanswered questions rising in this context.   In chapter 2, we explain how some of these problems can be solved naturally if one assumes a phase of inflation in the early Universe's evolution.   The homogeneous dynamics of 1-field inflationary models is described and the slow-roll approximation is introduced.   By using the linear theory of cosmological perturbations, observable predictions are derived in the slow-roll approximation.  Then, the dynamics of multi-field inflation models is described and we explain how to calculate the exact primordial power spectrum of scalar and tensor perturbations in this context.  Finally, the theory of the reheating after inflation is introduced.
   
In the second part, we study the exact multi-field dynamics of some hybrid models.  In chapter 3, these models are introduced and motivated.  In chapter 4, the effects of slow-roll violations during the field evolution along the valley are determined and discussed for the original hybrid model.  In chapter 5 we study, for a flat Universe, the set of the initial conditions leading to a sufficient amount of inflation, for all the hybrid models we have considered.  Chapter 6 is dedicated to the end of inflation in the original hybrid model.  In particular, we show that in a large part of the parameter space, inflation only ends after more than 60 e-folds of expansion are realized along the waterfall trajectories.  In chapter 7, we study the case of a closed Universe, in which the initial singularity is replaced by a classical bounce.  

The third part of the thesis is dedicated to 21cm forecasts.  In chapter 8, the theory of the 21cm cosmic background from the dark ages and the reionization is introduced.  In chapter 9, we analyze the ability of a typical FFTT radio-telescope to detect the 21cm power spectrum.   For two configurations of the experiment and ideal foreground removal, we calculate the forecasts on the cosmological parameters.  

The perspectives to this work are discussed in the conclusion.  In annexes, the bayesian MCMC methods and the Fisher matrix formalism are described.   The concepts of the FFTT radio-telescope are given and its advantages over the standard interferometers are explained.

%% file: FLRWmodel.tex
\chapter{Standard cosmological model}

\section{Introduction}

The standard cosmological scenario accurately describes the evolution and the structure of the Universe.  It relies on three major hypothesis:
\begin{enumerate}
\item The gravitational interaction obeys to the General Relativity (GR) theory.
\item At very large scales, the Universe can be considered as isotropic and homogeneous
\item The Universe is of trivial topology
\end{enumerate}
A dynamical cosmological model based on these assumptions was first proposed independently by Alexander Friedmann~\cite{Friedmann1922,Friedmann1924} and Georges Lema\^itre~\cite{Lemaitre1927}, respectively in 1922 and 1927.  
In 1929, Hubble first measured the expansion rate of the Universe~\cite{Hubble1929}, by interpreting the observation of redshifted spectral lines for the nearest galaxies~\cite{Slipher1915}.  
Related to the expansion, the idea that the Universe was born in a \textit{Big-Bang}, from an extremely dense initial state, has emerged and is today a cornerstone of the standard cosmological model.   This scenario is today confirmed by the observation of the Cosmic Microwave Background (CMB), relic of the period when free electrons recombined to atomic nuclei.  It relies also on the measurements of light element abundances.  These have been formed during a phase of primordial nucleosynthesis in the early Universe.  

In the next section the equations governing the space-time expansion are given.  In section~\ref{sec:fluids} we will apply these equations to various types of fluid filling the Universe.  We will find the corresponding expansion laws and energy density evolutions.  A brief description of the thermal history of the Universe will be given in section~\ref{sec:history}.  Section~\ref{sec:observations} is dedicated to the observations of astrophysical and cosmological signals that have permitted to measure precisely the various cosmological parameters.   Some unresolved questions rising from this cosmological scenario will be developed in section~\ref{sec:questions}.   

\section{The homogeneous FLRW model}


In comoving spherical coordinates $(r,\theta,\phi)$, imposing the hypothesis of isotropy and homogeneity leads to the Friedmann-Lema\^itre-Robertson-Walker (FLRW) metric\footnote{The metric signature $(-,+,+,+)$ is used and we work in the natural system of units $c=\hbar=k_{\mathrm B} =1$.  Greek indices go from 0 to 3.  Latin indices go from 1 to 3.   The Einstein convention of summing repeated indices is used.}
\begin{equation}
\dd s^2 = - \dd t^2 + a^2(t) \left[ \frac {\dd r^2}{1-K r^2} + r^2 \left( \sin^2{ \theta} \dd \phi^2 + \dd \theta^2 \right) \right] ~,
\end{equation}
where $t$ is the cosmic time, $a(t)$ is the scale factor and $K$ is the spatial curvature normalized to unity, such that $K=0$ if the Universe is flat, $K=1$ if it is closed and $K=-1$ if it is open. 
The cosmological dynamics is given by the Einstein equations\footnote{$R_{\mu \nu} = R^\sigma _{\ \mu \sigma \nu}$ is the Ricci tensor, $R_{\mu \nu \sigma \tau} $ is the Riemann tensor and $R = R^\mu _{\ \mu} $ is the Ricci scalar.  $T_{\mu \nu} $ is the stress-energy tensor.  In the comoving frame of an isotropic and homogeneous Universe, it is diagonal.  For a perfect fluid, the (0,0) component is the energy density $\rho$ and (i,i) components are the pressure $P$.  $\mpl = 1.2209 \times 10^{19} \rr{GeV} \  \rr{c}^{-2} $ is the Planck mass.  The reduced Planck mass will be denoted $\Mpl \equiv \mpl / \sqrt{8 \pi }$.  $\Lambda$ is a possible cosmological constant.}
\begin{equation}
 R_{\mu \nu} - \frac 1 2 R g_{\mu \nu}  + \Lambda g_{\mu \nu} = \frac{8 \pi}{\mpl^2} T_{\mu \nu}~,
 \end{equation}
applied to the FLRW metric.  This gives the Friedmann-Lema\^itre (FL) equations
\begin{eqnarray} \label{eq:FL}
H ^2 & = & \frac {8 \pi }{3 \mpl^2 } \rho - \frac{K}{a^2} + \frac 1 3 \Lambda~, \\  \label{eq:FL2}
\frac {\ddot a} a & = & - \frac {4   \pi }{3 \mpl^2}  (\rho + 3 P ) + \frac 1 3 \Lambda~,
\end{eqnarray}
in which a dot denotes the derivative with respect to the cosmic time $t$ and where the Hubble expansion rate $H(t) \equiv \dot a / a $ has been introduced.     
Furthermore, the conservation of the energy-momentum tensor ($\nabla_\mu T^{\mu \nu} = 0$) leads to 
\begin{equation} \label{eq:conserv}
\dot \rho + 3 H (\rho + P) = 0~,
\end{equation}
which is not independent since it can be derived also from Eq.~(\ref{eq:FL}) and Eq.~(\ref{eq:FL2}).    

\section{Matter content of the Universe } \label{sec:fluids}

The expansion dynamics depends on the characteristics of the fluid(s) filling the Universe.  One can define the equation of state parameter $w$ as 
\begin{equation}
w \equiv \frac P \rho ~.
\end{equation}
From the energy-momentum tensor conservation equation Eq.~(\ref{eq:conserv}) one concludes that the energy density of a perfect fluid characterized by a constant $w$ behaves like
\begin{equation}
\rho \propto a^{-3(1+w)} ~.
\end{equation}
Combined with Eq.~(\ref{eq:FL}), if $K = 0$ and $\Lambda = 0$, it is straightforward to show that the scale factor evolves like
\begin{equation}
a \propto t^{\frac{2}{3 (1+w)}}~.
\end{equation}
It is also useful to define the redshift, corresponding to the spectral shift of photon wavelengths due to the expansion,
\begin{equation}
z(t) + 1 \equiv \frac{a_0}{a(t)}~.
\end{equation}
The dynamics and the evolution of the energy density for some typical fluids are given below:
\begin{itemize}
\item \textit{Cosmological constant, $w= -1$} :  the cosmological constant acts like a perfect fluid whose energy density is constant, that is $w=-1$.  If the Universe is filled with such a fluid, the Hubble expansion term $H$ is constant and the scale factor grows exponentially,  $a \propto \exp ( H t) $.       

\item \textit{Curvature-like fluid, $w= -1/3$} :   if $K = -1 $, the curvature term in the FL equations is equivalent to a perfect fluid with $w= -1/3$.   The energy density goes like $\rho \propto a^{-2}$ and the scale factor grows linearly with the cosmic time, $a \propto t$ .

\item \textit{Pressureless matter dominated Universe, $w = 0 $} :   the energy density decreases like $\rho \propto a^{-3} $, because of the volume growth $\propto a^3$ of any comoving region.  The scale factor grows like $a \propto t^{2/3}$.  A non-relativistic baryonic fluid belongs to this class.   

\item \textit{Radiation dominated Universe, $w= 1/3$} :  the energy density decreases like $a^{-4}$.  Compared to the non-relativistic matter case, the additional $1/a$ factor can be viewed as the decrease of a photon energy whose wavelength increases due to the expansion.   The scale factor evolves like $a \propto t^{1/2} $.   Relativistic species (like relativistic massive neutrinos), belong to this class.   

\end{itemize}

As long as $w \geq - 1  $ and $K \leq 0 $, the scale factor reaches $0$ in a finite past while the energy density tends to infinity~\cite{PeterUzan}.  All known kinds of matter verify this bound on the equation of state parameter.  If  the Universe is closed and if the curvature term was dominant in the past, the singularity is replaced by a bounce~\cite{Starobinsky:1980}.  A model belonging to this specific case will be described and discussed in chapter 7. 

In the standard cosmological scenario, the so-called \textit{$\Lambda$CDM model}, the Universe has been successively dominated by radiation, pressureless matter and finally by cosmological constant, or identically a fluid with $w = -1$.     It is usual to introduce the notion of critical energy density $\rho_{\rr c}$, corresponding to the energy density of a flat Universe, 
\begin{equation}
\rho_{\rr c } \equiv \frac{3 \mpl^2  H^2  }{8 \pi}~.
\end{equation}
For each fluid filling the Universe, one can define the ratio of its energy density to the critical density today, 
\begin{equation}
\Omega_f \equiv \frac{\rho_f}{ \rho_{\rr c}}~.
\end{equation}
Let us define also $\Omega_\Lambda \equiv \frac{\Lambda }{ 3 H^2} $ and $\Omega_{\rr K} \equiv - \frac {K}{a^2 H^2} $.  Then the adimensional first FL equation reads
\begin{equation}
\sum_f \Omega_f + \Omega_\Lambda + \Omega_{\rr K} = 1~.
\end{equation}
In the $\Lambda$CDM model, the species contributing to the energy density today are the cosmological constant ($\Omega_\Lambda $), a pressureless matter component ($\Omega_{\rr{c}}$), the so-called \textit{cold dark matter} (CDM), the non-relativistic baryonic matter ($\Omega_{\rr b} $),  the photons ($\Omega_{\gamma}$) and the relativistic neutrinos ($\Omega_\nu $).  One may also consider a possible curvature term ($\Omega_{\rr K}$).   The Hubble expansion rate therefore evolves like
\begin{equation} \label{eq:H_of_t}
\frac{H(t)}{H_{\rr 0}} = \sqrt{ ( \Omega_{\rr{c}} + \Omega_{\rr b}  ) \left(\frac{a}{a_{\rr 0}} \right)^{-3}   + (\Omega_{\gamma} + \Omega_\nu  )   \left(\frac{a}{a_{\rr 0}} \right)^{-4}  + \Omega_\Lambda + \Omega_{\rr K} \left(\frac{a}{a_{\rr 0}} \right)^{-2} }~,
\end{equation}
where $H_{\rr 0}$ and $a_{\rr 0}$ are respectively the value of the Hubble parameter and the scale factor today.  The six cosmological parameters describing completely the homogeneous evolution of the $\Lambda$CDM model are therefore\footnote{Instead of  $\Omega_\nu $, it is usual to consider as a cosmological parameter the number of relativistic species } $\Omega_{\rr{c}}$, $\Omega_{\rr b} $, $\Omega_{\gamma}$, $\Omega_\nu $ and $\Omega_{\rr K}$ or  $\Omega_\Lambda $ today, as well as $h \equiv H_{\rr 0} / 100 \ \rr{km s^{-1} Mpc^{-1}} $. These parameters have been measured by observations, as explained later in section~\ref{sec:observations}.   

In a realistic scenario, the universe is only nearly isotropic and homogeneous.  The theory of cosmological perturbations~\cite{Mukhanov:1990me} permits to describe how density and metric perturbations are growing at the linear level.  Therefore, some additional cosmological parameters are required to provide initial conditions for the perturbative quantities.

\section{Thermodynamics in an expanding space-time}

Let us assume that the Universe is filled with several cosmological fluids. 
The Fermi-Dirac or Bose-Einstein distribution function for the species $i$ in kinetic equilibrium can be used to define the species temperature $T_i$, 
\begin{equation} 
f_i(\mathbf{p},T_i) = \frac{g_i}{\rr e^{(E - \mu_i )  /T_i } \pm 1 }~,
\end{equation}
where $\mathbf{p}$ is the momentum, $g_i$ is the degeneracy factor, $\mu_i$ is the chemical potential and $E^2 = p^2 + m^2$ where $m$ is the particle mass.   Other macroscopic quantities such as the particle number density, the energy density and the pressure are defined from this distribution function, 
\begin{eqnarray}  \label{eq:ni}
n_i & = & \frac{1}{(2 \pi)^3}   \int f_i(\mathbf{p}, T_i) \dd ^3 \mathbf p~, \\  \label{eq:rhodef}
\rho_i & = &   \frac{1}{(2 \pi)^3}   \int f_i(\mathbf{p}, T_i) E(p)  \dd ^3 \mathbf p~, \\
P_i & = &   \frac{1}{(2 \pi)^3}   \int f_i(\mathbf{p}, T_i) \frac{p^2}{3 E(p)}  \dd ^3 \mathbf p~, 
\end{eqnarray}
Let us consider several fluids in thermal equilibrium.  On one hand, the pressure variation with respect to the temperature reads
\begin{equation}
\frac{\dd P_i}{\dd T} = \frac 1 {T} \left(  \rho_i + P_i \right) + n_i T \frac{\dd }{\dd T} \left( \frac{\mu_i}{T} \right)~.
\end{equation}
On the other hand, the energy-momentum tensor conservation can be rewritten
\begin{equation}
a^3 \frac{\dd P}{\dd t} =   \frac{\dd }{\dd t} \left[ a^3 \left(\rho + P \right) \right] ~.
\end{equation}
Then, let us define a quantity $S$ as
\begin{equation}
S \equiv \sum_i \frac{a^3 \left( \rho_i + P_i - n_i \mu_i  \right)}{T}~.
\end{equation}  
By combining the last two equations, one obtains that $S$ satisfies to 
\begin{equation}
\dd S   = - \sum_i \frac{\mu_i}{T} \dd ( n_i a^3)~.
\end{equation}
For a constant number of particles in each species, 
 one sees that $S$ is conserved.  For a relativistic fluid, one has $T \propto 1/a $ and $\rho \propto T^4 $.  
 One recognizes in $S$ the \textit{entropy}.

In the standard $\Lambda$CDM model, the early Universe was dominated by interacting relativistic species.   Because their interaction rates were much larger than the expansion rate $\Gamma_i \gg H $, each species had the same temperature.   As the Universe expands, the interaction rates can become lower than the expansion rate.  In such a case, the corresponding species decouples from the other fluids and becomes a \textit{relic}.  


\section{Thermal history of the Universe} \label{sec:history}

\subsection{Early Universe}

At temperatures above 10 MeV ($z \sim 3 \times 10^{10}$)\footnote{In this section, the ratios $a / a_0 $ corresponding to the given temperatures are evaluated for the best fit values of the 
$\Lambda$CDM parameters.  These are given in section~\ref{sec:best_fits}.}, the standard model of particle physics predicts that the universe was filled with a mixture of photons, neutrinos, relativistic electrons and positrons, as well as non-relativistic protons and neutrons.  In the $\Lambda$CDM model, an additional non-relativistic component is assumed, the cold dark matter, that can be considered as interacting only gravitationally with the others species.  Due to the weak and electromagnetic interactions,
\begin{eqnarray}
\nu_{e} + n & \leftrightarrow & p + e~, \\
e^+ + n & \leftrightarrow & p + \bar \nu_{e}~, \\
n & \leftrightarrow & p + e + \bar \nu_e~,\\
e + e^+ & \leftrightarrow & \gamma + \gamma~,
\end{eqnarray}
these are in thermal equilibrium.  

Below 1 MeV ($z  \sim 3 \times 10^{9}$), neutrinos decouple and stop to interact with the other species.  A cosmic background of neutrinos is therefore expected.   

Below 511 keV ($z  \sim 2 \times 10^{9}$), electron-positron pairs annihilate into photons and thus the photon fluid is reheated compared to the neutrino fluid.


When the temperature decreases below 0.1 MeV ($z \sim 3 \times 10^{8}$), the photon energy becomes insufficient to photo-dissociate 
the eventually formed atomic nuclei.  Therefore light elements (deuterium, tritium, helium and lithium) can be formed~\cite{Alpher:1948,Gamow:1948,Alpher:1948b,Peebles:1966}.  This phase is called the \textit{primordial nucleosynthesis}, or \textit{Big-Bang nucleosynthesis} (BBN).   The resulting 
abundances in light elements depend on the baryon to photon ratio $\eta \equiv n_{\rr b} / n_\gamma $ and on the number of relativistic species $g_* $ (for a recent review, see~\cite{Mukhanov:2003xs}).  

The present light element abundances in the Universe can be evaluated with astrophysical observations  such that strong constraints have been established on the state of the Universe at the primordial nucleosynthesis epoch.   This will be explained in more details in the next section on observations.

\subsection{Matter dominated era}

Since radiation and relativistic matter energy densities decrease more slowly than the non-relativistic matter energy density, the Universe undergoes a transition from a radiation dominated era to a matter dominated era.  From Eq.~(\ref{eq:H_of_t}), if we neglect the cosmological constant, this happens when 
\begin{equation}
\frac{a}{a_{\rr 0}} = \frac{\Omega_{\gamma} + \Omega_\nu }{ \Omega_{\rr b} + \Omega_{\rr c} }~. 
\end{equation}
The time of matter-radiation equality thus depends on the energy density of neutrinos, itself depending on the effective number of neutrino species $N_\nu $.   For the best fit of the $\Lambda$CDM parameters, the matter-radiation equality occurs at a redshift $z_{\rr{eq}}= 3138$~\cite{Komatsu:2010fb}. 

\subsection{Recombination and cosmic microwave background} \label{sec:recomb}

When the photon energy goes below the binding energy of hydrogen atoms, $\epsilon_{0} = 13.6$ eV, 
free electrons can start to bind with protons and helium nuclei without being ionized anymore.   The mean free path of photons increases suddenly and becomes so large that they can propagate until today.   These photons can be observed as a black body, whose temperature was actually well below $T = 13.6 \rr{ eV} \simeq 158000$ K.  Indeed, since the number density of photons was much larger than the number density of electrons, $\eta^{-1} \sim 10^{10}$ (this ratio can be determined with the BBN theory and the observation of light element abundances), the recombination occurs only when the number density of photon with an energy $E > \epsilon_0$ is smaller than $n_\gamma / \eta $.
The corresponding temperature of the photon black body distribution is about  $T_{\rr{rec} } \sim 3100 \rr K $.    
Due to the expansion by a factor $ \sim 1100$, the photon black body spectrum is observed today with a temperature $T_{\rr{CMB}} = 2.725 \rr K$.   For an observer on Earth, it corresponds in the sky to an isotropic microwave radiation.   This was discovered accidently by Penzias and Wilson in 1964~\cite{Penzias:1965}. This is the so-called \textit{Cosmic Microwave Background} (CMB). 

\begin{figure}[ht]
\begin{center}
\scalebox{0.4}{\includegraphics{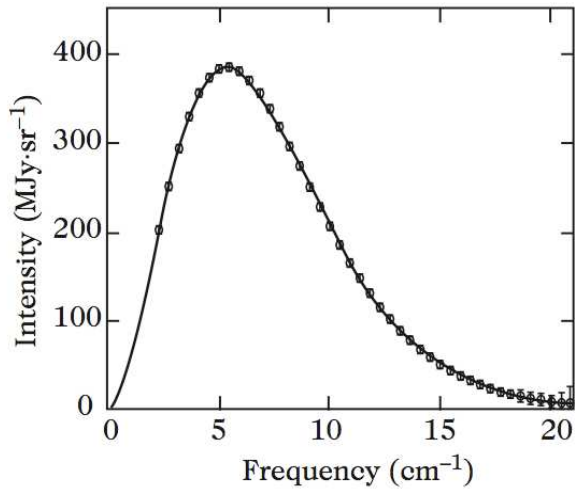}}
\caption{COBE measurements~\cite{Smoot:1998iq} of the CMB spectrum (error bars are multiplied by 200), in agreement with a black-body spectrum of temperature $T = 2.725$K.}
\end{center}
\end{figure}

Let us study more in details the recombination process.  The neutrality of the Universe imposes that $n_{\rr e} = n_{\rr p}$.  The free electron fraction is defined as $x_{\rr e} \equiv n_{\rr e} / (n_{\rr p} + n_{\rr H})  $.  Neglecting the Helium fraction, one has $n_{\rr b} = n_{\rr p} + n_{\rr H} $. As long as the interaction
\begin{equation}
p + e \leftrightarrow H + \gamma 
\end{equation} 
permits to maintain the equilibrium, one has from Eq.~(\ref{eq:ni}) (in the limit $m_{\rr e} \gg T$) 
\begin{equation}
n_{\rr {e}} = 2 \left(  \frac{m_{\rr e} T}{2 \pi}\right)^{3/2} \rr e^{-( m_{\rr e} - \mu_{\rr e} / T}~, 
\end{equation}
and the free electron fraction is given by the \textit{Saha equation}
\begin{equation} \label{eq:saha}
\frac{x_{\rr e } ^2}{1 - x_{\rr e}} = \frac{1}{n_{\rr b}  } \left( \frac{m_{\rr e } T}{2 \pi} \right)^{3/2} \rr e^{-\epsilon_0 / T  }~,
\end{equation}
where $\epsilon_0 = m_{\rr e} + m_{\rr p} - m_{\rr H} = 13.6$ eV.   Because $n_{\rr b } \ll n_\gamma$, when $T \sim \epsilon_{0} $, the right hand side is of the order of $10^{15}$ and the free electron fraction remains $x_{\rr e} \simeq 1$.  The recombination therefore only occurs at $T \ll \epsilon_0 $.

\begin{figure}[h!]
\begin{center}
\scalebox{1.}{\includegraphics{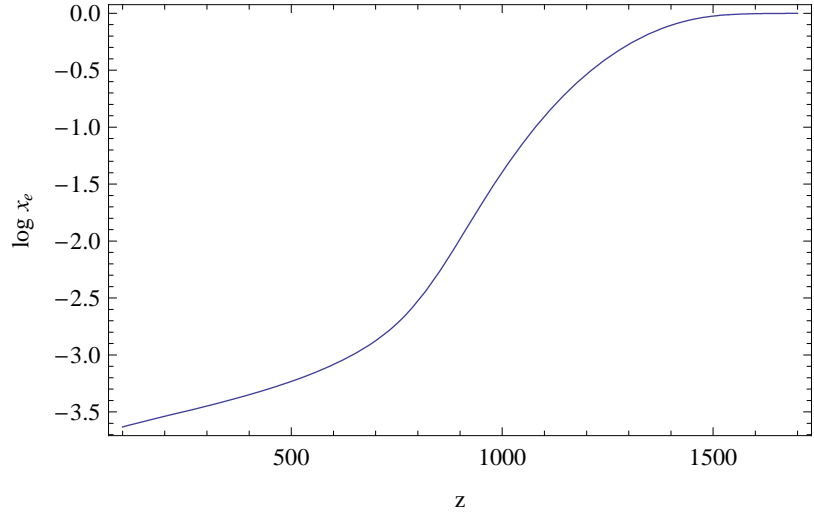}}
\caption{Free electron fraction as a function of redshift, calculated with Eq.~(\ref{eq:saha}) for $x_{\rr e} > 0.99 $ and by integrating numerically Eq.~(\ref{eq:recomb}) for $x_{\rr e} < 0.99 $.  After recombination at $z \sim 1100 $, it decreases and reach a value of the order $x_{\rr e} \sim 10^{-3} $ during the dark ages.  }  \label{fig:xe}
\end{center}
\end{figure}

At late time, the equilibrium is not maintained anymore and the Saha equation is not accurate.  To determine the free electron fraction, one needs to solve the Boltzmann equation for $x_{\rr e}$.  A good approximation\footnote{For more accurate results, the recombination can calculated numerically by using the RECFAST code~\cite{website:recfast}, taking account for additional effects like Helium recombination and 3-level atom. } is given in Refs.~\cite{Seager:1999bc,Seager:1999km,Dodelson},
\begin{equation} \label{eq:recomb}
\frac{\dd x_{\rr e}}{\dd t} = \left[ (1-x_{\rr e} ) \beta - x_{\rr e}^2 n_{\rr b} \alpha^{(2)} \right]~,
\end{equation}
where 
\begin{equation}
\beta \equiv \alpha^{(2)} \left( \frac{m_{\rr e} T }{2 \pi} \right)^{3/2} \rr e^{- \epsilon_{\rr 0} / T}
\end{equation}
is the ionization rate, and where
\begin{equation}
\alpha^{(2)} = 9.78 \frac{\alpha^2}{m_{\rr e}^2} \left( \frac{\epsilon_0}{T} \right)^{1/2} \ln \left(  \frac{\epsilon_0}{T} \right)
\end{equation}
is the recombination rate.  The superscript $^{(2)} $ indicates that recombination at the ground state is not relevant.  Indeed, this process leads to the production of a photon that ionizes immediately another neutral atom and thus there is no net effect.  
The free electron fraction evolution as a function of the redshift, calculated with the Saha equation Eq.~(\ref{eq:saha}) for $x_{\rr e} > 0.99$, and with Eq.~(\ref{eq:recomb}) for  $x_{\rr e} > 0.99$, is represented in Fig.~\ref{fig:xe}.

\subsection{The baryon decoupling of photons - the dark ages } \label{sec:darkages}

Between $z \sim 1100$ and $z \sim 10$, the first luminous objects are not yet formed.  
Because photons only interact weakly with the remaining small free electron fraction, no astrophysical or cosmological signal has been observed from this era.  This period is called the \textit{dark ages}.  Nevertheless,  because of the collisions between atoms, spin-flip transitions between the first hyperfine states of the neutral hydrogen atoms are possible.  This results in an absorption of 21-cm CMB photons, a signal in principle observable.  The interest of this signal and its ability to constrain cosmology will be studied and detailed in the last chapters of this thesis.   

Between $z \sim 1100$ and $z \sim 200$, even if photons interact weakly, the remaining free electron fraction, coupled to the baryons through Coulomb interaction, is sufficient for the gas temperature to be driven to the photon temperature.   
The energy transfer between photons and electrons is due to the Compton interaction.   The rate of energy transfer per unit of comoving volume between photons and free electrons is given by (see e.g.~\cite{Seager:1999bc,Seager:1999km})
\begin{equation}
\frac{\dd E_{\rr e, \gamma}}{\dd t} =    \frac{4 \sigma_{\rr T} \rho_\gamma  n_{\rr e} k_{\rr B} }{m_{\rr e} c} (T_\gamma - T_{\rr g} )~,
\end{equation}
where $ \sigma_{\rr T}$ is the Thomson cross section, $n_{\rr e} \sigma_{\rr T}$ is the scattering rate, $T_{\gamma} $ is the photon temperature and $T_{\rr g}$  is the gas temperature.  This energy transfer influences the baryon gas temperature. After using Eq.~(\ref{eq:rhodef}), it results that the gas temperature evolves according to 
\begin{eqnarray}
\frac{\dd T_{\rr g} }{\dd t} & = & -2 H T_{\rr g} + \frac {8 \sigma_{\rr T} \rho_{\gamma}  \left( T_{\gamma} - T_{\rr g} \right) } {3 m_{\rr e} c}\frac{n_{\rr e}} {n_{\rr b} } \\
& = &  -2 H T_{\rr g} + \frac {8 \sigma_{\rr T} \rho_{\gamma}  \left( T_{\gamma} - T_{\rr g} \right) } {3 m_{\rr e} c}   \frac{x_{\rr e}} {1+f_{\rr{He}} + x_{\rr e}} ~,
\end{eqnarray}
where $f_{\rr{He}} $ is the Helium fraction.  The first term on the right hand side is due to the volume expansion.   The second term accounts for the energy injection due to the Compton scattering between CMB photons and the residual free electrons.   Its last factor accounts for its distribution over the ionized fraction.   When the photon energy density $\rho_{\gamma} $ and the ionized fraction  $ x_{\rr i} = x_{\rr e}$ are sufficiently large, the Compton heating drives $T_{\rr g} \rightarrow T_{\gamma} $ such that the gas and the CMB photons have the same temperature.   With the expansion the photon energy density decreases, together with the ionized fraction.  At a redshift $z \sim 200 $, the gas temperature decouples from the radiation. Past this point it cools like $T_{\rr g} \propto 1/a^2 $, as expected for an adiabatic non-relativistic gas in expansion.  

\begin{figure}[h!]
\begin{center} \label{fig:TgTgamma}
\scalebox{1.}{\includegraphics{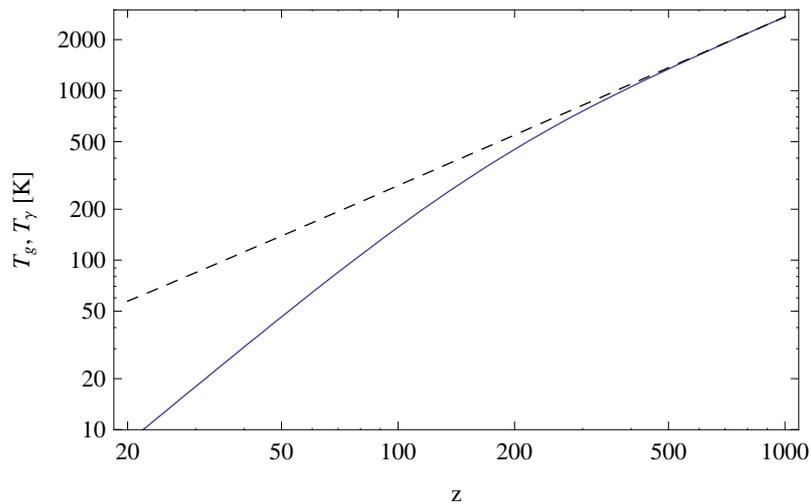}}
\caption{Evolution of the gas temperature $T_{\rr g}$ (plain line) and the photon temperature $ T_\gamma$ (dashed line), in the standard $\Lambda$CDM model.  The gas temperature is driven to the photon temperature until $z\sim 200$, due to the Compton interaction between photons and the remaining fraction of free electrons.  }
\end{center}
\end{figure}

\subsection{Reionization}

Around $z\sim 10 $ the first luminous astrophysical objects are formed.   These inject a large amount of radiation in the intergalactic medium (IGM) such that all the Universe is reionized.   

The reionization process is until now weakly known.  During reionization, free electrons can diffuse CMB photons and thus affect the optical depth of the signal.  
Our current knowledge about the reionization history relies on one hand on the measurement of the optical depth of CMB photons that can be used to determine the reionization redshift, for a given reionization model.    For instantaneous reionization, one has $z \sim 11 $~\cite{Komatsu:2010fb}.  

On the other hand, Gunn and Peterson~\cite{1965ApJ...142.1633G} have predicted in 1965 that the high-redshift quasar spectra must be suppressed at wavelengths less than that of the Lyman-$\alpha$ line at the redshift of emission, due to absorption by the neutral hydrogen in the IGM.   A Gunn-Peterson trough has been observed in the spectrum of quasars at $z > 6.28 $~\cite{Becker:2001ee}.   This method is used to fix a lower bound ($z\gsim 6$) on the reionization redshift.  

The reionization process itself can be investigated using complex numerical and semi-numerical methods (see e.g.~\cite{Santos:2009zk,Thomas:2010mz}), simulating the growth of structures and the energy transfer to the IGM.    

A 21-cm signal in absorption/emission against CMB from reionization is also in principle observable.   However, because collision rates are much lower than during the dark ages, the physical process generating hyperfine transitions is different.  Spin-flip transitions are induced by absorption and re-emission of Lyman-$\alpha$ photons emitted by the first stars.   Here again, details about the reionization 21cm signal and its ability to probe cosmology will be given in the last chapters of this thesis.


\section{Precision observational cosmology} \label{sec:observations}

With the measurements of temperature anisotropies in the CMB, the cosmology has entered into an era of high precision.  Combined with light element abundances, the observations of large scale structures and type Ia distant supernovae, they have permitted to determine with accuracy the  standard cosmological parameter values.  
The combination of several signals is important, for breaking the degeneracies between parameters that can affect a given observable in the same way.

The aim of this section is to give a brief review of these observations and to describe qualitatively how they can be used to constraint the cosmological parameters of the $\Lambda$CDM model.  At the end of this section, we give the current bounds  on these parameter values.  

\subsection{Hubble diagram}

\begin{figure}[!h]
\begin{center}
\scalebox{0.3}{\includegraphics{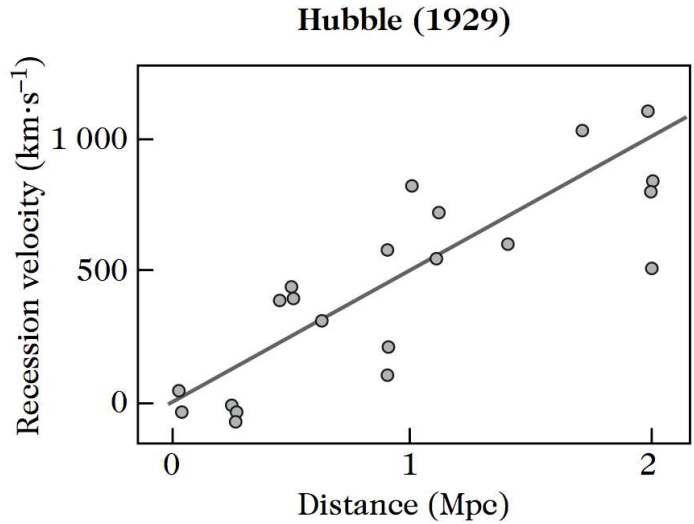}}
\caption{Original Hubble diagram~\cite{Hubble1929}, showing that the relative velocities of the nearest galaxies increase linearly with their distance.  Hubble found 
$H_0 \simeq 500 \ \rr{km \ s^{-1}   Mpc^{-1} }$.  } \label{fig:hubbleoriginal}
\end{center}
\end{figure}

\begin{figure}[!h]
\begin{center}
\scalebox{0.5}{\includegraphics{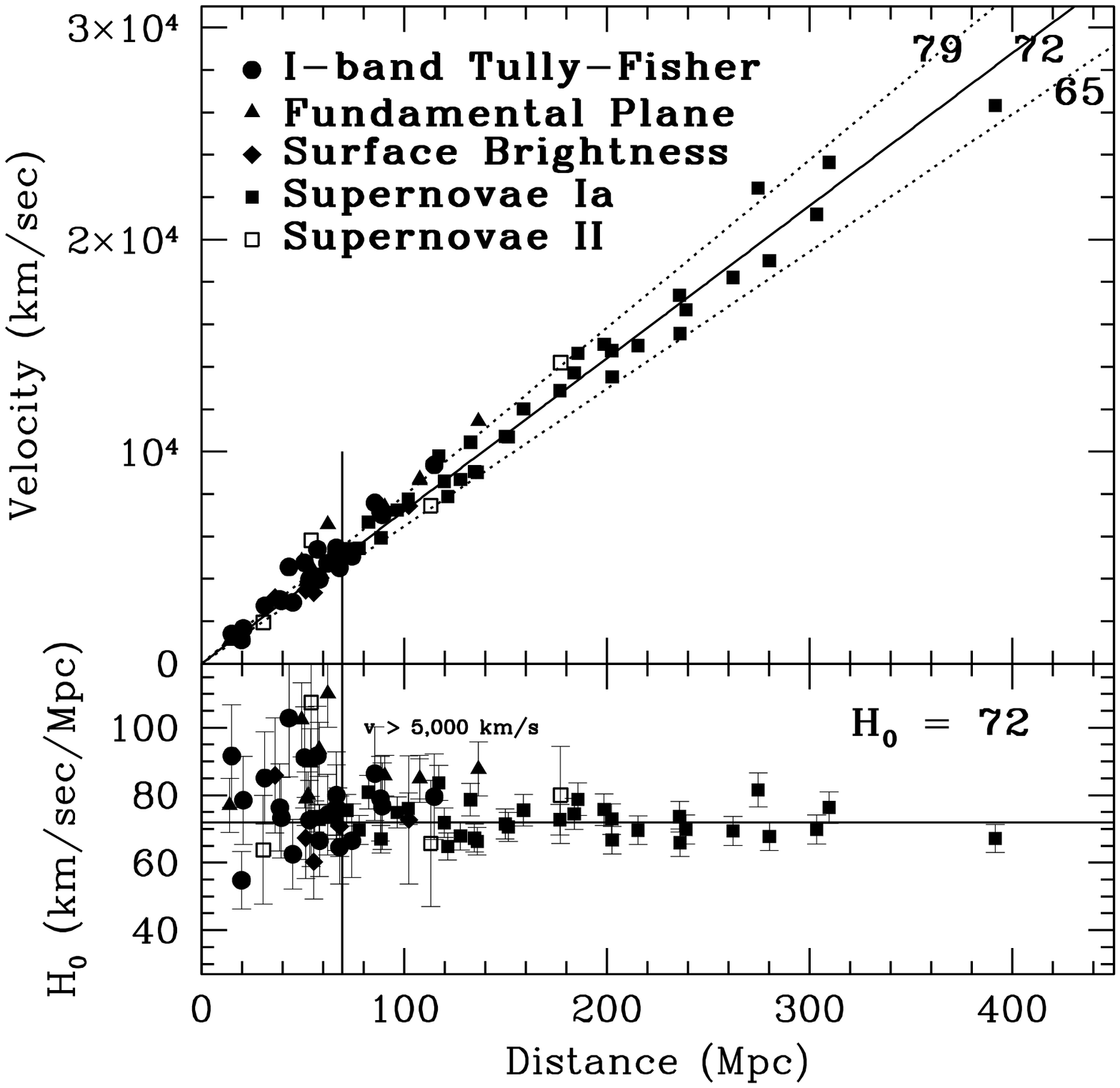}}
\caption{Hubble diagram~\cite{Freedman:2000cf}, from several observational techniques:  Tully-Fisher, Fundamental plane (analogue of the Tully-Fisher method for elliptic galaxies), Surface Brightness and supernovae.  The best constraints come from the type Ia supernovae, whose relative velocity can be measured up to distances of a few hundreds of Mpc.   Measurements are best fitted by the expansion rate value $H_0 = 72 \ \rr{km} \ \rr{ s^{-1} Mpc^{-1} }$.   } \label{fig:hubble}
\end{center}
\end{figure}

The Hubble expansion rate today $H_{\rr 0} = h \times 100 \ \rr{km \ s^{-1} Mpc^{-1}} $ can be determined by measuring the relative velocity of a large number of astrophysical objects as a function of their distance.   Velocities are calculated by measuring the spectral shift of the distant objects.  The original Hubble diagram~\cite{Hubble1929} (see Fig.~\ref{fig:hubbleoriginal}) put in evidence for the first time the expansion rate of the Universe by determining the velocity and the distance of 18 near galaxies.  Hubble found that $H_{\rr 0} \simeq 500  \rr{km \ s^{-1} Mpc^{-1}} $.  This value is far from the present measurement, $ H_{\rr 0} = 71   \pm 2.5  \rr{km \ s^{-1} Mpc^{-1}} $~\cite{Komatsu:2010fb} (see Fig.~\ref{fig:hubble}).  The difference between the original and the present values is due to inadequate methods for determining how distant the galaxies are~\cite{Webb:1999}.  

Today, distance measurements are much more accurate and several methods have permitted to calculate how distant extremely far objects are.    The present relative errors on the Hubble parameter are less than 5\%.   The methods for measuring distances include:

\begin{itemize} 
\item cepheids:  variable stars whose luminosity period has been empirically shown to be linked to their intrinsic luminosity~\cite{Udalski:1999pc}.
\item The Tully-Fisher relation:  this technique uses the correlation between the total luminosity of spiral galaxies with their maximal rotation velocity~\cite{Tully:1977fu}.  An analogous technique can be used for elliptic galaxies~\cite{PeterUzan}. 
\item  Type Ia supernovae:  they correspond to the explosion of white dwarf stars.  They are so luminous that they can be observed at a few hundreds of Mpc.  Their distance can be measured due to the correlation between their characteristic evolution time and their maximal luminosity~\cite{1979ApJ...232..404C}.  This technique can be used to estimate the Hubble expansion rate as a function of redshift.    

In 1998, the present acceleration of the Universe's expansion has been detected using type Ia supernova observations~\cite{Riess:1998cb}.   Combined with other signals, they have permitted to measure the value of $\Omega_\Lambda$ and to put constraints on the equation of state parameter $w$ of the dark energy fluid.    

\item Other techniques~\cite{Freedman:2000cf}, like type II supernovae (by measuring their angular size and their spectral Doppler shifts), or the fluctuations of the surface brightness of galaxies on the pixels of a CCD camera, that depend on the distance of the galaxy.     

\end{itemize}

\subsection{Abundances of light elements}

During the history of the Universe, the primordial abundances in light elements have been modified by various nuclear processes, like nuclear interactions in the stars.   
Nevertheless, these can be measured in sufficiently primitive astrophysical environments to be connected to the relative abundances at the end of the primordial nucleosynthesis.    From these measurements, and by using the theory of the primordial nucleosynthesis in an expanding space-time~\cite{Alpher:1948,Gamow:1948,Alpher:1948b,Peebles:1966,Mukhanov:2003xs}, it is possible to determine a range of
acceptable values of the parameter $\eta \equiv n_{\rr b} / n_{\rr \gamma} $ at the time of BBN.  

\begin{figure}[h!]
\begin{center} 
\scalebox{0.7}{\includegraphics{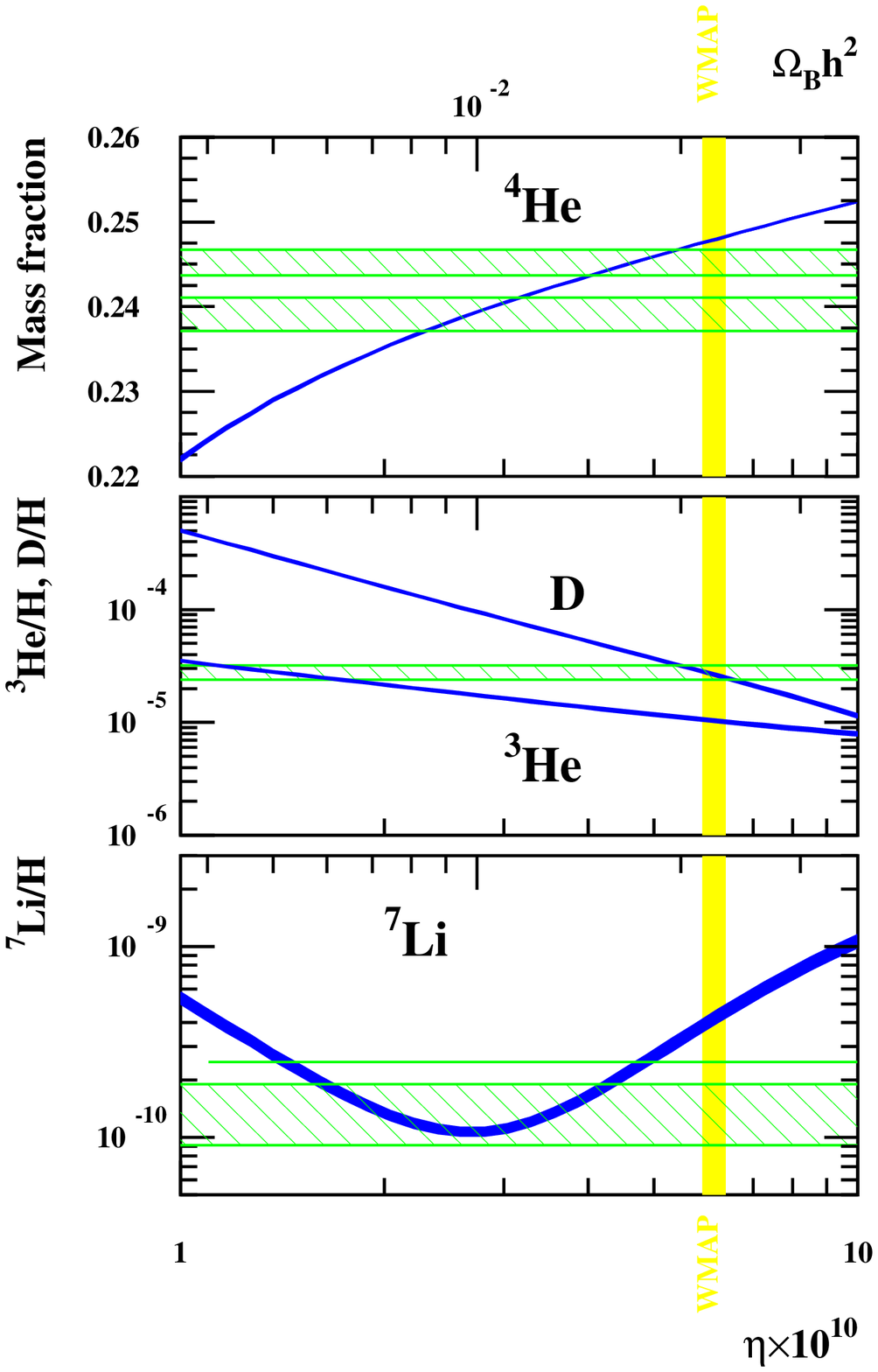}}
\caption{Abundances of Helium-3, Helium-4~\cite{Izotov:1999wa,Luridiana:2003jy,Izotov:2003xn}, Deuterium~\cite{Kirkman:2003uv}  and \hbox{Lythium-7~\cite{Ryan:2000zz}}, as a function of the baryon to photon ratio $\eta$.  The curves are the theoretical expectations.  Horizontal bands correspond to spectroscopic measurements, dark grey vertical band is the constraint from CMB observations by the WMAP satellite~\cite{Peiris:2005dt}.  Even if a slight difference is observed between astrophysical and CMB observations, they are in agreement with a value of $\eta$ of the order of $\eta \simeq 5 \times 10^{-10}$.  The figure is from Ref.~\cite{Coc:2003ce}.}  \label{fig:BBN}
\end{center}
\end{figure}

Fig.~\ref{fig:BBN} illustrates the observational status for the primordial abundances \hbox{Helium-4}~\cite{Izotov:1999wa,Luridiana:2003jy,Izotov:2003xn}, Deuterium~\cite{Kirkman:2003uv} and Lithium~\cite{Ryan:2000zz}.  One can see that observations are all compatible with a parameter $\eta \simeq 5 \times 10^{10}$.   This value corresponds today to a fractional energy density for baryons $\Omega_{\rr b} h^2\simeq 0.018 $.   Before CMB anisotropy observations, the light element abundances have been for a long time the only indirect measurement of the baryon energy density.  Today, observations also help to constraint other parameters, like the number of neutrino species $N_\nu$~\cite{Shimon:2010ug}.  They also constrain a possible variation of the fundamental constants~\cite{Dent:2007zz}.  

Finally, it must be noticed that the value of the parameter $\Omega_{\rr b} h^2 $ can be determined independently by CMB observations (see next point).   This results in a stronger constraint on the parameter $\eta$~\cite{Peiris:2005dt,Larson:2010gs}, as illustrated in Fig.~\ref{fig:BBN}.   

\subsection{The CMB anisotropies}

Before recombination, photons were tightly coupled to electrons and protons via Compton scattering.  The small inhomogeneities were prevented to collapse due to the pressure of photons.   Therefore, instead of growing like they do after recombination, the density perturbations have performed acoustic oscillations.
These left imprints in the cosmic microwave background, on the form of temperature anisotropies of the order of $10^{-5} $K.  These anisotropies have been put in evidence for the first time by the COBE (COsmic Background Explorer) satellite~\cite{1992ApJ...396L...1S} in 1992.  
The statistical properties of these temperature anisotropies have proved to be the an efficient tool to constrain the cosmological parameters.




\subsubsection{The angular power spectrum}


For gaussian temperature fluctuations, the statistical properties of the CMB sky map are encoded in the so-called \textit{angular power spectrum}.  
  
Let us define $\Theta_{\rr{obs}}(\mathbf{e})$, the temperature fluctuation compared to the average sky temperature in a specific sky direction $ \mathbf{e}$.   Then let us decompose these fluctuations in spherical harmonics $Y_{lm}(\mathbf e)$ with coefficients $a_{lm}^\Theta$, 
\begin{equation}
\Theta_{\rr{obs}}(\mathbf{e}) = \sum _l \sum_{|m| \le l }  a_{lm}^\Theta Y_{lm}(\mathbf e )~.
\end{equation}
The statistical properties of the sky map are encoded in the two-point correlation function 
\begin{equation}
\langle \Theta_{\rr{obs}}(\mathbf{e}) \Theta_{\rr{obs}}(\mathbf{e'})  \rangle = \frac 1 {4 \pi} \sum_l C_l^{\rr{obs}} P_l(\cos \theta )  ~,
\end{equation}
where we have used the isotropy hypothesis and the property of the Legendre polynomials,
\begin{equation}
P_l (\mathbf e \cdot \mathbf e' \equiv \cos \theta ) = \frac{4 \pi}{2 l +1} \sum_{m=-l}^l  Y_{lm}(\mathbf e )  Y_{lm}^* (\mathbf e' )~.
\end{equation}
Thus the $C_l^{\rr{obs}}$ coefficients are related to the $a_{lm}$ through the relation
\begin{equation}
C_l^{\rr{obs}} = \frac{1}{2 l +1} \sum_m |a_{lm}^\Theta |^2 ~.
\end{equation}
The next step is to compare the $C_l^{\rr{obs}} $ to the theoretical $C_l $.   The theoretical temperature field is an homogeneous and isotropic random field. The theoretical $a_{lm} $ coefficients are also independent stochastic fields, with vanishing mean value,
\begin{equation}
\langle  a_{lm}   \rangle = 0~, \hspace{20mm} \langle a_{lm} a^*_{l'm'} \rangle = \delta_{l-l'} \delta_{m-m'} C_l~.
\end{equation}
The theoretical $C_l$ are not directly observable, but an \textit{estimator} $\hat C_l$ can be obtained by summing over $m$,
\begin{equation}
\hat C_l = \frac{1 }{2 l +1} \sum_{|m| \le l } a_{lm} a^*_{lm}~.
\end{equation}
 If initial fluctuations follow a gaussian statistic, the $a_{lm} $ probability distribution function is also gaussian and reads
 \begin{equation} 
 P (a_{lm} )  = \frac{1}{\sqrt{2 \pi} C_l} \rr e^{- \frac{a_{lm}^2 } {2 C_l^2}}~.
 \end{equation}
 It results that the $\hat C_l$ follow a $\chi^2 $ distribution with $2 l + 1$ dof.  One sees that the $C_l^{\rr{obs}}$ can be used to \textit{estimate} the theoretical $C_l$.  This estimation can not be perfect, because it is obtained by averaging over the finite ($2 l +1$) set of the $a_{lm}^\Theta $ of the unique CMB sky map, whereas we want to estimate an ensemble average.   The intrinsic variance of the estimators $\hat C_l$ reads
\begin{equation}
\rr{Var}(\hat C_l ) \equiv \langle \hat C_l ^2 \rangle - \langle \hat C_l \rangle ^2 = \frac {2}{2 l +1} C_l^2~.
\end{equation}
As expected, the error is smaller if one has a larger number of $a_{lm} $ for the estimation.   Actually, it is possible to show that the $\hat C_l$ are the best estimators, and that the resulting variance is the smallest possible~\cite{Grishchuk:1997pk}.  This is called the \textit{cosmic variance}. 

\begin{figure}[h!]
\begin{center} 
\scalebox{1.3}{\includegraphics{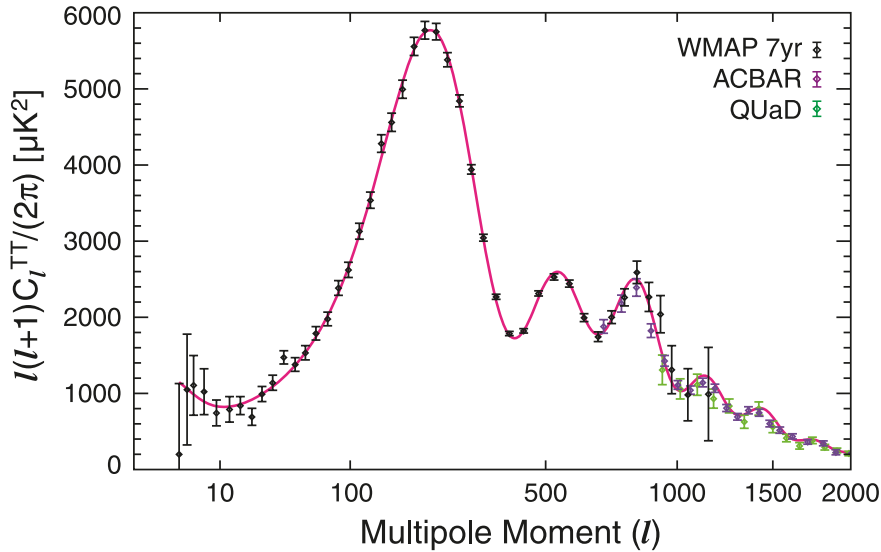}}
\caption{Angular power spectrum of CMB temperature fluctuations, showing the acoustic peaks and their damping at high multipoles, from WMAP7 data~\cite{Komatsu:2010fb,Larson:2010gs} and other recent CMB experiments (ACBAR~\cite{Reichardt:2008ay}, QUaD~\cite{Pryke:2008xp}). The curve is the best-fit for the $\Lambda$CDM model.  
}\label{fig:Cls}
\end{center}
\end{figure}

The present best measurements of the CMB angular power spectrum are shown in Fig.~\ref{fig:Cls}.  
As already mentioned in Sec.~\ref{sec:LSS}, prior to recombination the inhomogeneities in the tightly coupled baryon and photon fluid are prevented to collapse and perform acoustic oscillations.  These oscillations result from the competition between gravitational attraction and photon pressure.   At recombination, the density perturbations reaching for the first time a maximal amplitude lead to a maximum of temperature fluctuation for the emerging CMB photons, and induce a first peak in the CMB angular power spectrum.   The following peaks can be seen as its harmonics.  For instance, the second peak corresponds to perturbations having performed one complete oscillation at time of recombination.  These peaks are damped at high multipoles $l$.  This so-called \textit{Silk damping}~\cite{1968ApJ...151..459S} occurs because the acoustic waves can not propagate for perturbation modes whose wavelength is smaller than the mean free path of photons.   Large angular scale temperature fluctuations ($l\lesssim 20 $) are sourced at recombination by perturbations larger than the Hubble radius $1/H$.  In the super-Hubble regime, perturbations remain constant in time, and thus they conserve at recombination their initial amplitude.   

The complete evolution of the perturbations before the recombination can be determined in the context of the theory of cosmological perturbations~\cite{Mukhanov:1990me}.  This is done by solving both the perturbed Einstein equations and the first-order Boltzmann equations for all the species.  Since in the next chapters of the thesis we will focus mainly on models of inflation and on the post-recombination evolution, this calculation has not been reproduced here.  A detailed description of the evolution of perturbations prior to recombination and their effect on the angular power spectrum of CMB temperature anisotropies can be found in most textbooks on modern cosmology (see e.g.~\cite{Dodelson,PeterUzan}).   

\subsubsection{Dependance on cosmological parameters}

In this section, we give a qualitative description of the effects of the $\Lambda$CDM cosmological parameters on the shape of the CMB angular power spectrum, and more particularly on the positions and magnitudes of the acoustic peaks.  In the section~\ref{sec:best_fits}, the best fits of these parameters are given.  
   

\paragraph{The density of baryons $ \Omega_{\rr b} h^2$: }

The fractional energy density of baryons $  \Omega_{\rr b} h^2$ at fixed total matter density modifies the shape of the angular power spectrum in three ways:
\begin{itemize}
\item It fixes the sound velocity and thus the frequency of oscillations in the primordial baryon-photon plasma.  An increase of $ \Omega_{\rr b} h^2 $ leads to a reduction of the sound velocity, and thus a reduction of the oscillation frequency.  The acoustic peaks are therefore induced by perturbations of smaller wavelengths, entering earlier inside the Hubble radius, and they are thus shifted to higher multipoles $l$.  

\item It fixes the relative amplitude of odd and even peaks.   At constant total matter energy density, a reduction of the baryon energy density means that more dark matter can accumulate and dig deeper gravitational wells.  This induces a reduction of the relative magnitudes of the odd peaks, because the amplitude of the baryon perturbations are reduced each time they climb the more steep gravitational wells.  

\item The mean free path of photons due to the Compton scattering depends on the electron number density, and thus is affected by the baryon density (since the Universe is neutral, $n_{\rr b} = n_{\rr e}$).  The Silk damping is therefore affected by $ \Omega_{\rr b} h^2 $.  For an augmentation of $ \Omega_{\rr b} h^2 $, the diffusion length is reduced and the damping is less efficient, inducing a higher magnitude for the peaks at high multipoles in the angular power spectrum.
\end{itemize}

\paragraph{The total matter density $ \Omega_{\rr m} h^2=  \Omega_{\rr b} h^2 + \Omega_{\rr c} h^2$:}

The total matter energy density, for a fixed ratio $ \Omega_{\rr b} / \Omega_{\rr c}$, 
has two main effects on the angular power spectrum.
\begin{itemize}
\item When CMB photons emerge from an over-density, their wavelength is affected by the gravitational Doppler effect.  Increasing $ \Omega_{\rr m} h^2 $  affects the Doppler effect on the CMB photons and change the contrast between maxima and minima in the angular power spectrum. 
\item Increasing $ \Omega_{\rr m } h^2 $ also shifts acoustic peaks to higher multipoles, because it affects
the Hubble expansion rate.  At fixed value of\footnote{$ \Omega_{\rr r} = \Omega_\gamma + \Omega_\nu $ denotes the fractional energy density for the radiation} $ \Omega h^2$ and $ \Omega_{\rr r} h^2 $,  a higher value of $ \Omega_{\rr m} h^2$ reduces the time of matter/radiation equality and the moment of the last scattering.
\end{itemize}

\paragraph{The cosmological constant $\Omega_\Lambda $ and the curvature $\Omega_K $: }

At fixed values of $h$, $\Omega_{\rr m}$ and $\Omega_{\rr r} $, fixing a value of $ \Omega_K $ is equivalent to  fix $\Omega_\Lambda = 1 - \Omega_K - \Omega_{\rr m} - \Omega_{\rr r} $.  Their respective effect on the CMB angular power spectrum can thus be considered simultaneously.   
\begin{itemize}  
\item The Hubble expansion rate, and thus the relation between angular distances in the sky and the corresponding distance at a given redshift is modified with $\Omega_{\rr K}$ (as well as $\Omega_\Lambda $).
\item Spatial curvature induces geodesic deviations for CMB photons.  If the Universe is open, the positions of the acoustic peaks is shifted to higher multipoles, if it is closed, to lower multipoles.  
\end{itemize}

\paragraph{The neutrino density $\Omega_\nu $: }

The neutrino energy density, characterized by the effective number of relativistic neutrinos $N_\nu $, affects the total radiation energy density, and thus the time of the radiation/matter equality as well as the recombination time~\cite{riazuello}.  Thus the position of the acoustic peaks also depends on $\Omega_\nu $.



\subsection{The matter power spectrum}  \label{sec:LSS}

\begin{figure}[h!]
\begin{center}
\scalebox{1.8}{\includegraphics{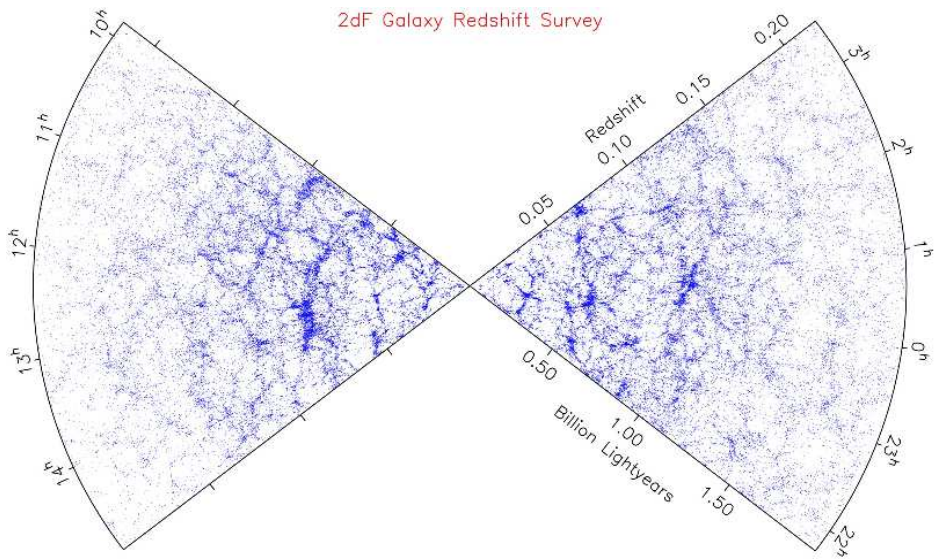}}
\caption{Distribution of 82821 galaxies in a 4 degree wide range as a function of the redshift, from the 2dF Galaxy Redshift Survey~\cite{Colless:1998yu,website:2dF}.  Galaxies are organized in a large scale structure of "filaments" and "walls".}   \label{fig:2df}
\end{center}
\end{figure}

\begin{figure}[h!]
\begin{center}
\scalebox{0.7}{\includegraphics{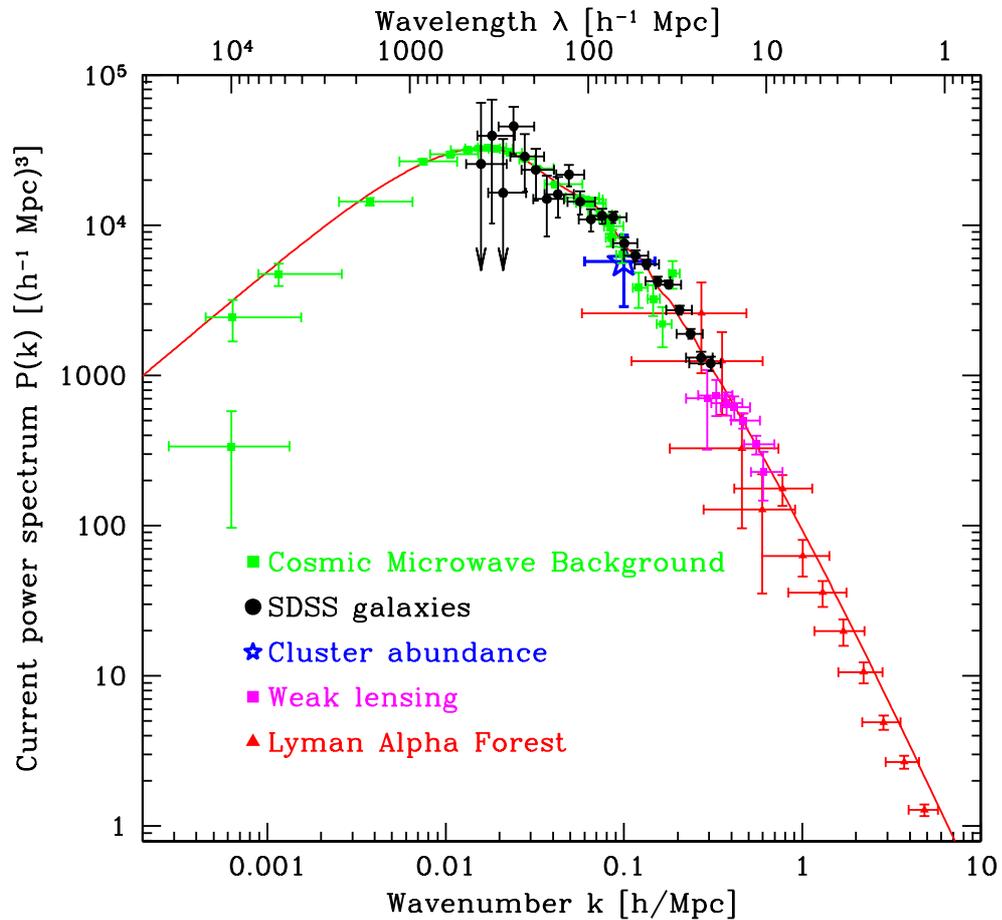}}
\caption{Matter power spectrum $P(k)$ measurements~\cite{Tegmark:2001jh} from various techniques of observations and theoretical expectation for the best fit of the $\Lambda$CDM model.  The best constraints are given by CMB and Large Scale Structures (SDSS).  The figure is from Ref.~\cite{PeterUzan}.}  \label{fig:Pk}
\end{center}
\end{figure}

With the recent surveys of galaxies and quasars (e.g. 2dF~\cite{Percival:2001hw, website:2dF}, SDSS~\cite{website:SDSS}), the large scale distribution of structures has been probed.   Galaxies are observed to be arranged in a complex structure of "walls" and "filaments" (see Fig.~\ref{fig:2df}).  The statistical properties of the matter distribution in the Universe are encoded in the \textit{matter power spectrum}.   If the mean density of galaxies is denoted $\bar n_{\rr {gal}} $, the fractional inhomogeneity $ \delta (\mathbf x )= [ n_{\rr {gal}} (\mathbf x ) - \bar n_{\rr {gal}}  ]  / \bar n_{\rr {gal}}  $ can be expanded in Fourier modes $\mathbf k$.  The power spectrum $P(k) $ is defined via\footnote{The adimensional form of the power spectrum $\mathcal P(k) \equiv k^3 P(k) / (2 \pi^2)$  is also often used. }
\begin{equation}
\langle \hat \delta (\mathbf k) \hat \delta (\mathbf k') \rangle = ( 2 \pi )^3 P(k) \delta^3({\mathbf{k - k'}})~,
\end{equation}
where $ \hat \delta (\mathbf k) $ is the 3-dimensional Fourier transform of $\delta ( \mathbf x) $, and where the brackets denote an average over the whole distribution.   The measurements of the matter power spectrum today are plotted in Fig.~\ref{fig:Pk}.  

\begin{figure}[h!]
\begin{center} 
\scalebox{0.7}{\includegraphics{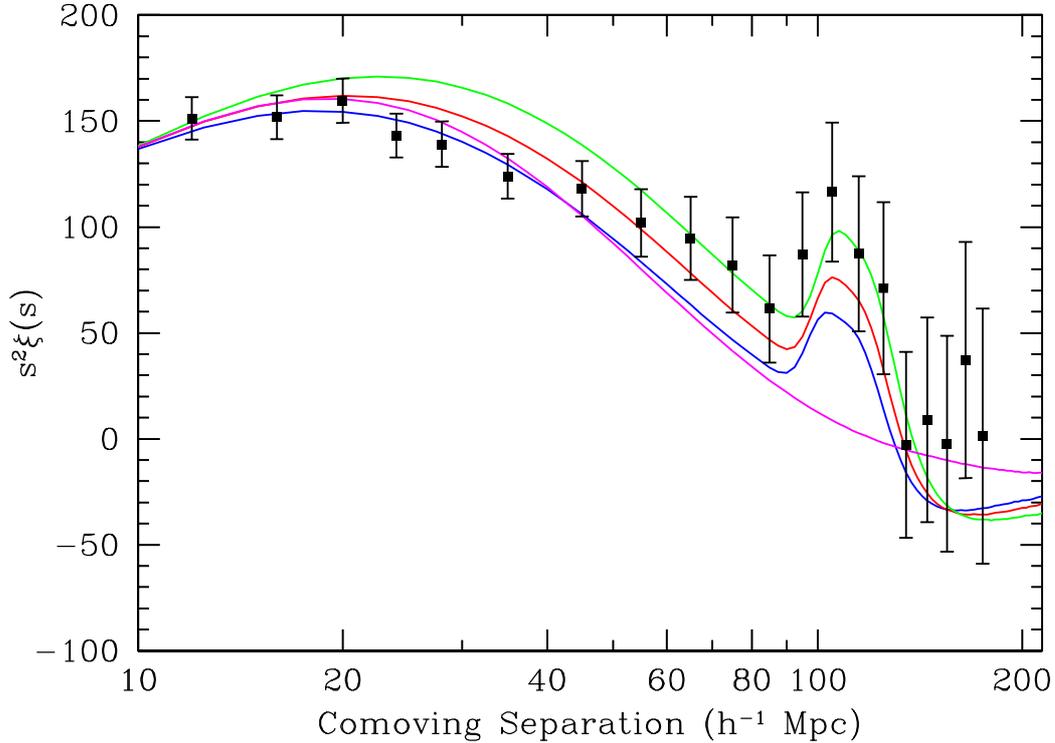}}
\caption{Detection of Baryon Acoustic Oscillations (BAO) by the SDSS experiment~\cite{Eisenstein:2005su}. From top to bottom, the curves are theoretical expectations for a $\Lambda$CDM model with $\Omega_{\rr b} h^2 = 0.024 $, and respectively $\Omega_{\rr m} h ^2 = \Omega_{\rr b} h^2 + \Omega_{\rr c} h ^2 = 0.12, 0.13, 0.14$.  The bottom curve is for a pure CDM model with $\Omega_{\rr m } h^2 = 0.105 $.  The vertical axis is the correlation function $\zeta(s) $ times $s^2$, where $s$ is the distance separation.  The BAO in the matter power spectrum are the imprint of the acoustic waves in the tightly coupled photon-baryon plasma prior to recombination.  }\label{fig:BAO}
\end{center}
\end{figure}

Some oscillations have been detected in the matter power spectrum~\cite{Eisenstein:2005su}, for perturbation wavelengths of a few Mpc (see Fig.~\ref{fig:BAO}).   These have been identified as the relic of the \textit{Baryon Acoustic Oscillations} (BAO) that took place in the early Universe, the same that are observed in the CMB.   


The shape of the matter power spectrum and the BAO are sensitive to the cosmological parameter values.  For instance, the largest possible wavelength for perturbation modes to oscillate is referred as the sound horizon.  It can be measured at recombination with CMB observations and compared to its present value measured with the matter power spectrum.   The ratio is sensitive to the expansion history, and thus to the cosmological parameters [see Eq.~(\ref{eq:H_of_t})].  This method can be used to determine the late-time acceleration of the Universe's expansion and to put a bound on the dark energy equation of state, independently of the type Ia supernova measurements.  

\subsection{Other signals}

To break the degeneracy between the cosmological parameters, it is necessary to combine data from several cosmological and astrophysical signals.   Besides the main signals described above, one could also mention:  Gravitational weak lensing~\cite{Bartelmann:1999yn}, Galaxy clusters~\cite{Voit:2004ah} , Ly-$\alpha$ forest~\cite{Viel:2006yh} and rotation curves of galaxies~\cite{Battaner:2000ef}.  


\subsection{Current bounds}  \label{sec:best_fits}

The best fits for the cosmological parameter values for the $\Lambda$CDM model are given in the table below~\cite{Komatsu:2010fb}.   The mean values of the probability distributions of these parameters and the corresponding 1-$\sigma$ errors are also given. 

\begin{table}[h!]
\begin{center}
\begin{tabular}{|c|c|c|}
\hline
Parameter & Best fit & Mean value and 1-$\sigma $ errors \\
\hline 
$\Omega_{\rr b} h^2$ & $2.253 \times 10^{-2}$ & $(2.255 \pm 0.054) \times 10^{-2} $ \\
$\Omega_{\rr c} h^2 $ & $0.1122$ & $0.1126 \pm 0.0036$ \\
$\Omega_\Lambda  $& $0.728$ & $0.725 \pm 0.016$ \\
$\Omega_{\rr K} $ &  & $-0.0111 \pm 0.006$ \\
$ N_{\rr{eff}} $ & & $4.34 \pm 0.88$ \\
$\tau $ & $0.085$ & $0.088 \pm 0.014$ \\
\hline 
$P_\zeta (k_*)$ & $2.42 \times 10^{-9}$ & $(2.430 \pm 0.091 ) \times 10^{-9}$ \\
$n_{\rr s}$ & $0.967$ & $0.968 \pm 0.012$ \\
$r $ & & $< 0.24 $ \\ 
\hline
$w$ &  & $-1.10 \pm 0.14$ \\
\hline
\end{tabular}
\caption{Best fit values of the $\Lambda $-CDM cosmological parameters, from WMAP7+BAO+$H_0$ data~\cite{Komatsu:2010fb}. The third column corresponds to the mean value of the marginalized posterior distributions with the corresponding 1-$\sigma$ errors.  $N_{\rr{eff}} = N_\nu + 1 $ is the effective number of relativistic species and is related to the energy density of neutrinos  $\Omega_\nu$. The parameter $\tau$ is the optical depth of the CMB photons.
$P_\zeta (k_*) $, $n_{\rr s}$ are respectively the amplitude and the spectral index of the primordial power spectrum of comoving curvature perturbations, $r$ is the tensor to scalar ratio (see Chapter 2 for further details).  The last line is the current bound on the equation of state parameter for the dark energy. }
\end{center}
\end{table}

\section{Unresolved problems}  \label{sec:questions}

The hot Big-Bang standard cosmological model has raised several questions and problems that remain today unresolved.   Some of them are described in this section.

\subsection{Nature of dark matter}

The nature of the cold dark matter component remains unknown.  Since the SM does not contain any dark matter candidate that is in agreement with all the observations, dark matter is a strong indication for new physics beyond the standard model.  A large number of models and dark matter candidates in agreement with cosmological and astrophysical observations have been proposed (for a review, see~\cite{Bertone:2004pz}).  

In the next few years, a major challenge will consist in identifying the nature of dark matter.  Dark matter particles could be produced and detected directly in particle accelerators, e.g. in the Large Hadron Collider (LHC) at CERN~\cite{DiCiaccio:2011zz}.   Other direct detection experiments in laboratories attempt to measure dark matter interactions with nuclei, e.g. in cryogenic detectors (CDMS~\cite{website:CDMS}, CRESST~\cite{Angloher:2008jj}, Edelweiss~\cite{website:Edelweiss},...).   Indirect detection experiments attempt to measure the decay/annihilation products of dark matter particles, that may lead to positron, antiproton, neutrino or gamma excesses.   

Recently, an excess of positrons has been reported by the PAMELA experiment~\cite{Adriani:2008zr}.  But this could be due to astrophysical sources~\cite{Delahaye:2008ua}.  
The DAMA experiment has measured an annual modulation~\cite{Bernabei:2000qi} that could be due to weakly interacting massive particles (WIMP's).  These results are subject to intense discussions in the community~\cite{Angle:2007uj}.

\subsection{Nature of dark energy}

The dark energy component is today the main contribution to the energy density of the Universe,  representing about 71\% of the total energy density.   Its energy density is thus comparable to the total matter energy density,  and the epoch from which the Universe became dominated by the dark energy coincides approximatively with the epoch of structure formation.  
In the $\Lambda$CDM model, dark energy is identified with a cosmological constant.  

But the dark energy could be also a dynamical quantity, due to an unknown fluid or a modification of gravity at cosmological scales.  A large number of models have been proposed in this context (for a recent review, see~\cite{Sapone:2010iz}).  
But because dark energy is not expected to be related to the matter content of the Universe, several model are said to suffer from a so-called \textit{coincidence problem}.    Recently, it has been proposed that the current cosmic acceleration can be due to an almost massless scalar field experiencing quantum fluctuations during a phase of cosmic inflation close to the electroweak energy 
scale~\cite{Ringeval:2010hf}.

A possible contribution to the cosmological constant could be the vacuum fluctuations.  However, when it is estimated using quantum field theories, it is found to be larger than the energy of the electro-weak breaking scale $ \rho_\Lambda ^{1/4} \sim 1 \rr{TeV}$.   But the measured value of the cosmological constant is $ \rho_\Lambda ^{1/4} \sim 10^{-3} \rr{eV}$. Its energy density is therefore at least 60 orders of magnitude smaller than expected~\cite{Weinberg:2000yb}.



\subsection{Horizon problem}  \label{sec:horizon}

\begin{figure}[ht] 
	\begin{center}
	\includegraphics[height=70mm]{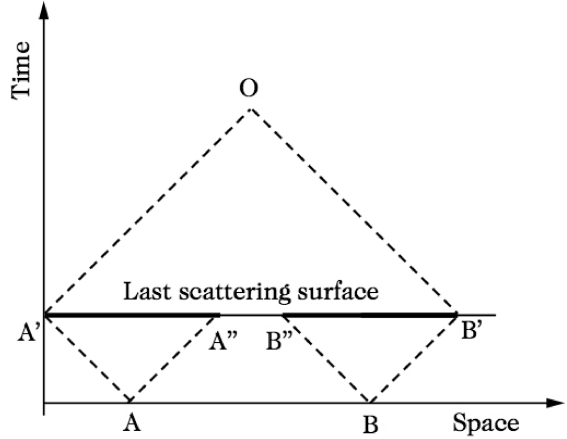}
	\caption{Scheme~\cite{PeterUzan} illustrating the horizon paradox.   The CMB is observed from the hypersurface $t=t_{\mathrm 0}$. The $AB'$ region at last scattering appears to be isothermal in the CMB sky, although it is constituted of patches causally disconnected if the Universe's expansion prior to recombination is assumed to be dictated by the $\Lambda$CDM model alone.}
	  \label{fig:horizon}
	\end{center}
\end{figure}

It is convenient to define the conformal time
\begin{equation}
\eta(t) = \int_{t_{\mathrm i}} ^{t} \frac {\dd t'}{a(t')} ,
\end{equation}
  that is the maximal comoving distance covered by the light between an initial hyper-surface at time $t_{\mathrm i}$ and the hyper-surface at time $t$.   Two points separated by a comoving distance larger than the conformal time $\eta$ do not have a causal link if one consider that the Universe's evolution begins at $t_{\rr i}$.   Usually, the initial hyper-surface is identified with the Planck-time, and points separated by a comoving distance larger than $\eta$ are said to be causally disconnected.  For an observer in $O$ at a time $t_{\mathrm 0}$  (see Fig.~\ref{fig:horizon}),  $\eta(t_{\mathrm 0})$ is the comoving radius of the sphere centered in $O$ separating particles causally connected to the observer of particles causally disconnected.  $\eta(t)$  is called the \textit{comoving horizon} or the \textit{particle horizon}.   It is important to distinguish between the \textit{particle horizon} and the \textit{event horizon}, which is, for the observer, the hypersurface separating the universe in two parts, the first one containing events that have been, are or will be observable, the second part containing events that will be forever unobservable.    Mathematically, the \textit{event horizon} exists only if the integral 
 \begin{equation} \label{evenements}
 \int_{t_{\mathrm i}} ^\infty \frac{\dd t'}{a(t')} 
 \end{equation}
 converges.  Finally, it is useful to define the \textit{comoving Hubble radius}, $ 1/(aH)$. 
 It is smaller than the conformal time, that is the logarithmic integral of the Hubble radius.

The \textit{horizon problem}  is linked to the isotropy of the CMB.   Indeed, how to explain that regions in the sky have the same temperature whereas their angular size is too large to correspond to causally connected patches at the time of last scattering, if the $\Lambda$CDM model alone is assumed to describe the whole Universe's expansion?

In the standard cosmological model, the early Universe is dominated by the radiation and the chemical potentials can be neglected most of the time.   One has therefore \hbox{$ a T = \mathrm{constant} $}, and in a comoving coordinate system, any physical distance growths like
\begin{equation} d (t) = \frac {T(t_{\mathrm 0})}{T(t)} d(t_{\mathrm 0})~. \end{equation}
The temperature of CMB photons today is $T_0 \approx 2.7 K \approx 2.3 \times 10^{-13} \mathrm{GeV}$. 

On the other hand, assuming that the expansion rate is dictated by the $\Lambda$CDM model at every time, the radius of the observable universe, that is the radius of the spherical volume in principle observable today by an observer at the center of the sphere, is $d_{H_0}(t_0) \approx 10^{26} 
\mathrm{m} 
$.   At the time corresponding to the last scattering surface $t_{\rr{LSS}}$, the radius of the observable universe was
 \begin{equation} 
 d_{H_0}(t_{\rr{LSS}}) \approx 7 \times 10^{22} \rr{m}~. 
 \end{equation}
Under the same assumption, at recombination, the maximal distance between two causally connected points would roughly be
\begin{equation} 
d_{\rr H_{\rr{LSS}}} (t_{\rr{LSS}}) \approx 2 \times 10^{21} \rr{m}~. 
\end{equation}
At last scattering, our observable Universe would therefore have been constituted of about $ 10^5$ causally disconnected regions.  But CMB photons emerging from these regions are observed to have all the same temperature, to a $10^{-5}$ accuracy.   
At the Planck time, the number of causally disconnected patches would have been much larger, about $10^{89} $.

\subsection{Flatness problem}

From the FL equations (\ref{eq:FL}), one can write the equation for the evolution of the curvature.  If we neglect the cosmological constant\footnote{This is a good approximation because $\Lambda$ dominates the energy density only at late times.}, we have
\begin{equation}
\frac{\dd \Omega_{\rr K} } { \dd \ln a} = (3 w + 1 ) ( 1 - \Omega_{\rr K} ) \Omega_{\rr K}~,
\end{equation}
This equation is easily integrated when $w$ is constant.  One has 
\begin{equation}
\frac{ \Omega_{\rr K 0} }{\Omega_{\rr K} (a) } = ( 1 - \Omega_{\rr K 0} ) \left( \frac{a}{a_0} \right)^{(-1 - 3 w)} + \Omega_{\rr K 0} ~,
\end{equation}
where $\Omega_{\rr K 0} $ is the curvature today.  Since it is constrained by observations ($|\Omega_{\rr K 0} - 1| \lesssim 0.01$~\cite{Komatsu:2010fb}) one has roughly at radiation-matter equality
\begin{equation}
|\Omega_{\rr K } (a_{\rr{eq}}) - 1|  \lesssim 3 \times 10 ^{-6}~,
\end{equation}
and at the Planck time,
\begin{equation}
|\Omega_{\rr K }(a_{\rr{p}}) - 1|  \lesssim 10^{-60}~.
\end{equation}
If the Universe is not strictly flat, the $\Lambda$CDM model does not explain why the spatial curvature is so small.  

\subsection{Problem of topological defects}

\begin{figure}[ht]
 	\begin{center}
	\includegraphics[height=90mm]{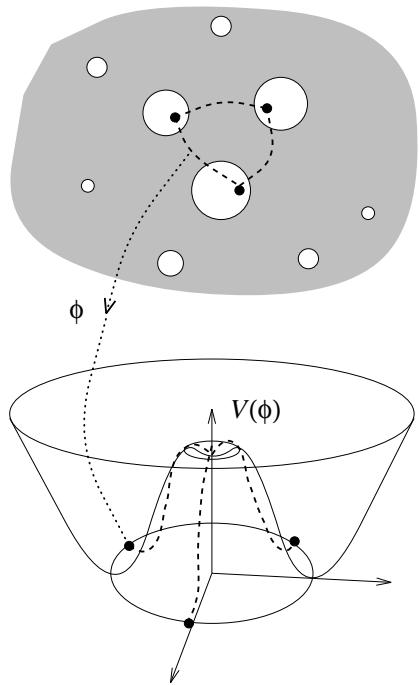}
	\caption{Illustration~\cite{Hindmarsh:1994re} of the formation of cosmic strings due to the breaking of the group $U(1)$ into $\{ \rr{id} \}$.   After the transition, the Higgs field $\phi$ takes a different value at each point in space.   When the Higgs field makes a complete loop in the field space along a closed path in the real space, there exists a point inside the path for which the phase is not defined.  At this point, the Higgs field vanished, the symmetry is restored and the resulting string configuration contains energy.  This process is called \textit{ the Kibble mechanism} (for a review, see~\cite{Hindmarsh:1994re}).}    \label{fig:kibble}
	\end{center}
\end{figure}

In Grand Unified Theories (GUT), the standard model of particle physics results from several phase transitions induced by the spontaneous breaking of symmetries.  Such symmetry breakings are triggered during the early Universe's evolution due to its expansion and cooling, and they can lead to the formation of topological defects like domain walls, cosmic strings and monopoles.   These defects correspond to configurations localized in space for which the initial symmetry remains apparent (see Fig.~\ref{fig:kibble}).

Let us consider the symmetry breaking of a group $\mathcal G$ resulting to an invariance under the sub-group $\mathcal H$:  $ \mathcal G \rightarrow \mathcal H $.  The vacuum manifold $\mathcal M$ is isomorphic to the quotient group $ \mathcal G / \mathcal H$~\cite{Nakahara:1990th}.  \textit{Domain walls} are formed when the 0th-order homotopy group of $\mathcal M$ is not trivial.  They can be due to the breaking of a $Z_{2}$ symmetry, or if the resulting vacuum contains several distinct elements.   \textit{Cosmic strings} are formed when the first homotopy group of $\mathcal M$ is not trivial, for instance for the breaking scheme $U(1) \rightarrow \{ \rr{Id} \}$.   \textit{Monopoles} are formed when the second homotopy group $\pi_2 (\mathcal M)$ of the vacuum manifold is not trivial.  This is the case for the breaking of a $SO(3)$ symmetry into $\mathcal H = \{ \rr{Id} \} $.  For higher homotopy groups, the resulting topological defects are called \textit{textures}.  

Groups involved in GUT are such that the first and second homotopy groups are trivial, $ \pi_1 (\mathcal G) \sim \pi_2 (\mathcal G) \sim \rr{Id}$.  In the SM, there remains a U(1) invariance corresponding to electromagnetism.  The first homotopy group of U(1) is $\pi_1 \left[ U(1)\right] \sim Z $.  Therefore, by using the property of homotopy groups~\cite{PeterUzan}
\begin{equation} \pi_n(\mathcal G) \sim \pi_{n-1} (\mathcal G) \sim {Id.} \Rightarrow \pi_n (\mathcal M) \sim \pi_{n-1} (\mathcal H)~,
\end{equation}
one obtains that the second homotopy group of the vacuum manifold corresponding to the breaking of a GUT group is not trivial.  That induces necessarily the formation of monopoles~\cite{Langacker:1980kd}.

However, monopole annihilation has been found to be very slow \cite{Zeldovich:1978wj,PhysRevLett.43.1365}.  As a consequence, their energy density today should be 15 orders of magnitude larger than the current energy density of the universe.   Domain walls can also lead to catastrophic scenarios, but they can be avoided in the schemes of symmetry breaking in GUT.   Cosmic strings are observationally allowed, but their contribution to the CMB angular power spectrum~\cite{Kaiser:1984iv} is constrained~\cite{Battye:2010xz}.  

\subsection{Why is the primordial power spectrum scale-invariant?}

The density perturbations at the origin of the CMB temperature fluctuations start to oscillate when their size becomes smaller than the Hubble radius.   On the contrary, the perturbations whose wavelength is much larger at recombination have remained constant and thus conserve their initial amplitude.  In the CMB angular power spectrum, these super-Hubble perturbations correspond to temperature fluctuations at low multipoles  ($l \lesssim 20$).   The CMB temperature fluctuations at large angular scales therefore directly probe the initial state of those density perturbations.   

With CMB observations, it has been established that the primordial power spectrum of density perturbations is (nearly) scale invariant.   The present measurements of the shape of the primordial power spectrum will be given in details in section~\ref{sec:observables}.   This constrains the possible physical processes at the origin of the initial density perturbations. 


\subsection{Contribution of iso-curvature modes}

There are two different kinds of primeval fluctuations:  the \textit{curvature} (or \textit{adiabatic}) and \textit{iso-curvature} (or \textit{entropic}).  

The adiabatic density fluctuations are characterized as fluctuations in the local value of the spatial curvature (hence the name of \textit{curvature} perturbations).  By the equivalence principle, all the species contribute to the density perturbation and one has for any fluid $f$,
\begin{equation}
\frac{\delta \rho}{\rho} = \frac{\delta n_f}{n_f} = \frac{\delta s}{ s}~,
\end{equation}
where $s \equiv S / a^3 $ is the entropy density.  Furthermore, one can write
\begin{equation}
\delta \left(  \frac{ n_f}{s} \right) = \frac{\delta n_f}{ s} - \frac{n_f \delta s} {s^2} = 0~.
\end{equation}
That means that the fluctuation in the local number density of any species relative to the entropy density vanished. 

The entropic fluctuations are perturbations for which $\delta \rho = 0 $ and therefore they are not characterized by fluctuations in the local curvature (hence the name \textit{iso-curvature}).   They correspond to fluctuations in the equation of state.   
 
 CMB observations have been used to determine that the temperature fluctuations are sourced by curvature perturbations, and the contribution of iso-curvature perturbation modes is constrained~\cite{Komatsu:2010fb}.  The mechanism leading to initial inhomogeneities therefore needs to generate (at least mostly) curvature perturbations.

\subsection{Why are the perturbations Gaussian?}

The statistical properties of the CMB anisotropies are encoded in the power spectrum of the temperature fluctuations, that is the two-point correlation function in the Fourier space.  Within a general framework, those are also encoded in the three-point, four-point, and higher order correlation functions.   But if the fluctuations follow a Gaussian statistic, these are all vanishing.  

The point is that the observations of the CMB have not detected a non-zero value neither for the three-point neither for higher-order correlation functions.  Since the temperature fluctuations in the CMB are induced by density perturbations, the mechanism generating the primordial density perturbations needs to be such that their statistics is Gaussian.








\subsection{Initial singularity}


As already mentioned in section~\ref{sec:fluids}, if the Universe has not been dominated by a positive spatial curvature, an initial singularity is generic for all known types of fluids.  In the $\Lambda$CDM model, the gravitation is assumed to be described correctly by GR at every time.  
However, some theories predict that GR is not valid anymore at the Planck energy scale.  Let us mention String Theories, Loop Quantum Gravity~\cite{Rovelli:1995ac} and Horava-Lifshitz theory~\cite{PhysRevD.79.084008}.    In some of these frameworks, the initial singularity is avoided and replaced by a bounce.  

It is nevertheless important to remark that all these theories are still highly hypothetic and not at all confirmed by observation.  


In the next chapter, the concept of inflation, that is an hypothetic phase of accelerated expansion in the early Universe, is introduced.  Models of inflation solve in an unified way several problems mentioned above.   They can provide Gaussian adiabatic primordial perturbations whose power spectrum is nearly scale invariant.  They solve also naturally the monopole, the horizon and the flatness problems.

%% file: The_inflationary_paradigm.tex
\chapter{The inflationary paradigm}
\label{chap:inflation}

\section{Motivations for an inflationary era}

Inflation is a phase of quasi-exponentially accelerated expansion of the Universe.  By combining the F.L. equations (\ref{eq:FL}) and (\ref{eq:FL2}), and assuming $K=0$, one obtains a necessary condition for inflation to take place, 
\begin{equation} \label{condition} 
\ddot a > 0 \Longleftrightarrow \rho + 3 P < 0~.
\end{equation}
The amount of expansion during inflation is measured in term of the number of \textit{e-folds}, defined as
\begin{equation}
N(t) \equiv \ln \left[ \frac {a(t)}{a_{\mathrm i}} \right],
\end{equation}
where $a_{\mathrm i} $ is the scale factor at the onset of inflation.   

The inflationary paradigm is motivated since it provides a solution to several problems of the standard cosmological model.  
\begin{itemize}
\item \textbf{The horizon problem}:  Inflation solves naturally this paradox if the number of e-folds of expansion is sufficiently large.  Indeed, isothermal regions in the CMB sky, appearing as causally disconnected at recombination if the $\Lambda$-CDM model alone is assumed, can actually be causally connected because of a primordial phase of inflation.  If the Universe's expansion was exponential during the inflationary era,
\begin{equation} a(t) = a_{\rr i} e^{H \Delta t }~, \end{equation}
(it will be shown later that this condition is nearly satisfied) one can evaluate the number of e-folds required to solve the horizon problem.   At the end of inflation, the size of the current observable Universe $d_{H_0}$ must have been smaller than the size of a causal region at the onset of inflation $d_{H_{\rr i}} $,
\begin{equation}
d_{H_0}(t_0)  \frac{a_{\rr{end}}  }{a_0}  <  d_{H_{\rr i}} \frac{a_{\rr{end} }  } {a_{\rr i}} = d_{H_{\rr i}}(t_{\rr i}) \rr e^{N}~,
\end{equation}
where $a_{\rr{end}}$ is the scale factor at the end of inflation.  If inflation ends at the Grand Unification scale ($T_{\rr{end}} \sim 10^{16}
$ GeV), one needs
\begin{equation}
N \sim \ln \left( \frac{T_0  d_{H_0}(t_0)  }{T_{\rr{end}}   d_{H_{\rr i}}(t_{\rr i}) }  \right) \gtrsim 57~,
\end{equation}
where $T_0$ is the photon temperature today, and for which we have assumed $ d_{H_{\rr i}}(t_{\rr i}) \sim l_{\rr{p}} T_{\rr p} / T_{\rr{end}}  $, where $l_{\rr p}$ and $T_{\rr p}$ are respectively the Planck length and the Planck temperature.   
If this condition is satisfied, the entire observable Universe can thus emerge out of the same causal region before the onset of inflation.  

\item \textbf{The flatness problem}:  During inflation, the Universe can be extremely flattened.    Indeed, if we assume $H$ to be almost constant during inflation, one has (see section~\ref{sec:fluids})
\begin{equation}
| \Omega_{\rr K} (a_{\rr{end} } ) | = | \Omega_{\rr K} (a_{\rr{i} } ) | \rr e^{- 2 N}~,
\end{equation}
With $N \gsim 70 $ and a curvature of the order of unity at the Planck scale, the flatness problem discussed in section~\ref{sec:questions} is naturally solved. 

\item \textbf{Topological defects}:  During inflation, topological defects are diluted due to the volume expansion and can have been "pushed" outside the observable Universe. 
\item \textbf{The primordial power spectrum}:  Models of inflation generically predict a nearly scale invariant power spectrum of curvature perturbations, and thus can provide good initial conditions for the perturbations in the radiation era. It will be explained later in this chapter how this power spectrum can be determined for a large class of models of inflation (single and multi-field models).    
\item \textbf{Gaussian perturbations}:  Inflation models predict that the classical perturbations leading to the formation of structures in the Universe are due to quantum metric and scalar field fluctuations.  As the Universe grows exponentially, the quantum-size fluctuations become classical, are stretched outside the Hubble radius, and source the CMB temperature fluctuations.  All the pre-inflationary classical fluctuations are conveniently stretched outside the Hubble radius today and can be safely ignored.   The Gaussian statistic of the perturbations therefore takes its origin in the Gaussian nature of the quantum field fluctuations.  
\item \textbf{Iso-curvature modes}:   Most models of inflation source only curvature perturbations. Nevertheless, for some models (like multi-field models), the iso-curvature mode contribution can be potentially important and eventually observable (e.g. in Ref.~\cite{Langlois:1999dw}).  In multi-field models, these are induced by field fluctuations orthogonal to the field trajectory, as explained more in details in section~\ref{sec:multifieldpert}.  
\end{itemize}

\section{Observables}  \label{sec:observables}

The CMB angular power spectrum is sensible to the initial conditions of the density and curvature fluctuations.   A model of inflation provide these initial conditions and it can therefore be confronted to CMB observations.  

Observations have permitted to measure and constrain the amplitude of the power spectrum of curvature perturbations at the end of inflation, its spectral tilt, as well as the ratio between curvature and tensor metric perturbations.   

\subsection{Power spectrum of primordial curvature perturbations} \label{sec:powerspectrum}

The primordial power spectrum of the curvature perturbation $\zeta$ is defined from,
\begin{equation}
\langle \hat \zeta(\mathbf k) \hat \zeta (\mathbf k' ) \rangle = (2 \pi)^3 P_\zeta(k) \delta^3 (\mathbf k - \mathbf k')~,
\end{equation}
where $\hat \zeta (\mathbf k)$ is the 3-dimensional Fourier transform of $\zeta (\mathbf x )$.  The spectral  index of this power spectrum $n_{\rr s}$ is defined as
\begin{equation}
n_{\mathrm s} \equiv 1+ \left. \frac {\dd \ln \left[ k^3 P_{\zeta} (k) \right] } { \dd \ln k } \right|_ {k_*} ~,
\end{equation}
where $k_*$ is a pivot scale in the observable range, e.g. $k_* = 0.002 \ \rr{Mpc}^{-1} $.  A power spectrum showing an excess at large angular scales ($n_{\rr s}< 1$) is called  \textit{red-tilted} while for an excess at small scales, it is called \textit{blue-tilted}. 
Deviation from a scale invariant primordial power spectrum have been detected by recent CMB experiments. The power spectrum is observed to be red-tilted, and the case $n_{\rr s } = 1$ is disfavored.   The present 1-$\sigma$ bound on the spectral index is~\cite{Komatsu:2010fb} $n_{\rr s} = 0.968 \pm 0.012 $, as illustrated in Fig.~\ref{fig:nsrplane}. 

On the other hand, measurements of the CMB temperature monopole and quadrupole have permitted to fix the amplitude of the curvature power spectrum~\cite{Komatsu:2010fb}, 
\begin{equation}  
\mathcal P_\zeta (k_*) \equiv \frac{k_*^3}{2 \pi^2} P_\zeta(k_*) = 2.43 \times 10^{-9} ~.
\end{equation}

\subsection{Tensor-to-scalar ratio}

\begin{figure}[ht] 
 	\begin{center}
	\includegraphics[height=100mm]{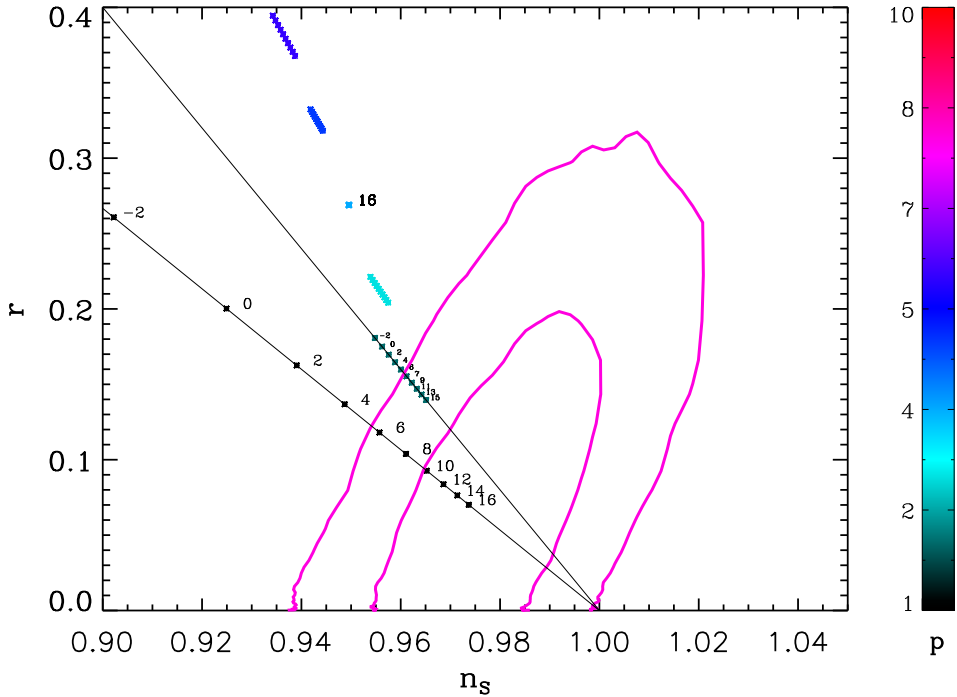}
	\caption{Reheating consistent 1-$\sigma$ and 2-$\sigma$ contours (in pink) in the plane $(n_{\rr s},r)$~\cite{Martin:2010kz}, from  WMAP7 data (marginalised over second order slow-roll).  Crosses are the predictions for the large field inflation model with the potential $V(\phi) = M^4 (\phi / \Mpl )^p $, for several values of the parameter $p$ (see the color scale).  The annotations correspond to $\log (30 \rho_{\rr{reh}} / \pi^2 )/4 $, where $\rho_{\rr{reh}} $ is the reheating energy in GeV$^4$, as discussed in section~\ref{sec:reheating}.  The two lines represent the locus of the $p \gtrsim 1 $ and $p=2$ models. }  
	 \label{fig:nsrplane}
\end{center} 
\end{figure}

The tensor metric perturbations, characterized by a power spectrum $P_{h}(k)$ at the end of inflation, can also affect the CMB angular power spectrum (for details, see e.g.~\cite{PeterUzan}).
CMB observations have permitted to put a significant limit on the primordial power spectrum of gravitational waves.  It is convenient to express this limit as an upper bound on the ratio $r$ between tensor and curvature power spectra.  At 2-$\sigma$, one has~\cite{Komatsu:2010fb}, 
\begin{equation} 
r \equiv \frac{P_{h}} {P_{\zeta}}  < 0.24~.
\end{equation}
The 1-$\sigma$ and 2-$\sigma$ bounds in the $(n_{\rr s},r)$ plane are shown in Fig.~\ref{fig:nsrplane}, as well as the predictions for some models of inflation.

\subsection{Other observables}

Other observable quantities are potentially interesting to improve the constraints on inflation models.  The observational bounds on these quantities are not yet sufficient to provide significant constraints on inflation. Nevertheless, the future data from the Planck satellite could change this perspective.   Some of  these observables are briefly described below:
\begin{itemize}
\item The running spectral index $\alpha_{\rr s}$:  it is defined as 
\begin{equation}
\alpha_{\rr s} \equiv \left. \frac{\dd n_{\rr s} }{\dd \ln k} \right|_{k=k_*}~.
\end{equation}
Present data are compatible with $\alpha_{\rr s} = 0$. 

Inflationary predictions can be compared to the constraints on $\mathcal P_\zeta (k_*) $, $n_{\rr s}$, $\alpha_{\rr s}$ and $r$.  However, it is more efficient to constrain directly the parameter space of a given model of inflation by confronting it directly to the $C_l$'s measurements, and by using Bayesian analysis to obtain the posterior probability density distributions of its parameters marginalized over all the cosmological parameters.  

\item The $f_{\rr{NL}} $ parameter:   this parameter characterizes the amplitude of the so-called local form bispectrum of $\zeta$, 
\begin{equation}
B_\zeta = \frac 6 5 f_{\rr{NL}}  \left[ P_\zeta(k_{\rr 1}) P_\zeta(k_{\rr 2}) + (\rr{2 \ perm.}) \right]~,
\end{equation}
defined as the Fourier transform of the three-point correlation function, 
\begin{equation}
\left\langle \prod_{i=1}^3 \zeta(\mathbf{ k}_i)   \right\rangle = (2 \pi )^3 \delta^3 \left(\sum_{i=1}^3 \mathbf{ k}_i \right) B_\zeta(k_{\rr 1},k_{\rr 2},k_{\rr 3})~.
\end{equation}
A non-zero bispectrum results from non-Gaussian curvature perturbations.  Inflation can be a source of small non-Gaussianities, but also the reheating phase, eventual cosmic strings, and various astrophysical processes.  In the \textit{squeezed limit}, corresponding to $ k_{\rr 3} \ll k_{\rr 1} \simeq k_{\rr 2} $, it has been shown that all single-field models of inflation yield to $f_{\rr{NL}} = \frac 5 {12} (1-n_{\rr s} ) \simeq 0.02 $~\cite{Gangui:2002qc,Creminelli:2004yq}.   For multi-field models, like hybrid models, the  $f_{\rr{NL}}$ value can be higher, possibly in the observable range of the Planck experiment.  For the other processes, the amplitude should be $f_{\rr{NL}} \sim \mathcal O (1)$ (see~\cite{Komatsu:2010hc} for a review), thus a convincing detection of $f_{\rr{NL}} \gg 1$ would rule out most\footnote{However, non-Gaussianities in single field models could be generated by trans-planckian effects~\cite{Collins:2009pf} or slow-roll violation~\cite{PhysRevD.83.103511}} single field inflation models.   The current best limit is~\cite{Sugiyama:2011jt} 
\begin{equation}
f_{\rr{NL}}  = 32 \pm 21 \ \rr{(68 \% C.L. )}~.
\end{equation}
The Planck satellite is expected to reduce the error bars by a factor of four.  

\item The  $\tau_{\rr{NL}} $ parameter:   this parameter characterizes one of the amplitudes of the local-form trispectrum of $\zeta$, 
\begin{equation}
T_\zeta = \tau_{\rr{NL}}  \left[ P_\zeta(\mathbf {k}_{\rr 1} + \mathbf{ k}_{\rr 3}) P_\zeta(k_{\rr 3})  P_\zeta(k_{\rr 4})+ (\rr{11 perm.}) \right]~,
\end{equation}
which is the Fourier transform of the four-point correlation function, 
\begin{equation}
\left\langle \prod_{i=1}^4 \zeta(\mathbf{ k}_i)   \right\rangle = (2 \pi )^3 \delta^3 \left(\sum_{i=1}^4 \mathbf{ k}_i \right) T_\zeta(k_{\rr 1},k_{\rr 2},k_{\rr 3},k_{\rr 4})~.
\end{equation}
The 2-$\sigma$ sensitivity of Planck is expected to be $\tau_{\rr{NL}} \sim 700 $~\cite{Sugiyama:2011jt}.   For multi-field inflation models, an interesting generic inequality between $f_{\rr{NL}} $ and $\tau_{\rr{NL}} $ have been established recently in Ref.~\cite{Sugiyama:2011jt},
\begin{equation}
\tau_{\rr{NL}} > \frac 1 2 \left( \frac 6 5 f_{\rr{NL}}  \right)^2
\end{equation}
A consequence of this inequality is the possibility to rule out most models of inflation, if a significant value of $f_{\rr{NL}} \gsim 30 $ is detected together with no detection of $\tau_{\rr{NL}}$.
\end{itemize}

\section{1-field models of inflation}

A period of inflation can be obtained by assuming that the early Universe was filled by one (or more) nearly homogeneous scalar field(s) slowly rolling along its (their) potential.  In the first part of this section, the equations governing the homogeneous 1-field background dynamics are derived and the slow-roll approximation is introduced.  The second part is dedicated to the theory of cosmological perturbations for such a scalar field.  The perturbation mode evolution equations are derived and it is explained how the observable power spectra of scalar curvature and tensor metric perturbations can be calculated in the slow-roll approximation.   As an example, the slow-roll predictions for the large field model are given at the end of the section.  

\subsection{Background dynamics}

The easiest realization of the condition (\ref{condition}) is to assume that the Universe is filled with an unique homogeneous scalar field $ \phi $, called the \textit{inflaton}. The lagrangian reads
\begin{equation}
 \mathcal L = - \sqrt {-g} \left[ \frac 1 2 \partial_{\mu} \phi \partial^{\mu} \phi   + V(\phi) \right] ~,
 \end{equation}
where $V(\phi)$ is the scalar field potential and $g$ is the determinant of the FLRW metric.  The equation of motion (e.o.m.) for this lagrangian is the Klein-Gordon equation in an expanding spacetime,
\begin{equation} \label{KGtc}
\ddot \phi + 3 H \dot \phi + \frac {\dd V}{\dd \phi} = 0~. 
\end{equation}
On the other hand, the energy momentum tensor reads
\begin{equation}
T_{\mu \nu} = - \frac 2 {\sqrt{-g}} \frac {\delta  \mathcal L }{\delta g_{\mu \nu} }~.
\end{equation}
The energy density and the pressure are therefore
\begin{eqnarray}  \label{eq:rho_inf}
 \rho & = & \frac{\dot \phi ^2}  2 + V(\phi)~, \\  \label{eq:P_inf}
 P & = & \frac {\dot \phi ^2} 2 - V(\phi)~.
 \end{eqnarray}
The condition (\ref{condition}) is satisfied if the scalar field evolves sufficiently slowly, so that
 $\dot \phi ^2 \ll V(\phi) $.
The expansion is governed by the Friedmann-Lema\^itre equations  
\begin{equation} \label{eq:FLtc1}
H^2 = \frac {8\pi }{3 \mpl^2}  \left[ \frac 1 2 \dot
\phi^2    + V(\phi) \right] ~, 
\end{equation}
\begin{equation} \label{eq:FLtc2}
\frac{\ddot a }{a} = \frac {8\pi}{3 \mpl^2} \left[ -  \dot \phi^2
 + V(\phi ) \right]~.
\end{equation}
These equations can be rewritten using the conformal time or the number of e-folds as the time variable.  For the conformal time, one has
\begin{eqnarray}  
\mathcal H^2 & = &   \frac {8\pi }{3 \mpl^2} \left[ \frac 1 2 {\phi'} ^2 + a^2 V(\phi) \right]~, \\
2 \mathcal H' + \mathcal H^2 & = &  \frac {8\pi }{ \mpl^2} \left[ - \frac 1 2 {\phi'} ^2 + a^2 V(\phi) \right]~,\\
\phi'' + 2 \mathcal H  \phi' + a^2 \frac {\dd V}{\dd \phi} & = & 0 ~,
 \end{eqnarray}
where a prime denotes the derivative with respect to the conformal time, and where $\mathcal H \equiv a' / a = a H $.  For the number of e-folds,
\begin{eqnarray}
H^2 & = &  \dfrac {8\pi }{ \mpl^2}  \frac{V(\phi)}{3 - \dfrac {4 \pi}{\mpl^2} \left( \dfrac{\dd \phi}{\dd N}  
\right)^2}~,\\
 \frac{1}{H}  \frac {\dd H}{\dd N} & = & - \frac{4 \pi}{\mpl^2} \left(  \frac{\dd \phi}{\dd N}\right)^2~,\\
\frac 1 {3- \frac{4 \pi}{\mpl^2} \left( \frac{\dd \phi}{\dd N} \right)^2 }\frac{\dd^2 \phi}{\dd N^2}  + \frac{\dd \phi}{\dd N} & = & -  \frac { \mpl^2} {8\pi } \frac{\dd \ln V}{\dd \phi}~,
\end{eqnarray}
and the field evolution is decoupled from the space-time dynamics.  

 \subsection{Slow-roll approximation}

For inflation to be very efficient, the kinetic terms in the F.L. equations must be very small compared to the potential.   The \textit{slow-roll approximation} consists in neglecting the kinetic terms and the second time derivatives of the field, 
\begin{equation}
 \dot \phi ^2 \ll V(\phi)~, \hspace{30mm} \ddot \phi \ll 3H\dot \phi~.
 \end{equation}
In the slow-roll regime, one has therefore 
\begin{eqnarray}\label{sr1} 
H^2 & = & \frac {8 \pi }{3m_{\mathrm p}^2} V(\phi)~,\\
\label{sr2}
 3H \dot \phi & = & - \frac {\dd V}{\dd \phi}~.
\end{eqnarray}
Using the number of e-folds as a time variable, the field evolution is governed by
\begin{equation}
\frac {\dd \phi}{\dd N} = -\frac { \mpl^2}{8\pi } \frac 1 V \frac {\dd V}{\dd \phi}~.
\end{equation}
One sees that a large number of e-folds is realized in a small range of $\phi$ when the logarithm of the potential is very flat.

The slow-roll regime is an attractor~\cite{Ringeval:2005yn} such that typically a few e-folds after the onset of inflation, the slow-roll approximation is valid.   It is convenient to study inflationary models in the slow-roll regime since observable predictions are easily determined in this regime, as it will be shown and discussed later.  

From this point, let us introduce the Hubble-flow functions~\cite{Leach:2002ar}, 
\begin{eqnarray}
 \epsilon_{1}  & \equiv &  - \frac {\dot H}{H^2} < 1 \Longleftrightarrow \ddot a > 0 ~,\\
 \epsilon_{n+1} & \equiv & \frac {\dd \ln |\epsilon_{n}| }{\dd N}~. 
\end{eqnarray}
Using these functions,  the F.L. and K.G. equations can be rewritten
\begin{eqnarray} H^2 & = & \frac {8 \pi}{m_{\mathrm p} ^2} \frac {V}{3-\epsilon_1}~, \\
\dot \phi & = & \dfrac {-1} { \left(3+\dfrac 1 2 \epsilon_2 - \epsilon_1 \right) H } \frac {\dd V}{\dd \phi}~, 
\end{eqnarray}
and one sees that the slow-roll regime is recovered when 
\begin{equation} \epsilon_1 \ll 3~, \hspace{2cm} \epsilon_2 \ll 6 - 2 \epsilon_1 ~. \end{equation}
One sees also that $\epsilon_1 < 3 $ is required for satisfying the condition $H^2 >0 $.  In  the slow-roll approximation, they can be expressed as a function of the potential and its derivatives.  For the first and second Hubble-flow functions, one has~\cite{Liddle:1994dx}
\begin{equation} \begin{split}
\epsilon_{\rr 1} (\phi) & \simeq  \frac{\mpl^2}{16 \pi} \left( \frac{1}{V} \frac{\dd V}{\dd \phi}  \right)^2 + \mathcal O (\epsilon_i ^2) ~,\\
\epsilon_{\rr 2} (\phi) & \simeq \frac{\mpl^2}{4 \pi} \left[\left( \frac{1}{V} \frac{\dd V}{\dd \phi}  \right)^2 - \frac{1}{V} \frac{\dd^2 V}{ \dd \phi^2}  \right] + \mathcal O (\epsilon_i ^2) ~.
\end{split} \label{eq:slowrollparams}
\end{equation}
The Hubble flow functions are usually referred as the \textit{slow-roll parameters}. 
Finally, let remark that many references use other slow-roll parameters, $ \epsilon $ and $\delta$, defined as
\begin{eqnarray} 
\epsilon & \equiv & \epsilon_1~, \\
 \delta & \equiv & - \frac{\ddot \phi}{H \dot \phi} = \epsilon - \frac{\dot \epsilon}{2 H \epsilon}  ~,
 \end{eqnarray}
such that the relation $\epsilon_2 =  2 ( \delta - \epsilon) $  is verified.  

\subsection{Cosmological perturbations}

The success of inflation is to provide the initial conditions for the density perturbations leading to the formation of structures in the Universe.  The classical density perturbations originate from quantum fluctuations.

The theory of cosmological perturbations permits to describe how the scalar field and metric fluctuations evolve during inflation.  At the linear level, the homogeneous metric is considered to be perturbed by $\delta g_{\mu \nu} $,
\begin{equation} g_{\mu \nu} (\mathbf x) = g_{\mu \nu} ^{\mathrm{FLRW}} +  \delta g_{\mu \nu} (\mathbf x)~.
\end{equation}
The 10 degrees of freedom (d.o.f.) associated to the metric perturbation $\delta g_{\mu \nu} $ can be decomposed in 
\begin{itemize}
\item 4 scalar d.o.f. $A,B,C,E$
\item 4 vector d.o.f. $ B_i $ et $ E_i $ resulting from two space-like vectors of null divergence
\item 2 tensor d.o.f. $h_{ij}$ resulting from a space-like tensor with vanishing trace and divergence.  
\end{itemize}
One can rewrite the perturbed metric as
\begin{equation} \begin{split}
 ds^2 = & a^2 (\eta) \left\{  -( 1+2A ) \dd \eta^2 + 2 (\partial_{i}  B + B_i) \dd x^i \dd \eta \right. \\
 & \left. + \left[  ( 1 +2 C ) \delta_{ij} + 2  \partial_{i} \partial_{j} E + 2 \partial_{(i} E_{j)} + 2 h_{ij} \right] \dd x^i \dd x^j  \right\} .
\end{split}
 \end{equation}

\subsubsection{The gauge problem}

A local perturbation in a quantity $Q$ can be defined as 
 \begin{equation}
 \delta Q (\mathrm x,t) = Q (\mathrm x, t ) - \bar Q (t )~,
 \end{equation}
 where $\bar Q (t)$ is this quantity in the un-perturbed space-time.  Any perturbation depends therefore on how are chosen the coordinate systems on each manifold.   In other words, if a coordinate system is fixed for the un-perturbed space-time, one needs to define an isomorphism identifying the points of same coordinates in the two space-times.   The liberty in this choice implies that four d.o.f. are non-physical and only linked to the choice of the coordinate systems on the two manifolds.  

Let us consider a transformation of the coordinate system
\begin{equation}
x^\mu \rightarrow x^\mu + \xi ^\mu~,
\end{equation}
where $\xi^\mu$ is a space-time like vector.   $\xi^\mu$ can be decomposed in two scalar ($T$ and $L$) and two vector ($L_i$) d.o.f.  via
\begin{equation} 
\xi^0 = T~, \hspace{2cm} \xi^i = D^i L + L^i~, \hspace{2cm} D^i L_i = 0~,
\end{equation}
where $D_i $ is defined as the spatial part of the covariant derivative.  Fixing this transformation is thus equivalent in fixing 4 d.o.f.. 

Under this modification of the coordinate system, the metric perturbation transforms as 
\begin{equation}
\delta g_{\mu \nu} \rightarrow \delta g_{\mu \nu} + \mathcal L _\xi g_{\mu \nu},
\end{equation}
where $\mathcal L_\xi$ is the Lie derivative along $\xi$.  The Lie derivative evaluates the change of a tensor field along the flow of a given vector field.   It is defined as
\begin{equation} 
\mathcal L_\xi T^{\mu_1 \ldots \  \mu_p}_{\nu_1 \ldots \ \nu_q } = \xi^\sigma \partial_\sigma T^{\mu_1 \ldots \  \mu_p}_{\nu_1 \ldots \ \nu_q }
- \sum_{i=1} ^p T^{\mu_1 \ldots \ \sigma \ldots \ \mu_p}_{\nu_1 \ldots \ \nu_q } \partial_\sigma \xi^{\mu_i} 
+ \sum_{j=1}^q T^{\mu_1 \ldots \ \mu_p}_{\nu_1 \ldots  \ \alpha \ldots \  \nu_q } \partial^\alpha \xi_{\nu_j} .
\end{equation} 
Applied to the symmetric metric, the Lie derivative gives
\begin{equation}
\mathcal L_\xi g_{\mu \nu} = \nabla_\mu \xi_\nu + \nabla_\nu \xi_\mu~,
\end{equation}
where $\nabla_\mu $ is the covariant derivative associated with $g_{\mu \nu} $.   It results that scalar, vector and metric perturbations transform as\cite{PeterUzan}
\begin{eqnarray}
A & \rightarrow & A + T' + \mathcal H T~, \\
B & \rightarrow & B - T + L'~, \\
C & \rightarrow & C + \mathcal H T~, \\
E & \rightarrow & E' + L~, \\
E^i & \rightarrow & E^i + L^i~,\\
B^i & \rightarrow & B^i + {L^i} '~, \\
h_{ij} & \rightarrow & h_{ij}~.
\end{eqnarray}
In the same way, the perturbation $\delta Q$ becomes
\begin{equation}
\delta Q \rightarrow \delta Q + \mathcal L_\xi Q~.
\end{equation}
A quantity is called \textit{gauge invariant} when it is independent of the coordinate system transformation, that is if its Lie derivative vanishes\footnote{This result is known as the Stewart-Walker lemma.}. Gauge invariants are for instance the Bardeen variables\footnote{$\Psi$ is also called the Kinney potential by V. Mukhanov. }~\cite{PhysRevD.22.1882}
\begin{eqnarray}
\Phi & \equiv & A + \mathcal H (B-E') + (B-E')'~, \\
\Psi & \equiv & -C - \mathcal H (B-E')~.
\end{eqnarray}
If we fix $T = B- E'$, $L = -E $ and $L_i' = -B_i$, one has
\begin{equation}
B=E=0~, \hspace{15mm} B_i = 0~,
\end{equation}
and the scalar metric perturbations are identified with the Bardeen variables
\begin{eqnarray}
 A & = & \Phi ,\\
 C & = & -\Psi~.
\end{eqnarray}
This is called the \textit{longitudinal gauge}.

 \begin{figure}[htbp] 
	\begin{center}
	\includegraphics[height=70mm]{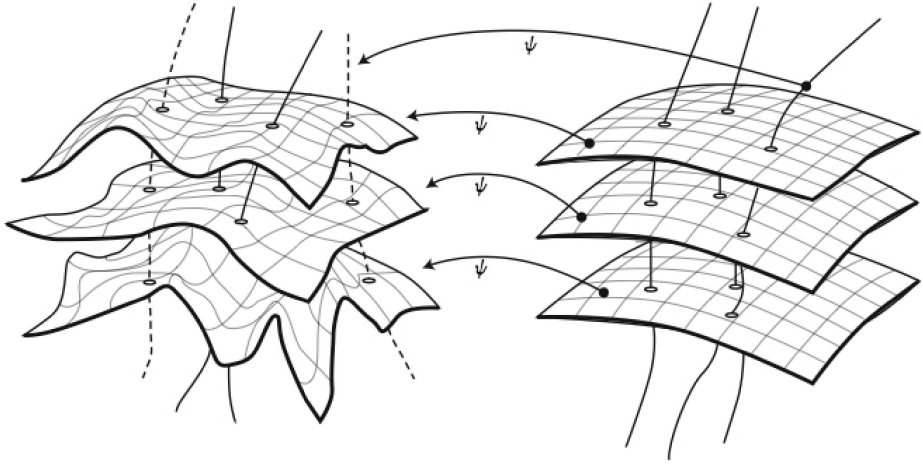}
	\caption{To define a perturbed quantity, one needs to fix an isomorphism $\psi$ identifying each point on the FLRW manifold with a point on the perturbed manifold.  The freedom in the choice of this isomorphism is called a \textit{gauge freedom}.  Fixing the gauge is equivalent to fix the coordinate system on the perturbed manifold.  The figure is taken from~\cite{PeterUzan}.   }
	\label{pertmetric}
	\end{center}
\end{figure}

\subsubsection{Scalar perturbations}

In the longitudinal gauge, the homogeneous metric is thus perturbed by the scalar Bardeen potentials, 
\begin{equation}
\dd s^2 = a^2(\eta) \left[ -(1+2 \Phi ) \dd \eta^2 + (1 - 2 \Psi ) \delta_{ij} \dd x^i \dd x^j \right] .
\end{equation}
The scalar field filling the Universe at a given spacetime point is given by its homogeneous part $ \bar \phi $ plus a small perturbation $\delta \phi \ll \bar \phi $,
\begin{equation}
\phi (\mathrm x,t) = \phi  (t) + \delta \phi(\mathrm x,t)~.
\end{equation}
In the longitudinal gauge, it is identified to the gauge invariant variable
\begin{equation}
\delta \phi_{\rr{g.i.}} = \delta \phi + \phi' ( B - E')~.
\end{equation}
After perturbing the energy momentum tensor, the $(0,0)$ and $(i,i)$ first order perturbed Einstein equations read
 \begin{eqnarray} \label{eq:ep1} 
 - 3 \mathcal H (\Psi'+\mathcal H \Phi ) + \nabla ^2 \Psi & = & \frac {4 \pi}{m_{\mathrm p}^2} \left( \phi' \delta \phi' - \phi'^2 \Phi +
a^2 \frac{\dd V}{\dd \phi} \delta \phi \right) ~, \\  \label{eq:ep2}
 \Psi' + \mathcal H \Phi & = & \frac{4\pi}{m_{\mathrm p}^2 } \phi' \delta \phi ~, \\ \label{eq:ep3}
 \Psi'' + 2 \mathcal H \Psi' + \mathcal H \Phi' +& &  
  \Phi \left( 2\mathcal H ' + \mathcal H ^2 \right) + \frac 1 2 \nabla ^2 (\Phi - \Psi ) \nonumber \\
& = & \frac {4 \pi}{m_{\mathrm p}^2}  \left( \phi' \delta \phi' - \phi'^2 \Phi - a^2 \frac {\dd V}{\dd \phi} \delta \phi \right) ~.
\end{eqnarray}
Moreover, because $ \delta T_{i}^{j} \propto \delta_{i}^{j} $ in absence of vector perturbations, one has  $ \Phi = \Psi $.  
On the other hand, the first order Klein-Gordon equation for the scalar field perturbation reads
\begin{equation}\label{kgp}
\delta \phi '' + 2 \mathcal H \delta \phi' - \nabla^2 \delta \phi + a^2 \delta \phi \frac {d^2V}{\dd\phi^2}
= 2 (\phi'' + 2 \mathcal H \phi' ) \Phi + \phi' ( \Phi' +3 \Psi')~.
\end{equation}
One sees that $ \delta \phi $ is directly related to $\Phi $ and its derivative, so there remains only one scalar d.o.f..  
By combining Eq.~(\ref{eq:ep1}) and Eq.~(\ref{eq:ep3}), and by using Eq.~(\ref{eq:ep2}) as well as the background equations, an unique second order evolution equation for scalar perturbations can be derived,
\begin{equation} \label{eq:Phievol}
\Phi'' + 2 \left( \mathcal H - \frac {\phi''}{\phi'} \right) \Phi' - \nabla^2 \Phi + 2 \left( \mathcal H' - \mathcal H \frac {\phi''}{\phi'} \right) \Phi = 0~.
\end{equation}
It is convenient to work in Fourier space, because in the linear regime each mode evolves independently and it is sufficient to follow their time evolution.  After a Fourier expansion, we define
\begin{eqnarray}
\mu_{\mathrm s} & \equiv &  -  \frac {4 \sqrt \pi}{  m_{\mathrm p}}  a (\delta \phi + \phi' \Phi / \mathcal H )~,\\
 \omega_{\mathrm s} ^2 & \equiv & k^2 - \frac {(a\sqrt\epsilon) '' }{a\sqrt \epsilon}~,
\end{eqnarray}
where $k$ is a comoving Fourier wavenumber, and equation (\ref{eq:Phievol}) can be rewritten in a simpler form, 
\begin{equation} \label{pertscal}
 \mu_{\mathrm s}'' + \omega_{\mathrm s} ^2 (k,\eta) \mu_{\mathrm s} = 0~.
\end {equation}
It is similar to an harmonic oscillator with a varying frequency.  Apart in some specific cases, this equation cannot be solved analytically.  However, it can be solved either numerically or analytically after a first order expansion in slow-roll parameters.  

Instead of $\Phi$ or $\mu_{\rr s} $, it is a common usage to calculate the mode evolution and the power spectrum of the curvature perturbation $\zeta $\footnote{$ \zeta$ can be identified to the spatial part of the perturbed Ricci scalar in the comoving gauge, in which the fluids have a vanishing velocity ($\delta T^0_{\ i} = 0 $).} defined as
\begin{equation} \label{zeta}
\zeta \equiv = \Phi - \frac{\mathcal H}{\mathcal H' - \mathcal H^2 } (\Phi' + \mathcal H \Phi )= - \mu_{\mathrm s} \frac 1 {2a\sqrt {\epsilon_{\rr 1}} }~.
\end{equation} 
Its power spectrum thus reads
\begin{equation}
 \mathcal P_{\zeta} (k) = \frac {k^3}{8\pi ^2 } \left| \frac {\mu_{\mathrm s}}{a\sqrt { \epsilon_1} } \right| ^2~.
 \end{equation}
By using Eq.~(\ref{kgp}), one can determine that $\zeta$ evolves according to
\begin{equation}
\zeta' = \frac{- 2 \mathcal H}{3 (1+w)}  \left( \frac{k}{\mathcal H}  \right)^2 \Phi~,
\end{equation}
and as long as the modes are super-Hubble ($ k/\mathcal H \ll 1 $), $\zeta (k) $  remains constant in time.   Therefore, observable modes re-entering into the Hubble radius during the matter dominated era have kept the value they had during inflation, when they exit the Hubble radius, independently of the details of the reheating phase\footnote{Let notice that a non linear growth of density perturbations during preheating is expected in some models, possibly affecting the linear curvature perturbations on very large scales~\cite{Jedamzik:2010dq,Jedamzik:2010hq}.} and the transition between inflation and the radiation dominated era.  For 1-field inflationary models, they can be used to probe directly the inflationary era.  

 \begin{figure}[htbp] 
	\begin{center}
	\includegraphics[height=80mm]{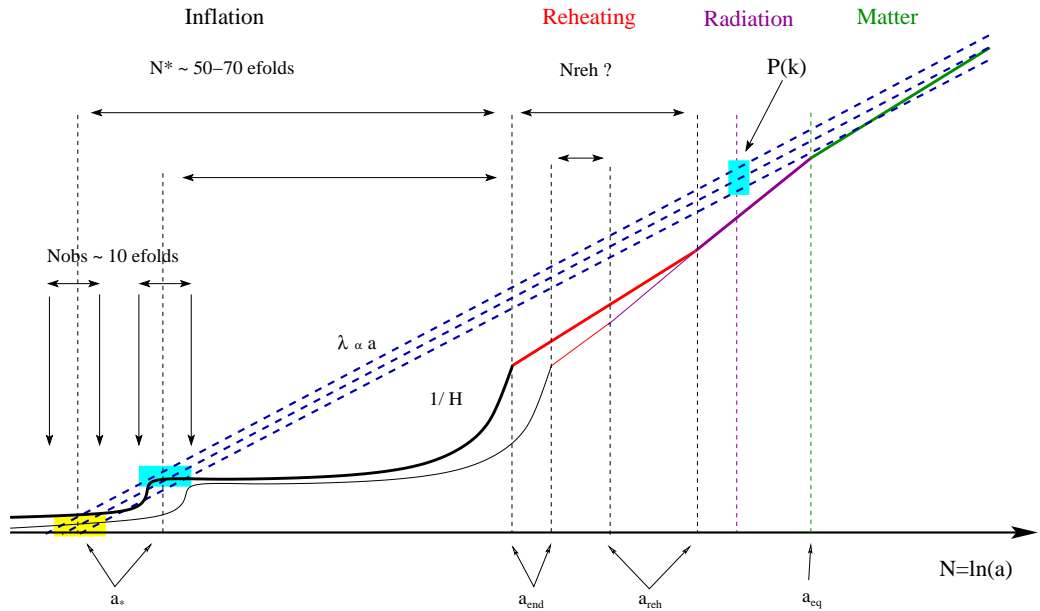}
	
	\caption{This scheme form Ref.~\cite{Ringeval:2007am} illustrates how observable perturbation modes evolve during and after inflation.  The horizontal axis represents the number of e-folds generated from the onset of inflation.  Observable modes exit the Hubble radius on a range of about ten e-folds.  From this time, inflation still lasts from 50 to 70 e-folds, depending on the energy scale of inflation~\cite{Liddle:2003as} and on the duration of the reheating phase (see section~\ref{sec:reheating}).  As explained in the text, the curvature and tensor perturbations remain constant for super-Hubble wavelengths, until they re-enter into the Hubble radius during the matter/radiation dominated era. }
	\label{modes}
	\end{center}
\end{figure}

\subsubsection{Tensor perturbations}

The metric for the tensor perturbations reads
\begin{equation}
 \dd s^2 = a^2(\eta) \left[ - \dd \eta^2 + (1 + h_{ij})  \dd x^i \dd x^j \right]~,
\end{equation}
and the metric perturbation $h_{ij}$ is gauge invariant.  It is convenient to express the two d.o.f. in $h_{ij}$ as
\begin{equation}
 h_{ij} = a^2 \left(
\begin{array}{lll}
h_+ & h_\times & 0 \\
h_\times & h_+  & 0 \\
0 & 0 & 0\\  \end{array} \right).
\end{equation}
As for the scalar perturbations, one can then write the first order perturbed Einstein equations,
\begin{equation}
h_{\alpha}'' + 2 \mathcal H h_{\alpha} + \nabla^2 h_{\alpha} =0,
\end{equation}
where $ \alpha = +,\times $.  By defining 
\begin{equation}
 \mu_{\mathrm t} \equiv \frac 1 2 ah_{ij} \delta^{ij} ,
 \end{equation}
after Fourier expansion, these two equations reduce to
\begin{equation} \label{tensoriel}
 \mu_{\mathrm t} '' + \omega_{\mathrm t} ^2  (k,\eta) \mu_{\mathrm t} = 0 ~,
 \end{equation}
where
 \begin{equation}
  \omega_{\mathrm t}^2 (k,\eta) \equiv k^2 - \frac {a'' }{a} ~.
  \end{equation}
The variable $ h = h_{ij} \delta^{ij}, $ is the analogous of $\zeta$ for the tensor perturbations and has similar properties.  Its power spectrum reads
\begin{equation}
\mathcal P_{h} (k) = \frac { 2 k^3 }{ \pi^2} \left| \frac {\mu_{\mathrm t} }{a} \right| ^2 .
 \end{equation}
 
 \subsubsection{Vector perturbations}
 
 The metric for the vector perturbations in the longitudinal gauge reads
 \begin{equation}
 \dd s^2 = a^2 (\eta) \left[  - \dd \eta^2 + 2 \partial_{(i} E_{j)} \dd x^i \dd x^j  \right]~,
 \end{equation}
and the vector perturbations $E_j $ can be identified in this gauge to the gauge invariant variable
\begin{equation}
 \Phi_i = E_i - B_i~.
\end{equation}
The perturbed energy-momentum tensor for a scalar field does not contain any source of vector perturbations and the first-order perturbed Einstein equations read
\begin{equation}
\Phi_i'' + 2 \mathcal H \Phi_i' = 0~.
\end{equation}
Vector perturbations therefore decay quickly, since $\Phi_i' \propto a^{-2} $ and because $a$ grows nearly exponentially with the cosmic time.  That is why vector perturbations are therefore usually neglected.

\subsubsection{Quantification of perturbations}

In the context of inflation, quantum fluctuations are responsible for large scale structures of the Universe observed today.   The canonical commutation relations are the basis of the quantization process.  But to define them, one needs the canonical momenta, and thus the action.  It is incorrect to interpret directly the classical equations of motion [Eqs.~(\ref{pertscal}) and~(\ref{tensoriel})] quantum mechanically, because it leads in general to an incorrect normalization of the modes~\cite{Mukhanov:1990me}. 


\paragraph{Scalar perturbations:}
If we perturb the total action of the system up to the second order in the metric and scalar field perturbations, one finds~\cite{Mukhanov:1990me}
\begin{equation}
^{(2)} \delta S = \frac 1 {2} \int \dd^4 x \left[ (v')^2 - \delta^{ij} \partial_i v \partial_j v + \frac{z_{\rr s}''}{z_{\rr s}} v^2 \right]~.
\end{equation}
where $ v \equiv a ( \delta \phi_{\rr{g.i.}} + \phi' \Phi) / \mathcal H $ can be identified to $- \mu_{\rr s} \mpl / 4 \sqrt \pi $ in the longitudinal gauge, and is the so-called \textit{Mukhanov-Sasaki variable}.    The quantity $z_{\rr s}$ is defined as $z_{\rr s} \equiv \sqrt{4 \pi} a \phi' / \mathcal H = a \sqrt \epsilon_1 $.  As expected, the e.o.m for this lagrangian reads
\begin{equation} \label{eq:class_evolu}
v'' - \left( \nabla^2 + \frac{z''}{z}  \right) v = 0~.
\end{equation}
The first step of the quantization process is to determine $\pi $, the conjugate of $v$, 
\begin{equation} 
\pi = \frac{\delta \mathcal L}{\delta v'} = v'~.
 \end{equation}
Then the Hamiltonian reads 
\begin{equation}
H = \int \dd x ^4 \left( \pi^2 + \delta^{ij} \partial_i v \partial_j v - \frac{z_s''}{z_s}  v^2  \right) ~.
\end{equation}
In a quantum description, the classical variables $v$ and $\pi$ are promoted as quantum operators $\hat v$ and $ \hat \pi $, satisfying the commutation relations 
\begin{equation}
[\hat v (\mathbf x, \eta ) ,\hat v (\mathbf y, \eta ) ] = [\hat \pi (\mathbf x, \eta ) ,\hat \pi (\mathbf y, \eta ) ] = 0~,
\end{equation}
\begin{equation}
[\hat v (\mathbf x, \eta ) ,\hat \pi (\mathbf y, \eta ) ] = i \delta^{(3)} (\mathbf x - \mathbf y) ~.
\end{equation}
In the Heisenberg picture, the operator  $\hat v$ can be expanded over a complete orthonormal basis of the solution of the field equation Eq.~(\ref{eq:class_evolu}).  If one takes a basis of plane waves, one has
\begin{equation}
\hat v (\mathbf x, \eta) = \frac{1}{(2 \pi)^{3/2}}\int \dd^3 \mathbf k \left( v_k \rr e^{i  \mathbf {k\cdot x} } \hat a_{ \mathbf k} + v_k^* \rr e^{-i  \mathbf {k\cdot x} } \hat a_{ \mathbf k}^+   \right)~,
\end{equation}
and the equation for the $v_k(\eta)$ is 
\begin{equation} \label{eq:vk}
v_k'' (\eta) + \left(k^2 -   \frac{z''}{z} \right) v_k =0~. 
\end{equation}
If the normalization condition 
\begin{equation}
v_k' (\eta) v_k^* (\eta) - {v_k^*}' (\eta) v_k (\eta) = 2 i
\end{equation}
is satisfied, the creation and annihilation operators $  \hat a_{ \mathbf k} $ and $ \hat a_{ \mathbf k}^+ $ satisfy the standard commutation relations 
\begin{equation}
[\hat a _{\mathbf k },\hat a _{\mathbf k' } ]=[\hat a _{\mathbf k }^+ ,\hat a _{\mathbf k' }^+ ] = 0~, \hspace{15mm} [\hat a _{\mathbf k },\hat a _{\mathbf k' }^+ ] = \delta^{(3)} (\mathbf {k - k'} )~.
\end{equation}
At a time $\eta_{\rr i} $, the vacuum $| 0 \rangle $ can now be defined, such that for all $\mathbf k$ one has
\begin{equation}
 \hat a_{\mathbf k} | 0 \rangle = 0~.
\end{equation}
From Eq.~(\ref{eq:vk}), in the sub-Hubble regime, we have
\begin{equation} \label{condinit_scalar}
\lim_{k/aH \rightarrow + \infty} v_k (\eta) = \frac { \mathrm e^{-ik(\eta-\eta_{\mathrm i}) } }{\sqrt{2k} } .
\end{equation}
This can be used to give consistent initial conditions to Eq.~(\ref{pertscal}).  

\paragraph{Tensor perturbations:}
 The quantification of the tensor perturbations is analogous.  One can first determine the second order perturbed action
 \begin{equation}
 ^{(2)} \delta S = -  \frac{\Mpl^2}{2} \sum_{\alpha=+,\times} \int \dd^4 x \left[ (h_\alpha')^2 - \delta^{ij} \partial_i h_\alpha \partial_j h_\alpha + \frac{a''}{a} h_\alpha^2 \right]~.
 \end{equation}
 The perturbations $h_\alpha (\eta, \mathbf x ) $ are the canonical variables.  They are promoted as quantum operators and are expanded in plane waves, 
 \begin{equation}
\hat h_j (\mathbf x, \eta) = \frac{1}{(2 \pi)^{3/2}}\int \dd^3 \mathbf k \left( h_{k,j} \rr e^{i  \mathbf {k\cdot x} } \hat a_{ \mathbf k , j} + h_{k,j}^* \rr e^{-i  \mathbf {k\cdot x} } \hat a_{ \mathbf k,j}^+   \right)~.
\end{equation}
The e.o.m. are
\begin{equation}
h''_{\mathbf k,j} + \left(  k^2 - \frac{a''}{a} \right) h_  {\mathbf k,j} = 0~,
\end{equation}
similar to Eq.~(\ref{tensoriel}).  
The quantification process can be used to determine the sub-Hubble tensor perturbation evolution,
\begin{equation} \label{condinit_tensor}
\lim_{k/aH \rightarrow + \infty} h_{k,j} (\eta) =\frac { \mathrm e^{-ik(\eta-\eta_{\mathrm i}) } }{\sqrt{2k} }~.
\end{equation}

\subsubsection{Expansion in slow-roll parameters} \label{sec:sr_expand}

Eqs.~(\ref{pertscal}) and~(\ref{tensoriel}) can be solved analytically if these are expanded at first order in the Hubble flow-functions around some pivot scale.
To do so, let us first rewrite 
\begin{equation} \label{eq:etasr}
\eta = \int \frac {\dd t}{a } = \int \frac{\dd a}{H a^2} = - \frac 1 {aH} + \int \dd a \frac {\epsilon_1}{a^2 H}~.
\end{equation}
In the slow-roll approximation, one has $|\epsilon_i| \ll 1$.  By definition, the derivative of the first and second Hubble-flow functions with respect to the number of e-folds are second order in $|\epsilon_i|$,
\begin{equation}
\frac{\dd \epsilon_1}{\dd N} = \epsilon_1 \epsilon_2~,\hspace{10mm} \frac{\dd \epsilon_2}{\dd N} = \epsilon_2 \epsilon_3~.
\end{equation}   
One can therefore neglect their variation over the time taken for observable modes to exit the Hubble radius (it corresponds typically to $\Delta N \sim 10$~\cite{Liddle:2003as}).   In this approximation, and by using Eq.~(\ref{eq:etasr}), one thus has
\begin{eqnarray}
a H & = & - \frac 1 \eta + \frac{a H} {\eta} \int \frac{\epsilon_1 } {a^2 H} \dd a \\
       & = & - \frac 1 \eta + \frac{a H} {\eta} \int \frac{\epsilon_1 } {a} \dd t \\
       & \simeq & - \frac 1 \eta + a H \epsilon_1 \\ 
       & \simeq & \frac{-1 + \epsilon_1} {\eta}~.  \label{eq:aH}
 \end{eqnarray}
By integrating the last equation, the scale factor is found to behave like
\begin{equation} \label{eq:aeta}
a(\eta) \simeq l_0 | \eta | ^{-(1+\epsilon_1)} \simeq  \frac{l_0}{|\eta|} \left( 1 - \epsilon_{1} \ln |\eta| \right) ~,
\end{equation}
where $l_0$ is an arbitrary parameter.  
Instead of choosing an arbitrary scale, it is more convenient to chose an arbitrary conformal time $\eta_* $, and to relate it to the scale $l_0 $ via
\begin{equation} \label{eq:Hetastar}
H(\eta_*) = - \frac{1 + \epsilon_{1*}}{a \eta_*} \simeq \frac 1 {l_0} \left[ 1 + \epsilon_{1*} (1+ \ln | \eta_* | ) \right] \equiv H_*~,
\end{equation}
where a star subscript denotes the evaluation at the time $\eta_*$.  We have fixed  $\eta_* $ such that the following relation is verified
\begin{equation}
k_* = a(\eta_*) H(\eta_*)~,
\end{equation}
where $k_*$ is the comoving pivot mode introduced in section~\ref{sec:powerspectrum}.  

\vspace{2mm}
The scalar and tensor perturbations evolve according to Eq.~(\ref{pertscal}) and Eq.~(\ref{tensoriel}).  These equations can now be expanded at first order in slow-roll parameters and then solved analytically.  The first step is to use Eq.~(\ref{eq:aH}) to expand
\begin{eqnarray}
 \frac { (a \sqrt \epsilon)''}{a \sqrt \epsilon } & \simeq & 
 \frac { 2 + 3 \epsilon_1 + \frac  3 2 \epsilon_2 }{ \eta ^2 } ~, \\
\frac { a'' }{a} & \simeq & \frac { 2 + 3 \epsilon_1 } {\eta^2} ~.
\end{eqnarray}
In the approximation that the slow-roll parameters are constant in time, a general solution to  (\ref{pertscal}) and (\ref{tensoriel}) can be found 
\begin{equation}
 \mu_{\mathrm s,\mathrm t} (k\eta ) = \sqrt {k \eta } \left[ A J_{\nu_{\mathrm s,\mathrm t}} (k\eta ) + B J_{-\nu_{\mathrm s,\mathrm t}} (k\eta) \right] ~,
 \end{equation}
where $ \nu_{\mathrm s} = - \frac  3 2 - \epsilon_1 - \frac 1 2 \epsilon_2  $ and $ \nu_{\mathrm t} = - \frac 3 2 - \epsilon_1 $.
It is convenient to express the Bessel function $J_{\nu} (k\eta ) $ in terms of the Hanckel functions of first and second kind $H^{(1)} _\nu (k\eta)$ et $ H^{(2)}_\nu(k\eta)$.
The quantification of the perturbations provide the initial conditions.  By using the asymptotic behavior of the Hanckel functions,
\begin{eqnarray}
H_\nu ^{(1)} (z \rightarrow \infty ) & = & \sqrt{ \frac 2 {\pi z} } \rr e^{i(z-\frac 1 2 \nu \pi - \frac 1 4 \pi)} , \\
H_\nu ^{(2)} (z \rightarrow \infty ) & = & \sqrt{ \frac 2 {\pi z} } \rr e^{-i(z-\frac 1 2 \nu \pi - \frac 1 4 \pi)} ,\\
\end{eqnarray}
and by comparing with the Eqs.~(\ref{condinit_scalar}) and (\ref{condinit_tensor}), $A$ and $B$ can be determined.  For scalar perturbations, one has
\begin{eqnarray} A & = & 2 i \frac \pi {\mpl} \sqrt k \sin (\pi \nu_s) \rr e^{i\left( \frac 1 2 \nu_s - \frac \pi 4 + k \eta_{\rr i} \right)}~,\\
B & = & - A \rr e^{-i \pi \nu_s}~.
\end{eqnarray}
On the other hand, one can use the limit condition
\begin{equation}
H^{(1)} _{1/2-\nu} (z \rightarrow 0) = - \frac i \pi \Gamma \left( \frac 1 2 - \nu \right) \left( -\frac z 2 \right)^{\nu - \frac 1 2} ,
\end{equation}
as well as the recurrence relation
\begin{eqnarray}
\Gamma (z+\epsilon) & = & \epsilon \psi(z) \Gamma(z) + \Gamma(z)~,\\
\psi(1/2) & = & - \gamma_{\mathrm{Euler}} - 2 \ln 2~, 
\end{eqnarray}
with $  \gamma_{\mathrm{Euler}} \simeq 0.5772 $ and where $\psi(z)$ is the polygamma function, to obtain the super-Hubble behavior of the perturbation modes. 
Since observable modes are super-Hubble at the end of inflation, one obtain the power spectrum expanded at first order in slow-roll parameters around $\eta_*$, by using Eqs.~(\ref{eq:aeta}) and~(\ref{eq:Hetastar}).  For scalar perturbations, one obtains   
\begin{eqnarray}
\mathcal P_{\zeta}(k) & = &  \frac {k^3} {8\pi ^2 } \left| \frac {\mu_{\mathrm s}}{a\sqrt { \epsilon_{1*}} } \right| ^2 \\
& = & \frac {H_* ^2} {\pi m_{\mathrm p} ^2 \epsilon_{\rr 1 *} } \left[ 1 - 2 (C+2) \epsilon_{1*} + C \epsilon_{2*}  - (2 \epsilon_{1*} + \epsilon_{2*} )  \ln \left( \frac {k}{k_*} \right)  \right]~.
\end{eqnarray}
For tensor perturbations,
\begin{eqnarray}
 \mathcal P_{h}(k) & = & \frac { 2  k^3 }{\pi^2} \left| \frac {\mu_{\mathrm t} }{a} \right| ^2 \\
 & = & \frac { 16 H_* ^2 }{\pi m_{\mathrm p}^2 } \left[ 1- 2(C+1) \epsilon_{1*} - 2 \epsilon_{1*} \ln \left( \frac k {k_*} \right) \right]~,
\end{eqnarray}
with $C = \gamma_{\mathrm{Euler}} + 2 \ln 2 - 2 $.  All the slow-roll parameters are evaluated at $\eta_*$.
At first order in slow-roll parameters, the scalar spectral index is therefore
\begin{equation}
n_{\mathrm s} -1 = - 2 \epsilon_{\rr 1*} - \epsilon_{\rr 2*}~.
\end{equation}
For the tensor perturbations, it is 
\begin{equation}
n_{\mathrm t}  = - 2 \epsilon_{1*}~.
\end{equation}
Finally the ratio between the tensor and scalar power spectrum is given by
\begin{equation}
r = 16 \epsilon_{\rr 1 *}~.
\end{equation}
The amplitude and the spectral tilt of the scalar and tensor power spectra can thus be derived easily in the slow-roll approximation, at first order in slow-roll parameters, for a given scalar field potential. 

\subsection{Example of 1-field potential:  large field models}

There exists a large variety of 1-field models of inflation, from the simplest power-law potentials to more complicated potentials originating from various high energy frameworks.  
In this section, the simplest model of large field inflation is presented in order to illustrate how the scalar and tensor power spectra are determined in the slow-roll approximation.  

The scalar field potential for large field models of inflation reads
\begin{equation} \label{eq:largefield}
V(\phi) = M^4 \left( \frac { \phi}{M_{\mathrm p}}  \right) ^{p}~.
\end{equation}
The background dynamics in the slow-roll approximation is given by Eqs.~(\ref{sr1}) and (\ref{sr2}).   In this approximation, the number of e-folds realized from an initial field value $ \phi_{\mathrm i} $ can be determined analytically, 
\begin{equation} N(\phi) = \frac {1 }{ 2 p} \left[ \left( \frac {\phi_{\mathrm i}}{M_{\mathrm p}} \right)^2 - \left(\frac {\phi}{M_{\mathrm p}} \right) ^2 \right]~. \end{equation}
The first and second slow-roll parameters read
\begin{equation}
\epsilon_1 (\phi) = \frac {p^2 M_{\mathrm p}^2 }{2 \phi^2 } ~,
\end{equation}
\begin{equation}
 \epsilon_2 (\phi)= \frac { 2 p M_{\mathrm p} ^2 }{  \phi^2 } ~.
 \end{equation}
Inflation stops when the first slow-roll parameter reaches $ \epsilon_1 = 1$.  This corresponds to the inflaton value
\begin{equation} 
\frac {\phi_{\mathrm {end}}}{M_{\mathrm p}} = \frac p {\sqrt 2} ~. 
\end{equation} 
For a given number of e-folds $N_*$ between the Hubble exit of the pivot mode and the end of inflation, the inflaton value and the slow-roll parameter values at Hubble exit can be obtained.  For $N_* = 60$ and $p=2$, they read
\begin{eqnarray}
\phi_{*} & = &  \sqrt{ 2 p \left(N_* + \frac p 4  \right)} M_{\rr{p}} \simeq 15.5 \Mpl \simeq 3.1 m_{\rr p} ~,\\
 \epsilon_{1*} & \simeq & 0.0083~, \hspace{10mm} \epsilon_{2*} \simeq 0.0166~.
\end{eqnarray}
It is then straightforward to derive the scalar power spectrum spectral index and the scalar to tensor ratio,
\begin{equation}
 n_{\rr s} =  1- 2 \epsilon_{1*} - \epsilon_{2*} \simeq  0.967~,     \hspace{10mm} r = 16 \epsilon_{1 *} \simeq 0.13~.
\end{equation}
These predictions are independent of the mass of the field and correspond to a point in the $(n_{\rr s}, r)$ plane, inside the 2-$\sigma$ contours.   Nevertheless, these depend on the reheating history via $N_*$ (see Section~\ref{sec:reheating}).  The mass scale is fixed by the scalar power spectrum amplitude given in section~\ref{sec:observables}.  One has $M \simeq 10^{-3} \mpl $.    For large field models, inflation takes therefore place close to the GUT scale.  
Let remark that the inflaton must be initially super-Planckian in order for inflation to last at least 60 e-folds.  Nevertheless, the energy density remains much smaller than the Planck scale. GR is thus valid and no effect of quantum gravity is expected\footnote{In supersymmetric models, SUGRA corrections are nevertheless expected (see Chapter 2 for further details).}.

 \begin{figure}[htbp] 
	\begin{center}
	\includegraphics[width=150mm]{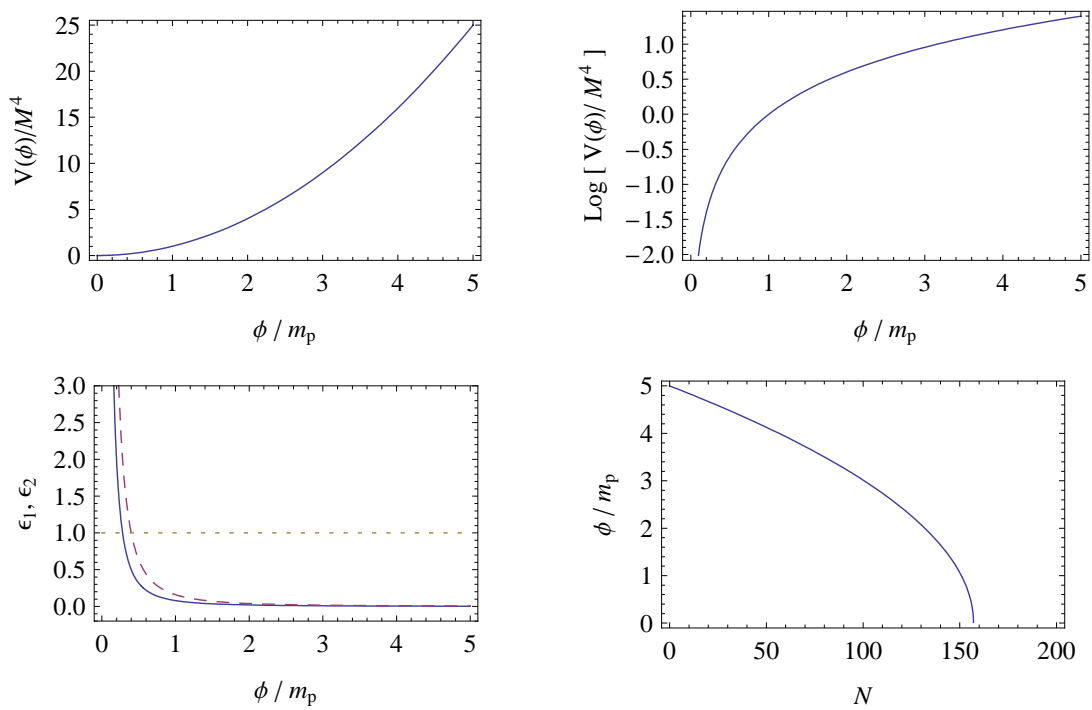}
	
	\caption{Top:  potential and logarithm of the potential of the large field model, for $p=2$.  Bottom-left:  evolution of slow-roll parameters  $\epsilon_1$ (plain line) and $\epsilon_2$ (dashed line).  Inflation stops when $\epsilon_1 = 1$ (dotted line).  Bottom-right: evolution of $\phi(N)$, for $ \phi_{\mathrm i}= 5 \mpl $}
	\label{largefield}
	\end{center}
\end{figure}

\newpage

\section{Multi-field inflation}

Inflation can be also realized in a multi-scalar field scenario.  Several models have been proposed, e.g. the double inflation~\cite{PhysRevD.35.419} model.   Hybrid models that will be studied in this thesis belong to the class of multi-field models.  In the first part of this section, the equations governing the homogeneous dynamics for a model with an arbitrary number $n$ of real scalar fields are given.  The second part concerns the evolution of the perturbations in the linear regime.  

\subsection{Background dynamics} \label{sec:multifield}

Let us assume that the Universe was filled with $n$ nearly homogeneous real scalar fields $\phi_{i=1,2,...,n} $.   The background dynamics is given by the Friedmann-Lema\^itre equations that read
\begin{equation} \label{eq:FLtc12field}
H^2 = \frac {8\pi }{3 \mpl^2}  \left[ \frac 1 2 \sum_{i=1}^n  \dot
\phi _i ^2 + V(\phi_{i=1,...,n})  \right] ~, 
\end{equation}
\begin{equation}
\frac{\ddot a }{a} = \frac {8\pi}{3 \mpl^2} \left[ - \sum_{i=1}^n  \dot \phi_i^2
 + V(\phi_{i=1,...,n}) \right]~.
\end{equation}
They are coupled to $n$ Klein-Gordon equations
\begin{equation} \label{KGtc2fieldd}
\ddot \phi_i + 3 H \dot \phi_i + \frac {\partial V}{\partial \phi_i} = 0~. 
\end{equation}
Then, let us introduce the velocity field 
\begin{equation} \label{eq:adiabfield}
\dot \sigma = \sqrt{\sum_{i=1}^n \dot \phi_i^2}~.
\end{equation}
$\sigma $ is the so-called \textit{adiabatic field}~\cite{Gordon:2000hv} and describes the collective evolution of all the fields along the classical trajectory.   
Its equation of motion reads 
\begin{equation} \label{KGtcadiab}
\ddot \sigma + 3 H \dot \sigma + V_\sigma = 0 ~, 
\end{equation}
where 
\begin{equation}
V_\sigma \equiv \sum_{i=1}^n u_i \frac{\partial V}{\partial \phi_i}~,
\end{equation} 
with $u_i$ being the components of an unit vector along the field trajectory $u_i \equiv \dot \phi_i / \dot \sigma $. 

\subsection{Multi-field perturbations }  \label{sec:multifieldpert}

\subsubsection{Scalar linear perturbations}

In the longitudinal gauge, the Einstein equations perturbed at first order, after using $\Phi = \Psi$ from $i\neq j $ equations, read
 \begin{eqnarray} \label{Eq:pert_multifield1} 
 - 3 \mathcal H (\Phi'+\mathcal H \Phi ) + \nabla ^2 \Phi & = & \frac {4 \pi}{m_{\mathrm p}^2} \sum_{i=1}^n \left( \phi_i' \delta \phi_i' - \phi_i'^2 \Phi +  a^2 \frac{\partial V}{\partial \phi_i} \delta \phi_i \right) ~, \\  \label{Eq:pert_multifield2} 
 \Phi' + \mathcal H \Phi & = & \frac{4\pi}{m_{\mathrm p}^2 } \sum_{i=1}^n \phi_i' \delta \phi_i ~,
  \\ 
  \Phi'' + 3 \mathcal H \Phi' +
  \Phi \left( 2\mathcal H ' + \mathcal H ^2 \right)  
& = & \frac {4 \pi}{m_{\mathrm p}^2}  \sum_{i=1}^n  \left( \phi_i' \delta \phi_i - \phi_i'^2 \Phi - a^2 \frac {\partial V}{\partial \phi_i} \delta \phi_i \right) ~, \label{Eq:pert_multifield3}
\end{eqnarray}
where $\delta \phi_i $ is the perturbation of the scalar field $\phi_i$.  On the other hand, the $n$ perturbed Klein-Gordon equations read
\begin{equation}\label{Eq:KGpert_multifield}
\delta \phi_i '' + 2 \mathcal H \delta \phi_i' - \nabla^2 \delta \phi_i + \sum_{j=1}^n a^2 \delta \phi_j \frac {\partial^2V}{\partial \phi_i \partial \phi_j}
= 2 (\phi_i'' + 2 \mathcal H \phi_i' ) \Phi + 4 \phi_i'  \Phi' ~.
\end{equation}
Because of the fourth term, the field perturbations are coupled to each other and only evolve independently if the cross-derivatives of the potential vanish.    By adding Eq.~(\ref{Eq:pert_multifield1}) to Eq.~(\ref{Eq:pert_multifield3}), and by using Eq.~(\ref{Eq:pert_multifield2}), one obtains the evolution equation for the Bardeen potential,
\begin{equation} \label{Eq:Bardeen_multifield}
\Phi'' + 6 \mathcal H \Phi' + ( 2 \mathcal H' + 4 \mathcal H^2 ) \Phi - \nabla^2 \Phi = - \frac {8 \pi}{m_{\mathrm p}^2} a ^2  \sum_{i=1}^n 
\frac{\partial V}{\partial \phi_i} \delta \phi_i~.   
\end{equation}
What we need is the comoving curvature perturbation $\zeta$, defined in Eq.~(\ref{zeta}) as
\begin{equation}\label{eq:defPhi}
\zeta \equiv = \Phi - \frac{\mathcal H}{\mathcal H' - \mathcal H^2 } (\Phi' + \mathcal H \Phi )~.
\end{equation}
By using the background dynamics, one has $ \mathcal H' - \mathcal H^2 = - 4 \pi \sigma'^2 / \mpl^2 $.  By using Eq.~(\ref{Eq:pert_multifield2}), the comoving curvature can thus be rewritten,
\begin{equation}
\zeta = \Phi + \frac{ \mathcal H}{\sigma'^2 }    \sum_{i=1}^n \phi_i' \delta \phi_i ~.
\end{equation}
By using the background and perturbed Einstein equations, one obtains that $\zeta$ evolves according to~\cite{Ringeval:2007am}
\begin{eqnarray}
\zeta' & = &  \frac{2 \mathcal H}{\sigma'^2} \nabla^2 \Phi - \frac {2 \mathcal H}{\sigma'^2} \left[ a^2  \sum_{i=1}^n \phi_i' \frac{\partial V}{\partial \phi_i}   - \frac{a^2}{\sigma'^2} \left( \sum_{i=1}^n \phi_i' \frac{\partial V}{\partial \phi_i} \right) \left(  \sum_{i=1}^n  \phi_i' \delta \phi_i\right)  \right] \\
 & = & \frac{2 \mathcal H}{\sigma'^2} \nabla^2 \Phi - \frac {2 \mathcal H}{\sigma'^2} \bot_{ij} a^2 \frac{\partial V}{\partial \phi_i} \delta \phi_j ~,
\end{eqnarray}
where the orthogonal projector $\bot_{ij} \equiv \rr{Id} - u_i u_j$ have been introduced.  
For a single field model, the second term vanishes and one finds back the 1-field evolution of $\zeta$.  For the multi-field case, one sees that entropy perturbations orthogonal to the field trajectory can source curvature perturbations, even after Hubble exit.   In a multi-field scenario, the second term is sourced by the field perturbations and the slope of the potential orthogonal to the trajectory.  

\subsubsection{Numerical integration}


In Ref.~\cite{Ringeval:2007am} the background and the perturbation dynamics has been integrated numerically.  The exact numerical method has the advantages to be robust and general, whereas the approximations required for an analytical treatment can sometimes break down for some exotic models.  We give here the guidelines proposed in Ref.~\cite{Ringeval:2007am}  for the calculation of the exact power spectrum of curvature perturbations in a multi-field scenario.    

It is convenient to use the number of e-folds as the time variable.  Since the perturbed FL and KG equations are redundant, one can reduce the number of equations to integrate.  For instance, the Bardeen potential can be expressed directly in terms of the field perturbations $\delta \phi_i$.   However, it is singular in the limit $k \rightarrow 0 $ and $\epsilon_1 \rightarrow 0$ and it is therefore more convenient to integrate simultaneously Eq.~(\ref{Eq:KGpert_multifield}) and Eq.~(\ref{Eq:Bardeen_multifield}).   In the longitudinal gauge, after expanding in Fourier modes, they read~\cite{Ringeval:2007am}
\begin{eqnarray} 
\frac{\dd^2 \delta \phi_i}{\dd N^2} + (3 - \epsilon_{\rr 1} ) \frac{\dd \delta \phi_i}{\dd N}+ \sum_{j=1}^n \frac{1}{H^2} \frac{\partial^2 V}{\partial \phi_i \partial \phi_j}  \delta \phi_j + \frac{k^2}{a^2 H^2} \delta \phi_i & = & 4 \frac{\dd \Phi}{\dd N} \frac{\dd \phi_i}{\dd N} - \frac {2 \Phi}{H^2} \frac{\partial V}{ \partial \phi_i} ~,\\
\frac{\dd^2 \Phi }{\dd N^2} + (7- \epsilon_{\rr 1} ) \frac{\dd \Phi}{\dd N} + \left( \frac{2 V}{H^2} + \frac{k^2}{a^2 H^2}  \right) \Phi & = & - \frac{1}{H^2} \frac{\partial V}{\partial \phi_i} \delta \phi_i~.
\end{eqnarray}

\subsubsection{Initial conditions}

The initial conditions (i.c.) on the Bardeen potential and its derivative are given by the constraint equations Eq.~(\ref{Eq:pert_multifield1}) and Eq.~(\ref{Eq:pert_multifield2}),
\begin{eqnarray}
\Phi_{\rr{i.c.}} & = & \sum_{j=1}^n  \dfrac{  \dfrac{\dd {\phi_i} _{,\rr{i.c.}}}{ \dd N} \dfrac{\dd \delta {\phi_i} _{,\rr{i.c.}}}{ \dd N} + 3  \dfrac{\dd {\phi_i} _{,\rr{i.c.}}}{ \dd N}  \delta {\phi_i} _{\rr{i.c.}} + \dfrac{1} {H^2 _{\rr{i.c.}} } \left. \dfrac{\partial V}{\partial \phi_i}\right|  _{\rr{i.c.}}  }{2 \left( \epsilon_{\rr {1,i.c.}} - \dfrac{k^2}{a_{\rr{i.c.}}^2 H_{\rr{i.c.}}^2}   \right)}~,\\
\left.  \frac{\dd \Phi}{ \dd N} \right|_{\rr{i.c.}} & = & - \Phi_{\rr{i.c.}}  + \frac 1 2 \sum_{j=1}^n  \frac{\dd {\phi_i} _{\rr{i.c.}}}{ \dd N}  \delta {\phi_i} _{\rr{i.c.}}~. 
\end{eqnarray}
The quantification of the field perturbations in the limit $k \gg a H$ provides initial conditions for the $\delta \phi_i $.   In the multi-field scenario, the normalized quantum modes are defined by 
\begin{equation} 
v_i = \sqrt 2 a k^{3/2} \delta \phi_i~.
\end{equation}
They obey to $v_i'' + k^2 v_i = 0$ and in the regime  $k \gg a H$,  one has
\begin{equation} \label{eq:condinit_multifield}
\lim_{k/aH \rightarrow + \infty} v_{k,i} (\eta) = \frac{\sqrt{8 \pi}}{m_{\mathrm p}} k  \mathrm e^{-ik(\eta-\eta_{\mathrm i}) }~.
\end{equation}
In terms of the field perturbations, the initial conditions therefore read, up to a phase factor,
\begin{eqnarray}
\delta \phi_{i,\rr{i.c.}} & = & \sqrt{\frac{8 \pi}{ \mpl^2 k} } \frac{1}{a_{\rr{i.c.}}}~, \\
\left[ \frac{\dd \delta \phi_i}{\dd N} \right]_{\rr{i.c.}} & = & -   \sqrt{\frac{8 \pi}{ \mpl^2 k} } \frac{1}{a_{\rr{i.c.}}}   \left( 1 + i \frac{k}{a_{\rr{i.c.} } H_{\rr{i.c.}}}   \right)~.
\end{eqnarray}

It is not convenient to integrate the perturbations from the onset of inflation, since the total number of e-folds of can be much larger than $N_*$.  In order to avoid the time consuming numerical integration of sub-Hubble modes behaving like plane waves, it is convenient to start the numerical integration of perturbations later, but when the following condition is still satisfied, 
\begin{equation} \label{eq:Cq}
\frac{k}{ \mathcal H(n_{\rr{i.c.}} )}  = C \gg 1 
\end{equation}
where $C$ is a constant characterizing the decoupling limit.  Practically, the numerical integration process follows the algorithm described below
\begin{enumerate}
\item The background dynamics is integrated until the end of inflation, such that $N_{\rr{end}} $ and $ N_{\rr{end}}  - N_* $ are obtained.  
\item The background dynamics is integrated again, until $N_{\rr{i.c.}} $ is reached.  Initial conditions for the perturbations are fixed at this time.
\item For each comoving mode $k$, the background and the perturbation dynamics are integrated simultaneously from $N_{\rr{i.c.}} $ to $ N_{\rr{end}} $. 
\item The last step is the determination of the observable scalar power spectrum, from the perturbation modes of $\Phi$ at the end of inflation.   By definition, for each mode the curvature perturbations are related to $\Phi$ through  Eq.~(\ref{eq:defPhi}).   Due to the contribution of entropic modes, the conservation of $\zeta$ at super-Hubble scale may be broken.   

\end{enumerate}

\begin{figure}
\begin{center}
\includegraphics[width=12cm]{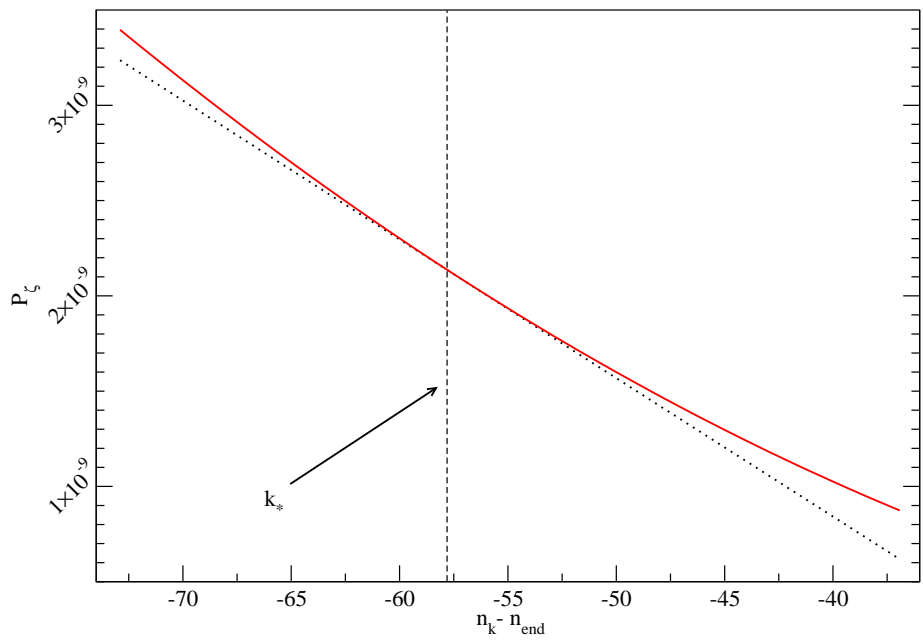}
\caption{Power spectrum~\cite{Ringeval:2007am} of the comoving curvature perturbation $\zeta$, at the end of inflation, for the large field model of Eq.~(\ref{eq:largefield}), with $p=2$, from the numerical exact integration (plain line) and the 
first order slow-roll approximation (dotted line).  The wavenumbers are expressed as the e-fold time $n_k - n_{\rr{end}} $ at which they exit the Hubble radius, $k = \mathcal H(N_k) $. $N_{\rr{end}}$ is the number of e-folds at the end of inflation.  }
\label{fig:compsrnum}
\end{center}
\end{figure}

\subsubsection{Tensor perturbations}

The numerical integration of the tensor modes doe not present any major difficulty.  Indeed, the perturbation modes evolve according to Eq.~(\ref{tensoriel}), identically to the 1-field case.  Following the previous algorithm, initial conditions on $\mu_{\rr t} $ are fixed deep inside the sub-Hubble regime, when the condition of Eq.~(\ref{eq:Cq}) stands, and then $\mu_{\rr t} (k) $ is integrated numerically simultaneously with the background dynamics.  The tensor power spectrum at the end of inflation is calculated for each comoving mode $k$.

\section{Reheating} \label{sec:reheating}

\subsection{Reheating for large field models}

Inflation stops when the first Hubble flow function reaches unity, $\epsilon_1 = 1 $.  From this point, in many models of inflation, the field(s) starts to oscillate around the minimum of the potential.   The oscillation frequency is much larger than the Hubble rate, because $V'' \gg H $.   

Let us follow~\cite{Martin:2004um} and consider for simplicity the class of large field potentials.  
By using Eqs.~(\ref{eq:rho_inf}) and~(\ref{eq:P_inf}), the conservation equation Eq.~(\ref{eq:conserv}) can be rewritten
\begin{equation}
\dot \rho = - 3 H \dot \phi^2~ = - 6 H (\rho - V)~.
\end{equation}
Because the Hubble rate only varies marginally during one oscillation, one can time average this equation, 
\begin{equation} \label{eq:meanrho}
\left\langle \dot \rho \right\rangle =  - 6 H \left\langle \rho - V \right\rangle~.
\end{equation} 
One has also
\begin{equation}
\langle \rho - V \rangle = \frac 1 T \int_0^T (\rho - V ) \dd t~,
\end{equation}
where $T$ is the oscillation period.  After using 
\begin{equation}
\frac{\dd \phi}{\dd t} = \sqrt{2 ( \rho - V)}~,
\end{equation}
one obtains
\begin{equation}
\langle \rho - V \rangle = \dfrac{\int_{-\phi_{\rr{max}}} ^{\phi_{\rr{max}}}  \sqrt{\rho - V} \dd \phi }
{\int_{-\phi_{\rr{max}}}^{\phi_{\rr{max}}}  \dfrac{1}{\sqrt{\rho - V}} \dd \phi}~,
\end{equation}
where $ \phi_{\rr{max}} $ is the maximum value of the scalar field during one oscillation.  
Over one period $T$, the energy density $\rho \simeq V(\phi_{\rr{max}}) $ remains nearly constant.  For large field models $V(\phi ) \propto \phi^p $, one has
\begin{equation}
\langle \rho - V \rangle \simeq \rho \dfrac{\int_{-\phi_{\rr{max}}}^{\phi_{\rr{max}}} \sqrt{1  - \dfrac{\phi^p}{\phi_{\rr{max}}^p}} \dd \phi }
{\int_{-\phi_{\rr{max}}}^{\phi_{\rr{max}}} \dfrac{1}{\sqrt{1 - \dfrac{\phi^p}{\phi_{\rr{max}}^p }}} \dd \phi}
 = \frac{p \rho}{p + 2}~.
\end{equation}
On the other hand, the left hand side of Eq.~(\ref{eq:meanrho}) can be considered as $\Delta \rho / T $, and if we now consider time scales much larger than $T$, it can be rewritten $\dot \rho$.  Therefore, one has
\begin{equation} \label{eq:reheat}
\dot \rho = - \frac{6 p } {p + 2} H \rho~,
\end{equation}
and the energy density thus evolves according to $\rho \propto a^{-6 p / (p+2)} $.  For $p=2$, this corresponds to a matter dominated epoch.  This is expected since $\langle \dot \phi^2 / 2 \rangle \simeq \langle V(\phi) \rangle$, resulting in a vanishing pressure.  

In a realistic scenario, the scalar field must decay into other particles, since one may recover particles of the standard model and the radiation era.  Phenomenologically, the particle creation can be described by adding a friction term in the KG equation,
\begin{equation}
\ddot \phi + 3 H \dot \phi + \Gamma \dot \phi + \frac{\dd V}{\dd \phi} = 0~.
\end{equation}
Adding a friction term induces a modification of Eq.~(\ref{eq:reheat}) that reads
\begin{equation}
\dot \rho =  - \frac{6 p } {p + 2} \left(H + \frac \Gamma 3 \right) \rho~.
\end{equation}
Integrating this equation, one obtains that the energy density is exponentially decreasing with time,
\begin{equation}
\rho(t) = \rho_{\rr{end}} \left( \frac{a}{a_{\rr{end}}} \right)^{-6 p / (p+2)} \exp \left[ - \Gamma \frac{2p}{p+2}(t-t_{\rr{end}}) \right] ~,
\end{equation}
where $  \rho_{\rr{end}} $, $a_{\rr{end}} $ and $t_{\rr{end}}$ are respectively the scalar field energy density, the scale factor and the cosmic time at the end of inflation, when oscillations are triggered.  

The total energy density must be conserved, and thus one has for the energy density of radiation (assuming that the particles produced are very light compared to the inflaton mass, so that they are relativistic),
\begin{equation}
\dot \rho_{\rr r} = - 4 H \rho_{\rr r} + \Gamma \rho~.
\end{equation}
As solution to this equation in the limit $ t \ll \Gamma^{-1} $ is given in~\cite{Martin:2004um}.  In~\cite{Martin:2006rs}, this solution is compared to numerical results and is shown to remain a good approximation until the reheating is completed, at $t \approx t_{\rr{reh}} \equiv  \Gamma^{-1} $, and one has
\begin{equation}
\rho_{\rr{reh}} \approx \rho_{\rr r} (t_{\rr{reh}} ) \simeq \left(\frac{3 p }{p+ 8}\right) \Gamma^2 \mpl^2~.
\end{equation}
It does not depend on the energy scale of inflation, but only on the decay rate $\Gamma $ of the inflaton.  

\subsection{Parametrization of the reheating}

In general, the perturbation mode evolution outside the Hubble radius does not depend on the detailed physics between the end of inflation and the radiation era.  However, the relation linking the physical scales today and those scales during inflation depends on the amount of expansion during these eras.  Indeed, the physical length of the comoving pivot scale $k_*$ at $\eta_* $ is 
\begin{equation} 
\frac{k_*}{a_*}  = \frac{k_*}{a_0} \frac{a_0}{a_{\rr{end} }  }  \frac{a_{\rr{end} } } {a_*}~,
\end{equation}
where $ k_* / a_0 $ is the physical pivot scale today and $a_{\rr{end}}$ is the scale factor at the end of inflation. $a_0 / a_{\rr{end}} $ is known if the radiation era is triggered instantaneously after inflation, but in a realistic scenario it will depend on the reheating history.  $a_{\rr{end} } / a_* $ can thus been fixed once given the Universe's evolution after inflation.   Since the determination of the primordial power spectra requires the evaluation of the Hubble rate and the slow-roll parameters $N_* $ e-folds before the end of inflation, they will therefore be affected by the reheating history.  Since we are also interested in the energy scales of the various phases in the early Universe, it is convenient to rewrite this relation 
\begin{equation}
\frac{k_*}{a_*}  = \frac{k_*}{a_0} \left(  \frac{\rho_{\rr{end}} }{\rho_{\gamma 0}  }\right)^{1/4} R_{\rr{rad}}^{-1}  \frac{a_{\rr{end} } } {a_*}~,
\end{equation}
 where $\rho_{\rr{end}} $ is the energy density at the end of inflation and where $\rho_{\gamma 0} $ is the present radiation energy density.  This relation defines a new parameter $R_{\rr{rad}}$.  This parameter actually records the expansion history between the end of inflation and the onset of the radiation era.  Indeed, let us consider the equation of state
 \begin{equation}
 w_{\rr{reh}} (N)  \equiv \frac{P(N)}{\rho(N)} ~,
 \end{equation}
during the reheating.  To account for a general reheating history, it can depend on time.   From Eq.~(\ref{eq:conserv}), it is straightforward to show that 
 \begin{equation}
 \rho(N) = \rho_{\rr{end}} \rr e ^{-3 \int_{N_{\rr{end} } }^ {N} \left[ 1 + w(n) \right] \dd n }~.
 \end{equation}
 Let us introduce 
 \begin{equation}
 \bar w \equiv \frac{1}{\Delta N} \int_{N_{\rr{end}}} ^{N_{\rr{reh}}} w_{\rr{reh}} (n) \dd n ~,
 \end{equation}
  that is the mean equation of state parameter, with $\Delta N \equiv N_{\rr{reh}} - N_{\rr{end}} $ being the total number of e-folds of reheating, between the end of inflation and the onset of the radiation era.  
It follows that the new parameter $R_{\rr{rad}}$ can be rewritten as
\begin{equation}
 \ln R_{\rr{rad}} = \frac{\Delta N}{ 4} \left( -1 + 3 \bar w  \right)~.
 \end{equation}
This parameter therefore only depends on the reheating history.  If reheating is instantaneous ($\Delta N = 0 $) or if the overall reheating behaves like a radiation era ($\bar w = 1/3$), one has $R_{\rr{rad}} = 1$ and thus the relation between the physical observable modes today and during inflation is not affected.   

For a given model of inflation, $R_{\rr{rad}} $ must be ideally added to the inflation potential parameters in order to be constrained by CMB observations.   If in addition a model of reheating is assumed (and thus $\bar w$ is given), the relation relating $ R_{\rr{rad}} $ and the energy of reheating: 
\begin{equation}
\ln  R_{\rr{rad}} = \frac{1 - 3 \bar w_{\rr{reh}}}{12 (1+ \bar w_{\rr{reh}} )} \ln \left( \frac{\rho_{\rr{reh}}}{\rho_{\rr{end}}} \right)~,
\end{equation}
can be exploited to put constraints on the reheating energy~\cite{Martin:2010kz}.    

For instance, for large field power-law models (keeping $p$ as a real potential parameter), a 2-$\sigma$ lower bound on $  R\equiv R_{\rr{rad}}  \rho_{\rr{end}}^{1/4}  / M_{\rr p}$ was found in Ref.~\cite{Martin:2010kz}, 
\begin{equation}
 \ln  R_{\rr{rad}} > - 28.9 ~.
 \end{equation}
If reheating is assumed to proceed only by parametric oscillations around the minimum of the potential, we have shown above that
\begin{equation}
\bar w = \frac{p-2}{p+2}~,
\end{equation}
and a 2-$\sigma$ lower bound on the reheating energy can be found, 
\begin{equation}
\rho_{\rr{reh}}^{1/4} > 17.3 \rr{TeV}~.
\end{equation}
This parametrization of the reheating history should be used to determine the ability of 21cm observations from the dark ages and the reionization to put strong constraints on inflation and reheating.

%% file: hybrid.tex
\chapter{Hybrid models of inflation}
\label{chap:hybridmodels}

\begin{center}
\textit{based on}\\
S. Clesse, J. Rocher, \\Avoiding the blue spectrum and the fine-tuning of initial conditions\\ in hybrid inflation\\
Phys.Rev.D79:103507, 2009, arXiv:0809.4355\\
\vspace{3mm}  S. Clesse, C. Ringeval, J. Rocher, \\ 
Fractal initial conditions and natural parameter values in hybrid inflation\\
Phys.Rev.D.80:123534, 2009, arXiv:0909.0402 
\end{center}
Among the zoo of inflation models, the hybrid class is particularly motivated and promising.  Hybrid models are easily embedded in various high energy frameworks like supersymmetry, supergravity~\cite{Halyo:1996pp,Binetruy:1996xj,Dvali:1994ms,
Kallosh:2003ux}, grand unified theories \cite{Jeannerot:1997is,
Jeannerot:2003qv} and extra-dimensional theories~\cite{Davis:2008dj,Brax:2007fe,Brax:2007xq,Dvali:1998pa,Koyama:2003yc,Fukuyama:2008dv,Fairbairn:2003yx,Linde:2005dd,Brax:2006ay}.   Contrary to large field models \cite{Martin:2010kz}, the energy scale of hybrid inflation can be low and such models do not  need super-plankian field values.   Hybrid inflation is realized along a nearly flat direction of the potential and it ends due to a Higgs-type tachyonic instability.   However, for the original hybrid model~\cite{Linde:1993cn,Copeland:1994vg}, the scalar power spectrum calculated in the 1-field slow-roll approximation exhibits a slight blue tilt, which is disfavored by CMB experiments~\cite{Martin:2006rs}.   

In this chapter, the original hybrid model is introduced, as well as some other hybrid models from various frameworks that will be studied in the next chapters.  

\section{The original hybrid model} \label{sec:original_hybrid}

\subsection{2-field potential}
The original hybrid model of inflation was proposed by A. Linde~\cite{Linde:1993cn} as a new way to stop inflation, when a symmetry is spontaneously broken.  Its potential reads
\begin{equation} \label{eq:potenhyb2dNEW}
V(\phi,\psi) = \Lambda^4 \left[  \left( 1 - \frac{\psi^2}{M^2} \right)^2 
+ \frac{\phi^2}{\mu^2} + \frac{ 2 \phi^2 \psi^2}{\phi_{\rr c}^2 M^2}\right] .
\end{equation}
The field $\phi$ is the inflaton, $\psi$ is an auxiliary transverse field and
$M,\mu,\phi_{\rr c}$ are three mass parameters. 
Inflation is assumed to be realized in the
false-vacuum~\cite{Copeland:1994vg} along the valley $\langle\psi\rangle=0$.  In the usual description, inflation ends when the transverse field develops a Higgs-type tachyonic instability soon after the inflaton reaches a critical value $ \phi_{\rr c}  $.  From this point, the classical system is assumed to evolve quickly toward one of 
its true minima $\langle\phi\rangle=0$, $\langle\psi\rangle=\pm
M$, whereas in a realistic scenario one expects the
instability to trigger a tachyonic preheating era~\cite{Kofman:1997yn,
  Garcia-Bellido:1997wm, Felder:2000hj, Felder:2001kt, Copeland:2002ku, Senoguz:2004vu, Micha:2004bv,
  Allahverdi:2007zz}.

Let remark that in many papers, the 2-field potential is written as
\begin{equation} \label{pothybride} 
 V(\phi,\psi )= \frac 1 2 m^2 \phi^2 + \frac {\lambda}4 (\psi^2 - L^2 )^2 + \frac {\lambda'} 2 \phi^2 \psi^2 ,
 \end {equation}
where $\lambda'$ and $\lambda$ are two coupling constant.   The relations between the potential parameters of Eq.~(\ref{eq:potenhyb2dNEW}) read  
\begin{eqnarray}
\phi_c^2 & = & \frac{\lambda L^2 }{ \lambda'}~,\\
 M & = &  L~, \\
 \Lambda & = &  \frac {\lambda^{1/ 4} L}{\sqrt 2}~,\\ 
 \mu & = & \sqrt { \frac \lambda 2 } \frac {L^2}{m}~.
\end{eqnarray}

 \begin{figure}[htbp]
	\begin{center}
	\includegraphics[height=80mm]{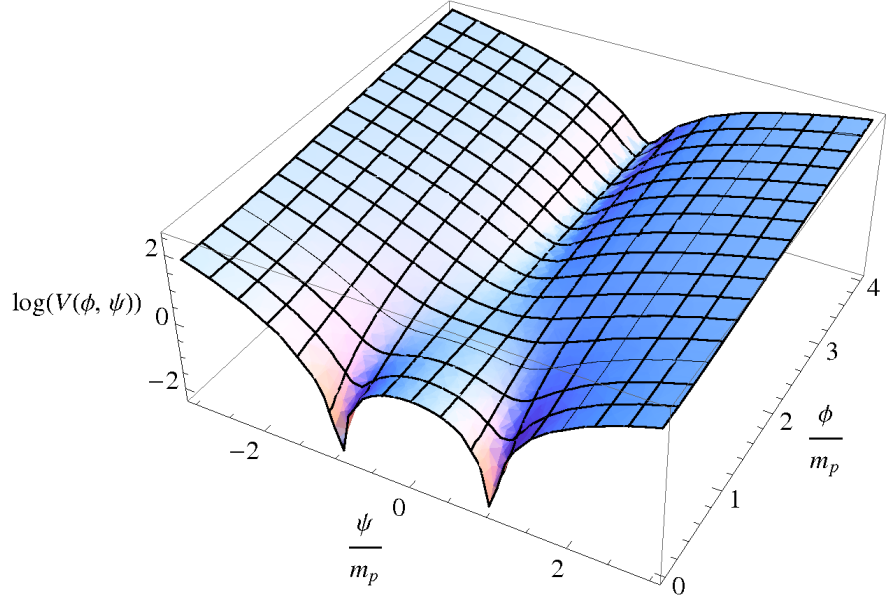}
		\caption{ Logarithm of the original hybrid potential, for $\Lambda = M=\phi_{\rr c} = \mpl$, $\mu = 100 \ \mpl$.   Inflation can take place along the valley in the $\psi=0$ direction.  The two global minima are in $\phi = 0, \psi = \pm M$. } 
	\end{center}
\end{figure}

\subsection{Effective 1-field potential}

Observable predictions can be derived in the slow-roll approximation by approximating the 2-field dynamics to the evolution of $\phi$ along the valley $\psi = 0$.   The effective one-field potential therefore reads
\begin{equation} \label{eq:effpot}
  V_{\rr{eff}} (\phi) = \Lambda^4 \left[ 1 + \left( \frac \phi \mu \right)^2  \right] ,
\end{equation}   
and inflation is assumed to end abruptly once the critical instability point $\phi_{\rr c} $ is reached.   For this effective potential, the Hubble flow functions in the slow-roll approximation read~\cite{Martin:2006rs}:
\begin{equation} \begin{split} \label{eq:eps1hybrid}
\epsilon_{\rr 1} & = \frac {1} {4 \pi} \left( \frac{\mpl}{\mu}  \right)^2 \dfrac{\left(\dfrac{\phi }{ \mu} \right)^2}{\left[1 + \left( \dfrac{\phi}{ \mu} \right)^2  \right]^2} \ ,  \\ 
\epsilon_{\rr 2} & =  \frac {1} {2 \pi} \left( \frac{\mpl}{\mu}  \right)^2 \dfrac{\left(\dfrac{\phi }{ \mu }\right)^2-1}{\left[1 + \left( \dfrac{ \phi}{ \mu} \right)^2  \right]^2} ~.
\end{split} 
\end{equation}
$\epsilon_{\rr 1} $ is maximum when $\phi = \mu$.  Thus two phases of inflation can be identified.   The first one occurs at \textit{large field} values ($\phi > \mu$), where the potential behaves like Eq.~(\ref{eq:largefield}).   The second phase takes place at \textit{small field} values ($\phi < \mu $).    It is important to remark that there exists a critical value of $\mu$ under which inflation is interrupted between these two phases,
\begin{equation}
\exists \phi \ | \  \epsilon_{\rr 1}(\phi) > 1 \Leftrightarrow \frac{\mu}{m_{\mathrm p}} < \frac 1 {4 \sqrt \pi} ~.
\end{equation} 
During the transition, slow-roll is violated and the subsequent dynamics is affected. The effects of such slow-roll violations will be discussed in the next chapter.  

On the other hand,  the number of e-folds generated in the slow-roll regime reads
\begin{equation}
N(\phi) = \frac{ 2 \pi \mu^2} {m_{\mathrm p}^2 } \left[ \left( \frac {\phi_{\mathrm i}}{\mu} \right)^2 - 
\left( \frac \phi \mu \right)^2 - 2 \ln \left( \frac{\phi}{\phi_{\rr i}}  \right) \right]~.
\end{equation}
In the small field phase, the number of e-folds is typically much larger than the $60 $ e-folds required.   Therefore the observable modes are usually considered to leave the Hubble radius at small field values.   As a consequence, $ \epsilon_{\rr 1}$ is extremely small, $\epsilon_{\rr 2}$ is negative and drives the scalar spectral index towards blue value, $n_{\rr s} = 1 - 2 \epsilon_{1*} - \epsilon_{2*} > 1 $, which is disfavored by WMAP7 observations~\cite{Larson:2010gs}.  

\section{F-term  and D-term (SUGRA) hybrid model}
\label{section:fsugra}

\subsection{Motivations}
The minimal supersymmetric versions of hybrid inflation are
known as the F-term and D-term inflationary
models~\cite{Dvali:1994ms,Binetruy:1996xj,Halyo:1996pp}, for which the
slope of the valley is generated by radiative corrections. The F-term 
model we focus on is compatible with the present CMB data because it exhibits a red spectrum
of the scalar perturbations~\cite{Dvali:1994ms,Senoguz:2003zw,Jeannerot:2005mc}. In
addition, this model is more predictive and testable than its non-SUSY 
version since it contains only one coupling constant and one mass 
scale as free parameters.

\subsection{F-term Potential}

The supersymmetric F-term hybrid
model is based on the superpotential~\cite{Dvali:1994ms}
\begin{equation}\label{superpotFterm}
W_{\rm infl}^{\rm F}=\coupling S(\bar \Phi  \Phi - M^2)\,.
\end{equation}
$\kappa $ is a coupling constant\footnote{$\kappa$ is not $1/\Mpl^2$ like in many references}. The inflaton is contained in the superfield $S$. The Higgs pair
$\bar \Phi,\Phi$ is charged under a gauge group $G$, that is broken at
the end of inflation when the Higgs pair develops a non-vanishing
expectation value (\vev) $M$. The superpotential leads in global SUSY
to a tree level potential~\cite{Dvali:1994ms}
\begin{equation}
V^{\rm SUSY}_{\rm tree}(s,\psi)=\coupling^2\left(M^2-\frac{\psi^2}{4}\right)^2 
+\frac{1}{8}\coupling^2 s^2 \psi^2~,
\end{equation}
where the effective inflaton $s$ and the Higgs field $\psi$ can be made
real and canonically normalized [$s \equiv \sqrt{2}\,\Re e(S)$,
$\psi=2\Re e(\bar \Phi)=2 \Re e(\Phi)$]. The local minima
of the potential at large $S$ provide a flat direction for the
inflaton $s$: $V_0=\coupling^2M^4$.

This tree level flat direction is lifted by two effects. Firstly,
radiative corrections induced by the interactions between the fields. In addition, if the field values are close to the reduced
Planck mass $\Mpl$, one should expect supergravity corrections
to the tree level potential.  Assuming that radiative corrections along
the inflationary valley are given by the Coleman-Weinberg
formula~\cite{Coleman:1973jx}, they reduce to~\cite{Dvali:1994ms}
\begin{equation}\label{VFterm1dim}
\begin{split}
V^{\rm cw}_{\rm 1-loop}(s)=\frac{\coupling^4M^4\mathcal{N}}{32\pi^2}&\left[2\ln\frac{s^2
\coupling^2}{\Lambda^2}+(z+1)^2\ln(1+z^{-1})\right.\\
&+(z-1)^2\ln(1-z^{-1})\Big],
\end{split}
\end{equation}
where $z=s^2/M^2$, $\mathcal{N}$ stands for the dimensionality of the
representations to which $\Phi$ and $\bar \Phi$ belong and $\Lambda$
is a renormalization mass. Realistic
values of $\mathcal{N}$ can be derived from the embedding of the model in
realistic SUSY Grand Unified Theories (GUT) as shown in
Ref.~\cite{Jeannerot:2003qv}. For example, in the case of an embedding
of the model in SUSY SO(10), $\Phi$ and $\bar \Phi$ belong to the representation
$\mathbf{16},\mathbf{\overline{16}}$ or
$\mathbf{126},\mathbf{\overline{126}}$. However, as pointed out in
Ref.~\cite{Jeannerot:2006jj}, it is possible that only some components
of $\Phi$ and $\bar \Phi$ take a mass correction of order $M$ so that
effectively\footnote{This depends on the mass spectrum of the assumed
GUT model.} $\mathcal{N}=2,3$. For the sake of generality, we will 
assume that $\mathcal{N}$ can take values in the range $[2,126]$. This 
model is also known to generically produce cosmic strings at the end of 
inflation \cite{Jeannerot:2003qv}.  Their contribution to the CMB angular power spectrum is constrained and depends on the energy scale of inflation, so that an upper limit can be fixed on~\cite{Rocher:2004et,Jeannerot:2005mc,Jeannerot:2006jj}
\begin{equation}
\label{contrainteparamFterm}
M\lesssim 2\times 15\,\mathrm{GeV}, \quad \coupling \lesssim 7\times 10^{-7}\,
\frac{126}{\mathcal{N}}\,.
\end{equation}

Secondly, SUGRA corrections also contribute to lifting the tree-level 
flat direction at field values of the order of, and larger than, the Planck
mass. 
We
  will restrict to minimal SUGRA corrections\footnote{It has been noticed in Ref.~\cite{Copeland:1994vg} that the F-term hybrid inflation model does not suffer from the $\eta$-problem
only when the K\"ahler potential is (close to) minimal}, neglecting
  SUSY breaking soft terms and the non-renormalizable corrections to
  the superpotential\footnote{see \cite{Senoguz:2003zw,Jeannerot:2005mc,Jeannerot:2006jj} 
  for an analysis of their effects.}
\begin{equation}
K\simeq |S|^2+|\bar \Phi|^2+|\Phi|^2~,
\end{equation}
In terms of the canonically
normalized effective inflaton $s$ and waterfall field $\psi$, the
SUGRA corrected potential reads~\cite{Copeland:1994vg} 
\begin{equation}\label{VFSUGRA}
\begin{split}
V_{\rm tree}^{\rm{F-sugra}}(s,\psi) & = \coupling ^2 \exp \left(\frac{s^2 + \psi^2}{2 \Mpl^2}\right) \\
& \times \Bigg\{ \left( \frac {\psi^2}{4} - M^2 \right)^2  
\left( 1 - \frac{s^2}{2 \Mpl^2} + \frac{s^4}{4 \Mpl^4} \right) \\
&+ \frac{s^2 \psi^2}{4 } \left[ 1+ \frac 1 {\Mpl^2} \left(\frac 1 4 \psi^2 - M^2\right)\right]^2\Bigg\}~.
\end{split}
\end{equation}

The dynamics along the inflationary valley is driven by the radiative
corrections and by the SUGRA corrections. 
The radiative corrections
play a major role in the last e-folds of inflation (thereby generating
the observable spectral index), whereas most of the dynamics takes
actually place for field values dominated by the SUGRA corrections. We
have calculated the amplitudes for both corrections and found that
 the radiative corrections may dominate over the SUGRA corrections
 only at field values along the valley near the critical instability point (for $s\in [M,8M]$ if
$\mathcal{N}=3$ and $s\in [M,3.5M]$ if $\mathcal{N}=126$). In chapter 5, we will investigate for the F-term model the set of initial field values leading to a sufficient number of e-folds.    
The number of e-folds generated during inflation along the valley is generically much larger than $60$, so this set does not depend on the very last part of the field evolution inside the valley.  Outside the inflationary valley, we expect the tree
level dynamics to dominate over the radiative corrections, especially
for small coupling $\coupling$. There also, in addition to the tree
level, at large fields, SUGRA corrections are expected to be
important.

As a result, we have neglected radiative
corrections and kept only the potential of 
Eq.~(\ref{VFSUGRA}).

\subsection{D-term Potential}

Let us consider a theory invariant under the transformation of the $U(1)$ group.  The D-term supersymmetric hybrid model is obtained by considering the superpotential 
\begin{equation}
W^{\rr D} = \kappa S  \Phi_+ \Phi_-~,
\end{equation}
as well as a Fayet-Iliopoulos D-term in the potential~\cite{Dvali:1994ms,Binetruy:1996xj,Halyo:1996pp}.
  The superfields $S$, $\Phi_+ $ and $\Phi_- $ have respectively the charges $0$, $+1$ and $-1$ under $U(1)$, and the inflaton is identified with the radial part of the superfield $S$.  In global SUSY this leads to a scalar field tree-level potential that is the sum of F-term and D-term,
\begin{equation}
V_{\rr{tree}} ^{\rr D}(s,\psi) = \kappa^2 \left( | \phi_+ \phi_- |^2 + | s \phi_+ |^2 + | s \phi_- |^2 \right) + \frac{g^2}{2} \left( \xi + | \phi_+ |^2 - | \phi_- |^2 \right)^2~,
\end{equation}
where $g$ a coupling constant, and where the parameter $\xi > 0 $.   This potential owns a flat direction $\phi_+ = \phi_- = 0$, along which inflation can occur.   The Fayet-Iliopoulos term allows the $U(1)$ symmetry to be broken at the end of inflation, after reaching the critical instability point 
\begin{equation}
s_{\rr c} = \sqrt{\frac{g^2 \xi}{2 \kappa^2}}~.
\end{equation}
 The global minima are at $s= \phi_+ = 0$ , $\phi_ -  = \pm \sqrt \xi $.   A red primordial scalar power spectrum is generic and the model can be in agreement with CMB observations~\cite{Battye:2010hg} when radiative corrections are taken into account.  One obtains the effective potential along the direction $\phi_+ = \phi_- = 0$ \cite{PeterUzan},
 \begin{equation}
 V_{\rr{1-loop}} ^{\rr D} (s) = \frac 1 2 g^2 \xi^2 \left[ 1+ \frac{g^2}{16 \pi^2} \ln \left( \frac{s^2}{\Lambda^2} \right) \right]~.
 \end{equation}
 As for F-term inflation, supergravity corrections can be quite important at field values of the order of the Planck mass.  For a minimal K\"ahler potential, the tree level potential in SUGRA reads~\cite{Mazumdar:2010sa}
 \begin{equation} \begin{split}
 V_{\rr{tree}} ^{\rr{D-SUGRA}} & = \kappa^2 \exp \left( \frac{| \phi_ - |^2 + | \phi_+ |^2 + | s |^2}{\Mpl^2}   \right)  \left[   | \phi_ + \phi_- |^2 \left( 1 + \frac{|s|^4}{\Mpl^4}  \right) \right. \\ & + \left. 
 | \phi_+ s |^2 \left( 1 + \frac{| \phi_- |^4}{\Mpl^4} \right) + | \phi_- s |^2 \left( 1 + \frac{| \phi_+ |^4}{\Mpl^4}  \right) + 3 \frac{| \phi_- \phi_+ s |^2}{\Mpl^2} \right] \\ & + \frac{g^2}{2} \left( \xi^2 + | \phi_+ |^2 - | \phi_- |^2 \right)^2~.
\end{split}
 \end{equation}
 In the next chapters, we have not considered the D-term (SUGRA) model.  However, our results are expected to be generic and should apply also to the D-term hybrid model.   

\section{Smooth and Shifted hybrid inflation}

\subsection{Motivations}

In F-term inflation, the field develops a non-vanishing vev which leads to the breaking of
a group $G$. Topological defects can be produced during this breaking,
depending on $G$ and the subgroup  $H$ in which it is broken. They can be cosmic strings~\cite{Jeannerot:2003qv}
which would be in agreement with CMB
data~\cite{Rocher:2004et,Jeannerot:2005mc,Fraisse:2006xc},
provided that their effect on the CMB is subdominant~\cite{Bevis:2007gh}.
But they could also be monopoles or domain walls and then be in
contradiction with observations~\cite{Vilenkin:1994}.
It is possible to implement hybrid inflation in such a way that the topological defect problem is avoided, for any symmetry breaking scheme.
Two extensions of the F-term
model have been proposed in this context: the smooth~\cite{Lazarides:1995vr} and the
shifted~\cite{Jeannerot:2000sv} hybrid inflation. They are both
based on the idea of shifting the inflationary valley away from
$\psi=0$. As a consequence the symmetry group $G$ is broken \emph{during
or before} inflation, and thus any topological defect formed
during this breaking are diluted away by inflation. This is
achieved by introducing non-renormalizable terms in the
potential~\cite{Lazarides:1995vr,Jeannerot:2000sv} and imposing an
additional discrete symmetry for the
superpotential~\cite{Lazarides:1995vr}.

If these models are considered
realistic, that is if the scalar potential is assumed to be
originated from SUSY or SUGRA, one cannot consider
super-planckian fields.  However,  one may study these models
beyond super-planckian fields if they originate from other
frameworks where non-renormalizable corrections can be controlled or
prevented by other mechanisms.

\subsection{Smooth Inflation}
\subsubsection{The potential in SUSY}
\begin{figure}
	\begin{center}
	\includegraphics[height=80mm]{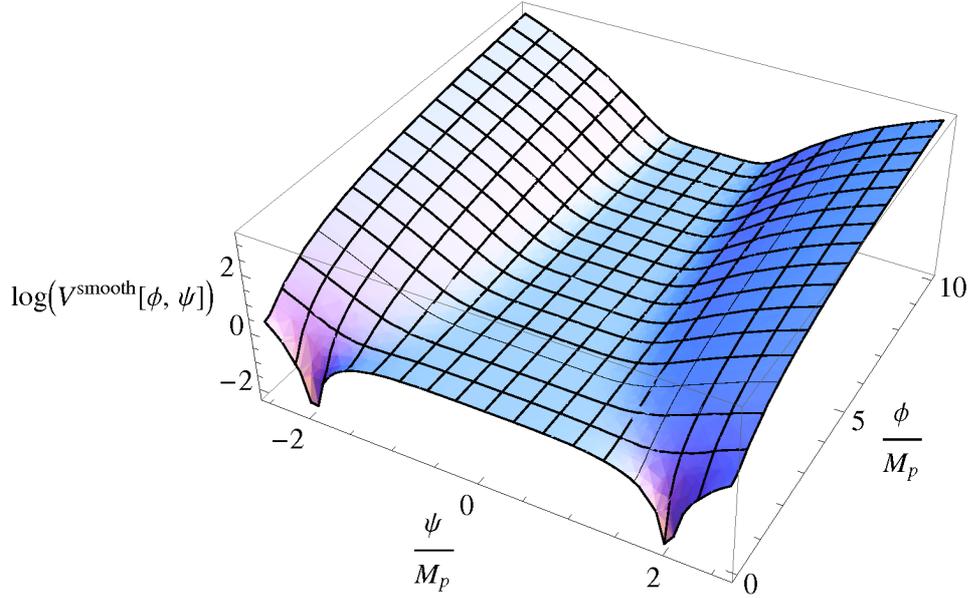} 
		\caption{ Logarithm of the smooth hybrid potential of Eq.~(\ref{eq:smoothpot}), for $\kappa =1$, $ M=  \Mpl$.   Inflation can take place along one of the two valleys, given by Eq.~(\ref{eq:smoothvalleys}).  The two global minima are in $\phi = 0, \psi = \pm 2 M$. } \label{fig:smoothpotb}
	\end{center}  
\end{figure}

Smooth inflation has been introduced by Lazarides and
Panagiotakopoulos~\cite{Lazarides:1995vr}.  It assumes that the
superpotential is invariant under a $Z_2$ symmetry under which
$\Phi\bar{\Phi}\rightarrow -\Phi\bar{\Phi}$. This forbids the
first term in the F-term superpotential of Eq.~(\ref{superpotFterm}) 
but allows for one non-renormalizable term\footnote{Note that our
choice of setting the renormalization scale to the reduced Planck
mass is arbitrary. In general, we can write $W^{\mathrm{sm}}=\kappa S
\left[-M^2+(\bar{\Phi}\Phi)^2/\Lambda^2\right]$, $\Lambda$
corresponding to the scale of new physics.}~\cite{Lazarides:1995vr}
\begin{equation}
W^{\rr{smooth}}=\kappa S \left[-M^2+\frac{(\bar{\Phi}
\Phi)^2}{\Mpl^2}\right]~.
\end{equation}
In the context of global supersymmetry, the scalar potential
has been derived in Ref.~\cite{Lazarides:1995vr},
\begin{equation}
\begin{aligned}
V^{\rr{smooth}}(S,\Phi,\bar{\Phi}) & = \kappa^2 \left| -M^2 + \frac{(\bar
\Phi \Phi)^2 }{\Mpl^2 } \right|^2\\
& + 4\kappa^2 | S| ^2 \frac {|\Phi|^2 |\bar \Phi|^2  }{\Mpl^4}
\left( |\Phi|^2 + |\bar \Phi|^2  \right).
\end{aligned}
\end{equation}
Here again, two real scalar fields $\phi$ and $\psi$ can be defined as the
relevant components of $S$, $\Phi$, $\bar \Phi$ fields such
that the fields are canonically normalized
\begin{equation}\label{eq:defininflatwaterfall}
\phi\equiv \sqrt{2} \mathrm{\Re e}(S)~,\quad
\psi\equiv 2\mathrm{\Re e}(\Phi)=2\mathrm{\Re e}(\bar{\Phi})~,
\end{equation}
and the potential becomes~\cite{Lazarides:1995vr}
\begin{equation} \label{eq:smoothpot}
V^{\rr{smooth}}(\phi,\psi)=\kappa^2\left(M^2-\frac{\psi^4}{16\Mpl^2}
\right)^2  + \kappa^2 \phi^2 \frac{\psi^6}{16\Mpl^4}\,.
\end{equation}
This potential contains a flat direction along $\psi=0$, but it is a
local maximum. The global minima are obtained for non-vanishing
values of $\psi$: they define two distinct inflationary valleys,
along
\begin{equation} \label{eq:smoothvalleys}
\psi=\pm\sqrt{-6\phi^2+6\sqrt{\phi^4+\frac{4}{9}M^2 \Mpl^2}}~.
\end{equation}
Note that these inflationary valleys, shown in Fig.~\ref{fig:smoothpotb}, progressively shift away from
$\psi=0$ as $\phi$ evolves towards $0$.

\subsubsection{Supergravity corrections}

As for the F-term model, we have considered SUGRA corrections to the smooth potential.  
Assuming supergravity with a minimal K\"ahler potential,
\begin{equation}
K_{\rr{min}}= |\Phi|^2+|\bar{\Phi}|^2+|S|^2,
\end{equation}
the scalar potential reads,
\begin{equation}
\begin{aligned}
V_{\rm SUGRA}^{\rr{sm}}(S,\Phi,\bar{\Phi})  & = \kappa^2\mathrm{exp}
\left[\frac{K_{\rm min}}{\Mpl^2}\right]\left\{ \left| \frac{(\bar
\Phi \Phi)^2 }{\Mpl^2 } -M^2 \right|^2\left(1-\frac{|S|^2}{\Mpl^2}
+\frac{|S|^4}{\Mpl^4}\right)\right.\\
+&\left. \frac {|S|^2}{\Mpl^4}\left[\left(\left| \frac{(\bar
\Phi \Phi)^2 }{\Mpl^2 }-M^2  \right|^2 + 4 |\Phi|^2 |\bar \Phi|^2 \right)
\left( |\Phi|^2 + |\bar \Phi|^2 \right) \right. \right. \\
& \left. \left.+ 4\Phi^2 \bar{\Phi}^2\left( \frac{\bar
\Phi^{*2} \Phi^{*2} }{\Mpl^2 }-M^2\right)+\rr{c.c.}\right]\right\}~.
\end{aligned}
\end{equation}
This potential is in agreement with~\cite{Yamaguchi:2004tn}.
We define again the
inflaton and waterfall fields as in
Eq.~(\ref{eq:defininflatwaterfall}), and we obtain the full potential
in SUGRA,
\begin{equation} \label{eq:smoothsugra}
\begin{split}
V_{\rm SUGRA}^{\rr{sm}}&(\phi,\psi)=\kappa^2\mathrm{exp}\left(\frac{\phi^2+\psi^2}
{2\Mpl^2}\right)\left[  \left(M^2-\frac{\psi^4}{16\Mpl^2}\right)^2\right.\\
&\times\left(1-\frac{\phi^2}{2\Mpl^2}+\frac{\phi^4}{4\Mpl^4}
+\frac{\phi^2\psi^2}{4\Mpl^4}\right)  \\
&\left.+ \frac{\phi^2\psi^6}{16\Mpl^4}-\frac{M^2\phi^2\psi^4}{4\Mpl^4}
+\frac{\phi^2\psi^8}{64\Mpl^6} \right]~.
\end{split}
\end{equation}

\subsection{Shifted Inflation} \label{sec:shifted}

\subsubsection{The potential}

\begin{center}
\begin{figure}
\begin{center}
\scalebox{1.}{\includegraphics{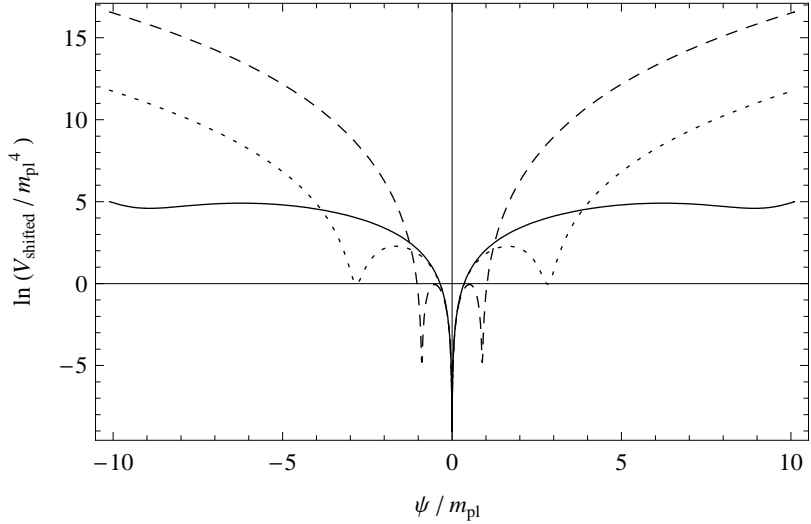}}
\caption{Cut of the logarithm of the shifted potential of Eq.~(\ref{eq:shiftedpot}), at $\phi=2 \mpl$, for $M=0.1 \mpl$, $\kappa=1$, and $\beta = 10^{-3}
\mpl^{-2} $ (plain line), $\beta = 10^{-2} \mpl^{-2}$ (dotted line),
$\beta = 10^{-1} \mpl^{-2}$ (dashed line). Notice the appearance of
two inflationary valleys, whose positions depend
on the parameter $\beta$. 
}
\label{fig:shifted_potential_cut}
\end{center}
\end{figure}
\end{center}

The shifted hybrid inflation model, proposed by Jeannerot et
al.~\cite{Jeannerot:2000sv}, is similar to the smooth inflation
model, but the additional $Z_2$ symmetry of smooth inflation is
not imposed anymore.
Thus the superpotential reads
\begin{equation}
W^{\rr{shifted}} = \kappa S \left[- M ^2 + \bar{\Phi}\Phi - \beta \frac{(\bar
\Phi \Phi)^2}{\Mpl^2}\right]~,
\end{equation}
where $\beta$ is a new dimensionless parameter.  This gives rise to the following F-terms contributions to the
scalar potential, in the context of global
supersymmetry
\begin{equation}
\begin{aligned}
V^{\rr {shifted}}&(S,\Phi,\bar{\Phi}) =\kappa^2\left[ \left| -M^2 + \bar{\Phi} \Phi - \beta
\frac{(\bar{\Phi} \Phi)^2}{\Mpl^2} \right|^2 \right.\\
&\left.+ \left| S \right|^2 \left( \left|  \bar \Phi \right|^2 +
\left| \Phi \right|^2\right) \left| 1 - 2
\beta \frac{\bar{\Phi} \Phi}{\Mpl^2} \right|^2\right]~.
\end{aligned}
\end{equation}
We can define the
relevant inflaton and waterfall fields as in
Eq.~(\ref{eq:defininflatwaterfall}) so as to cancel the D-term
contributions of the potential and to have canonical kinetic terms.
The effective scalar potential then becomes~\cite{Jeannerot:2000sv},
\begin{equation}\label{eq:shiftedpot}
\begin{aligned}
V^{\rr{shifted}}(\phi,\psi) = \kappa^2 & \left( \frac{\psi^2}{4}-M^2-
\beta\frac{\psi^4}{16\Mpl^2} \right)^2 \\ &+ \frac{\kappa^2}{4}
\phi^2 \psi^2 \left( 1- \beta\frac{\psi^2}{2\Mpl^2}\right)^2.
\end{aligned}
\end{equation}
In the limit of negligible $\beta$, one recovers the
same potential as for the original hybrid model, that is with a valley of local minima at
$\psi=0$. As $\beta$ increases, two symmetric valleys appear, parallel
to the central one as represented in Fig.~\ref{fig:shifted_potential_cut}.
These
new inflationary valleys get closer to the central one as $\beta$
gets larger.   Inflation can be realized along one of the three valleys.



\subsubsection{Supergravity corrections}
Let us discuss, as for the smooth hybrid model, the effects of
embedding the shifted model in supergravity.
As mentioned for the F-term and smooth models, supersymmetry
 is not a valid framework for describing
super-planckian fields and in this regime, the model is considered as an effective one. However supergravity corrections
allow to extend the domain of validity up to Planckian like field
values.

The supergravity corrections to the shifted potential are computed
assuming again a minimal K\"ahler potential and we obtain,
\begin{equation}
\begin{split}
V_{\rm SUGRA}^{\rr{shifted}}(S,\Phi,\bar{\Phi}) &= \kappa^2\mathrm{exp}
\left(\frac{K_{\rm min}}{\Mpl^2}\right) \left\{ \left| \bar
\Phi \Phi -M^2  -\beta\frac{(\bar \Phi \Phi)^2 }{\Mpl^2} \right|^2\left(1-\frac{|S|^2}{\Mpl^2}
+\frac{|S|^4}{\Mpl^4}\right)\right.\\
&+|S|^2\left( |\Phi|^2 + |\bar \Phi|^2 \right)\left[\left|1-2\beta \frac{\bar
\Phi \Phi }{\Mpl^2}\right|^2+\frac{1}{\Mpl^4}\left| \bar
\Phi \Phi -M^2 -\beta\frac{(\bar \Phi \Phi)^2 }{\Mpl^2} \right|^2 \right]\\
&\left. + 2\frac{|S|^2}{\Mpl^2}\left[ \Phi \bar{\Phi}\left(1 - 2\beta\frac{\Phi
\bar{\Phi}}{\Mpl^2}\right) \left(\bar{\Phi}^* \Phi^*-M^2-\beta\frac{(\bar{\Phi}^* \Phi^*)^2}{\Mpl^2 }\right)+\rr{c.c.}\right]\right\}~.
\end{split}
\end{equation}
By defining the inflaton and the waterfall field to be the
canonically normalized real part of the fields $S$, $\Phi$ and
$\bar{\Phi}$, we obtain the effective 2-field
potential,
\begin{equation}
\begin{split}
&V_{\rm SUGRA}^{\rr{shifted}} = \kappa^2 \mathrm{exp}\left(\frac{\phi^2+\psi^2}{2\Mpl^2}\right)
\left\{\left(\frac{\psi^2}{4}-M^2-\beta \frac{\psi^4}{16\Mpl^2} \right)^2\right.\\
&\phantom{V_{\rm eff}^{\rm sh} = \kappa^2 \mathrm{Exp}^{\frac{\psi^2+\psi^2}{\Mpl^2}}
}~~~~~~~~~~~~\times\left(1-\frac{\phi^2}{2\Mpl^2}+\frac{\phi^4}{4\Mpl^4} \right)\\
&\left. + \frac{\phi^2 \psi^2}{4} \left[ 1- \beta \frac{\psi^2}{2 \Mpl^2}
+\frac{1}{\Mpl^2}\left(\frac{\psi^2}{4}-M^2-\beta \frac{\psi^4}{16\Mpl^2}\right)
\right]^2\right\}.
\end{split}
\end{equation}
SUGRA corrections affect the dynamic
of inflation at large fields.  For super-planckian field values, the exponential term
dominates and the potential becomes too steep for inflation to be
automatically realized, as for the F-term and the smooth models.

\section{Radion Assisted Gauge Inflation}
\subsection{Motivations}
The radion assisted gauge inflation
model~\cite{Fairbairn:2003yx} belongs to the class of \textit{gauge inflation} (or \textit{extra-natural inflation})
models~\cite{ArkaniHamed:2003wu,ArkaniHamed:2003mz,Kaplan:2003aj}.   This class of models is based on the concept of \textit{natural inflation}~\cite{Freese:1990rb}, for which the inflaton is assumed to be a Pseudo Nambu Goldstone Boson (PNGB), parametrized by an angular variable $ \theta \sim \theta + 2 \pi$.  The flat potential obtained in the limit of exact symmetry is lifted up by explicit symmetry breaking terms in the Lagrangian
\begin{equation}  \label{eq:potengauge}
\mathcal L = \frac{f^2}{2}  (\partial \theta)^2 - V_0  \left[1- \cos (\theta)\right]~,
\end{equation} 
where $f$ is the spontaneous breaking scale.  The canonically normalized field is $\phi = f \theta$, so the potential is flat for large $f$.  For having a scenario compatible with CMB observations, one needs typically $f \gg \Mpl$.  The spontaneously breaking scale is therefore presumably outside the range of validity of an effective field theory description.  Moreover, higher dimensional operators from quantum gravity corrections~\cite{Kallosh:1995hi}, usually suppressed by powers of $f / \Mpl$, could destroy the flatness of the potential. In the extra-natural version of the model~\cite{ArkaniHamed:2003wu,ArkaniHamed:2003mz,Kaplan:2003aj}, these problems are naturally solved by assuming an effective 5-dimensional
Universe, one of the dimension being compactified on a circle of radius\footnote{The effective 4-dimensional (reduced)
Planck mass is related to the 5-dimension Planck mass $M_5$ by
$\Mpl^2=2\pi R M_5^3$.} $R$.  In this model, a gauge
symmetry is assumed together with a gauge field $(A_\mu,A_5)$. The
inflaton field is proportional to the phase $\theta$ of a
Wilson-loop wrapped around the compact dimension,  
\begin{equation}
\theta=\oint \dd x^5
A_5~.
\end{equation}
Its
potential is flat at tree level but at one-loop, takes the form of
an axion-like potential with an effective
\begin{equation}
f_{\rr{eff}} = \frac{f_{\rr{4D}} }{2 \pi R}~.
\end{equation}
Observations require $f_{\rr{eff}} \gg \Mpl$ but allow $f_{\rr{4D}} < \Mpl$.  The potential is protected from non-renormalizable operators,
suppressed by powers of $1/R$, and it becomes safe to consider super-planckian values of the canonically normalized inflaton field $\phi = f_{\rr{eff}} \theta$.  Finally, since the
inflaton is a phase, one can show~\cite{Freese:1990rb} that the
probability to have a sufficiently homogeneous distribution of the
field is quite large.

\subsection{Potential}

The ``radion assisted'' gauge inflation differs from standard
gauge inflation by assuming a varying radius of the
extra-dimension $R$, around a central value $R_0$. The ``radion''
field is defined by
$| \psi | \equiv (2\pi R)^{-1}$ and is subject to a
potential for which $R_0$ is assumed to be the minimum (for the
late time stability of the extra-dimension). The simplest way to
implement this stabilization is to use a Higgs-type potential for
$\psi$. By expanding, at first order, the potential of
Eq.~(\ref{eq:potengauge}), and by adding the Higgs-type sector,
the full scalar potential reads~\cite{Fairbairn:2003yx}
\begin{equation}\label{eq:potenradion}
V(\phi, \psi ) = \frac{1}{4}  \frac{\phi^2}{f^2} \psi^4 + \frac{\lambda}{4}
\left( \psi^2 - \psi_0 ^2 \right)^2~,
\end{equation}
where $\psi_0=(2\pi R_0)^{-1}$. This potential is similar to
the hybrid potential discussed in the last section. It is flat
for $\psi=0$ which corresponds to a global maximum. For a given
$\phi$, the minima of the potential are located in the valleys
\begin{equation}
\langle\psi\rangle^2 = \frac{\psi_0^2}{1+\phi^2/(\lambda f^2)}~.
\end{equation}
More than $60$ e-folds of inflation can take place in these throats.


\vspace{1cm}


\section{Conclusion}

In this chapter, six models of hybrid inflation, from various frameworks, have been introduced and their 2-field potential have been given:  the non-supersymmetric original model, the supersymmetric F-term, D-term, smooth and shifted models, both in they SUSY and SUGRA versions, and the radion assisted gauge inflation model.  

Their exact 2-field dynamics will be studied in the next chapters.   More particularly, in chapter 4, the effects of slow-roll violations during the field evolution along the valley, for the original hybrid model, will be studied.   Then, in chapter 5, for all the considered hybrid models, the space of initial conditions leading to more than 60 e-folds of inflation will be determined.  We will focus especially on the question of the necessity or not to fine-tune the initial field values.  For the original hybrid model, it will be shown in chapter 6 that the exact 2-field dynamics at the end of inflation reveals regions in the parameter space for which the last 60 e-folds are realized classically along waterfall trajectories.  The modifications of the observable predictions will be discussed.   
Finally, in chapter 7, for the original hybrid model, the analysis of the space of initial conditions will be extended to the case of a closed Universe in which the initial singularity is replaced by a classical bounce.  There again the existence or not of a fine-tuning problem of initial conditions (in the contracting phase) will be analyzed.

%% file: Slow-roll_violations.tex
\chapter{Slow-roll violations in hybrid inflation}
\label{chap:slowrollviol}

\begin{center}
\textit{based on}\\
S. Clesse, J. Rocher, \\Avoiding the blue spectrum and the fine-tuning of initial conditions\\ in hybrid inflation\\
Phys.Rev.D79:103507, 2009, arXiv:0809.4355\\
\end{center}

\section{Introduction}

In the original hybrid inflation scenario, two regimes can be identified for the field evolution along the valley (see section~\ref{sec:original_hybrid}).  The first one at large field values, where the potential is similar to the large field model.   The second one at small field values, where inflation is driven by the false vacuum.     Assuming that the tachyonic instability is developed in the small field regime, the scalar power spectrum in the slow-roll approximation is found to be generically blue, which is now disfavored by CMB data.   

However, during the transition between the two regimes, the slow-roll conditions can be violated.  In this chapter, we use the exact field dynamics to determine how such slow-roll violations can affect the dynamics during and after the transition, as well as the observable predictions of the model.   In particular, we give a condition on the potential parameter to generate a red spectrum of scalar perturbations, independently of the position of the critical instability point, and discuss the field values that this condition require.  



\section{Effective one-field potential}

To study the original hybrid model dynamics along the valley $\psi=0$, one may consider the 1-field effective potential of Eq.~(\ref{eq:effpot}),
\begin{equation} \label{eq:potenhybeffectif}
V(\phi) =  \Lambda^4 \left[1+ \left( \frac{\phi}{\mu}
\right)^2\right]~.
\end{equation}
Inflation occurs for $\epsone=-\dot{H}/H^2 < 1$ and slow-roll
conditions are satisfied when $|\epsilon_n|
\ll 1$. For the effective hybrid potential, the analytical
expression of the first and second Hubble-flow functions in the slow-roll approximation are given in Eqs.~(\ref{eq:eps1hybrid}),
\begin{equation}\label{eq:epshybrid}
\begin{split}
\epsilon_1 (\phi) & =\frac{1}{4 \pi} \left( \frac {\mpl}{\mu} \right)^2
\frac{(\phi/\mu)^{2}}{\left[ 1 + (\phi/\mu)^2 \right]^2}~, \\
\epsilon_2(\phi)&=\frac{1}{2\pi}\left(\frac{\mpl}{\mu}\right)^2
\frac{(\phi/\mu)^2 -1}{\left[1+(\phi/\mu)^2\right]^2}~.
\end{split}
\end{equation}
It is clear from these expressions that the slow-roll conditions
are satisfied for large field ($\phi \gg \mu$) and small field ($\phi \ll \mu$) values. As illustrated in
Fig.~\ref{fig:eps1_srviolated}, $\epsilon_{\rr 1} $ is maximum at $\phi_{\rr{max}} = \mu $, at which 
$\epstwo(\phi)$ changes its sign.
As mentioned in section~\ref{sec:original_hybrid}, there exists a critical value of $\mu$ under which inflation is interrupted between these two phases,
\begin{equation}
\exists \phi \ | \  \epsilon_{\rr 1}(\phi) > 1 \Leftrightarrow \frac{\mu}{m_{\mathrm p}} < \frac 1 {4 \sqrt \pi} ~.
\end{equation} 
The slow-roll conditions can therefore be violated during the transition, as it is
illustrated by the dashed line in
Fig.~\ref{fig:eps1_srviolated}. Thus the resolution of the exact
equations of motion for the field is required to study the
influence of slow-roll violations on the dynamics of inflation during and after the transition period.

\section{Exact field dynamics}\label{sec:1fielddynamics}

\begin{figure}[ht]
\begin{center}
\scalebox{1.}{\includegraphics{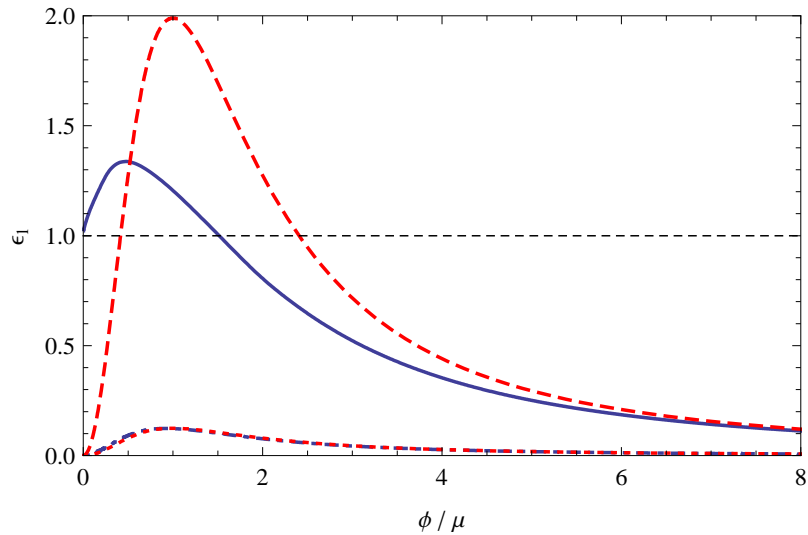}}
\caption{First Hubble flow function $\epsone$, as a function of the
inflaton field, in the slow-roll approximation (red dashed and dotted
lines) and for the exact dynamics  (blue solid and dot-dashed
lines) of field trajectories started in the slow-roll attractor in the large field phase, at $\phi_{\rr i} / \mu = 8$.  The curves correspond to $\mu=0.1
\mpl$ (two top curves), $\mu=0.4 \mpl$ (two bottom curves,
quasi-superimposed).   Top curves show that the slow-roll can be violated during the transition 
between the large field and the small field regimes, for sufficiently small values of $\mu$.  The exact trajectory shows that the field acquires a sufficient velocity for the slow-roll attractor to not be reached at small fields, whereas it is in the slow-roll treatment.}
 \label{fig:eps1_srviolated}
\end{center}
\end{figure}

The dynamics of the one-field effective hybrid inflation, without
assuming slow-roll, is described by Eqs.~(\ref{KGtc}) and (\ref{eq:FLtc1}).  These have been integrated numerically. The
function $\epsone(\phi)$ has been computed exactly. It
is represented in
Fig.~\ref{fig:eps1_srviolated}  and compared to the analytical
slow-roll expressions of Eqs.~(\ref{eq:epshybrid}).

The exact integration of field trajectories starting in the slow-roll attractor in the large field phase 
confirms the existence of the two regimes
before and after the maximum of $\epsone$, at which the slow-roll
conditions can be violated and inflation can even be interrupted
(when $\epsone \geq 1$)
depending on the value of the parameter $\mu$. But there are two
important novelties. Firstly, $\phi_\mathrm{max}$ is displaced
toward \emph{smaller} values in the exact treatment compared to
its slow-roll value $\mu$. Secondly, in the slow-roll
approximation, after the peak, $\epsone(\phi)$ decreases and
vanishes for vanishing field. One may think that inflation always
takes place for $\phi< \phi_\rr{max}$. However, exact numerical
results show that this conclusion is erroneous: $\epsone$ does not
necessarily become negligible when the field vanishes (see the
plain blue curve).  Actually, due to the slow-roll violations, 
the field velocity can increase sufficiently for 
the trajectory to not reach the slow-roll attractor a small field values.
As a consequence, \emph{inflation does not
necessarily produce the last $60$ e-folds in the small field
regime ($\phi<\phi_{\max}$)}.   \\

\begin{figure}
\begin{center}
\scalebox{1.}{\includegraphics{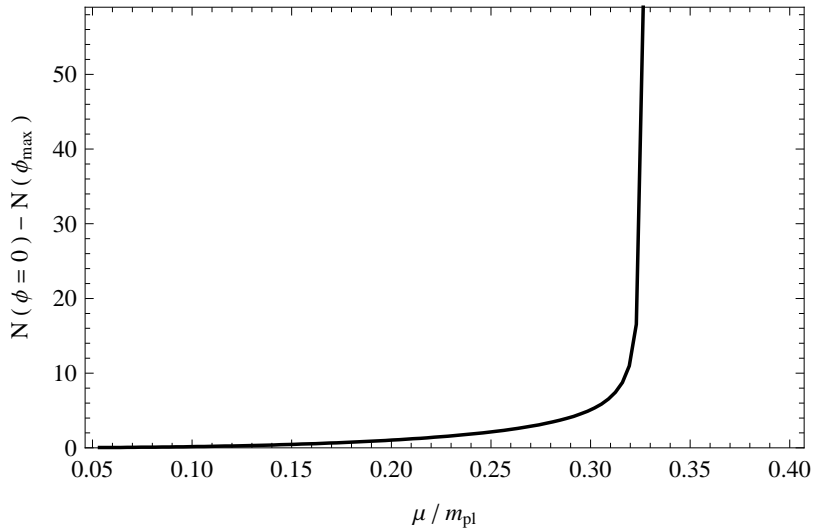}}
\caption{Number of e-folds created between $\phi_\rr{max}$ and
$\phi=0$ as a function of $\mu$, when slow-roll is not assumed, for trajectories started 
in the large field phase.  At large values of $\mu$, inflation at small field ($\phi < \phi_\rr{max}$) is very efficient in terms of the number of e-folds generated.   
But there exists a critical value of the parameter $\mu$ under which
the phase of inflation at small field values is not triggered and for which only a marginal number of e-folds is created between $\phi_\rr{max}$ and
$\phi=0$.}
\label{fig:criticalmu}
\end{center}
\end{figure}

From Fig.~\ref{fig:eps1_srviolated}, it is clear that the presence
or not of a small field phase of inflation depends on the parameter
$\mu$ (difference between the dashed and plain curves). In order
to measure the efficiency/existence of this second phase of
inflation, we have plotted in Fig.~\ref{fig:criticalmu} the number
of e-folds created between $\phi_\rr{max}$ and $\phi=0$ as a
function of $\mu$, for trajectories initially in the slow-roll attractor at large field values.
This shows that there exists a critical value
\begin{equation}\label{cond_redspectrum_2}
\mu_{\mathrm{crit}}\simeq 0.32 \,\mpl,
\end{equation}
under which the number of e-folds generated after $\phi_\rr{max}$
is reached is marginal. Therefore, observable modes become super-Hubble during the large field phase  ($\phi>\phi_{\max}$), provided
$\phi_\ui > \phi_{\max}$. In this case, the potential of hybrid
inflation is similar to the potential of the large field model, independently of the
way inflation ends. This has important consequences for the
generated spectral index.

\section{Scalar spectral index}

\begin{figure}
\begin{center}
\scalebox{0.9}{\includegraphics{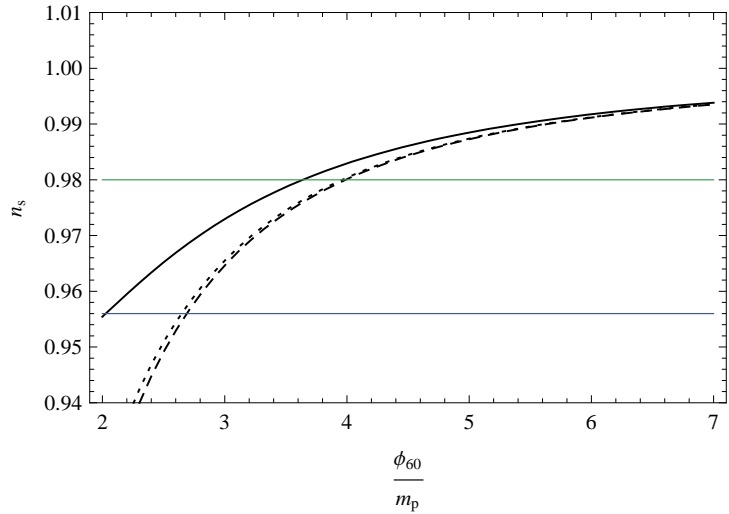}}\caption{Spectral
index $n_{\rr s}$ of the scalar power spectrum as a function of $\phi_{60}$, the
value of the field $60$ e-folds before the end of inflation.  This has
been computed for the effective hybrid potential for $\mu=\mpl$ (full
line), $\mu=0.3 \mpl$ (dotted line) and $\mu = 0.01 \mpl$ (dashed
line), in the slow-roll approximation (in any case the slow-roll regime is valid at Hubble exit of observable modes). One can see that almost any
value of the spectral index in the 1-$\sigma$ bounds of WMAP7~\cite{Komatsu:2010fb} (horizontal lines) can be accommodated  within hybrid
inflation.  If $\phi_{60}$ is pushed in the large field phase due to slow-roll violations, the model is in agreement with CMB experiments} \label{fig:ns}
\end{center}
\end{figure}

At first order in slow-roll parameters, we have determined in section~\ref{sec:sr_expand} that the scalar spectral index  can be expressed
as
\begin{equation}
n_\mathrm{s} -1  = - 2 \epsilon_{1*} -
\epsilon_{2*}~.
\end{equation}
If observable modes leave the Hubble radius during inflation at small field values, it is generically blue.  If this phase is avoided, either by instability, either due to slow-roll violations, observable modes leave the Hubble radius at large field values, when the slow-roll is valid, and the spectral index is generically red.  As illustrated in Fig.~\ref{fig:ns}, the scalar spectral index value can be accommodated 
to be inside the observational bounds of the WMAP7 data~\cite{Komatsu:2010fb}.

In that case, we would like to emphasize that the value of the inflaton at Hubble exit of the observable modes, about $60$ e-folds
before the end of inflation and denoted $\phi_{60}$, is necessarily
super-planckian,
independently of $\mu$.  If $\mu \geq \mu_{\rm crit}$, the slow-roll
approximation can be used and with $\phi_{\rm end}=
\phi_{\uc}=\phi_{\rr {max}}=\mu$, the minimum
value of $\phi_{60}$ is given by~\cite{Copeland:1994vg}
\begin{equation}
\frac{2\pi\mu^2}{\mpl^2}\left[2\ln\left(\frac{\phi_{60}}{\mu}\right)
+ \left(\frac{\phi_{60}}{\mu}\right)^2-1\right]=N_*=60,
\end{equation}
which is always around $3\mpl$ or greater. If $\mu \leq \mu_{\rm crit}$,
solving numerically the exact field dynamics is required, but we also
found that $\phi_{60}\gtrsim 3\mpl$.


Finally, notice that even for $\mu \le \mu_{\mathrm{crit}}$, one
could still have an inflationary period at small field if the trajectory start at $\phi_\ui \ll
\phi_{\max}$.  This would lead to a blue-tilted spectrum. 

\section{Conclusion}

In the original hybrid model, the inflaton
field is assumed to be coupled to a Higgs-type auxiliary field
that ends inflation by instability, when developing a
non-vanishing expectation value. We have reanalyzed the problem of the blue scalar power spectrum that was thought to be generic due to a very efficient phase of inflation at small field values.

Besides the trivial well-known possibility to have the
waterfall ending inflation in the large field phase, we have identified the possibility to generate a red spectrum due to slow-roll violations only.  

A new criteria on the mass scale $\mu$ has been found, so
that a violation of the slow-roll conditions ensures automatically the
\emph{non-existence of the small field phase of inflation}, independently of the position of the critical instability point. In
this case, the spectral index generated is less than unity (see
Fig.~\ref{fig:ns}). However, like for the large field model, this requires a large initial value of the inflaton, typically $\phi_{\rr i} \gtrsim 3\mpl$.

%% file: Initial_conditions.tex
\chapter{Initial field and natural parameter values in hybrid inflation}
\label{chap:initial_conditions}

\begin{center}
\textit{based on}\\
S. Clesse, J. Rocher, \\Avoiding the blue spectrum and the fine-tuning of initial conditions\\ in hybrid inflation\\
Phys.Rev.D79:103507, 2009, arXiv:0809.4355\\
\vspace{3mm}  S. Clesse, C. Ringeval, J. Rocher, \\ 
Fractal initial conditions and natural parameter values in hybrid inflation\\
Phys.Rev.D.80:123534, 2009, arXiv:0909.0402 
\end{center}

\section{Introduction}

Several fundamental questions about initial conditions for
inflation are still open (see for
example~\cite{Goldwirth:1991rj,Vachaspati:1998dy,Kaloper:2002cs,
Linde:1986fd,Tetradis:1997kp,Lazarides:1997vv,Mendes:2000sq}).
In this chapter, we will not address the important problem of spatial
homogeneity of the fields~\cite{Goldwirth:1991rj}, and we will assume that the field values do not enter the self-reproducing
inflationary regime~\cite{Linde:1986fd}. For the original hybrid model, even when restricting to
the classical approximation, the existence of a
fine-tuning on the initial field values has been found in Refs.~\cite{Tetradis:1997kp,Lazarides:1997vv,Mendes:2000sq}
(an opposite conclusion has been obtained~\cite{Lazarides:1996rk}
for the smooth hybrid inflation model. We will comment on this
model at section~\ref{sec:othermodels}). 

The space of initial
conditions is described by regions in the space
$(\phi_\ui,\psi_\ui, \dot \phi_\ui , \dot \psi_\ui)$, where $\phi_\ui$ and $\psi_\ui$ denote the
initial values of the inflaton and the waterfall field
respectively, and $\dot \phi_\ui$, $\dot \psi_\ui$ denote their initial velocities.  A field trajectory leading to the realization of more than 60 e-folds of inflation is said successful.   By fine-tuning of initial conditions, one means
that the successful trajectories need to be located initially~\cite{Tetradis:1997kp,Mendes:2000sq} either in an extremely thin
band around the valley $\psi_\ui=0$, either in a few very subdominant regions exterior to it. Uncertainties were remaining on whether these successful initial conditions exterior to the valley are
of null measure~\cite{Mendes:2000sq} or not~\cite{Tetradis:1997kp}.
The thin band was considered as fine-tuned
because $\psi_\ui$ had to be very close to 0 whereas the inflaton $\phi$ can take values of the order of the Planck scale. 
The problem of the fine-tuning of initial conditions for hybrid-type models of inflation is important because it means that these models would
not easily be the natural outcome of some pre-inflationary era
(see however~\cite{Calzetta:1992bp}).

Several papers have proposed some solutions to the fine-tuning
problem. It has been proposed to replicate many times identically
the inflaton sector~\cite{Mendes:2000sq}, even though no
motivations have been proposed for this replication. A similar
idea had been employed to construct the N-flation
model~\cite{Dimopoulos:2005ac} but the replication in this context
is no more natural~\cite{Easther:2005zr}. It has also been
proposed~\cite{Mendes:2000sq,Underwood:2008dh} to embed hybrid
inflation into a brane description. The induced modifications to
the Friedmann-Lema\^itre equations provide additional friction in
the evolution of scalar fields. Thus slow-rolling is favored and
more of the initial condition space gives rise to successful
inflation. This friction can also be efficiently played by
dissipative effects~\cite{Ramos:2001zw}, as in warm inflation~\cite{PhysRevLett.75.3218}, 
when couplings between
the inflaton and the waterfall field with a thermal bath of other fields
are assumed. Finally, it has been
proposed~\cite{Panagiotakopoulos:1997if} to solve this problem by
accepting a short ($N\sim 10$) phase of hybrid inflation and
implementing a second one responsible for the generation of the
primordial fluctuations, thus solving the standard FLRW problems.

However, to our knowledge, little has been done to explain the
properties of the \hbox{(un-)successful} space of initial conditions:
discreteness, sub-dominance, size and limits. In this chapter, we
provide a detailed analysis of the properties of the initial condition
space, explain why parts of this space were thought to be
discrete, and what are the field trajectories leading to these
apparently isolated points. In particular, we show that they can be
viewed as the ``anamorphosis'' (that is a deformed image) of the
thin successful band. We also give the area of successful initial
conditions  in the the plane $(\phi_\ui,\psi_\ui)$.  When
restricting the fields to sub-planckian values, instead of sub-dominant, we find up to $15\%$
of successful initial conditions, for the original hybrid scenario and for specific sets of potential parameters.  Most of these initial field values are exterior to the valley.  The fine-tuning problem is therefore absent in some parts of the parameter space. 
One may wonder whether
these features are robust with respect to the potential parameters.  Moreover, the effect of the initial field velocities could be important.  The purpose
of this chapter is also to quantify how the successful
inflationary regions are widespread in the higher dimensional space of
all the model parameters, i.e. by considering not only the initial
field values but also their initial velocities and the potential
parameters.  

In order to deal with a multi-dimensional parameter space, after
having discussed the fractal nature of the successful inflationary
regions, we introduce a probability measure and perform their
exploration by using Monte--Carlo--Markov--Chains (MCMC) methods. The
outcome of our approach is a posterior probability distribution on the
model parameters, initial velocities and field values such that
inflation lasts more than $60$ e-folds\footnote{Such probability
  distributions are almost independent of the chosen number of
  e-folds: once the field rolls down in a flat enough region of the
  potential, the total number of e-folds generated is always very
  large.}.  
  For the original hybrid model,   
this treatment allows us to establish 
natural bounds on the potential parameters.


To prove the robustness of the results, we explore the space of
initial conditions for all the hybrid models introduced in Chapter 3:  besides the
original hybrid model, we study the supersymmetric and supergravity F-term~\cite{Dvali:1994ms}, 
``smooth''~\cite{Lazarides:1995vr,Lazarides:2007fh,Yamaguchi:2004tn}
and ``shifted''~\cite{Jeannerot:2000sv,Jeannerot:2002wt} models,
as well as the ``radion assisted'' gauge
inflation~\cite{Fairbairn:2003yx}.

The chapter is organized as follows. 
 In the
next section, using exact numerical methods
applied on the two-field potential, we provide an extensive analysis
of the space of initial conditions and revisit the above-mentioned
fine-tuning problem. In Sec.~\ref{sec:othermodels}, we test the
robustness of our results on the three other models: SUSY/SUGRA smooth
hybrid inflation, SUSY/SUGRA shifted hybrid inflation and radion assisted gauge inflation.
In the following sections, we discuss
the fractal nature of the sub-planckian successful regions for the
original hybrid model and define a probability measure over the full
parameter space. In Sec.~\ref{section:mcmc}, the MCMC method is
introduced and we study step by step the effect of the initial field
velocities and the potential parameters on the probability of
obtaining more than $60$ e-folds of inflation. We then present the full
posterior probability distributions of these parameters for the original hybrid
scenario. In Sec.~\ref{section:fsugrab}, we perform the same analysis for 
the F-term SUGRA hybrid potential. Some conclusions and perspectives are 
finally presented in the last section.

\section{Exact two-field dynamics and initial conditions} \label{sec:ICoriginal}


In this section, we integrate numerically the exact multi-field dynamics, given by Eqs.~(\ref{eq:FLtc12field}) and (\ref{Eq:KGpert_multifield}), for the 
original hybrid model, characterized by the 2-field potential of Eq.~(\ref{eq:potenhyb2dNEW}).  
We explore the space of initial conditions and extend previous
studies to super-planckian initial values. 
We quantify the
amount of fine-tuning of the model by computing the ratio of
successful/unsuccessful area and study the effect of varying the
parameters of the potential on our results.

\subsection{Classical dynamics and stochastic effects}
Considering large values for the fields can induce stochastic
(quantum) effects which affect the field dynamics.  In this
chapter, the evolution remains purely classical~\cite{Linde:1993xx,Martin:2005ir}.
Since we also consider super-planckian field values, it is important to
check that for such values, the dynamics is still dominated by the
classical motion.   It is valid if the classical evolution dominates over the quantum fluctuation scale,
\begin{equation}
\Delta \sigma_{\rr{cl}} = \frac{\dot \sigma}{H} > \Delta \sigma_{\rr{qu}} \simeq \frac{H}{2 \pi}~,
\end{equation}
where $\sigma$ is the adiabatic field.
Because the potential can be rescaled without affecting the 2-field dynamics, the energy scale of inflation in this chapter can be adjusted such that the COBE normalization is satisfied.  

When slow-roll is realized along the valley, stochastic effects were found to not affect the classical dynamics~\cite{Martin:2005ir}, so that it can be safely considered.  Since we are not yet interested in the field dynamics during the waterfall phase, the stochastic effects during the waterfall are not relevant in this chapter. 



\subsection{Exploration of the space of initial conditions}
Let us now study the space of initial values [i.e. the $(\phi_{\rr
i},\psi_{\rr i})$ plane] of the fields that leads to successful
inflation. For simplicity, we have first assumed initial velocities to
be vanishing $\dot{\phi}_\mathrm{i}=\dot{\psi}_\mathrm{i}=0$ as
their effect can always be mimicked by starting in a different
point with vanishing velocities. Then for each initial conditions,
we have integrated the equations of motion and computed the field
values and the number of e-folds as a function of time. Choosing
to end simulations when inflation is violated would have not
allowed us to study trajectories where inflation is transiently
interrupted as it may happen (see Sec~\ref{sec:1fielddynamics}).
Therefore, we chose to end the numerical integration when the
trajectory is trapped by one of the two global minima,
because at that point, no more e-folds will be produced. This is
realized when the sum of the kinetic and potential energy of the
fields is equal to the height of the potential barrier between the
vacua, i.e. when
\begin{equation}
\Lambda^4=\frac{1}{2}\left(\dot\phi^2+\dot\psi^2\right)
+V(\phi,\psi).
\end{equation}

Throughout this chapter we will define a successful initial condition (i.c.) as a point
in field space that lead to a sufficiently long phase of inflation to
solve the horizon and flatness problem. We will assume that
$N=\ln(a/a_\uini)\simeq 60$ e-folds is the critical value required,
though this value can change by a factor of two depending on the
reheating temperature and the Hubble parameter at the end of
inflation~\cite{Liddle:2003as, Ringeval:2007am}. However, generically,
once inflation starts it lasts for much more than $60$ e-folds and our
results are not sensitive to the peculiar value chosen.


Let us mention that our aim here is not to provide the best fit to
the cosmological data but to explore the space of initial
conditions that lead to sufficient inflation within the hybrid
class of models. However, notice that the COBE normalisation can
always be achieved by a re-scaling of the potential without
affecting the inflaton dynamics.

\begin{figure} \begin{center}
\scalebox{.30}{\includegraphics{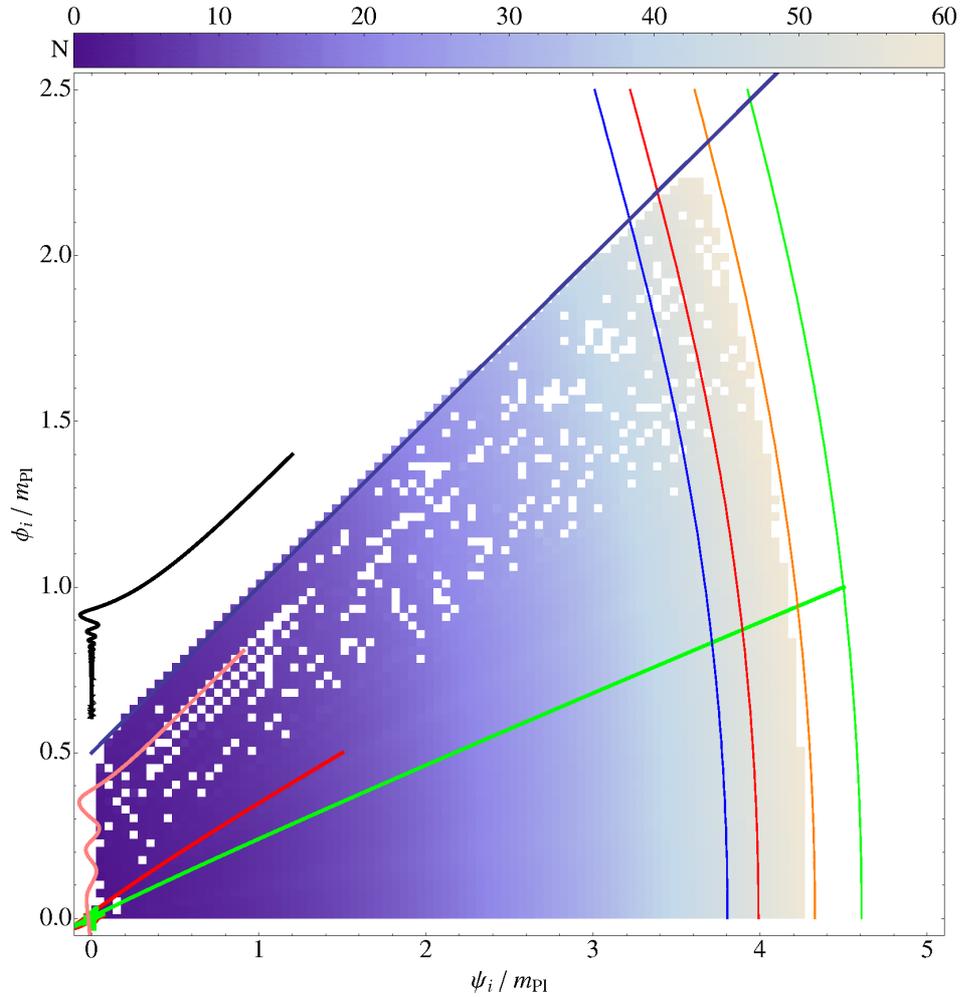}}
\caption{Grid of initial conditions leading to successful (white
regions) and unsuccessful inflation (colored region), for the
original hybrid inflation with $M=\phi_{\rr c}=0.03 \mpl$ and $\mu = 636 \mpl$.  The color code denotes the number of
e-folds realized. Three typical successful trajectories [in the
valley (black), radial (green), and from an isolated point (pink)] are added
as well as an unsuccessful trajectory (red).
Also plotted are the
iso-curves of $\epsone$, in the slow-roll approximation, for
$\epsone = 0.022$, $0.02$, $0.0167$ and $0.015$ (from left to
right), and the limit of the unsuccessful region (blue oblique line), obtained with Eq.~(\ref{eq:pentelimit})} \label{fig:completegrid-hybrid}
\end{center}
\end{figure}

\begin{figure}[h!]
\begin{center}
\scalebox{.27}{\includegraphics{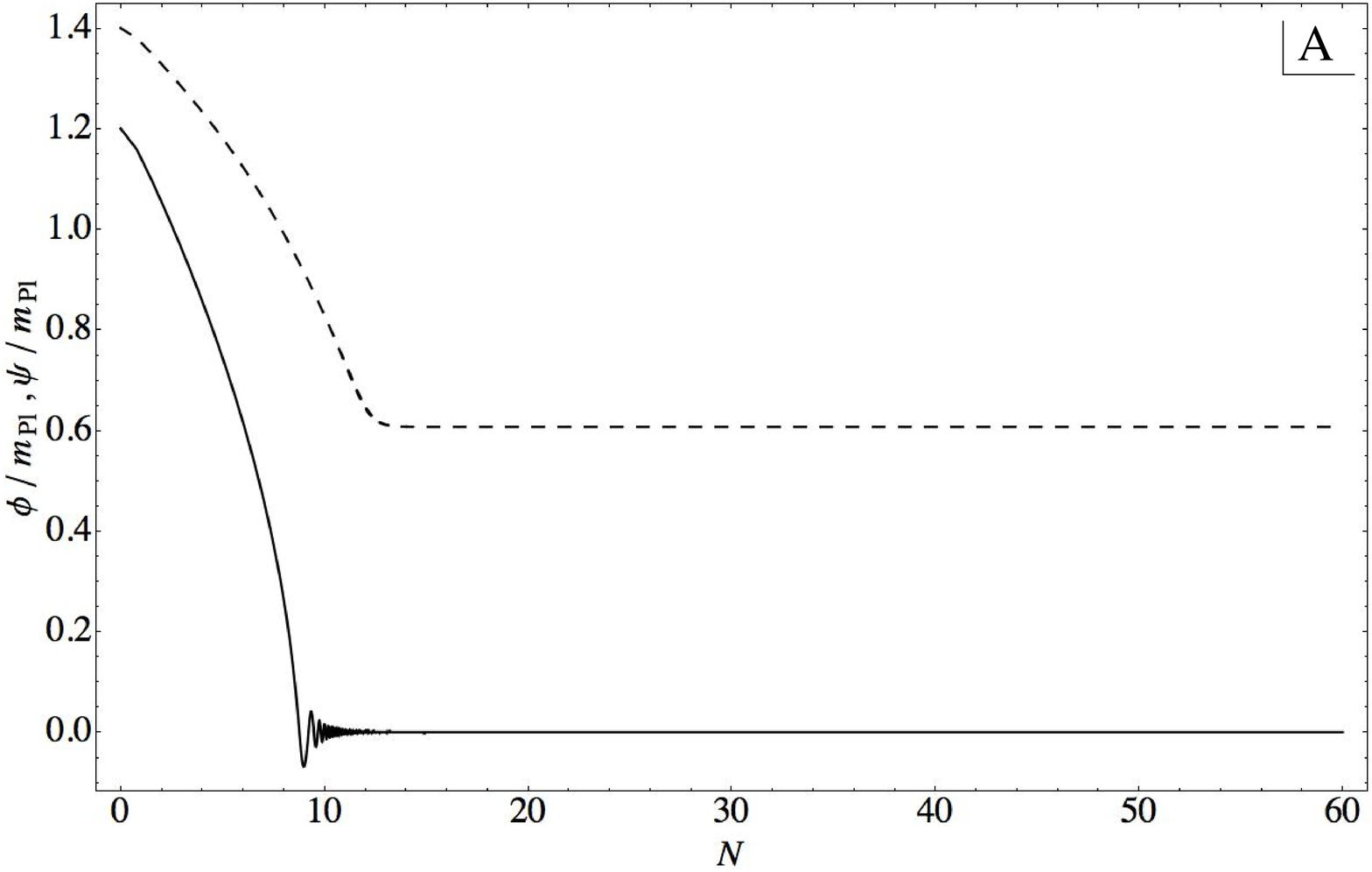}}

\scalebox{.27}{\includegraphics{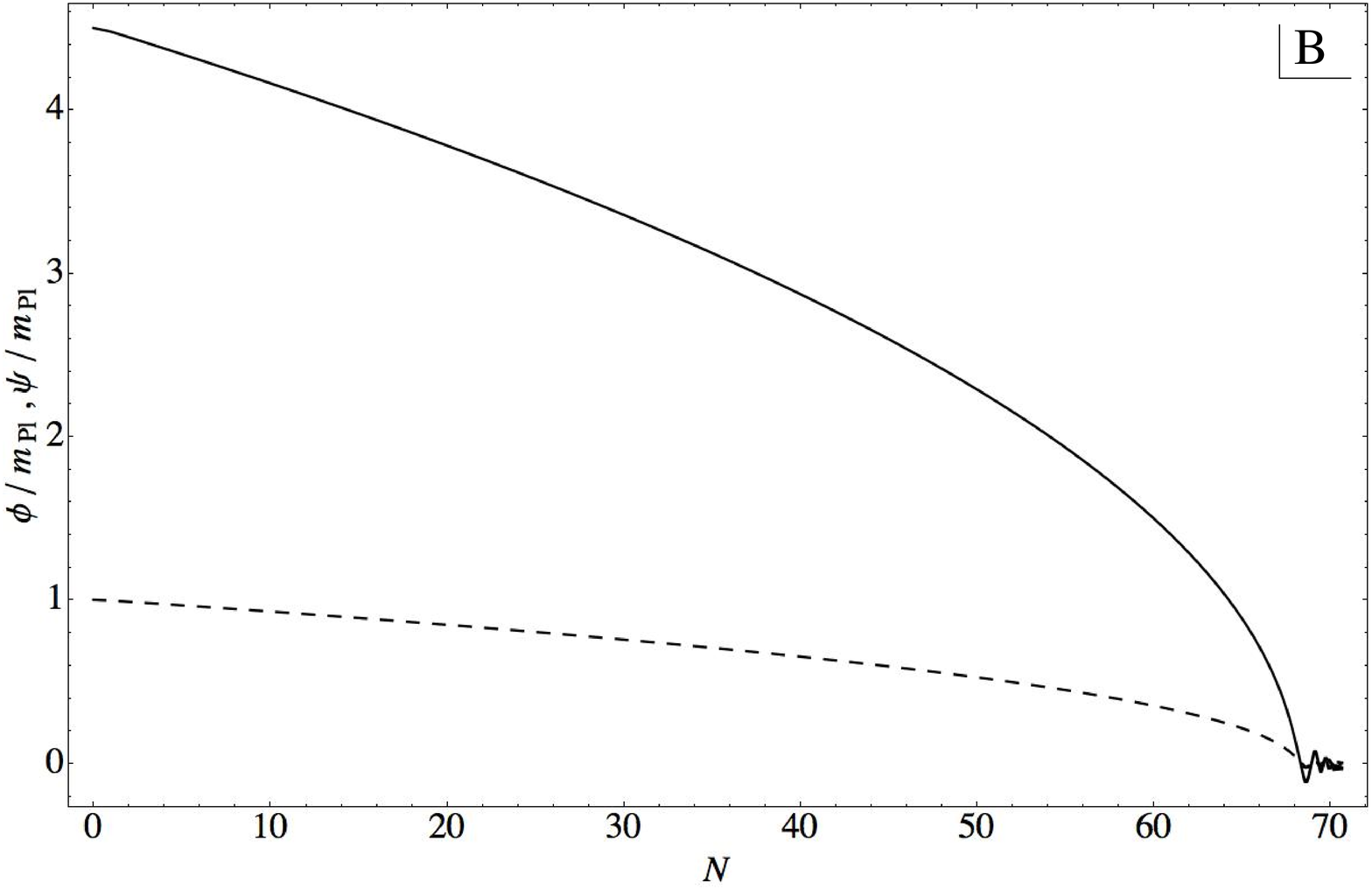}}

\scalebox{.27}{\includegraphics{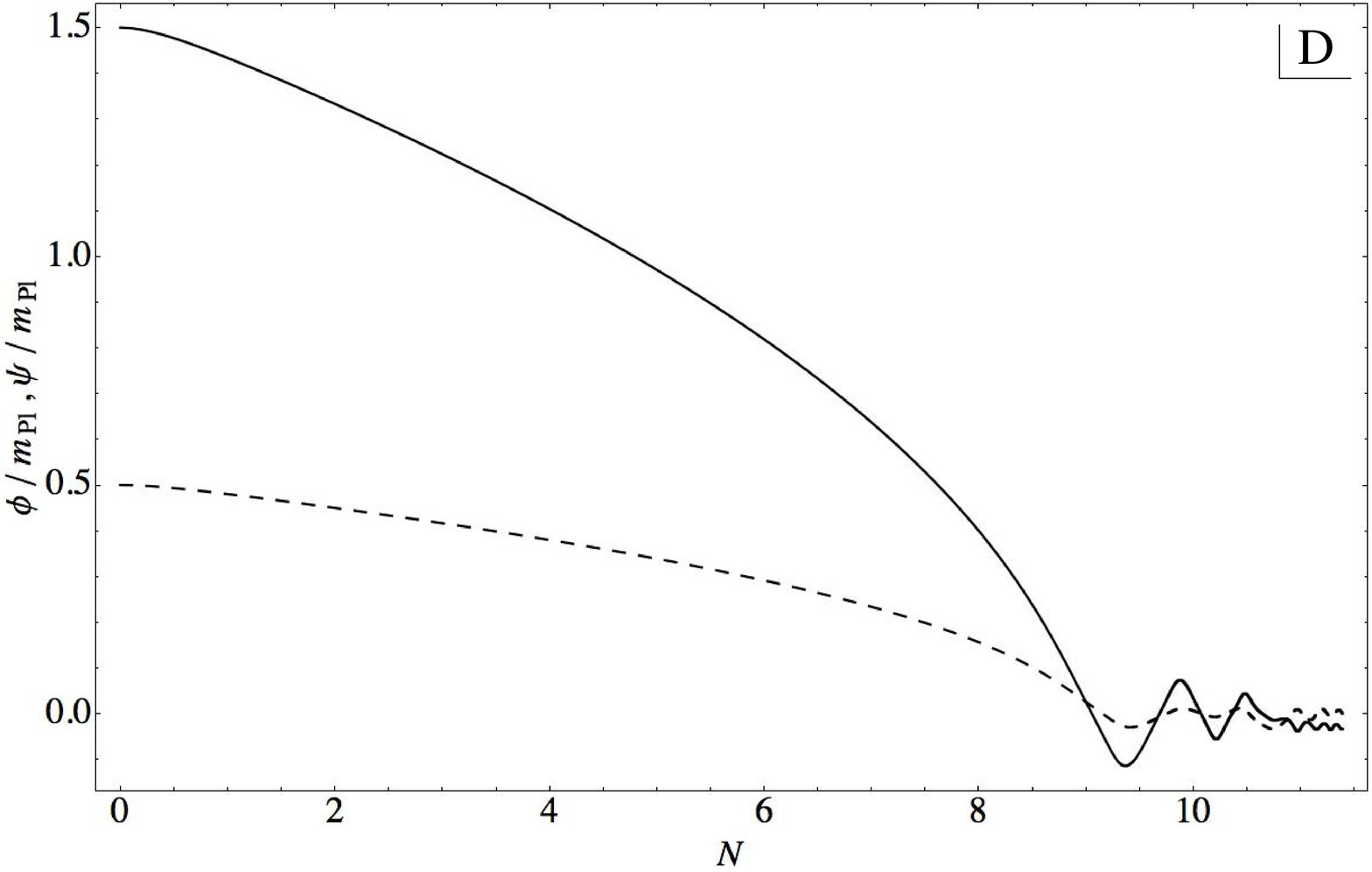}}
\caption{Evolution of the fields $\phi$ (dashed lines) and $\psi$
(plain lines) with the number of e-folds realized, for the
trajectories A, B, and D (from top to bottom) as represented in
Fig.~\ref{fig:completegrid-hybrid}. The more interesting type-C
trajectory is represented in Fig.~\ref{fig:TypeCtrajectory}.} \label{fig:efolds-hybrid}
\end{center}
\end{figure}

\begin{figure}[h] \begin{center}
\scalebox{.3}{\includegraphics{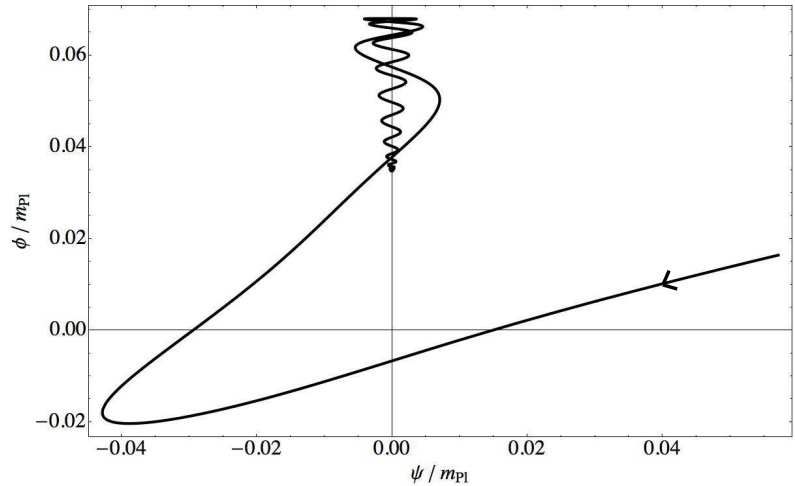}}
\caption{More detailed description of the field values during a
type-C trajectory as defined in
Fig.~\ref{fig:completegrid-hybrid}. This is a zoom of the
trajectory close to the bottom of the potential. One can notice
that the system quickly rolls down while few e-folds are produced
before ``accidently'' climbing up the valley. Then it starts a
second efficient phase of inflation like a type-A trajectory.}
\label{fig:TypeCtrajectory}
\end{center}
\end{figure}

In Fig.~\ref{fig:completegrid-hybrid} the grid of initial values
is presented for the original hybrid inflation model of
Eq.~(\ref{eq:potenhyb2dNEW}). For values of parameters comparable to
those used in~\cite{Tetradis:1997kp} and \cite{Mendes:2000sq}, we
have put in evidence three types of trajectories in the field
space to obtain successful inflation. An example of each has been
represented in Fig.~\ref{fig:completegrid-hybrid} and identified
by a letter A, B, or C whereas an example of a failed trajectory
is identified by a D.
The details of these trajectories are represented in
Fig.~\ref{fig:efolds-hybrid} where the values of the fields for
three trajectories are plotted as a function of the number of
e-folds. A more detailed description of the more interesting type-C
trajectory is represented separately in
Fig.~\ref{fig:TypeCtrajectory}. Each trajectory is described and
explained below.\\

\paragraph{Trajectory A: along the valley}
This region of successful inflation corresponds to a narrow band
along the $\psi = 0$ line and is the standard evolution.
Trajectories are characterized first by damped oscillations around
the inflationary valley which do not produce a significant
number of e-folds. However once the oscillations are damped, the
evolution is identical to the one for the effective one-field
potential and inflation becomes extremely efficient in terms of
e-folds created. This explains the abrupt transition between the
unsuccessful and the successful type-A regions observed in
Fig.~\ref{fig:completegrid-hybrid}. Indeed, unsuccessful points
with around 10 e-folds created can be found right next to the
white successful region where $N \gg 60$. The difference between
two close points in each region is that for the successful one,
the system just has the right amount of time for the oscillations
to become damped before entering the global minimum where
inflation ends.

For larger initial values of the $\phi$ field (around and above
the Planck mass), the narrow band of successful inflation opens up
and inflation is always successful (in agreement
with~\cite{Tetradis:1997kp,Lazarides:1997vv}. In this region (at
the top of Fig.~\ref{fig:completegrid-hybrid}), it is always
possible for the oscillations to become damped and for the
efficient regime of inflation to start before the end of
inflation: \emph{the fine-tuning on the initial conditions
disappears at large values of $\phi$ for any values of $\psi$}.

By comparing the time necessary for the expansion to damp the
oscillations and the time taken by the inflaton to reach the
critical point of instability, an analytical approximation of the
width $\psi_{\rr w}$ of the narrow successful band has been
proposed in~\cite{Tetradis:1997kp},
\begin{equation}\label{eq:largeurvallee}
\psi_{\rr w} \simeq \sqrt{\frac{3 \pi^3 \phi_{\rr c}}{4}} \frac{M}{\mpl}~.
\end{equation}
For the parameter values of the
Fig.~\ref{fig:completegrid-hybrid}, $\psi_{\rr w} \sim 4\times 10^{-3} \mpl$.
This provides a good fit of the width of the inflationary valley
at small $\phi \ll \mpl$. This successful band is so thin that
quantum fluctuations would have an amplitude large enough to shift
the field $\psi$ outside the successful
band~\cite{Tetradis:1997kp}.  For larger initial values of $\phi$,
it is also possible to provide an analytical fit of the limit
successful/unusuccessful. Fig.~\ref{fig:completegrid-hybrid}
suggests that the limit $\phi_{\rm lim}(\psi)$ is a linear
function.  From a given set of initial conditions $(\phi_\ui,
\psi_\ui)$, the total number of e-folds generated depends almost
only on the value $\phi=\phi_\uhit$ at which the oscillations in
$\psi$ become damped and the slow-roll starts in the valley. The
reason is that a type-A trajectory rolls faster before $\phi_\uhit$
and thus doesn't generate many e-folds before the valley. As a
consequence, the limit between successful and unsuccessful regions
necessarily follows the unique trajectory for which $\phi_\uhit$
becomes large enough to generate exactly $60$ e-folds by slow-roll
in the valley. As a result, using the slow-roll approximation, the
slope of the limit is simply given by the gradient of the
potential.  Above the instability point $\phi > \phi_{\rr c} $, for large values of the parameter $\mu$, it is well approximated by
\begin{equation}\label{eq:pentelimit}
\alpha = \frac{\partial V(\phi,\psi)/\partial \phi} {\partial
V(\phi,\psi)/\partial \psi}\simeq \dfrac{\psi \phi}{\phi_{\rr c}^2 \left( \dfrac{\psi^2}{M^2} + \dfrac{\phi^2}{\phi_{\rr c}^2 }  \right)}~.
\end{equation}
 Given one point of the
transition line, for example $(1. \mpl,1. \mpl)$, we can check that the slope
of the limit is $\alpha \simeq 0.5$ for the parameters of
Fig.~\ref{fig:completegrid-hybrid}.\\

\paragraph{Trajectory B: radial}
Enlarging the space of initial conditions to super-planckian values
shows another region where successful inflation is automatic. It is
observed for super-planckian initial values of the auxiliary field
$\psi$ beyond a few Planck mass,  in a way reminiscent to the double inflation scenario~\cite{PhysRevD.35.419}.
 In this case, the trajectory is radial and the $60$
e-folds are realized mostly before reaching the valley or the global
minima.  

From the $\phi$-axis to larger values of $\psi_{\rr i}$, the number of
e-folds realized increases slowly (see
Fig.~\ref{fig:completegrid-hybrid}). Therefore, this limit between
the two regions is smooth unlike the limit with A-type
trajectories described in the previous paragraph. Increasing
$\phi_{\rr i}$, the critical value of $\psi_{\rr i}$ leading to enough
inflation decreases slowly, because inflation is radial and the
trajectory longer. To describe this limit more precisely, we have
plotted the iso-curves of $\epsone$ in
Fig.~\ref{fig:completegrid-hybrid} in the two-fields slow-roll
approximation. We can see that this limit follows one of these
iso-curves, namely $\epsone\simeq 0.0167$. This observation can be
understood using a kinematic analogy~\cite{Ringeval:2007am} as
long as $\epstwo$ is negligible. This critical value of $\epsone$
can be computed analytically, by studying the easiest trajectory
of this kind at $\phi_{ \rr i} = 0 $. In this case, the effective
potential is dominated by $\Lambda^4 \psi^4 / M^4  $, and the critical
$\psi_{\rr i}$ is obtained by requiring a phase of inflation of exactly
$N_{\rr{suc}}=60$ e-folds. We find
\begin{equation}
\psi_{\rr {ic}} = \sqrt{\frac {\mpl ^2}{\pi} \,N_{\rr{suc}} }
\approx 4.37 \mpl.
\end{equation}
At this value, the corresponding first Hubble-flow parameter
$\epsilon_{\rr {1c}}$ reads
\begin{equation}
{\epsone}_{\rr c} \simeq \frac {1}{N_{\rr{suc}}} \approx 0.0167.
\end{equation}

\paragraph{Trajectory C and D: isolated successful points and
unsuccessful points.}
Previous works~\cite{Tetradis:1997kp,Mendes:2000sq} pointed out
the presence of unexplained successful isolated points in the
central unsuccessful region. In this paragraph, we justify their
existence, study their properties and quantify the area they
occupy.

Let us first describe the D-type trajectories that are
unsuccessful. As shown in Fig.~\ref{fig:efolds-hybrid}, in these
cases, the system quickly rolls down the potential to one of the
global minima of the potential during which only a few e-folds are
created. What is then the difference between the D-type and the
C-type trajectories  plotted in Fig.~\ref{fig:TypeCtrajectory}?
The fields roll towards the bottom of the potential with
sufficient kinetic energy and, after some oscillations close to
the bottom of the potential, the momentum is ``by chance''
oriented toward the inflationary valley. Thus the system goes up
the valley until it looses its kinetic energy and then starts
slow-rolling back down the same valley producing inflation with a
large number of e-folds.
Note that there are more of these points in a band under the
limit of type-A trajectories. This is because, at higher
$\phi_\ui$, there are more chances to find a trajectory where the
momentum at the bottom of the potential is oriented toward the
inflationary valley.

\begin{figure}[] \begin{center}
  \includegraphics[width=12cm]{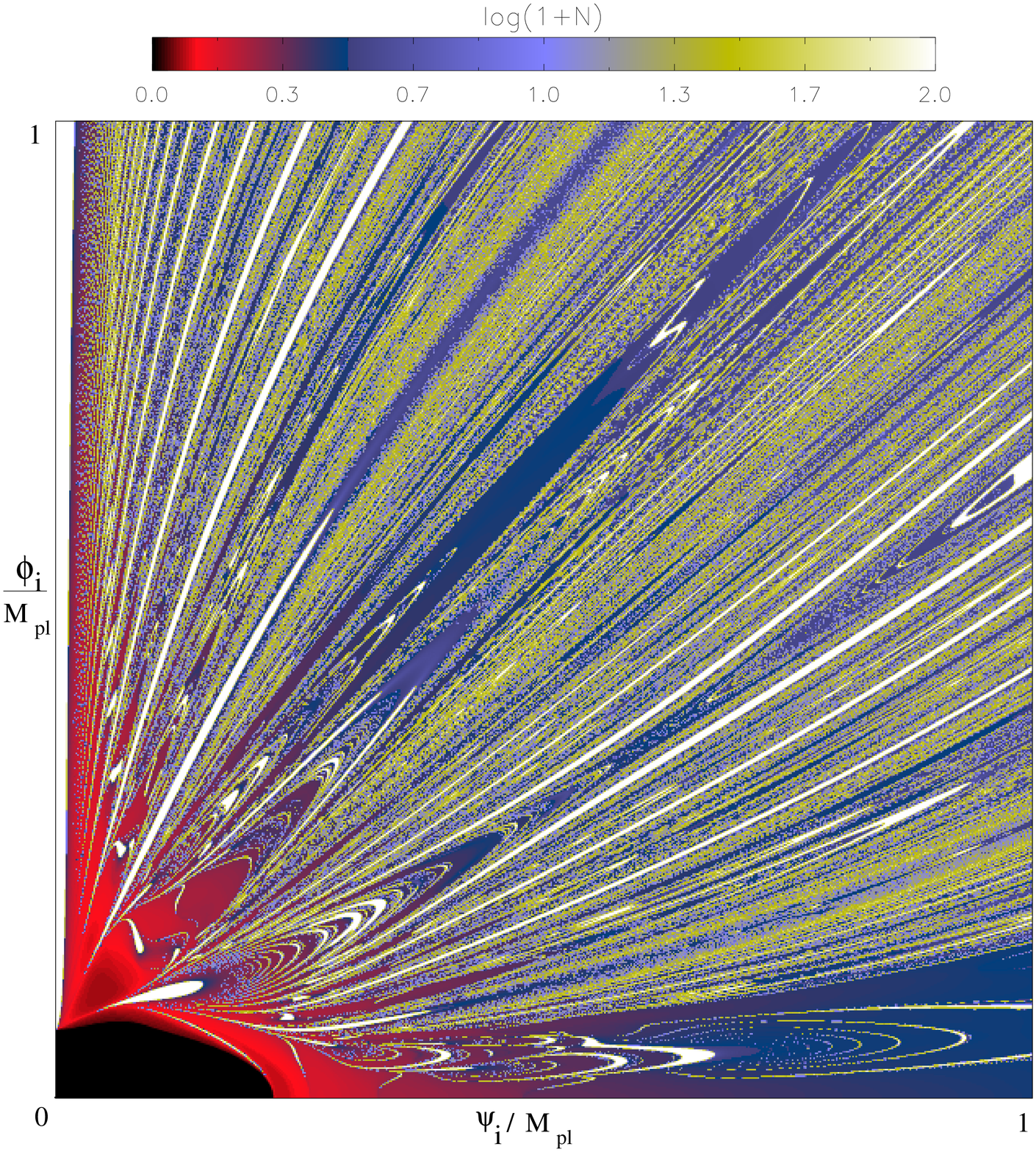}
  \caption{Mean number of e-folds obtained from $512^2$ initial
    field values in the plane $(\psi_\ui/\Mpl,\phi_\ui/\Mpl)$. This
    figure has been obtained by averaging the number of e-folds
    (truncated at $100$) produced by $2048^2$ trajectories down to
    $512^2$ pixels. The potential parameters have been set to
    $M= 0.03 \ \mpl$, $\phi_{\rr c} = 0.014 \ \mpl $, $\mu = 636 \  \mpl$~\cite{Clesse:2009ur}.} 
       \label{fig:grid}
\end{center}
\end{figure}

\begin{figure} \begin{center}
\scalebox{.30}{\includegraphics{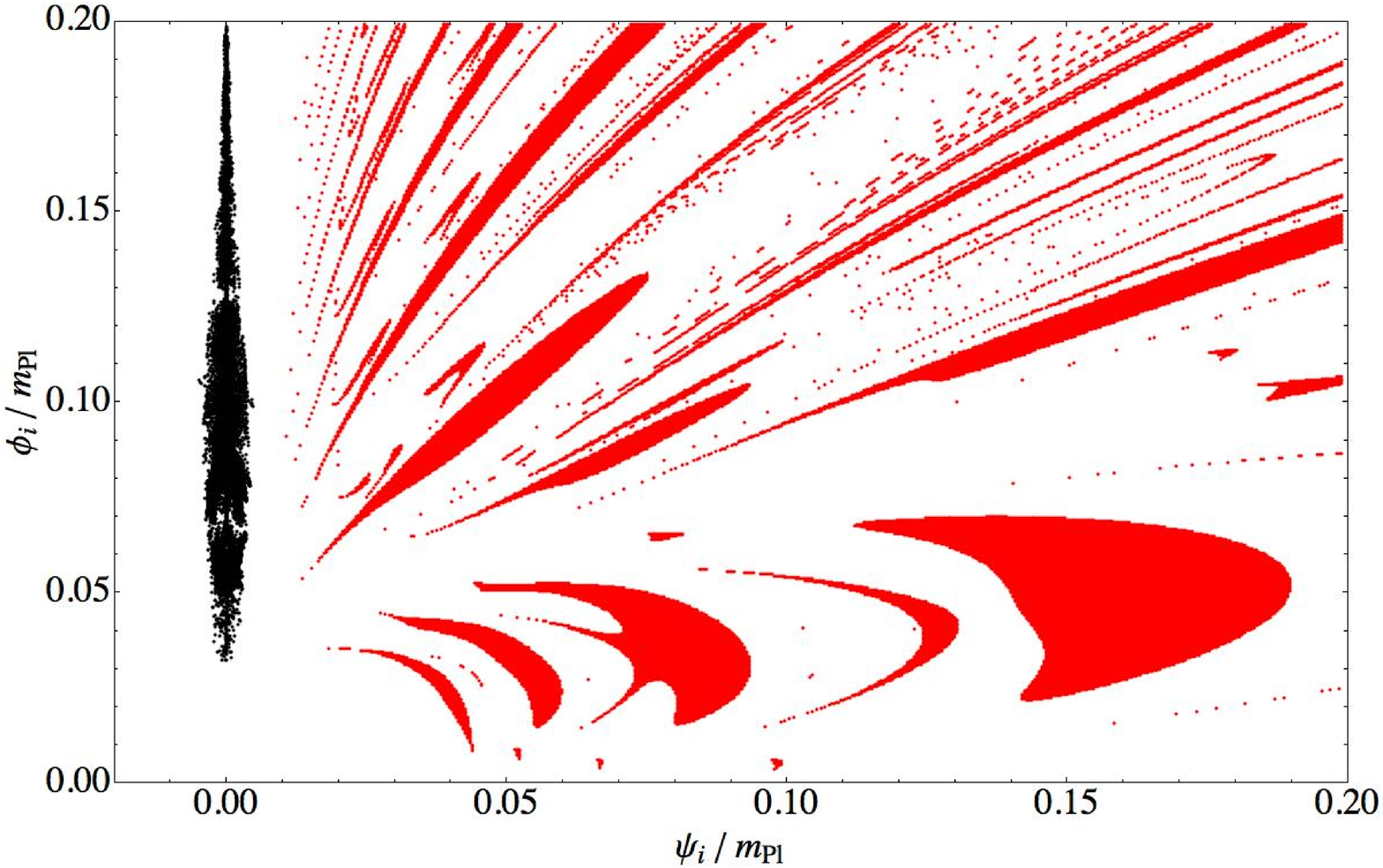}}
\caption{Structure of the successful initial field values exterior to the valley
(in red), 
together with their images (in black), defined as the points
of vanishing velocity on the field trajectories.  These image points populate the narrow successful band along the valley.   The structure in red can be seen as the ``anamorphosis'' of the inflationary valley. In this analogy the
field trajectories on the potential correspond to the trajectories of the light on the optic device.  The apparently senseless red patterns create a meaningful image along the inflationary valley. 
This is obtained for
$M=\phi_{\rr c} = 0.03 \mpl$ and $\mu= 636 \mpl$.}\label{fig:anamorphosis} \end{center}
\end{figure}

High resolution grids and zooms on peculiar regions of
Fig.~\ref{fig:completegrid-hybrid} show that these apparently
random isolated points form actually a complex structure. Some of
it, for small initial conditions, is visible in
Figs.~\ref{fig:anamorphosis} and~\ref{fig:grid}.

The points are organized in long thin lines, or crescents. The
points that seem isolated actually belong to structures that a
better resolution would show continuous. Some of our biggest
structures can be identified also in~\cite{Tetradis:1997kp} but
are not recovered in~\cite{Mendes:2000sq} where only isolated
points were found. This may be explained by the need of a higher
resolution to resolve the structures. A detailed analysis of
trajectories in a continuous successful patch shows that they all cross the axis $\phi = 0 $ the same number of times, before climbing up and going back down the
inflationary valley along the $\psi=0$ direction.

For each of these type-C trajectories, we can identify the point
(that we will call the ``image'') on the inflationary valley at
which the velocities of the fields become (quasi)null. We show the
robustness of the previous description of the type C trajectories,
by observing that all these images are in the successful narrow
band responsible for the type-A trajectories. More precisely, the
images obtained populate \emph{exactly} the narrow successful band. 
The
identification between the successful points exterior to the valley and their images along the
inflationary valley is represented in Fig.~\ref{fig:anamorphosis}, for positive initial field values. Using
the analogy with optical \emph{anamorphosis}, we can say that the
observed structures of type-C initial conditions is the anamorphosis, that is the deformed image, of the successful narrow band around the inflationary
valley. In this analogy, the potential plays the role of the
optical instrument used to create the meaningful image. The
trajectories of the light rays on the optic device are then
replaced by the field trajectories to create a meaningful image
(in the valley) from the apparently senseless patterns of
successful initial conditions.

Let us elaborate a little more on the properties of the images in
Fig.~\ref{fig:anamorphosis}. Since the potential is invariant
under $\phi\rightarrow -\phi$, there exist two
inflationary valleys, one going toward $\phi>0$ and one going
$\phi<0$. Some of these type-C initial conditions give rise to
inflation thanks to the first valley when the others will realize
inflation in the second. Obviously the two situations are
equivalent and symmetric. 


\subsection{Dependencies on the parameters}
The grid of initial conditions, and therefore the proportion of
successful points in a given range of initial values naturally
depend on the values of the parameters of the potential. 

\begin{figure}[h!]  \begin{center}
\scalebox{.25}{\includegraphics{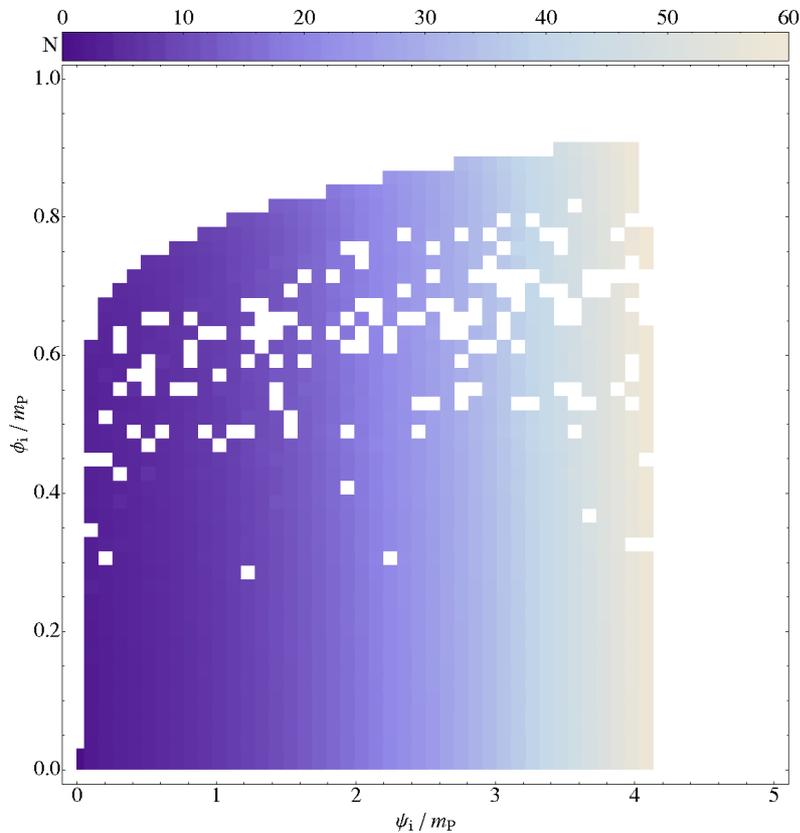}}
\caption{Grid of initial conditions, for hybrid potential with
$M=0.01 \ \mpl$, $\phi_{\rr c} = 0.03 \  \mpl$, $\mu = 636 \ \mpl$.}
\label{fig:hybridVaryLambda'}  \end{center}
\end{figure}

\paragraph{Evolution of the limit of A-type trajectories}
At small $\phi$,  reducing the parameters $\phi_{\rr c}$ and $M$ induces a narrower
inflationary valley.   At large $\phi$, the slope $\alpha$ in
Eq.~(\ref{eq:pentelimit}), is also mostly a function of  $\phi_{\rr c}$ and $M$.   
By reducing the ratio $M / \phi_{\rr c} $, this slope decreases, as seen in Fig.~\ref{fig:hybridVaryLambda'}. This effect is
due to the potential now dominated by the $\psi^4 / M^4 $ term and less 
dependent  on $\phi$. Thus the velocity in the $\psi$
direction is enhanced compared to the $\phi$ one.\\

\begin{figure} \begin{center}
\scalebox{.25}{\includegraphics{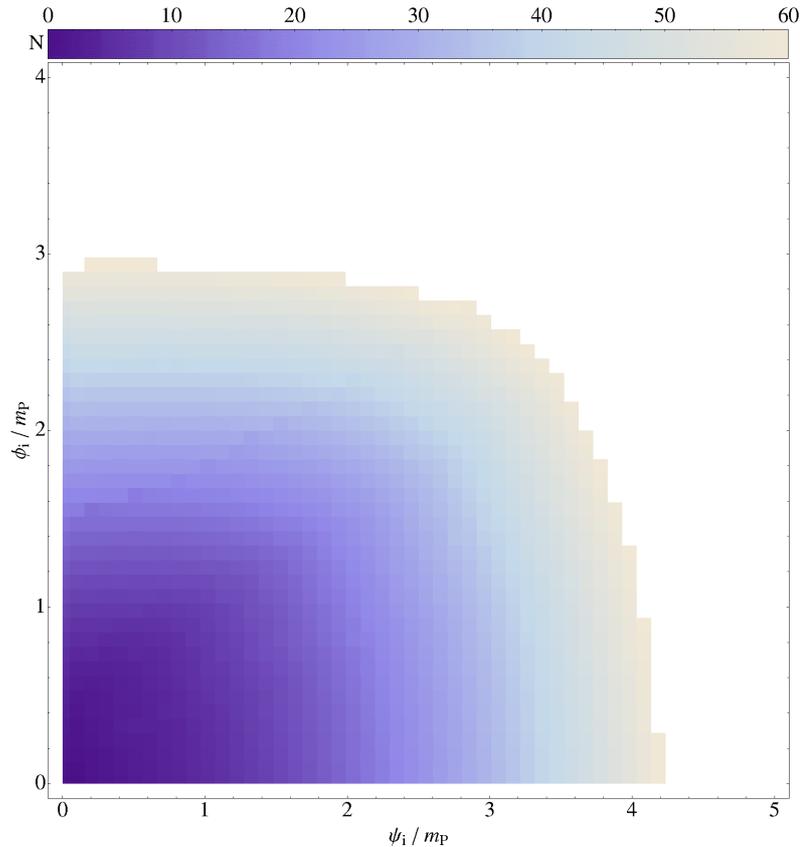}}
\caption{Grid of initial conditions for
the original hybrid model, with $\mu = 0.1 \ \mpl$, $M = \phi_{\rr c} = 0.03 \ \mpl$.  The small value of $\mu$ induces slow-roll violations along the valley, preventing the phase of inflation at small field values to take place (see Chapter 4).  As a consequence, the model require super-planckian initial field values.}
\label{fig:hybrid_redspectrum} \end{center}
\end{figure}

As long as $1 /\mu ^2$ is subdominant compared to
$  \psi^2/ (M^2 \phi_{\rr c}^2  )$, its variation does not affect the properties of the
initial condition plane.  But if $\mu $ is sufficiently reduced, the velocity in the $\phi$ direction increases and tends to spoil
the slow-roll evolution in the inflationary valley. As already
described in the section~\ref{sec:1fielddynamics}, this violation
of the slow-roll conditions in the valley imposes for inflation to
occur in the large field phase. In the space of initial
conditions, the narrow successful band then disappears together
with the type-C trajectories. Finally the unsuccessful region
takes an elliptic form as represented in
Fig.~\ref{fig:hybrid_redspectrum}, with a smooth transition
between successful and unsuccessful regions. The model becomes
comparable to double inflation~\cite{PhysRevD.35.419} and we
recover the feature of this model: it is almost unavoidable to
have super-planckian initial values of the fields to realize a
sufficiently long inflation.

\paragraph{Evolution of the amount of C-type trajectories}
A similar explanation can be given to justify the reduction of
the amount of successful points outside the valley, when the ratio  $ M / \phi_{\rr c}  $ is reduced (see
Fig.~\ref{fig:hybridVaryLambda'}). The potential is dominated by the $
\psi^4 / M^4 $ term and the $\phi$-component of the velocity becomes
small. Thus the chances for the system to climb up the valley are
suppressed.   This result is illustrated in Tab.~\ref{tab:anamorphhyborigin}
below.\\

\paragraph{Quantification of successful initial conditions}
We end this section by quantifying what proportion of the initial
condition space gives rise to inflation for hybrid inflation, for
some sets of parameters and for field values smaller than the reduced Planck mass.  Our results are represented in
Tab.~\ref{tab:anamorphhyborigin}.
\begin{table*}
\begin{center}
\begin{tabular}{|c|c|c|c|}
\hline Model & Parameters & Succ. points (\%) &
Out of the valley (\%)  \\\hline\hline
Hybrid & $M=\phi_{\rr c} = 0.03 \mpl$, $\mu = 636 \mpl$ & 17.4  & 14.8  \\
Hybrid & $M = 0.01 \mpl$, $\phi_{\rr c} = 0.03 \mpl$, $\mu  = 636 \mpl$ & 4.4 & 2.7 \\
Hybrid & $M = 0.03 \mpl$, $\phi_{\rr c} = 0.01 \mpl$, $\mu  = 636 \mpl$ & 15.6 & 13.6 \\

Hybrid & $M = 0.01 \mpl$, $\phi_{\rr c} = 0.01 \mpl$, $\mu  = 636 \mpl$ & 15.9 & 14.0 \\

Hybrid & $M = 0.03 \mpl$, $\phi_{\rr c} = 0.03 \mpl$, $\mu  = 0.1 \mpl$ & 0 & 0 
 \\ \hline
\end{tabular}
\caption{Percentage of successful points in grids of initial
conditions, for different values of parameters, when restricting to
$\phi_\ui,\psi_\ui\leq \Mpl$. The third column represents the area
of the whole successful initial condition parameter space over the
total surface. The fourth column represents the surface of the
successful space only located outside the valley, over the total
surface. This allows to visualize the importance of the anamorphosis regions. }  \label{tab:anamorphhyborigin}
\end{center}
\end{table*}
From this table, we can see that unless the ratio $M / \phi_{\rr c}$ is reduced, the hybrid model possesses
about $15\%$ of initial conditions that leads to successful
inflation. For this percentage to be translated into a probability
of realizing inflation, one would need a measure in the
probability space. The problem will be considered later in the chapter and the measure will be
shown to be flat, such that for the considered specific sets of parameters,  \emph{the
successful initial conditions should not be considered as
fine-tuned but simply sub-dominant when fields are restricted to
sub-planckian values}.   However, a full analysis of the parameter space is required to extend this result.  This will be done in Sec.~\ref{sec:mcmc_IC} by performing a bayesian MCMC analysis of the parameter space, including non-vanishing initial velocities. 

For field values sufficiently larger than the Planck mass, more than 60 e-folds of inflation become generic (see Fig.~\ref{fig:completegrid-hybrid}).   With the requirement $\phi_{\rr i}, \psi_{\rr i} \leq 5 \mpl$,
we found that the percentage of successful initial conditions
raise to $72\%$ for the parameter values of
Fig.~\ref{fig:completegrid-hybrid}.

\section{Initial conditions for extended models of hybrid inflation}\label{sec:othermodels}
In this section, we study the properties of initial
conditions leading to successful inflation for three hybrid-type
models of inflation and study how generic the properties observed
for the original model are. The models are the ``smooth'', and
``shifted'' hybrid inflation both in global SUSY and SUGRA, and
radion inflation.  These models are described in chapter~3.

\subsection{Space of initial conditions for Smooth Inflation}

In a previous study by Lazarides et al.~\cite{Lazarides:1996rk},
an exploration of the space of initial conditions leading to
sufficient inflation was performed, with a low resolution, for smooth inflation.  This
exploration led to a conclusion opposite to the one found for the
non-supersymmetric hybrid inflation model in Ref.~\cite{Mendes:2000sq}: most of the space
was found to be successful. Therefore, smooth hybrid inflation seems
a good laboratory to test the validity of the results we found in
the previous section. We performed the exploration of the space of
initial conditions, for a higher resolution, and for a larger
range of initial field values and parameter values. Imposing
$\phi_i,\psi_i \leq \Mpl$, we computed the proportion of successful
initial conditions and the proportion of isolated successful points
away from the inflationary valleys.

\begin{figure}[h!] \begin{center}
\scalebox{.25}{\includegraphics{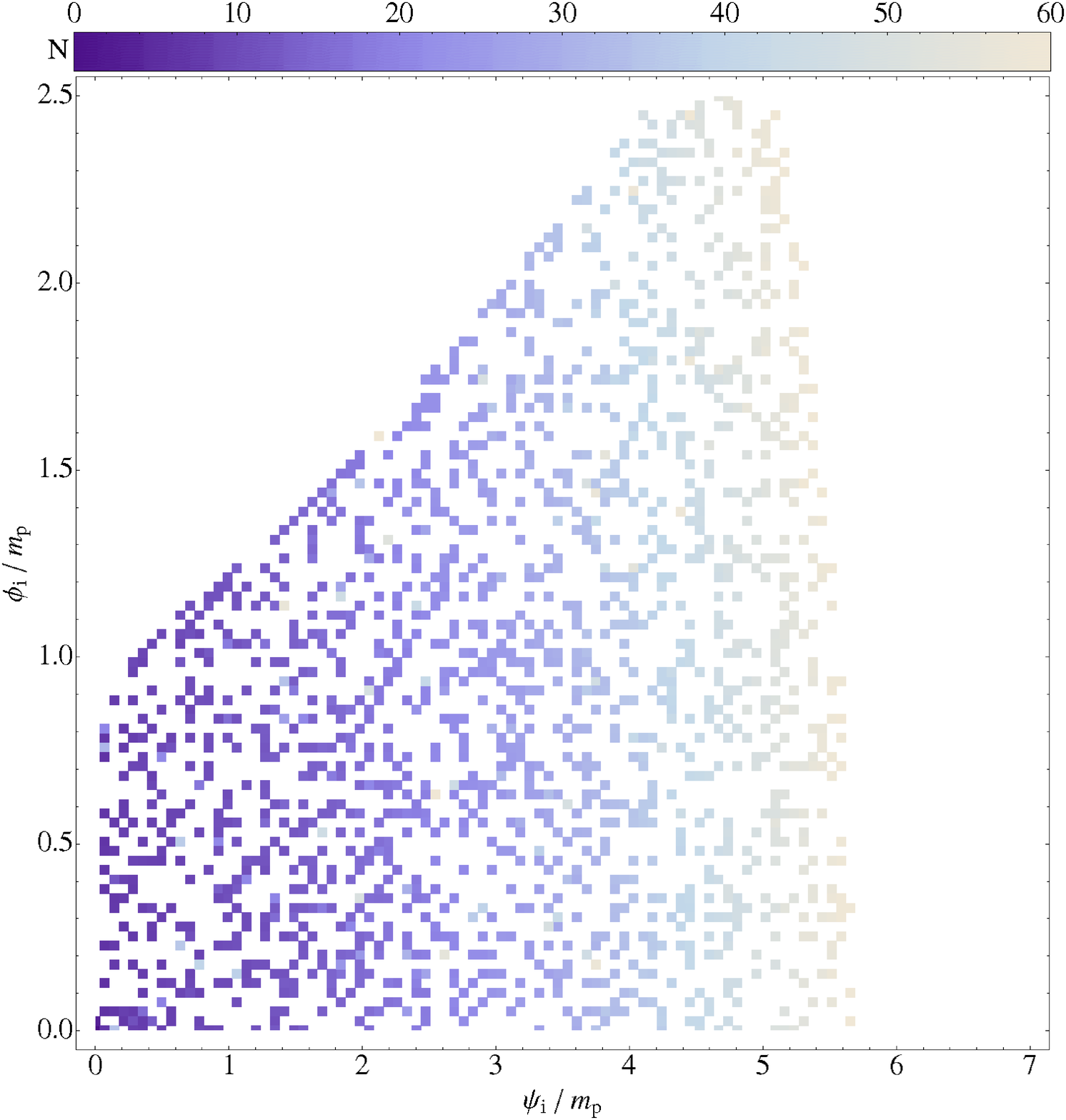}}
\caption{Grid of initial conditions for smooth inflation, using
the values of the parameters of \cite{Lazarides:1996rk}: $\kappa
\simeq 10$, $M\simeq 2.3\times 10^{-5} \mpl$.  A large proportion of successful initial conditions is observed in the unsuccessful region exterior to the valley.  A much higher resolution would reveal that these are organized in a complex connected pattern, as for the original hybrid inflation model.}
\label{fig:smoothVaryAllLazarides} \end{center}
\end{figure}

This analysis is also extended to super-planckian values of the fields.  It
always reveals a structure similar to that of the original model.
We observe (see for e.g. Fig.~\ref{fig:smoothVaryAllLazarides})
a narrow band of
fine-tuned successful initial conditions along $\psi=0$, a
triangular unsuccessful region, and successful areas for large
initial values of one or both of the fields. Anamorphosis is also
present, leading to successful patterns in the
unsuccessful region.  There again, these successful initial conditions are observed to be connected when zooming over particular regions of the field space.  For the values of the parameters quoted
in Ref.~\cite{Lazarides:1996rk}, that is with a mass scale of
order $10^{-5} \mpl$, they occupy most of the space of
initial condition as shown on Fig.~\ref{fig:smoothVaryAllLazarides}.
We find almost $80\%$ of initial conditions below the reduced
Planck mass to be successful.

We have also studied how this grid evolves with the parameters of
the potential. We first observe that the amount of successful
initial conditions is independent of the coupling constant
$\kappa$ (it only scales the potential or the CMB spectrum), but
only depends on the mass scale $M$. This analysis shows a strong
dependency with the value of $M$, the amount of successful initial
conditions ranging from $15\%$ to almost $80\%$ when $M$ ranges
from $10^{-2}$ and $10^{-5}$. For $M$ below the GUT scale,
$10^{16}$ GeV, the quantification of successful initial conditions
is larger than $50\%$, providing a good mechanism to produce
inflation without fine-tuning of initial conditions. As a
conclusion, we confirm the qualitative results
of Lazarides et al. in Ref.~\cite{Lazarides:1996rk}, and we note that they depend on the
values of potential parameters. We note also that most of the
successful initial conditions are exterior to the valleys and are the images of successful points along the valleys, like in the hybrid inflation model. These results are summarized
in the Tab.~\ref{tab:anamorph} at the end of this section.

\subsubsection{Supergravity corrections}

We
have studied for the potential in SUGRA, given by Eq.~(\ref{eq:smoothsugra}), the space of initial conditions
leading to enough inflation and compared the results to the SUSY case. We
observe two properties of the space of initial conditions.
First, at low initial field values, this space is mostly
unchanged.  This is expected since the SUGRA corrections are small. In particular, the patterns of
successful initial conditions exterior to the valleys exist.
We note that the relative area covered by these patterns
 is higher in SUGRA than in SUSY and can be as high as $70\%$ for small values of $M$.
Secondly, because SUGRA corrections increase exponentially for super-planckian fields, we do not observe automatic successful inflation at large
super-planckian initial field values.  Indeed, SUGRA corrections induce slow-roll violations preventing the fields to reach directly the slow-roll attractors along the valleys.  Nevertheless, they can be reached after some field oscillations around the global minima of the potential, and therefore the anamorphosis mechanism is still efficient.   Patterns of successful super-planckian initial conditions are therefore observed.

\subsection{Space of initial conditions for Shifted Inflation}

\begin{center}
\begin{figure}[ht]
\begin{center}
\scalebox{.25}{\includegraphics{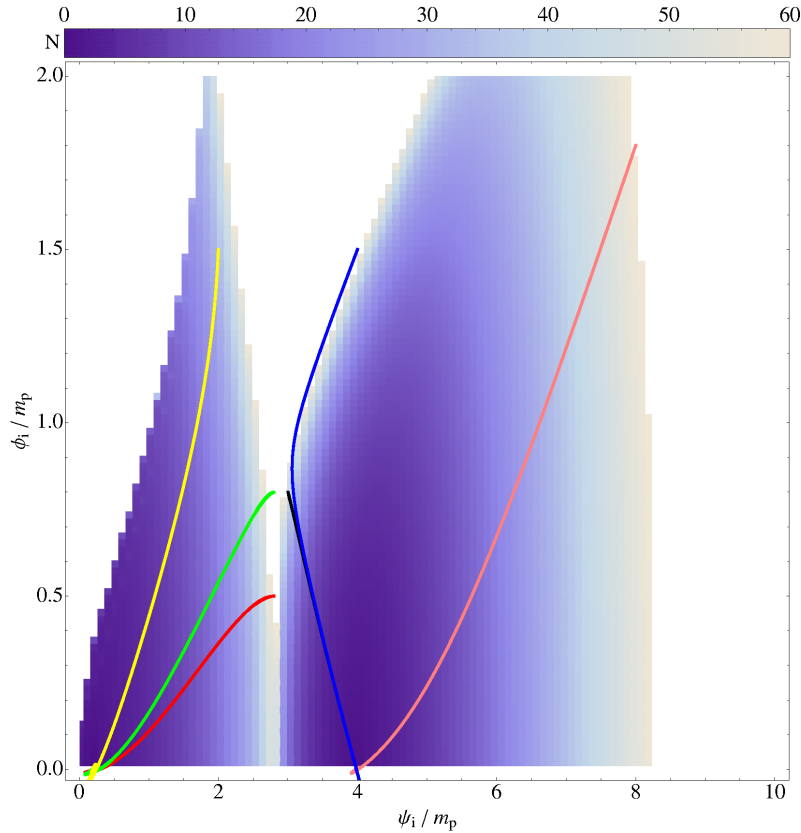}}
\caption{Grid of initial conditions leading or not to inflation,
for the shifted potential of Eq.~(\ref{eq:shiftedpot}), with $M=0.1 \mpl$, $\kappa=1$, and
$\beta = 10^{-2} \mpl^{-2}$. Some trajectories in field space have been
represented to identify where local maxima and minima are.} \label{fig:grid-shifted2}
\end{center}
\end{figure}
\end{center}

Grids of initial conditions leading or not to inflation have
been computed for the shifted inflation model, introduced in section~\ref{sec:shifted}; one of them is represented in Fig.~\ref{fig:grid-shifted2}
for one set of parameters. It corresponds to one cut of the potential
in Fig.~\ref{fig:shifted_potential_cut} (dotted line).

The shifted potential is given by Eq.~(\ref{eq:shiftedpot}),
\begin{equation}
\begin{aligned}
V^{\rr{shifted}}(\phi,\psi) = \kappa^2 & \left( \frac{\psi^2}{4}-M^2-
\beta\frac{\psi^4}{16\Mpl^2} \right)^2 \\ &+ \frac{\kappa^2}{4}
\phi^2 \psi^2 \left( 1- \beta\frac{\psi^2}{2\Mpl^2}\right)^2.
\end{aligned}
\end{equation}
  For a small coupling $\beta$ (say of order $10^{-3}$), if we restrict
ourselves to values of the waterfall field smaller than $5\mpl$,
we obtain a space of initial conditions similar to the original
hybrid case (see Fig.~\ref{fig:completegrid-hybrid}), with a
triangular shaped region of unsuccessful inflation surrounded by
successful regions at higher values of the fields.

At larger values of $\psi$, around the new inflationary valley (the
``shifted'' one), a second triangular
shaped unsuccessful region is observed.
Unlike the
central one, the shifted valley is too steep to generate inflation when
the fields start inside it. Thus no line of successful initial
conditions along the valley is observed. 

If we increase $\beta$, the shifted valley gets closer to
the $\psi=0$ one. As a consequence, the two unsuccessful regions
become closer as well, with interferences between them, as shown
in Fig.~\ref{fig:grid-shifted2}.
The shape of the first unsuccessful
region is modified because the presence of the second valley renders
some unsuccessful trajectories successful. We have represented some
examples of such trajectories in Fig.~\ref{fig:grid-shifted2}.
Finally, note that in the limit of small
$\beta$ and sub-Planckian field values, this model reduces to the original hybrid one. Conclusions
concerning the relative area of successful points are then identical.
These results are
summarized in Tab.~\ref{tab:anamorph}.

\subsubsection{Supergravity corrections}

We have computed for several sets of the parameters the percentage
of successful initial conditions taking into account SUGRA corrections.  Similarly to what is observed for smooth inflation, we
do not find significant modifications compared to the SUSY case except at
large mass scale, where the steepness of the potential prevent the field trajectories to reach directly the slow-roll attractor in the valleys.  Inflation can therefore be generated only if trajectories reach the central valley after oscillations around the global minima of the potential.  

\subsection{Space of initial conditions for Radion Assisted Gauge Inflation}

The set of initial field values leading to more than 60 e-folds of inflation has also been analyzed for the Radion Assisted Gauge inflation model.   The two-field potential is given by Eq.~(\ref{eq:potenradion}), 
\begin{equation}
V(\phi, \psi ) = \frac{1}{4}  \frac{\phi^2}{f^2} \psi^4 + \frac{\lambda}{4}
\left( \psi^2 - \psi_0 ^2 \right)^2~.
\end{equation}
The effective field $\psi$ is the inverse of the radius of an extra-dimension and
quantum gravity effects could dominate when it
gets larger than the five dimensional Planck mass. We have nevertheless considered 
super-planckian values of $\psi$ or $\psi_0$ [the potential of
Eq.~(\ref{eq:potenradion}) could be an effective model].  For the parameter values $\psi_0 = 10^{-2}\mpl, f=\mpl,
\lambda=10^{-5}$, the grid of initial conditions is very similar
to the hybrid case, with a triangular unsuccessful region, and a
generic successful inflation at larger values of the fields (see
Fig.~\ref{fig:grilleRadion} below).
\begin{figure}[ht]  \begin{center}
\scalebox{.25}{\includegraphics{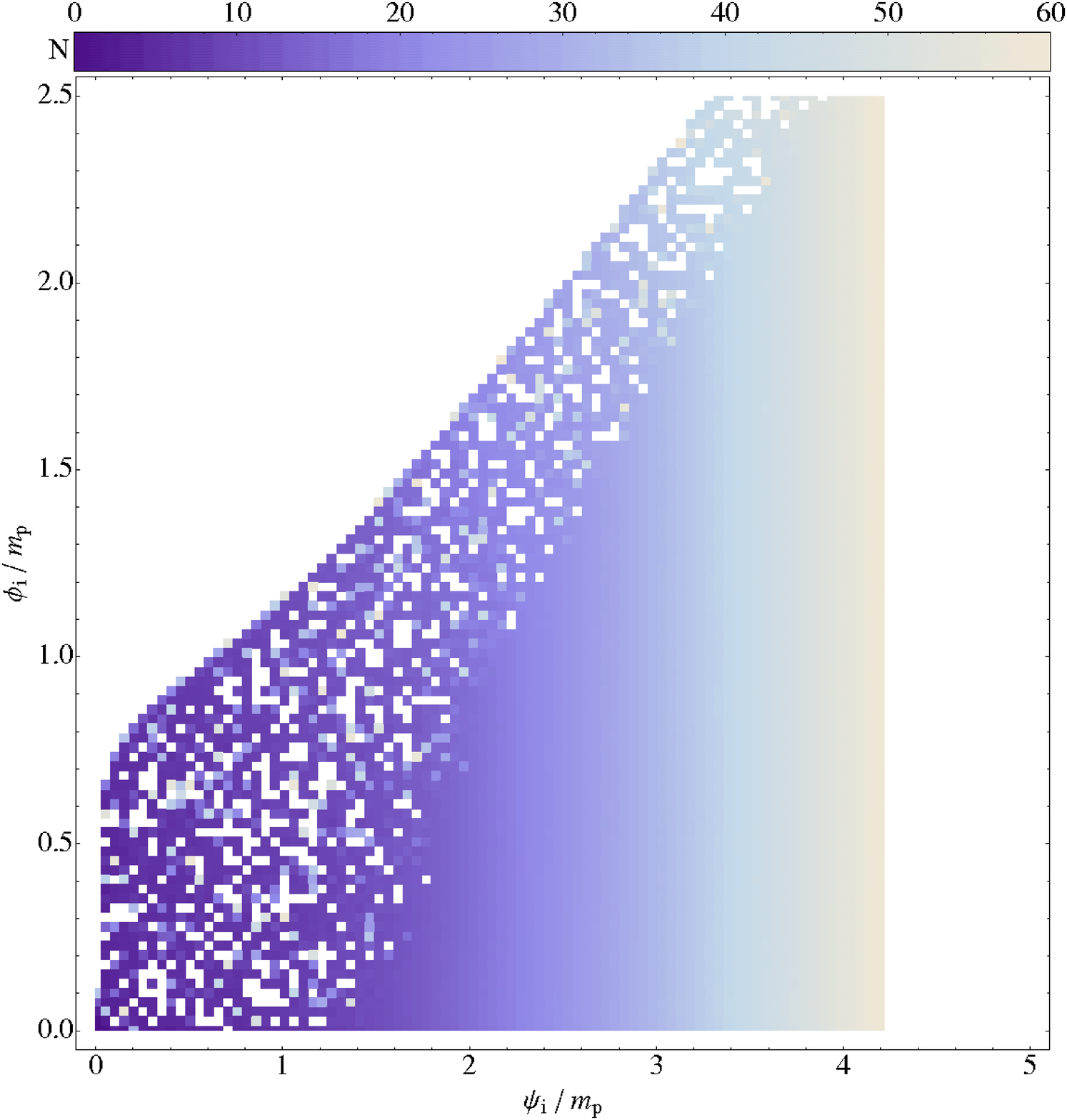}}
\caption{Grid of initial conditions for the radion assisted gauge inflation potential, with
$\psi_0=10^{-2} \mpl$, $f=1\mpl$, $\lambda=10^{-5}$.}
\label{fig:grilleRadion}  \end{center}
\end{figure}

Many successful trajectories also appear in the unsuccessful area
(type-C trajectories), for sufficiently small values of $\lambda
$. We observe a slightly higher successful area, compared to the
hybrid case: for $\phi_\ui,\psi_\ui< \Mpl$, more than $20\%$ of the
points are successful. Grids for different values of the parameter
$M$ show a behavior similar to the hybrid model. However, varying
$\lambda$ has a major impact on the amount of type-C trajectories
as shown in Tab.~\ref{tab:anamorph}. In particular we do not find a
significant amount of successful initial conditions
for the choice of parameters of~\cite{Fairbairn:2003yx}
($\psi_0=10^{-2} \mpl$, $f=1 \mpl$, $\lambda = 10^{-3}$).
We also observe a transition between the successful and
unsuccessful region less abrupt (see Fig.~\ref{fig:grilleRadion}).
This is due to the fact that at small inflaton field, the potential
slightly differs from the hybrid potential: the slope of the
potential is slightly more steep and the same amount of e-folds
requires a larger variation of field values.

Our results on the proportion of successful initial conditions for
all models are summarized in the Tab.~\ref{tab:anamorph} below,
when restricting to initial fields values below the reduced Planck
mass. For comparison the results for the
original hybrid model are recalled. Two percentages are
given: first the total number of successful initial field values
(column 3) and the number of initial conditions that are localized
in the initial condition space, outside of the inflationary valley(s)
(column 4). 

\begin{table*}
\begin{center}
\begin{tabular}{|c|c|c|c|}
\hline Model & Values of parameters & Successful &
out of the    \\
 & & points (\%)  & valley(s) (\%)  \\\hline\hline

Hybrid & $M=\phi_{\rr c} = 0.03 \mpl$, $\mu = 636 \mpl$ & 17.4  & 14.8  \\
Hybrid & $M = 0.01 \mpl$, $\phi_{\rr c} = 0.03 \mpl$, $\mu  = 636 \mpl$ & 4.4 & 2.7 \\
Hybrid & $M = 0.03 \mpl$, $\phi_{\rr c} = 0.01 \mpl$, $\mu  = 636 \mpl$ & 15.6 & 13.6 \\
Hybrid & $M = 0.01 \mpl$, $\phi_{\rr c} = 0.01 \mpl$, $\mu  = 636 \mpl$ & 15.9 & 14.0 \\
Hybrid & $M = 0.03 \mpl$, $\phi_{\rr c} = 0.03 \mpl$, $\mu  = 0.1 \mpl$ & 0 & 0 \\ \hline

Smooth & $M=10^{-2} \mpl$, $\kappa=1$  & 16 & 9  \\
Smooth & $M=10^{-3} \mpl$, $\kappa=1$  & 53 & 49  \\
Smooth & $M \approx 2.37\times 10^{-5}\mpl$, $\kappa\approx 10.3 $
& 78 & 60  \\
Smooth SUGRA & $M=10^{-2} \mpl$, $\kappa=1$  & 29 & 17  \\
Smooth SUGRA & $M=10^{-5}\mpl$, $\kappa=1 $ & 70 & 70  \\\hline
Shifted & $M=0.1 \mpl$, $\kappa^2=1$, $\beta = 0.1 \mpl^{-2} $ & 6 & $ < 1$ \\
Shifted & $M=10^{-2} \mpl$, $\kappa^2=1$, $\beta = 0.1 \mpl^{-2} $ & 15 & 14 \\
Shifted & $M=10^{-2} \mpl$, $\kappa^2=1$, $\beta=1 \mpl^{-2} $ & 14 & 13  \\
Shifted SUGRA & $M=0.1 \mpl$, $\kappa^2=1$, $\beta = 0.1 \mpl^{-2} $ & $ < 1$ & $ < 1$ \\
Shifted SUGRA & $M=10^{-2} \mpl$, $\kappa^2=1$, $\beta = 0.1 \mpl^{-2} $ & 13 & 12 \\
Shifted SUGRA & $M=10^{-2} \mpl$, $\kappa^2=1$, $\beta=1 \mpl^{-2} $ & 13 & 12  \\ \hline
Radion & $\psi_0 = 10^{-2} \mpl$, $\lambda=10^{-3}$, $f=1\mpl $ & $<0.1$ & $<0.1$  \\
Radion & $\psi_0 = 10^{-2} \mpl$, $\lambda=10^{-4}$, $f=1\mpl $ & 9.4 & 9.4  \\
Radion & $\psi_0 = 10^{-2} \mpl$, $\lambda=10^{-5}$, $f=1\mpl $ &
25.6 & 24.8  \\\hline
\end{tabular}
\caption{Area of the whole successful initial condition
parameter space over the total surface (column 3), from grids of initial
conditions, for different models and values of parameters, when
restricting to $\phi_\ui,\psi_\ui\leq \Mpl$.  The fourth column
represents the surface of the successful space exterior to the valley(s), over the total surface.} \label{tab:anamorph}
\end{center}
\end{table*}

The results for the range $\phi_i,\psi_i <5 \mpl$ are given in
Tab.~\ref{tab:superplanck} below. The relative area in this case have been
computed only to give an information about how fast the proportion
of successful initial conditions increases when the space of
allowed initial values is enlarged.
\begin{table}
\begin{center}
\begin{tabular}{|c|c|c|}
\hline Model & Values of parameters & Successful (\%)
\\\hline\hline
Hybrid &  $M=\phi_{\rr c} = 0.03, \mu = 636 \mpl$ &   72 \\
Smooth &  $M \approx 2.37\times 10^{-5}\mpl$, $\kappa\approx  10.3 $ &   92 \\
Shifted & $M=10^{-2} \mpl $, $\kappa^2 = 1$, $\beta=10^{-2} \mpl^{-2}$  &  73 \\
Radion &  $\psi_0=10^{-2} \mpl$, $\lambda=10^{-3}$, $f=1$ & 76
\\\hline
\end{tabular}
\caption{Percentage of successful points in grids of initial
conditions on a range $0 < \phi, \psi < 5\mpl$, for each model and some standard
values of the parameters.} \label{tab:superplanck}
\end{center}
\end{table}

\section{Fractal properties of sub-planckian initial field values}

\label{sec:fractalic}

\subsection{The set of successful initial field values}

In the last section, the space of successful initial field values has been found to be composed of an intricate
ensemble of points organized into continuous patterns. They are 
plotted in Fig.~\ref{fig:grid} for a \emph{fixed} set of
potential parameters and assuming vanishing initial velocities.

The white vertical narrow strip located along
$\psi=0$ correspond to trajectories joining directly the slow-roll attractor along the valley.    On the other hand, successful regions exterior to the valley exhibit a fractal looking
aspect.  One may wonder if the area of this
two-dimensional set of points is indeed well-defined? Equivalently, do
new successful regions appear inside unsuccessful domains and
conversely? In order to quantify how much the anamorphosis points are
a probable way to have inflation in the whole parameter space, we
 address the question of defining a measure on the initial field
values space. In particular, this requires to determine the dimension
of the set
\begin{equation}
\label{eq:Sdef}
\calS \equiv \left\{(\phi_\ui, \psi_\ui) \ \diagup \ N>60 \right\}.
\end{equation}

\subsection{Chaotic dynamical system}

\subsubsection{Phase space analysis}

As suggested by Fig.~\ref{fig:grid}, at fixed potential parameter
values, the dynamical system exhibits a chaotic behavior. In particular,
the sensitivity of the trajectories to the initial field values comes
from the presence of three attractors. Two of them are the global
minima of the potential, $\calM_\pm$ respectively at $(\phi=0,\psi =
\pm M)$, in which all classical trajectories will end, whereas the
less obvious is a quasi-attractor $\calI$ defined by the inflationary
valley itself ($\psi=0$, $\phi>\phi_\uc$). Indeed, whatever the initial field values,
as soon as the system enters slow-roll one has (in Planck
units)~\cite{Ringeval:2005yn},
\begin{equation}
v^2 \equiv \left(\dfrac{\ud \phi}{\ud N}\right)^2 + \left(\dfrac{\ud \psi}{\ud N}
\right)^2 = 2 \epsilon_1 \ll 1,
\end{equation}
where $\epsilon_1$ is the first Hubble flow
function~\cite{Schwarz:2001vv}. The system therefore spends an
exponentially long amount of time in the valley. The
sensitivity to the initial conditions comes from the presence of these
three attractors: either the trajectory ends rapidly into one of the
two minima, or it lands on the valley where it freezes.

\begin{figure}[]  \begin{center}
  \includegraphics[width=12cm]{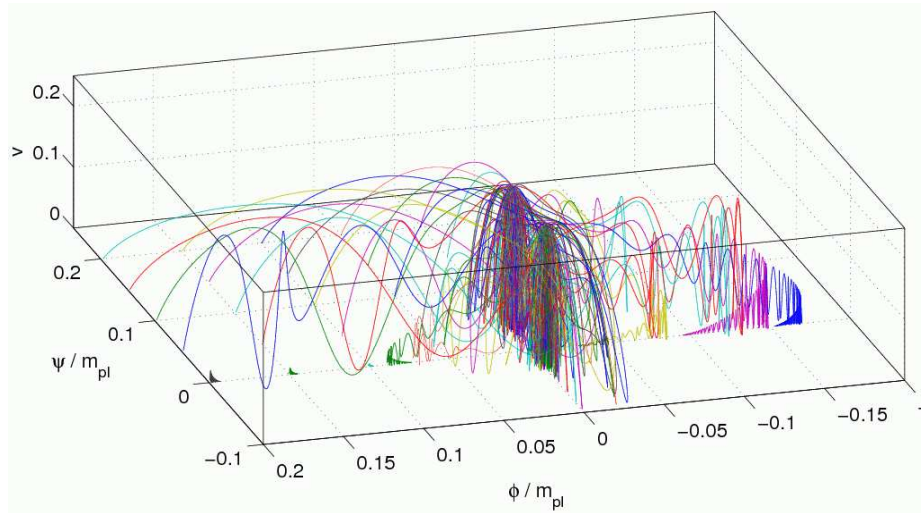}
  \caption{Phase space $v^2(\phi, \psi)$ for 25 trajectories and
    vanishing initial velocities. The potential parameters are fixed
    to the values $M=\phi_{\rr c} = 0.03 \mpl, \mu = 636 \mpl$. All trajectories end on one of the three attractors of the
    dynamical system: the two global minima of the potential, and the
    inflationary valley with almost vanishing slow-roll
    velocity. These three attractors induce the chaotic behavior.}
  \label{fig:phasespace} \end{center}
\end{figure}

A phase space plot is represented in Fig.~\ref{fig:phasespace} in
which we have computed $25$ trajectories from a grid of initial field
values. The inflationary valley clearly appears as the attractor with
quasi null velocity vector ($\epsilon_1 \ll 1$), while around the two
global minima, two ``towers'' appear due to the field oscillations
around them.

\subsubsection{Basins of attraction}

From the definition of $\cal S$ in Eq.~(\ref{eq:Sdef}), one has
\begin{equation}
\calS = F^{-1}(\calI),
\end{equation}
where $F(\phi,\psi)$ stands for the mapping induced by the
differential system of equations (\ref{eq:FLtc12field}) and
(\ref{KGtc2fieldd}). The set of successful initial field values
$\calS$ is therefore the basin of attraction of the attractor
$\calI$~\cite{Ott:fractals, Falconer:fracgeo}. Since the attractor
$\calI$ is a dense set of dimension 2 and $F$ is continuous, one
expects $\calS$ to contain a dense set of dimension
2~\cite{Falconer:fracgeo}. As can be intuitively guessed, the boundary
of $\calS$ can however be of intricate structure because of the
chaotic behavior of the system: two trajectories infinitely close
initially can evolve completely differently. As we show in the
following, $\calS$ is actually a set of dimension two having a fractal
boundary of dimension greater than one.

Finally, by the definition of a continuous mapping, all parts of $\calS$, 
boundary included, must be
connected together and to the inflationary valley $\calI$. The fractal
looking aspect of Fig.~\ref{fig:grid} is only induced by the
intricate boundary structure of $\calS$ which is exploring all the
initial field values space. The fractality of the boundaries of the
space of initial fields values was first mentioned in
Ref.~\cite{Ramos:2001zw}, but the study was restricted to a small
region of the field space and the model included dissipative
coefficients. As an aside remark, let us notice that the existence of
a fractal boundary may have strong implications in the context of
eternal chaotic inflation: there would exist inflationary solution close to
any initial field values.

In order to quantify the chaotic properties of the dynamical
system defined by the mapping $F(\phi,\psi)$, we turn to the
calculation of the Lyapunov exponents.

\subsubsection{Lyapunov exponents}

\begin{figure}[] \begin{center}
  \includegraphics[width=11cm]{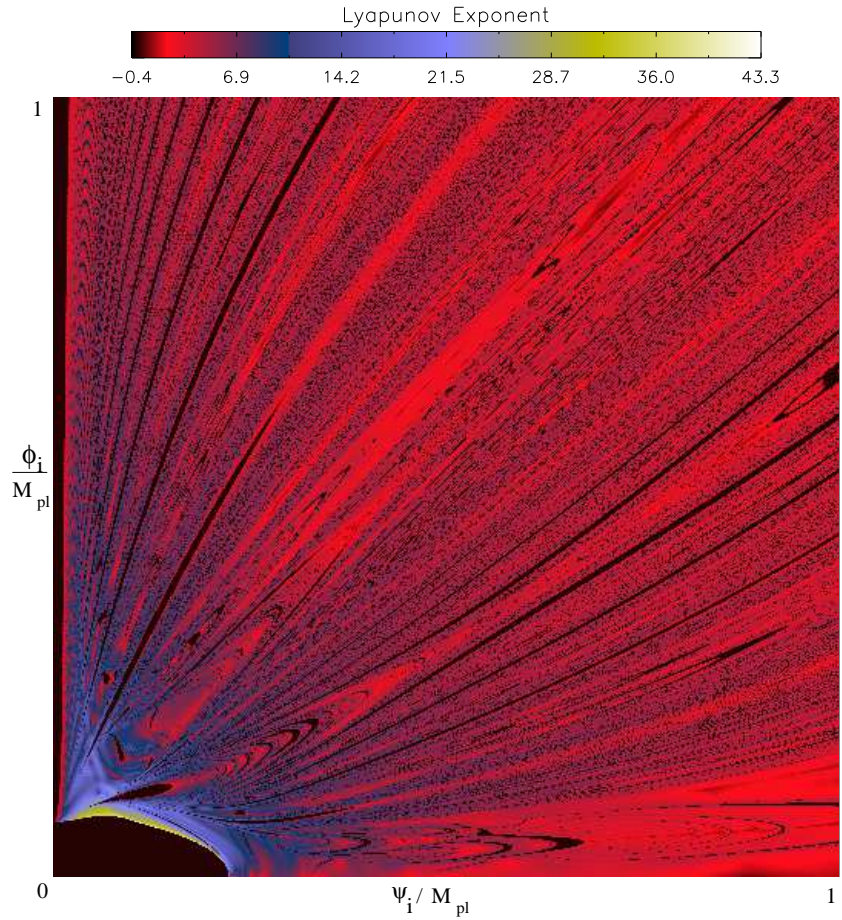}
  \caption{Highest Lyapunov exponent as a function of the initial
    field values in the original hybrid model. The potential parameters
    are the same as in Fig.~\ref{fig:grid}. The field evolution is
    therefore stable on the set $\calS$ of successful initial field
    values (black) but exhibits chaotic behaviour elsewhere.}
  \label{fig:lyap} \end{center}
\end{figure}

The Lyapunov exponents at an initial point
$\vect{\chi}_\ui=(\phi,\psi,\deriv{\phi}{N},\deriv{\psi}{N})|_\ui$
measures how fast two infinitely close trajectories mutually diverge
or converge. They give a mean to characterize the stretching and
contracting characteristics of sets under the mapping induced by the
differential system. A small perturbation $\vect{\delta \chi}$ around
the trajectory $\vect{\chi}(N)$ will evolve according to
\begin{equation}
\label{eq:pertevol}
\dfrac{\ud \vect{\delta \chi}}{\ud N} = \ud F \cdot \vect{\delta \chi}\,,
\end{equation}
where $\ud F$ stands for the Jacobian of the differential system
$F$. The Lyapunov exponents at the initial point $\vect{\chi}_\ui$ and
along the direction $\vect{\delta \chi}_0$ are the numbers defined
by~\cite{Ott:fractals}
\begin{equation}
  h(\vect{\chi}_\ui,\vect{\delta \chi}_0) = \lim_{N\to\infty} \dfrac{1}{N}\ln \dfrac{
    \left|\vect{\delta \chi}(N)\right|}{\left|\vect{\delta \chi}_0 \right|}\,,
\end{equation}
where $\vect{\delta \chi}(N)$ is the solution of
Eq.~(\ref{eq:pertevol}) with $\vect{\delta \chi}(0)=\vect{\delta
  \chi}_0$ and $\vect{\chi}(0)=\vect{\chi}_\ui$. If the considered set
is an attractor or an invariant set of the differential system having
a natural measure, one can show that the exponents do not depend on
the initial point $\vect{\chi}_\ui$. At fixed potential parameters,
there are four Lyapunov exponents associated with the differential
system of Eqs.~(\ref{eq:FLtc12field}) and (\ref{KGtc2fieldd}). If
the largest exponent is positive, then the invariant set is chaotic.

In Fig.~\ref{fig:lyap}, we have computed the largest Lyapunov exponent
at each point of the plane $(\phi_\ui,\psi_\ui)$. The numerical method
we used is based on Refs.~\cite{1997:Dieci, 2003:dieci} and uses the
public code $\texttt{LESNLS}$. Let us notice that since $\calI$ is
only a quasi-attractor, we have stopped the evolution at most when
$H_\uend^2=V/(3\Mpl^2)$, i.e. just before the fields would classically
enter either $\calM_+$ or $\calM_-$. As can be seen, all points
belonging to $\calS$ exhibit the same and small negative Lyapunov
exponent: the invariant set $\calS$ is therefore non-chaotic. On the
other hand, all the other initial field values associated with the
basins of attraction of $\calM_\pm$ have a positive Lyapunov
exponent. For those, the field evolution is chaotic and exhibits a
sensitivity to the initial conditions. Notice that these exponents may
slightly vary from point to point due to our choice to stop the
integration at $H_\uend$ instead of the classical attractors
$\calM_\pm$. This is particularly visible for the trajectories
starting close to $H_\uend$ (green shading): there is not enough
evolution to get ride of the transient evolution associated with the
initial conditions.

\subsection{Fractal dimensions of $\calS$ and its boundary}

\subsubsection{Hausdorff and box-counting dimension}\label{sec:boxcountingdim}

Since we suspect a set with fractal properties, the natural
measure over $\calS$, extending the usual Lebesgue measure, is the
Hausdorff measure. The $s$-dimensional Hausdorff measure of $\calS$ is
defined by~\cite{Falconer:fracgeo}
\begin{equation}
\calH^s(\calS) = \lim_{\delta \to 0} \inf \left\{ \sum_{i=1}^\infty
      |U_i|^s\ \diagup \ \calS \subset \bigcup_{i=1}^\infty
      U_i\ ; \ |U_i| \le \delta \right\}.
\end{equation}
In this definition, the sets $U_i$ form a $\delta$-covering of $\calS$
and the diameter function has been defined by $|U|\equiv \sup\{|x-y|\
\diagup \ x,y \in U\}$. As a result, $\calH^s(\calS)$ is the smallest
sum of the $s$th powers of all the possible diameters $\delta$ of all
sets covering $\calS$, when $\delta \to 0$. Having such a measure, the
fractal dimension of $\calS$ is defined to be the minimal value of $s$
such that the Hausdorff measure remains null (or equivalently the
maximal value of $s$ such that the measure is infinite). In practice,
measuring the Hausdorff dimension using this definition is not
trivial, due to the necessity of exploring all
$\delta$-coverings. However, in our case, we are interested in the
fractal properties of a basin of attraction associated with a
continuous dynamical system and one can instead consider the so-called
box-counting dimension~\cite{Falconer:fracgeo}. This method simply
restricts the class of the $U_i$ to a peculiar one, all having the
same diameter $\delta$. When the mapping $F$ is self-similar, one can
show that box-counting and Hausdorff dimensions are equal. In general,
the Hausdorff dimension is less or equal than the box-counting
one. Here, $F$ being a contracting continuous flow, we expect the
equality to also hold.

To define the box-counting dimension, we cover the set $\calS$ with
grids of step size $\delta$, and count the minimal number of boxes
$N(\delta)$ necessary for the covering. The box-counting dimension is
then given by
\begin{equation}\label{eq:boxcountingdim}
D_\uB= \lim _{\delta \rightarrow 0} \frac{\log N(\delta)}{\log (1/ \delta) }\,.
\end{equation}
This method has the advantage to be easily implemented numerically 
and, in the following, we will apply it to calculate the dimension of
$\calS$ and its boundary. 

\subsubsection{Fractal boundary of $\calS$}

For each randomly chosen point of the plane $(\phi_\ui,\psi_\ui)$, we compute
three trajectories. The first one starts from the point under
consideration while the two others have initial conditions modified by
$+\delta$ and $-\delta$ along one direction (for example along $\phi$,
but the chosen direction does not affect the result). For each of
these trajectories, we determine in which attractor ($\calM_\pm$ or
$\calI$) the flow ends. Since we are interested in the boundary of
$\calS$, we calculate the proportion $f(\delta)$ of points for which
at least one trajectory ends in $\calI$, and another in $\cal M_+$ or 
$\cal M_-$. The process is iterated for increasingly smaller values of 
$\delta$ and we evaluate how the area of the
$\delta$-grid covering of $\calS$ scales with $\delta$. So strictly
speaking, our evaluation of the box-counting dimension is made through
the determination of the Minkowski dimension of the boundary of
$\calS$~\cite{Falconer:fracgeo}. From Eq.~(\ref{eq:boxcountingdim}),
assuming that, at small $\delta$, 
\begin{equation}
f(\delta) \propto \delta^\alpha,
\end{equation}
the box-counting dimension of the $\calS$ boundary is then given
by~\cite{Ott:fractals}
\begin{equation}
D_\uB = 2 - \alpha.
\end{equation}

\begin{figure}[] \begin{center}
\includegraphics[width=12cm]{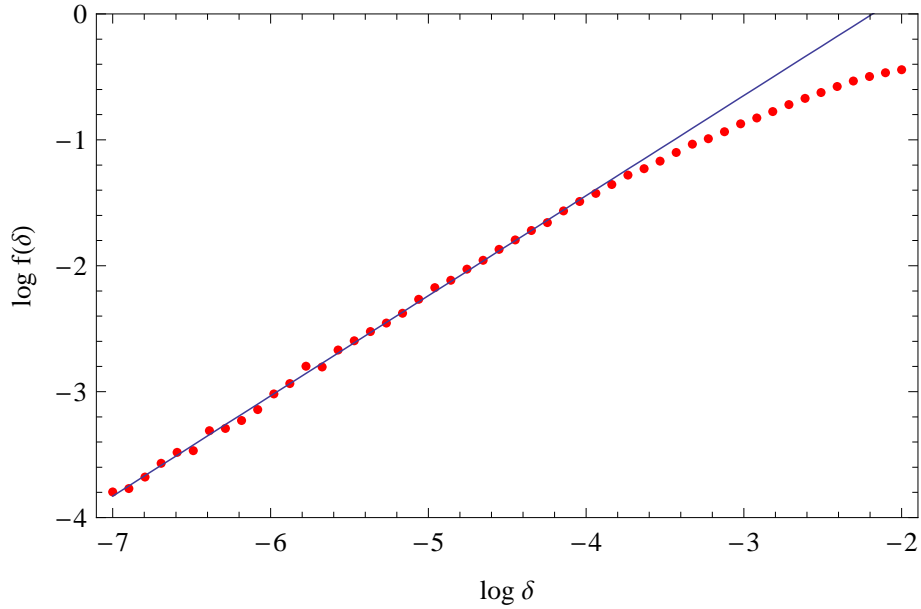}
\caption{Fraction of initial field values in a $\delta$-sized box
  intercepting the set $\calS$, as a function of $\delta$. The field
  has been restricted to sub-planckian values and the potential
  parameters are fixed to $\mu = 636 \mpl $ and
  $M= \phi_{\rr c} = 0.03 \mpl$. The exponent $\alpha$ of the power law dependency
  gives the box-counting dimension $D_\uB = 2 - \alpha \simeq 1.2$
  showing that $\calS$ possesses a fractal boundary.}
 \label{fig:dim_fractale}  \end{center}
\end{figure}

In Fig.~\ref{fig:dim_fractale}, we have plotted $f(\delta)$ as a
function of $\delta$ at fixed potential parameters. We recover the
expected power law behaviour, the slope of which is approximately
$\alpha \simeq 0.80$. As a result, the boundary of $\calS$ is indeed a
fractal of box-counting dimension
\begin{equation}
D_\uB \simeq 1.20.
\end{equation}
Notice that this value depends on the chosen set of potential
parameters, as one may expect since they affect the shape of $\calS$
and the typical size of the structures.

\subsubsection{Dimension of $\calS$}

\begin{figure}[] \begin{center}
\includegraphics[width=12cm]{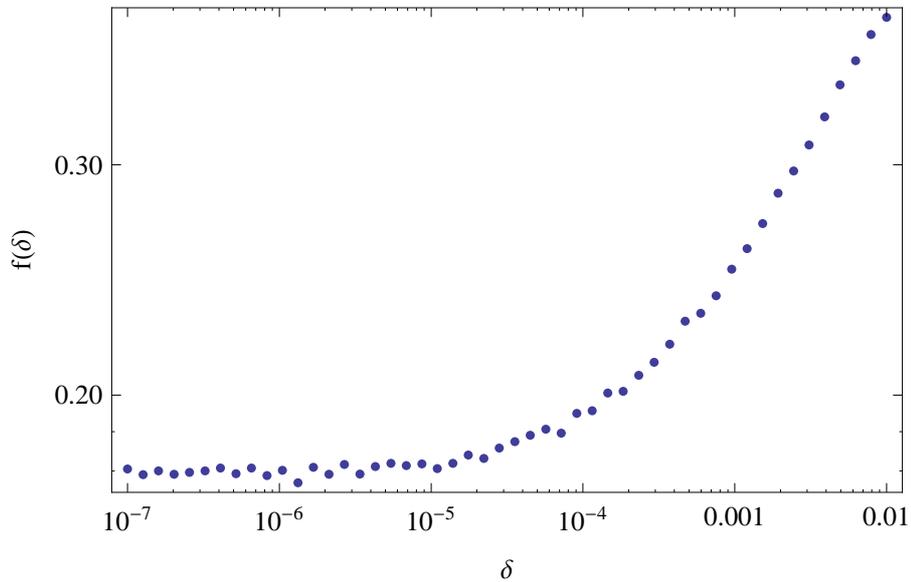}
\caption{Fraction of initial field values leading to inflation in a
  $\delta$-sized box as a function of $\delta$. The potential
  parameters are the same as in Fig.~\ref{fig:dim_fractale}. Once the
  box is small enough to be fully contained in $\calS$, $f(\delta)$
  remains constant. As a result, the box-counting dimension of $\calS$
  is $D_{\uB}=2$ and the interior of $\calS$ is not fractal.}
\label{fig:fractal_area}  \end{center}
\end{figure}

In order to determine the box-counting dimension of $\calS$ itself one
can apply a similar method than the one used for its boundary. Now
$f(\delta)$ denotes the proportion of points for which at least one of
the three trajectories end in the attractor $\calI$ (this condition
therefore includes also the points belonging the boundaries). The
resulting power-law is represented in Fig.~\ref{fig:fractal_area}.

For small enough values of $\delta$, the $\delta$-sized boxes are
small enough to be fully contained in $\calS$ and the function
$f(\delta)$ appears to be constant in that case. As a result, the
box-counting dimension of $\calS$ is $2$. We therefore conclude that,
like for the well-known Mandelbrot set~\cite{Mandelbrot:1980}, the 
boundary of $\calS$ is fractal but
the set of successful inflationary points is not and has the dimension 
of a surface. Consequently, although the boundary of $\calS$
has an infinite length ($D_\uB = 1.2$), it has a vanishing area: the
Haussdorf dimension of $\calS$ (boundary included) is therefore also
2. As a result, the two-dimensional Haussdorf measure on $\calS$
reduces to the usual two-dimensional Lebesgues measure and this will
be our choice for defining a probability measure in the rest of this chapter.

As previously
emphasized, the potential parameters and initial field velocities have
been fixed in this section and the set $\calS$ is actually the
two-dimensional section of an higher dimensional set, whose boundary
is also certainly fractal (and therefore of null measure). Since one 
can no longer use griding method to
explore such a high dimensional space, we move on in the next
section to a MCMC exploration of the full parameter space to assess
the overall probability of getting inflation in the hybrid model.

\section{Probability distributions in hybrid inflation} \label{sec:mcmc_IC}

\label{section:mcmc}

The aim of this section is to use Monte-Carlo-Markov-Chains (MCMC)
techniques in order to explore the whole parameter space, including the
initial field velocities and all the potential parameters. With
unlimited computing resources, we could have used a griding method to
localize the hypervolumes in which inflation occurs, as we have done
for the two-dimensional plane $(\phi_\ui,\psi_\ui)$ in the previous
section. For the original hybrid model, we have in total seven
parameters that determine a unique trajectory: two initial field
values, initial field velocities and the three potential parameters
$M$, $\mu$ and $\phi_{\rr c}$. To probe this seven-dimensional space, more than
just measuring the hypervolume of the successful inflationary regions,
we define a probability measure over the full parameter space. Using
Bayesian inference, one can assess the posterior probability
distribution of all the parameters to get enough e-folds of
inflation. Monte--Carlo--Markov--Chains (MCMC) method is a widespread
technique to estimate these probabilities, its main power being that
it numerically scales linearly with the number of dimensions, instead 
of exponentially.

Several algorithms exist in order to construct the points of a Markov
chain, the Metropolis\-Hastings algorithm being probably the
simplest~\cite{Metropolis53,Hastings70}. Each point $x_{i+1}$,
obtained from a Gaussian random distribution (the so-called proposal
density) around the previous point $x_i$, is accepted to be the next
element of the chain with the probability
\begin{equation}
\label{eq:mcmctrans}
P(x_{i+1}) = \min \left[1,\frac{\pi(x_{i+1})}{\pi(x_i)}  \right],
\end{equation}
where $\pi(x)$ is the function that has to be sampled via the Markov
chain. MCMC methods have been intensively used in the context of CMB
data analysis~\cite{Christensen:2001gj, Lewis:2002ah, Martin:2006rs,
  Lorenz:2007ze, Dunkley:2008ie} where the function $\pi(\theta|d)
\propto \calL(d|\theta)P(\theta)$ is the posterior probability
distribution of the model parameters given the data. In the context of
Bayesian inference, this one is evaluated from the prior distributions
$P(\theta)$ and the likelihood of the experiment
$\calL(d|\theta)$. After a relaxation period, one can show that
Eq.~(\ref{eq:mcmctrans}) ensures that $\pi$ is the asymptotic
stationary distribution of the chain~\cite{MacKayBook}. The MCMC
elements directly sample the posterior probability distribution
$\pi(\theta|d)$ of the model given the data.

In our case, we can similarly define a likelihood $\calL$ as a
binary function of the potential parameters, initial field values and
velocities. Either the trajectory ends up on $\calI$ and produces more 
than $60$ e-folds of inflation, or it does not. In
the former case we set $\calL=1$ whereas $\calL=0$ for no
inflation. The function $\pi$ we sample is then defined by $\pi =
\calL P(\theta)$ where $\theta$ stands for field values, velocities
and potential parameters and $P$ is our prior probability distribution
that we will discuss in the next section.

\subsection{Prior choices}
\label{sec:priors}
MCMC methods require a prior assumption on the probability
distributions of the fields, velocities and potential parameters.  As
we only consider in this work the initial condition and parameter
space leading to at least $60$ e-folds of inflation, the prior choices
are only based on theoretical arguments. These arguments can be linked
to the framework from which the potential is deducted. As discussed in the chapter 3, if one
considers the hybrid model to be embedded in supergravity, the 
fields have to be restricted to values less than the reduced Planck
mass. We adopt here this restriction for initial field values, not only 
because of this argument, but also because it has been shown in section~\ref{sec:ICoriginal} that if super-plankian fields are allowed,
trajectories become generically successful. On the other hand, the
model was considered to suffer some fine-tuning when one of the fields
has to be order of magnitudes smaller than the other. As inflation is
not possible for very small initial values of both fields (because of
the Higgs instability), we have considered a flat prior for initial
field values in $[-\Mpl,\Mpl]$ as opposed to a flat prior for the
logarithm of the fields. Note that one has to include
negative values of the fields in order to take into account the 
orientation of the initial velocity vector.

Concerning the initial field velocities, from the equations of motion,
one can easily show that there exist a natural limit\footnote{This is
  just the limit $\epsilon_1 < 3$ in Planck units \cite{Ringeval:2005yn}.}
\begin{equation}
  v^2 =  \left(  \frac{\dd \phi} {\dd N} \right)^2 + \left( \frac{\dd
      \psi} {\dd N} \right)^2 < 6 M_{\rr p}^2.
\end{equation}
Similarly, our prior choices are flat distributions inside such a
circle in the plane $(\deriv{\phi}{N},\deriv{\psi}{N})$, where ``$,N$''
denotes a partial derivative with respect to the number of e-folds.

In the absence of a precise theoretical setup determining the
parameters of the potential [given in Eq.~(\ref{eq:potenhyb2dNEW})], there are no a priori theoretical constraint on $M$, $\mu$ and $\phi_{\rr c}$. Let us just remind that for $\mu < 
0.3$, the dynamics of
inflation in the valley is possibly strongly affected by slow-roll
violations~\cite{Clesse:2008pf}. As a result, with the concern to not
support a particular mass scale, we choose a flat prior on the
logarithm of the parameters.  Notice that the dependancies in $\Lambda$ are not considered here because this parameter only
rescales the potential and thus does not change the dynamics.

In the next sections, we perform the MCMC exploration of the parameter
space from these priors. Firstly by reproducing the results of
Sec.~\ref{sec:fractalic} in the two-dimensional section
$(\phi_\ui,\psi_\ui)$, then by including the initial field velocities
and finally by considering all the model parameters.

\subsection{MCMC on initial field values}

\begin{figure}
\includegraphics[width=12cm]{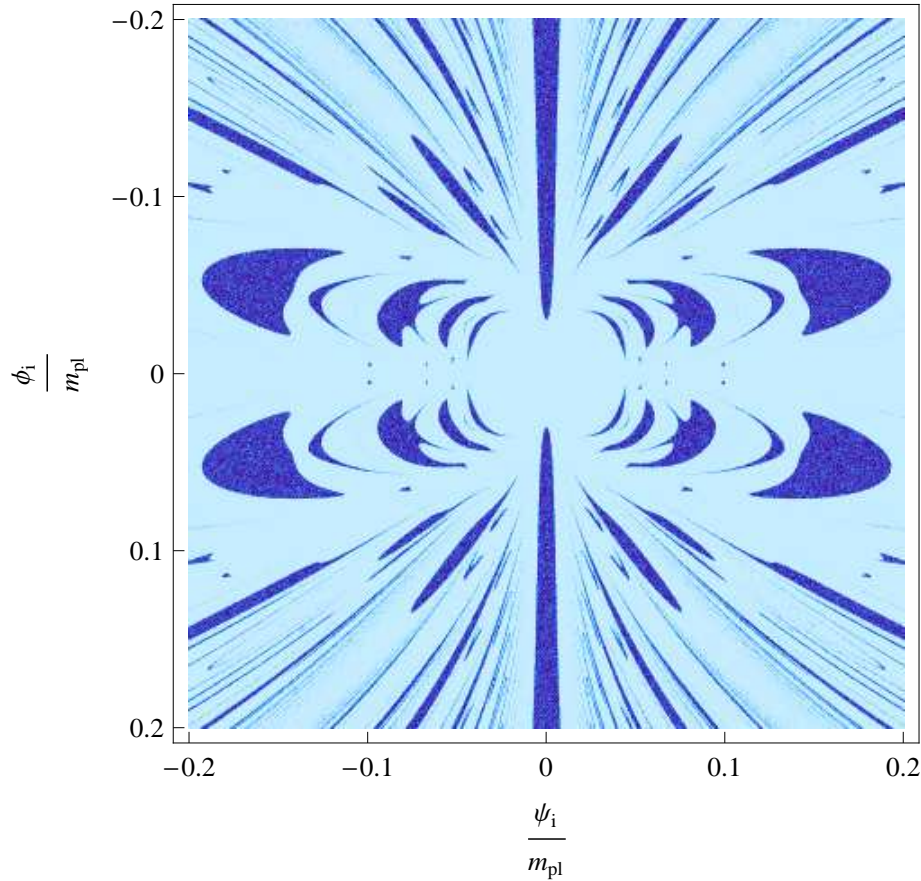}
\caption{Two-dimensional posterior probability distribution in the
  plane $(\phi_\ui,\psi_\ui)$ leading to more than $60$ e-fold of
  inflation in the hybrid model. The dark blue regions corresponds to
  a maximal probability whereas it vanishes elsewhere. Following~\cite{Tetradis:1997kp,Clesse:2008pf}, the potential parameters are set to $M=\phi_{\rr c } = 0.03 \ \mpl  , \ \mu=636 \  \mpl$. As expected, the MCMC
  exploration matches with the griding methods (see Fig.~\ref{fig:grid}).}
\label{fig:2Dfield_nov}
\end{figure}

In order to test our MCMC, we have first explored the space of initial
field values leading or not to more than $60$ e-folds of
inflation. The potential parameters have been fixed to various values
already explored by griding methods in Sec.~\ref{sec:fractalic} and
Ref.~\cite{Clesse:2008pf}, while the initial velocities are still
assumed to vanish. The MCMC chain samples have been plotted in
Fig.~\ref{fig:2Dfield_nov}. Notice that to recover the fractal
structure of the boundary of $\calS$, one has to adjust the choice of
the Gaussian widths of the proposal density distribution. If those are
too large, the acceptance rate will be small because the algorithm
tends to test points far away from the last successful point, and if
they are too small the chains remain stuck in the fractal structures
without exploring the entire space. Nevertheless, with an intermediate
choice, Fig.~\ref{fig:2Dfield_nov} shows that the intricate structure
of the boundary of $\calS$ can be probed with the MCMC. More than
being just an efficient exploration method compared to griding, the
MCMC also provides the marginalised probability distributions of
$\phi_\ui$ and $\psi_\ui$ such that one gets inflation. They have been
plotted in Fig.~\ref{fig:1Dfields} (top two plots), the normalisation
being such that their integral is unity. As one can guess from
Fig.~\ref{fig:2Dfield_nov}, with vanishing initial velocities and a
fixed set of potential parameters, inflation starting in the valley is
\emph{not} the preferred case since the area under the distribution of
$\psi_\ui$ outside of the valley is larger than inside. Moreover,
these distributions take non-vanishing values everywhere and there is
therefore no fine-tuning problem. Of course, one still have to
consider the other parameters and this is the topic of the next
sections.

\subsection{MCMC on initial field values and velocities}

\begin{figure}[]
  \includegraphics[width=12cm]{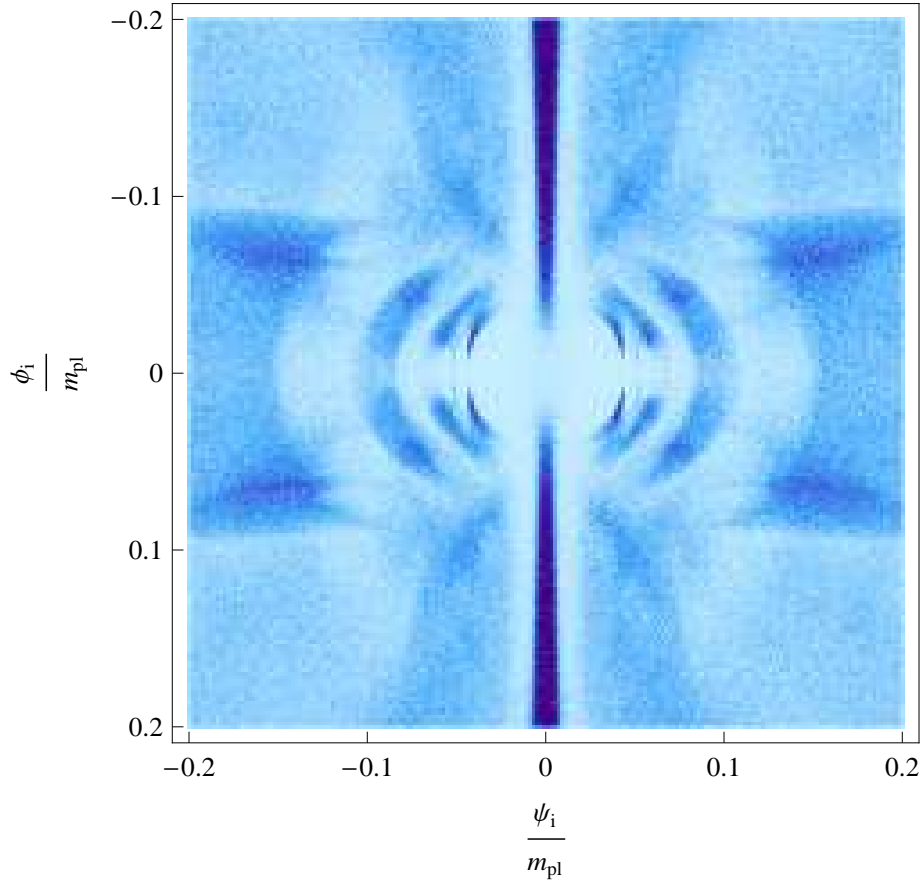}
  \caption{Two-dimensional marginalised posterior probability
    distribution for the initial field values. The marginalization is
    over the initial field velocities whereas the potential parameters
    are still fixed. The shading is proportional to the probability
    density value. Although the inflationary valley has the highest
    probability density, its area remains restricted such that the
    most probable initial field values to get inflation are still out
    of the valley (see Fig.~\ref{fig:1Dfields}).}
\label{fig:2Dfield_v}
\end{figure}

The initial
values of the field velocity are inside a disk of radius $\sqrt {6}$
in the plane $(\deriv{\phi}{N},\deriv{\psi}{N})$ (in reduced Planck mass
units). The marginalised two-dimensional posteriors for the initial
field values is plotted in Fig.~\ref{fig:2Dfield_v} whereas the
marginalised posterior for each field are represented in
Fig.~\ref{fig:1Dfields} (middle line). Even if non-vanishing velocities are
considered, the successful inflationary patterns remain. Notice that
they appear to be blurred simply because of the weighting induced by
marginalizing the full probability distribution over the initial
velocities.

\begin{figure}[] \begin{center}
\includegraphics[width=8.cm]{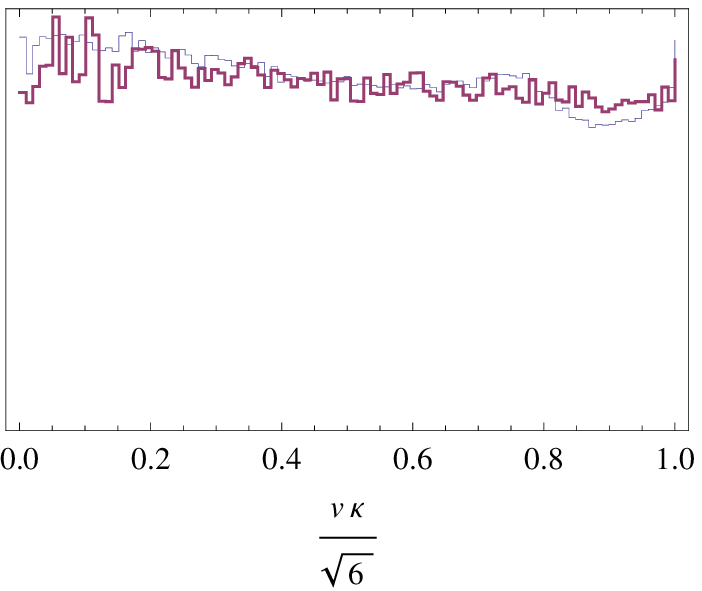}

\includegraphics[width=8.cm]{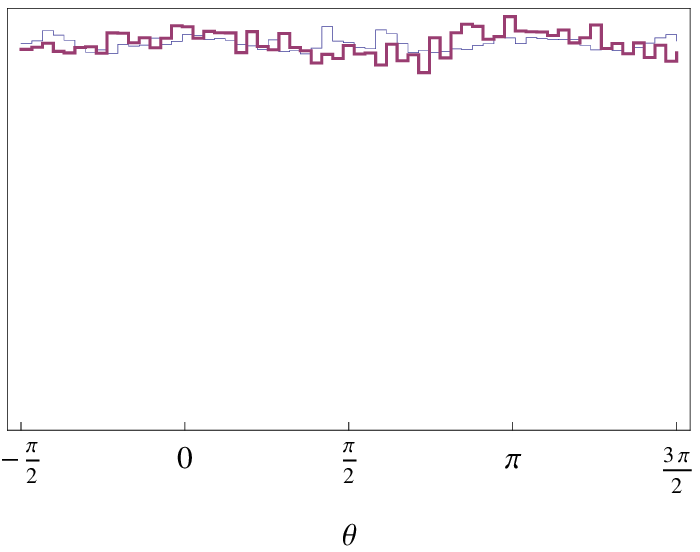}
\caption{Marginalised posterior probability distributions for the
  modulus (top) and angle (bottom) of initial field velocity. The thin
  superimposed blue (lighter) curves are obtained at fixed potential parameters, while the
  thick red are after a full marginalization over all the model
  parameters. As expected from Hubble damping, all values are
  equiprobable since the field do not keep memory of the initial
  velocity.}
\label{fig:1Dspeeds} \end{center}
\end{figure}

In Fig.~\ref{fig:1Dspeeds}, we have also represented the marginalised
posterior probability distribution for the modulus and direction of
the initial velocity vector. Their flatness implies that there are no
preferred values.  This is an important result because one could think
that large initial velocities could provide a way to kick trajectories
in or out of the successful regions.  But 
because of the Hubble damping term in the Friedmann equations, allowing 
only a generation of a small number of
e-folds before the trajectory falls in one of the three attractors, 
this effect does not affect the marginalized posterior distribution of the initial velocities.

\subsection{MCMC on initial field values, velocities and potential
  parameters}

\begin{figure}[]
  \includegraphics[width=12cm]{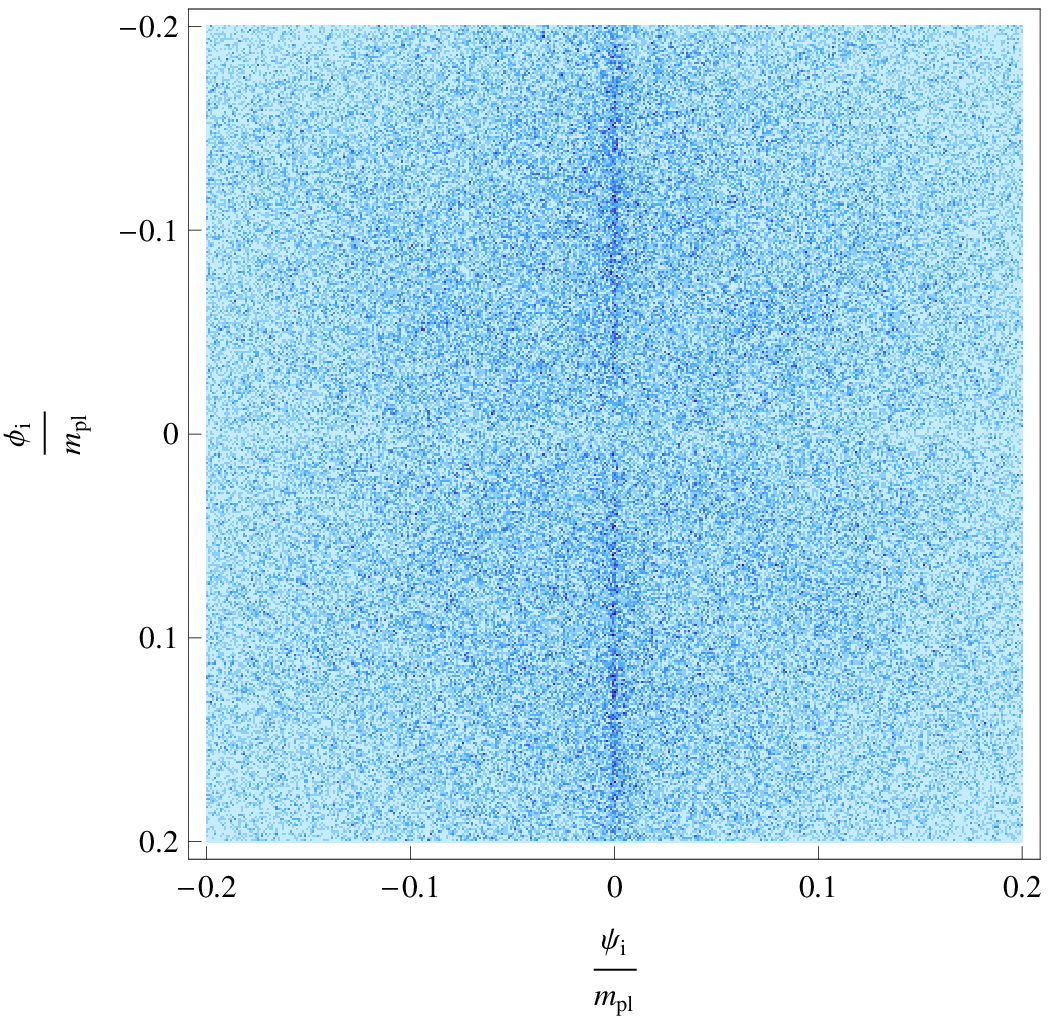}
  \caption{Two-dimensional marginalised posterior probability
    distribution for the initial field values. The marginalisation is
    over the initial field velocities and all the potential
    parameters. The shading is proportional to the probability density
    value. The inflationary valley is still visible around
    $\psi_\ui=0$ and the posterior takes non-vanishing values
    everywhere in the $(\phi_\ui,\psi_\ui)$ plane.  A sub-density is nevertheless observed (as well as in Fig.~\ref{fig:1Dfields}) along 
    the direction $\phi = 0$.  Field trajectories initially oriented along this direction indeed can not reach the inflationary valleys that are orthogonal to this direction.    }
\label{fig:2Dfield_all}
\end{figure}

\begin{figure*}[]
\includegraphics[width=15cm]{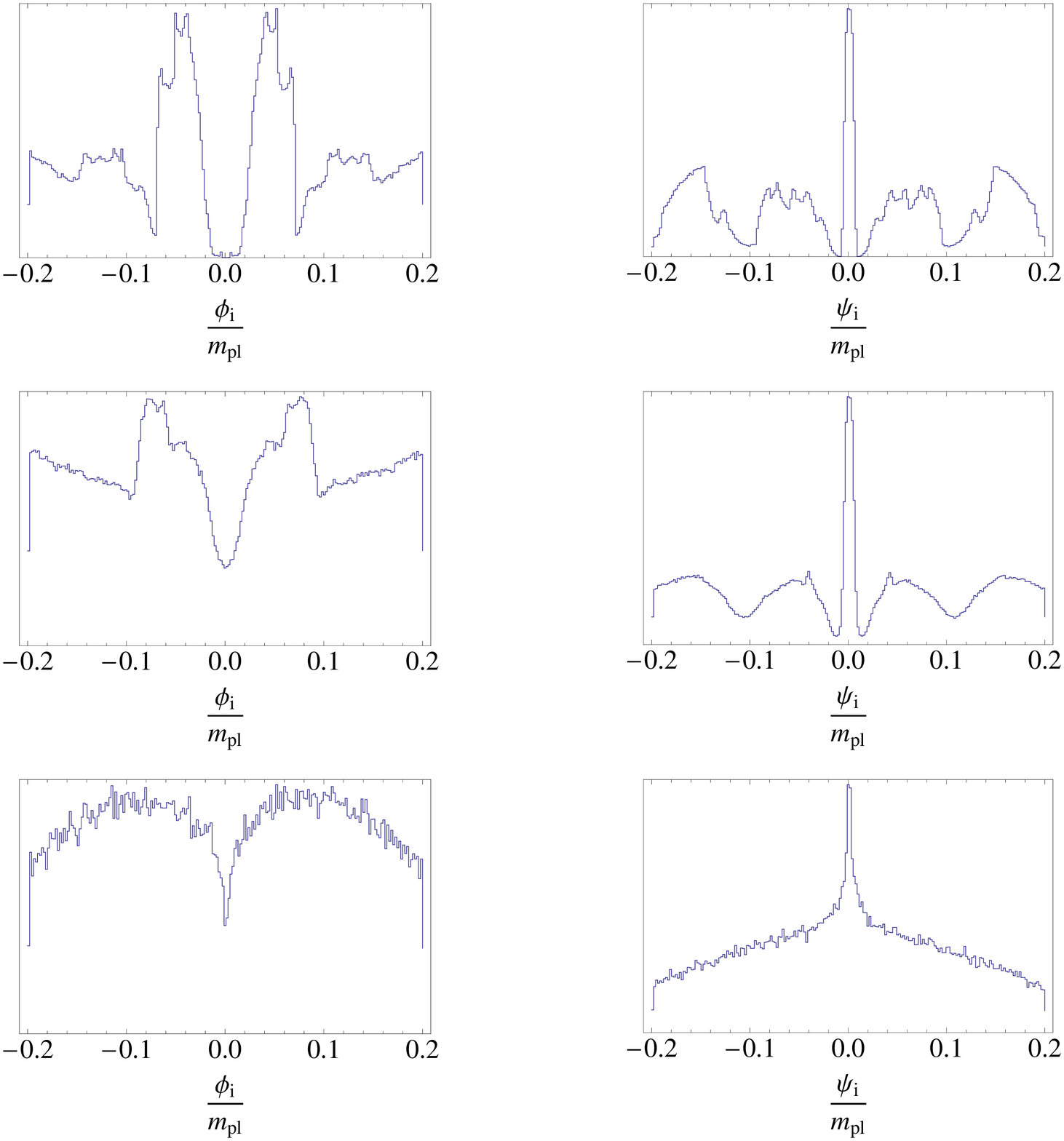}
\caption{Marginalised posterior probability distributions for the
  initial field values $\phi_\ui$ and $\psi_\ui$. The top panels
  correspond to vanishing initial velocities and fixed potential
  parameters, the middle ones are marginalised over velocities at
  fixed potential parameters, while the lower panels are marginalised
  over velocities and all the potential parameters.}
\label{fig:1Dfields}
\end{figure*}

The most interesting part of the exploration by MCMC technique
concerns the study of the full parameter space. The only
restriction being associated to the necessity of $M< \Mpl $ as
discussed in Sec.~\ref{sec:priors}.

We have plotted in Fig.~\ref{fig:2Dfield_all} the marginalised
two-dimensional posterior for the initial field values. In comparison
with Fig.~\ref{fig:2Dfield_nov} and \ref{fig:2Dfield_v}, the most
probable initial field values are now widespread all over the
accessible values; the intricate patterns that were associated with
the successful field values (at fixed potential parameters) are now
diluted over the full parameter space. The resulting one-dimensional
probability distributions for each field are plotted in
Fig.~\ref{fig:1Dfields} (bottom panels). One can observe that the
$\psi$ distribution is nearly flat outside the valley but remains
peaked around a extremely small region around $\psi=0$. Integrating
over the field values, initial conditions outside the valley are still
the preferred case.

Concerning the probability distributions of the modulus $v$ and the
angular direction $\theta$ of the initial velocity vector, results integrated
over the whole parameter space do not present qualitative differences
compared to the posteriors with fixed potential parameters, as one may
expect since the Hubble damping prevents the initial velocities to
influence the dynamics (see Fig.~\ref{fig:1Dspeeds}).

\begin{figure} [] \begin{center}
\includegraphics[width=8cm]{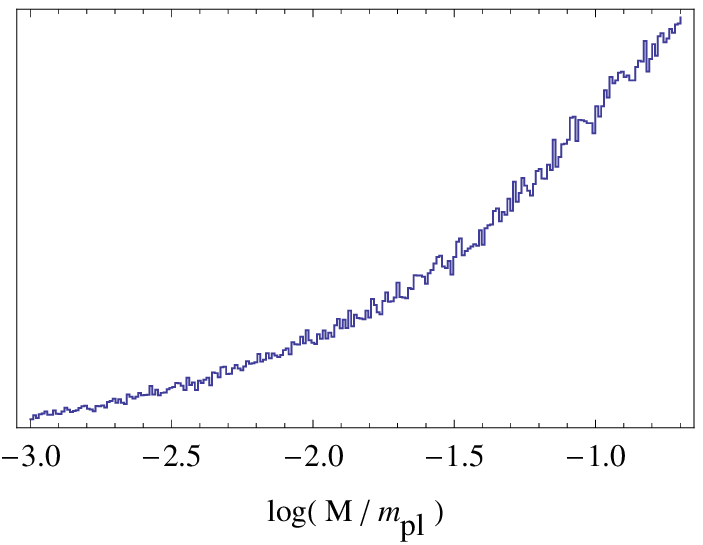}
\includegraphics[width=8cm]{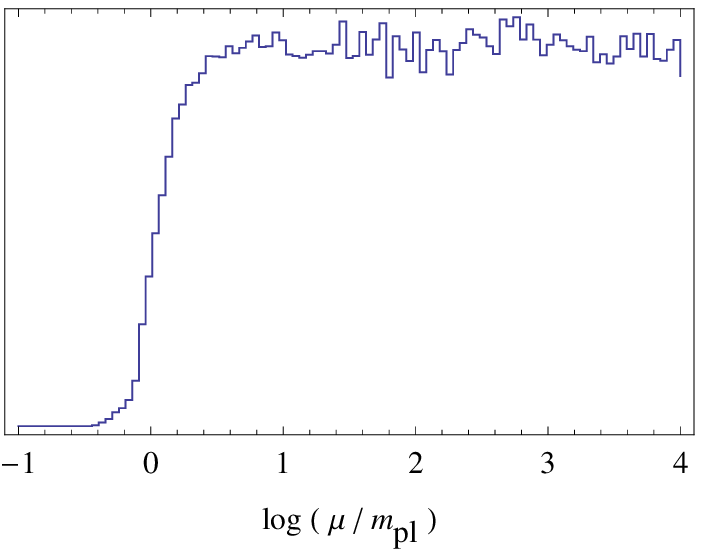}
\includegraphics[width=8cm]{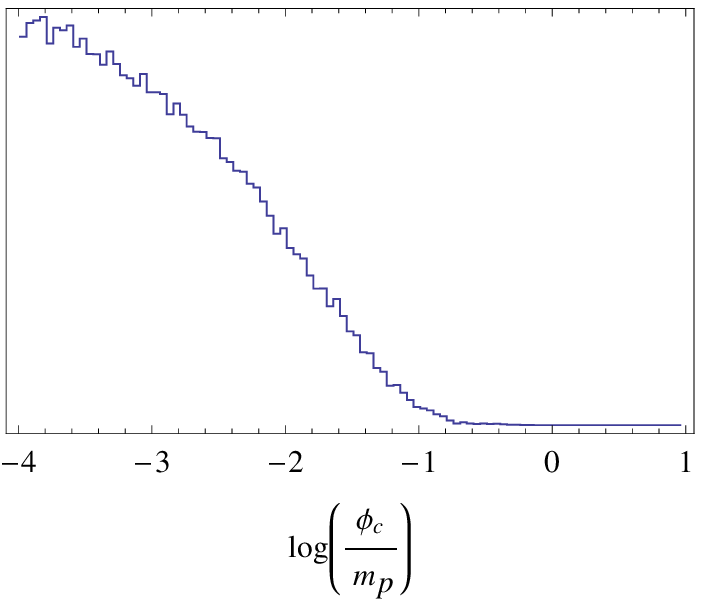}
\caption{Marginalised probability distribution for the potential
  parameters of the hybrid model.   Natural bounds on the parameters $\mu$ and $\phi_{\rr c}$ are observed.}
\label{fig:1Dpot} \end{center}
\end{figure}

The marginalised probability distributions for the potential
parameters are represented in Fig.~\ref{fig:1Dpot}. These
posteriors seem to indicate that two potential parameters are bounded.
The critical instability point $\phi_\uc$ should not be larger than the 
anamorphosis image of the i.c. (the point in the valley where slow-roll 
starts). Restricting initial fields to sub-Planckian values leads to 
an uper bound on the largest image, and thus an upper bound on the 
instability point. 
At $95\%$ of confidence level, we find
\begin{equation} 
\label{eq:critbound}
\phi_{\rr c } < 4\times 10^{-3} \ \mpl.
\end{equation}

The parameter $\mu$ is the other constraint that the MCMC exhibits. It
is explained by the appearance of slow-roll violations in the
valley, when $\mu$ becomes too small. These slow-roll violations
prevent the generation of an inflationary phase if the trajectory
climbs too high in the valley. At a two-sigma level, one has
\begin{equation}
\label{eq:mubound}
\frac{\mu}{\mpl} > 1.7\,.
\end{equation}
Let us stress that these constraints come
only from requiring enough inflation in the hybrid model whatever the
initial field values, velocities, and other potential parameters. In
this respect, the limits of Eqs.~(\ref{eq:critbound}) and
(\ref{eq:mubound}) can be
considered as ``natural''.

To conclude this section, we have shown that inflation is generic in
the context of the hybrid model and we have derived the marginalised
posterior probability distributions of all the parameters such that 60
e-folds of inflation occur. The
original hybrid model under scrutiny is however a toy model. In this respect, one may wonder
whether our results are peculiar to this model or can be generalized
to other more realistic two field inflationary models. This point is
addressed in the next section in which we have performed a complete
study of the SUGRA F-term hybrid inflation. In that model, the
dynamics depends on only one potential parameter; also constrained by
cosmic strings formation. The challenge will thus be to confront this
constraint to the natural bounds that can be deducted from MCMC
methods by requiring enough e-folds of inflation.

\section{Probability distributions in F-SUGRA inflation}
\label{section:fsugrab}

\subsection{Fractal initial field values}

\begin{figure}[] \begin{center}
\includegraphics[width=12cm]{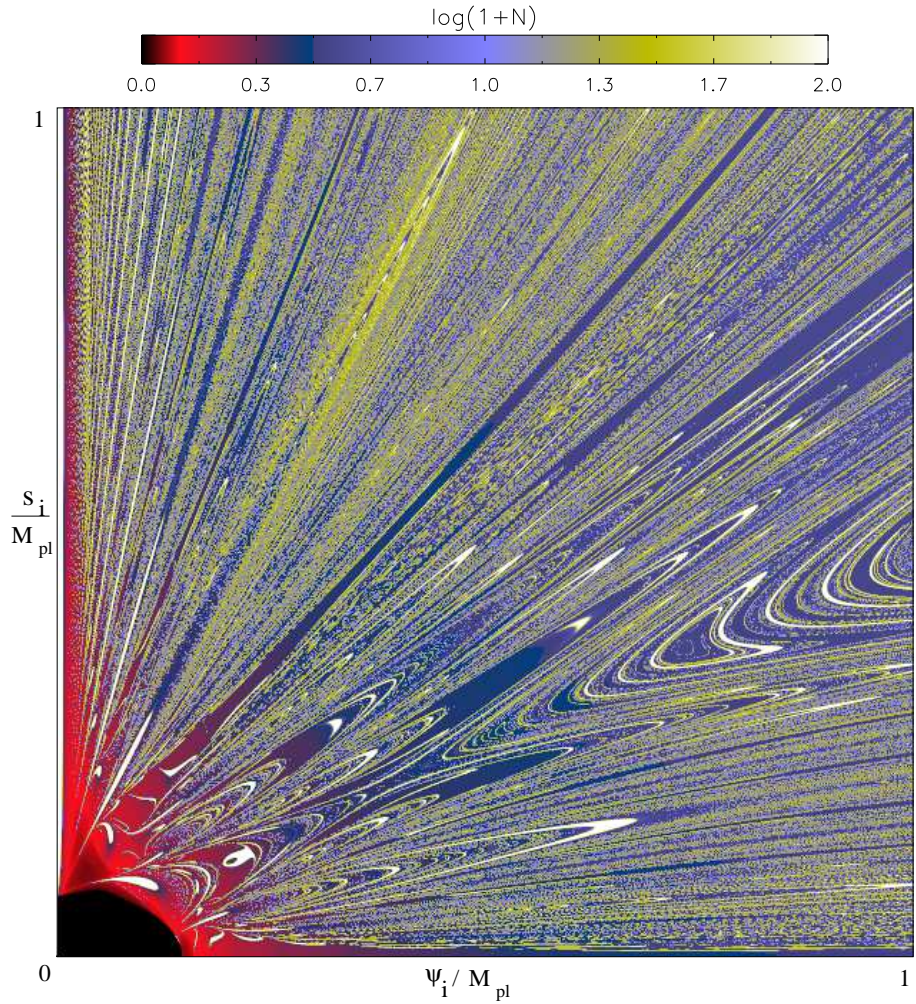}
\caption{Mean number of e-folds obtained from $512^2$ initial
    field values in the plane $(\psi_\ui/\Mpl,\phi_\ui/\Mpl)$, for
  the SUGRA F-term model. This
    figure has been obtained by averaging the number of e-folds
    (truncated at $100$) produced by $2048^2$ trajectories down to
    $512^2$ pixels. The initial field velocities are assumed to
  vanish and the relevant potential parameter is fixed at $M=10^{-2}\mpl$. As
  for the original hybrid model, we recover a set of dimension two
  with a fractal boundary.} \label{fig:gridSUGRA}
 \end{center}
\end{figure}

The analysis of the SUGRA F-term model of inflation, described in Chapter 3,  has been conducted
along the lines described in Sec.~\ref{sec:fractalic} and
Sec.~\ref{section:mcmc}. We have first verified that, at fixed
potential parameter $M$ and vanishing initial velocities, the set of
initial field values $\calS$ defined by Eq.~(\ref{eq:Sdef}) is
two-dimensional with a fractal boundary. In Fig.~\ref{fig:gridSUGRA},
we have represented the set $\calS$ of successful initial field
values for the mass scale $M=10^{-2}\mpl$. Notice that the coupling
constant $\coupling$ being an overall factor, it does not impact the
dynamics of the fields. Our study is therefore valid for any value of
$\coupling$ and of the dimensionality of the Higgs field $\mathcal{N}$, 
since the relationship $M(\coupling)$ depends only on $\mathcal{N}$.

\begin{table}
\begin{center}
\begin{tabular}{|l|l|}
\hline Values of $M$ & Area of $\calS$ (\%)
\\\hline\hline
$M=10^{-1}~\mpl$  &   0 (exact)\\
$M=10^{-2}~\mpl$ & $12.9 \pm 0.1$\\ 
$M=10^{-3} \mpl $ & $12.0 \pm 0.3$ \\ 
$M=10^{-4} \mpl $ & $10.3 \pm 0.5$ \\ 
\hline
\end{tabular}
\caption{Percentage of successful initial field values, at vanishing
  initial velocities, for various values of the potential parameter
  $M$. The error bars come from the finite numerical precision, 
  which decreases with $M$.}
\label{tab:successFterm}
\end{center}
\end{table}

For vanishing initial velocities, we have reported in
Table~\ref{tab:successFterm} the area occupied by the set $\calS$ in
the plane $(s_\ui,\psi_\ui)$ for various sections along the potential
parameter $M$. Like for the original hybrid model, we recover a
significant proportion of successful initial field values outside the
valley. This result holds even for $M\ll 1$ though at small M, the
potential becomes very flat and the number of oscillations of the
system before being trapped in the inflationary valley can exceed
$10^3$. Simulations become therefore more time-consuming and
error-bars in Tab.~\ref{tab:successFterm} increase. Reducing $M$ also
reduces the typical size of structures in the plane
$(s_\ui,\psi_\ui)$, which evolves from Fig.~\ref{fig:gridSUGRA} to a
more intricate space of thinner successful i.c.. As suggested by the
Tab.~\ref{tab:successFterm}, we will see below that this does not
affect the probability of getting inflation by starting the field
evolution outside the valley.

\begin{figure}[] \begin{center}
\includegraphics[width=12cm]{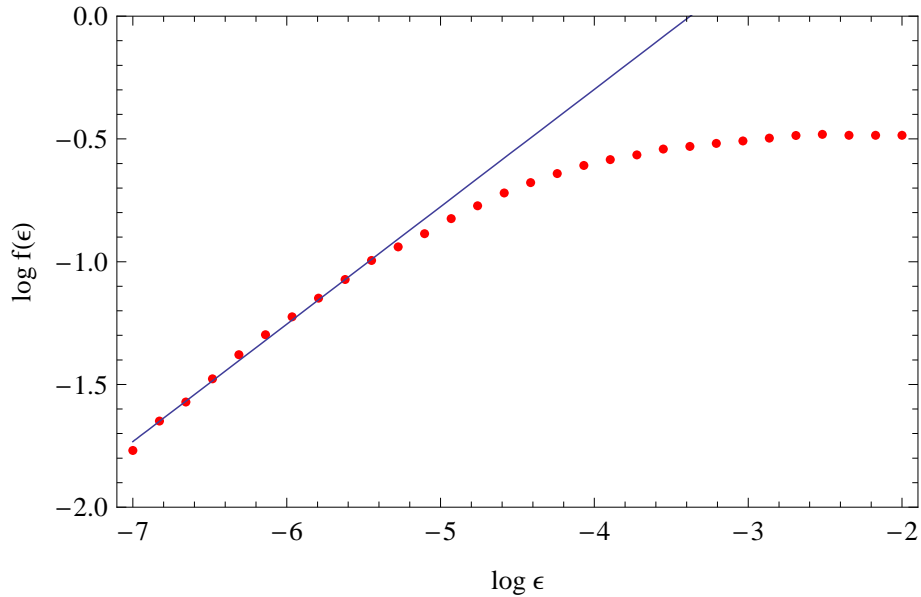}
\caption{Fraction of initial field values in a $\delta$-sized box
    intercepting the set $\calS$ as a function of $\delta$ for the
    SUGRA F-term model. The potential parameter has been fixed to
    $M=10^{-2} \mpl$. The box-counting dimension of boundary of 
    $\calS$ is
    given by the power law behaviour for small $\delta$ and found to
    be $D_\uB\simeq 1.5$.}
\label{fig:dim_fractale_SUGRA}  \end{center}
\end{figure}

Concerning the fractal properties of $\calS$, we have applied the same
method as in Sec.~\ref{sec:boxcountingdim} to compute the box-counting
dimensions of $\calS$ and its boundary. As expected, we recover that
$\calS$ is of box-counting dimension two whereas the function
$f(\epsilon)$ for its boundary is represented in
Fig.~\ref{fig:dim_fractale_SUGRA}. We obtain that, as in the non-SUSY
case, the boundaries are fractal with dimension
\begin{equation}
D_\uB \simeq 1.5~.
\end{equation}
These results allow us to use the usual Lebesgues measure to define
the probability distribution over the whole parameter space.

\subsection{MCMC on initial field values, velocities and potential 
parameters}

\begin{figure}
\includegraphics[width=16cm]{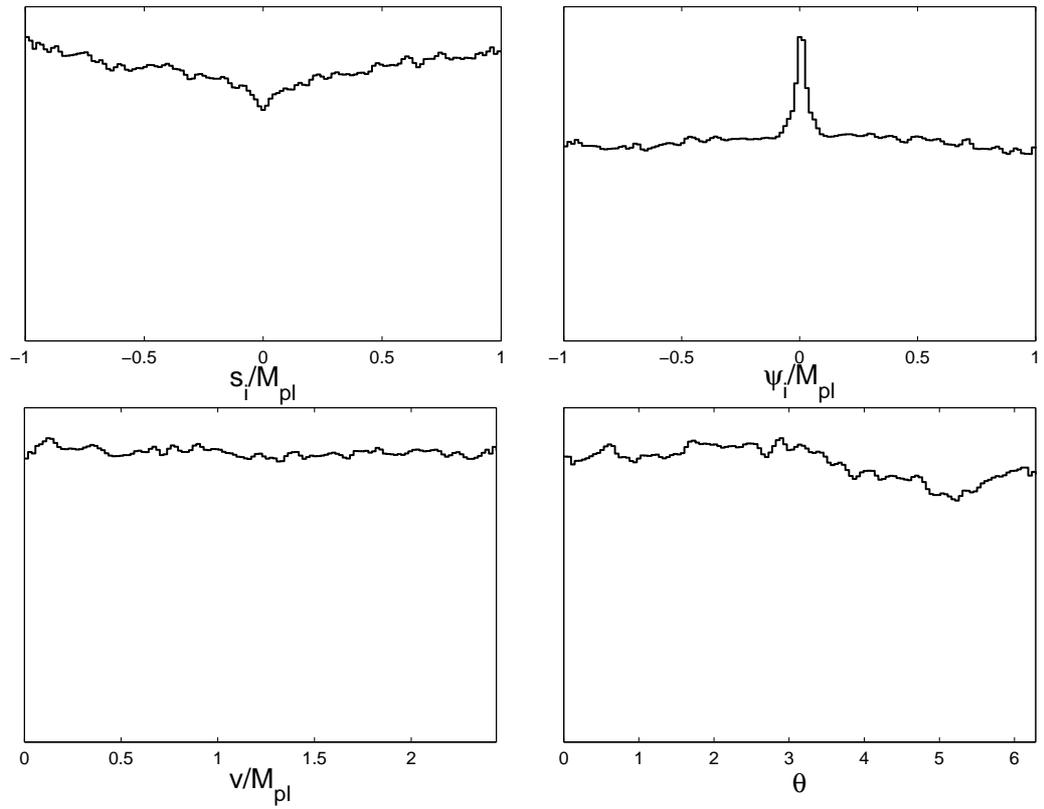}
\caption{Marginalised posterior probability distributions for the
  initial field values (upper panels) and the initial velocities,
  modulus $v$ and angle $\theta$. The F-SUGRA inflationary valley has
  a slightly higher probability density around $\psi=0$ but is
  extremely localized: as a result, inflation is more probable by
  starting out of the valley.}
\label{fig:fsugra1Dfields}
\end{figure}

As already mentioned, there is only one potential parameter $M$ in
F-term SUGRA model that may influence the two field dynamics. The goal
of this section is to evaluate the probability distributions of the
initial field values, velocities and of $M$ such that inflation lasts
more than $60$ e-folds. As for the original hybrid model, we have
performed a MCMC analysis on the five-dimensional parameter space
defined by $s_\ui$, $\psi_\ui$, $v$, $\theta$ and $M$ where
\begin{equation}
\left. \dfrac{\ud s}{\ud N}\right|_\ui = v \cos \theta, \quad
\left. \dfrac{\ud \psi}{\ud N}\right|_\ui = v \sin \theta.
\end{equation}

We have chosen the same sub-planckian priors for the initial field
values and initial velocities than in Sec.~\ref{section:mcmc}. Since
the order of magnitude of $M$ is not known, we have chosen a flat
prior on the $\log(M/\Mpl)$ over the range $[-2,0]$. The lower limit
on $M$ is motivated by computational rather than physical
considerations. The resulting marginalised posterior probability
distributions for each of the parameters are represented in
Fig.~\ref{fig:fsugra1Dfields}. The chains contain $400000$ samples
producing an estimated error on the posteriors around a few percent
(from the variance of the mean values between different chains).

\begin{figure}[h] \begin{center}
\includegraphics[width=10cm]{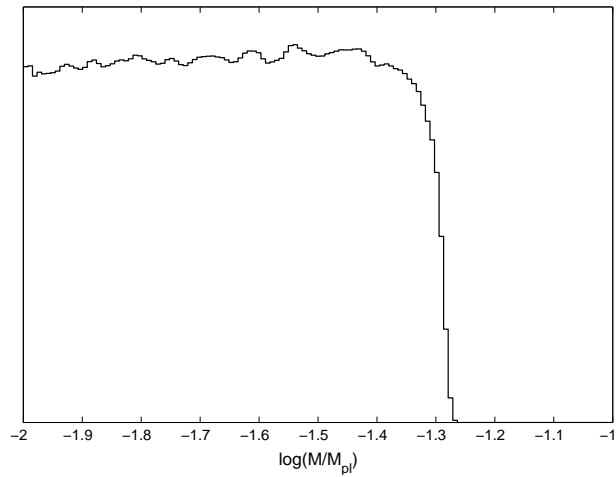}
\caption{Marginalised posterior probability distribution of the mass
  scale $M$ of F-SUGRA inflation.}
\label{fig:fsugraM} \end{center}
\end{figure}

The posteriors for the field velocities are flat showing that all
values are equiprobable to produce inflation. The initial field values
are also flat, up to a sharp peak of higher probability density around
$\psi=0$ corresponding to the inflationary valley. As for the hybrid
model of Sec.~\ref{sec:fractalic}, after integration of these curves
over the field values, inflation is clearly more probable by starting
out of the valley. Finally, only the posterior probability
distribution of $\log M$ is strongly suppressed at large values. We
find, at $95\%$ of confidence level
\begin{equation}\label{eq:M2sig}
\log(M) < -1.33\,.
\end{equation}
As for the original hybrid model, this limit comes from the condition
of existence of a sub-planckian inflationary valley which is related
to the position of the instability point. Indeed, from
Eq.~(\ref{VFSUGRA}), one finds
\begin{equation}
  \left. \dfrac{\ud^2 V^{\mathrm{SUGRA}}_{\mathrm{tree}}}{\ud
      \psi^2 }\right|_{\psi=0} = 0 \Rightarrow s=s_\uc=\pm \dfrac{M}{\Mpl}\sqrt{1-\sqrt{1-4\dfrac{M^4}{\Mpl^4}}}\,,
\end{equation}
where we have kept only the sub-planckian solutions. This expression
shows that there is an inflationary valley at $\psi=0$ only for $M/\Mpl
< 1/\sqrt{2}$, and for field values such that $s>s_\uc$. As a result of
the two-field dynamics, we find that a valley supporting at least $60$
e-folds of inflation require the more stringent bound of
Eq.~(\ref{eq:M2sig}). Let us finally notice that the most probable
values we obtain on $M$ to get inflation in Eq.~(\ref{eq:M2sig}) are
compatible with the existing upper bound coming from cosmic strings
constraint: $M \lesssim 10^{-3} \mpl$ (see
Ref.~\cite{Rocher:2004et,Jeannerot:2005mc}).

\section{Conclusion and discussion}

\label{sec:conclu}


In this chapter, we have numerically integrated the exact equations of motion of
the fields and analysed the space of initial
conditions in hybrid inflation.   The study has been conducted for the five different models
of hybrid-type inflation, introduced in Chapter 3: the original
non-supersymmetric model, its F-term supersymmetric version in supergravity, its
extensions ``smooth'' and ``shifted'' hybrid inflation in global
supersymmetry and supergravity and the ``radion assisted'' gauge
inflation.  



For the original hybrid model, instead of fine-tuned along the inflationary valley~\cite{Mendes:2000sq}, the set of the sub-planckian initial field values leading to more than 60 e-folds of inflation is found to occupy a non negligible part of the field space exterior to the valley.  
These initial conditions 
correspond to special trajectories for which the velocity in the field
space becomes oriented along the inflationary valley after some
oscillations at the bottom of the potential. Therefore the system
climbs up the valley before slow-rolling back down, generating
enough inflation.


In fact, the inflationary
valley, indeed of small extension in field space, is one of the three
dynamical attractors of the differential system given by the Einstein
and Klein--Gordon equations in a FLRW universe (the others being the
minima of the potential). As a result, any trajectory will end in one
of these three attractors and the set $\calS$ of successful initial
conditions therefore belongs to the basin of attraction of the
inflationary valley.  This set forms a complex
structure, as represented in Fig.~\ref{fig:anamorphosis}.  We have shown that such a set is connected and of
dimension two while exhibiting a fractal boundary of dimension greater
than one.  

For specific potential parameter values and vanishing initial field velocities, 
the relative area in the field space that this set
occupy is typically of order of $15\%$ for the original
hybrid model. This value can go up to $25\%$ for radion assisted gauge inflation
and even above $70\%$ for smooth inflation, even though these
results depend on the values of the parameters of the potential
(see Tab.~\ref{tab:anamorph}). Moreover, even when supergravity
corrections are included, these trajectories still exist and their
proportion stays similar.  We would like to note that these
percentages allow us to claim that the fine-tuning on hybrid
inflation is much less severe that what was found in the past. 

 In order to quantify what are the natural
field and parameter values to get inflation for both of these models,
we have introduced a probability measure and performed a MCMC
exploration of the full parameter space. It appears that the
inflationary outcome is independent of the initial field velocities,
is more probable when starting out of the inflationary valley, and
favors some ``natural'' ranges for the potential parameter values
that cover many order of magnitudes. The only constraints being that
the inflationary valley should at least exist.
Let us notice that the posterior probability distributions we have
derived are not sensitive on the fractal property of the boundary of
$\calS$. This is expected since, even if fractal, the boundary remains of
null measure compared to $\calS$. However, its existence may have
implications in the context of chaotic eternal
inflation~\cite{Linde:1986fd, Guth:2000ka}. Indeed, the boundary
itself leads to inflation and spawn the whole field space such that
its mere existence implies that inflationary bubbles starting from
almost all sub-planckian field values should be produced. Here, we
have been focused to the classical evolution only and our prior
probability distributions have been motivated by theoretical prejudice
(flat sub-planckian prior). In the context of chaotic eternal
inflation, our results are however still applicable provided one uses
the adequate prior probabilities which are the outcome of the
super-Hubble chaotic structure of the
universe~\cite{Linde:2008xf}. Provided the eternal scenario does not
correlate with the classical dynamics, one should simply factorise the
new priors with the posteriors presented here to obtain the relevant
posterior probability distributions in this context.

Finally, let remind that the inflationary trajectories have only been considered for the first 60 e-folds of inflation.  But when the fields reach the valley, inflation continues until the critical instability point $\phi_{\rr c}$ is reached.  From this point, they deviate from the valley direction and fall through one of the global minima of the potential.  To study such a waterfall phase, the 2-fields dynamics is also required to be integrated.  This is done in the next chapter, for the original hybrid scenario.


%% file: waterfall_phase.tex
\chapter{The waterfall phase }
\label{chap:waterfall}

\begin{center}
\textit{based on}\\
S. Clesse, \\
Hybrid inflation along waterfall trajectories \\
Phys.Rev.D83:063518, 2011, arXiv:1006.4522  \\
\end{center}

As already discussed earlier in this thesis, in the 1-field slow-roll approximation, the scalar power spectrum for the hybrid model, for inflation at small field values, exhibits a slight blue tilt, which is disfavored by WMAP7 observations ~\cite{Larson:2010gs}.   
Indeed, in the small field regime ($\phi \ll \mu$), $ \epsilon_{\rr 1}$ is extremely small and the effective potential curvature is positive such that $\epsilon_{\rr 2}$ is negative and is the dominant contribution to the spectral tilt.  This result also relies on the assumption that inflation stops nearly instantaneously in a waterfall phase whereas tachyonic preheating is triggered due to the exponential growth of the tachyonic field perturbations.  

In this chapter, we discuss the validity of the last assumption.  So we do not assume an instantaneous end of inflation at the critical instability point.  We show that in a large part of the parameter space, the last $60$ e-folds of inflation, relevant for the calculation of the observable scalar power spectrum, are actually realized in a non-trivial way during the waterfall phase, after crossing the instability point.  

Tachyonic preheating is not triggered during this phase because the effective mass of the auxiliary field is small compared to the expansion term.  
More precisely,  the exponential growth of tachyonic modes is avoided because the Hubble expansion term is dominating the equation governing the linear perturbations of $\psi$.

In this chapter, the classical dynamics is investigated both numerically and analytically by using the adiabatic field formalism \cite{Gordon:2000hv}.  The whole potential parameter space is explored using a Monte-Carlo-Markov-Chains (MCMC) method.  Regions for which much more than 60 e-folds are realized after instability are shown to be generic.    In such cases, observable modes leave the Hubble radius during the waterfall and a modification of the predicted scalar spectral index is expected.  For adiabatic perturbations, the power spectrum is actually generically red.  

The potential is very flat near the critical instability point and quantum backreactions could dominate the classical dynamics.  Therefore, a particular attention has been given to consider only trajectories not affected by quantum stochastic effects.   These comprise both the quantum backreactions of the adiabatic and the entropic transverse field.  The classical evolution is valid only if it is not affected by field quantum jumps in the longitudinal direction and if the quantum diffusion of the transverse field \cite{GarciaBellido:1996qt}  does not increase too much the spread of the transverse field probability distribution.



If inflation continues during the waterfall phase for a low number of e-folds ($N <  60$), some problems become evident \cite{GarciaBellido:1996qt}.   Inflating topological defects can induce large-scale perturbations and primordial black holes can be formed after inflation.  When the waterfall phase is much longer, we argue that these problems are naturally avoided.  Indeed, topological defects are so strongly diluted by inflation during the waterfall that they do not affect the observable Universe.  Primordial black holes are expected to form when fractional density perturbations occurring at the phase transition reenter the horizon\footnote{Let nevertheless mention that primordial black holes could be formed during the exponential growth of the waterfall field, as discussed recently in Ref.~\cite{Lyth:2011kj}.}.  Thus they affect the observable universe only if inflation after the critical instability point $\phi_{\rr c} $ lasts much less than typically $60$ e-folds.  This is not the case here.

The chapter is organized as follows:    Section 1 is dedicated to the dynamics inside the valley, before the critical instability point is reached.   It is shown that classical oscillations of the waterfall field are quickly dominated by its quantum fluctuations.  Section 2 concerns the waterfall phase description in the usual fast approximation and the resulting tachyonic preheating phase.  In section 3, we show for some sets of parameters that much more than $60$ e-folds can be realized classically along a waterfall trajectory, i.e. after crossing the critical instability point.   In section 4 the generic character of this effect is studied.  The dependences on the potential parameters and initial conditions are determined by using a MCMC method.  At the end of the chapter, important implications  for hybrid models (e.g. on the formation of topological defects) are discussed.


\section{Field dynamics before instability}

Given an arbitrary set of initial conditions, two classical behaviors are possible.  As shown in the previous chapter, either the trajectory falls through one of the global minima of the potential without inflating.  Either it reaches the nearly flat valley along the $\psi = 0 $ direction.  When the valley is reached, field trajectories are first characterized by damped oscillations in the transverse direction (orthogonal to the valley).  After some oscillations, the slow-roll regime is triggered and a large number of e-folds is realised along the valley.   

At the critical point of instability $\phi_{\rr c}$, only a small transverse displacement allows inflation to end with a waterfall phase.  The two competing processes able to cause this displacement from the $\psi = 0$ valley line are the remaining classical transverse oscillations and the quantum fluctuations of the auxiliary field.   In this section, it is shown that classical oscillations own generically an amplitude so small that the dynamics is dominated by the quantum fluctuations of the auxiliary field.

Quantum fluctuations are typically of the order $\Delta \psi_{\rr{qu}} \simeq H /2 \pi$.
The primordial nucleosynthesis fixes an extreme lower bound on the energy scale of inflation, and thus on the Hubble rate $H_{\rr{end}}$ at the end of inflation through the Friedmann-Lema\^itre equations Eqs.~(\ref{eq:FLtc12field})  ($H \gsim 10^{-35} \mpl $).  On the other hand, measurements of the primordial scalar power spectrum amplitude,
 \begin{equation}
\mathcal P(k=0.002/\rr{Mpc})\simeq 2.43 \times10^{-9} = \frac{H_* ^2} {\pi \mpl^2 \epsilon_{\rr 1 *}}~, 
\end{equation}
with $ \epsilon_{\rr 1} \lesssim 0.1 $~\cite{Martin:2010hh},  allow to fix a higher bound on $H_*$.  For the original hybrid model, if inflation is realized in the false vacuum ($\phi \ll \mu$), one has typically $H \simeq H_* \simeq H_{\rr{end}}$.   One can thus determine the range of transverse quantum oscillations, $10^{-35} \mpl \lesssim \Delta \psi_{\rr{qu}} \lesssim 10^{-6} \mpl $.


In the regime of small classical oscillations $\psi \ll M$, at inflaton values sufficiently larger than the critical one but still in the small field phase, that is $\phi_{\rr c} \ll \phi \ll \mu$, the potential is well approximated by
\begin{equation}
V(\phi,\psi) \simeq \Lambda^4 \left[ 1 + \frac{\phi^2}{\mu^2} + 2 \frac{\phi^2 \psi^2 }{\phi_{\rr c}^2 M^2}  \right] \ .
\end{equation}

We can assume that the inflaton field is slow-rolling along the valley $\psi = 0$, such that  
 \begin{equation} \label{eq:Hevol}
H \simeq \frac{1}{\sqrt 3 \Mpl} \sqrt{\Lambda^4 (1+ \phi^2 / \mu^2)} \ .
\end{equation}   
In a short time scale, the inflaton $\phi$, and thus the Hubble parameter $H$, can be assumed to be constant.
As a consequence, the Klein-Gordon equation for the auxiliary field now read
\begin{equation}
\ddot \psi + 3  \dot \psi \frac{1}{\sqrt 3 \Mpl} \sqrt{\Lambda^4 (1+ \phi^2 / \mu^2)} + \frac{4 \Lambda^4 \phi^2} {\phi_{\rr c} ^2 M^2} \psi = 0 \ .
\end{equation}
It has a simple oscillating solution with exponentially decreasing amplitude
\begin{equation}
\begin{aligned}
\psi(t)  & =  \ \rr e ^{-  \frac{3}{2 \sqrt 3 \Mpl} \sqrt{\Lambda^4 (1+ \phi^2 / \mu^2)} t}  \\ 
 & \left[ C_{\rr 1} \  \rr e^{-\frac 3 2 \frac{1}{\sqrt 3 \Mpl} \sqrt{\Lambda^4 (1+ \phi^2 / \mu^2)} t \sqrt{1-16 \Mpl^2 \phi^2 / (3 \phi_{\rr c}^2 M^2 ) } } \right.    \\
  & \left.   + C_{\rr 2} \  \rr e ^{ \frac{3}{2 \sqrt 3 \Mpl} \sqrt{\Lambda^4 (1+ \phi^2 / \mu^2)} t \sqrt{1- 16 \Mpl^2 \phi^2 / (3 \phi_{\rr c}^2 M^2) } }  \right] , 
\end{aligned}
\end{equation}
where $C_{\rr 1} $ and $C_{\rr 2}$ are two integrating constant fixed by initial conditions.  As an example, it takes about $ N \sim 40 $ e-folds
for initial oscillations of amplitude $A\simeq  10^{-3} \mpl $ to be reduced by a factor of $10^{32}$ at the minimal level of possible quantum fluctuations.   

Given this time scale, the assumption that $\phi$ is constant can be justified a posteriori.  Indeed, in the slow-roll approximation, straightforward manipulations give 
\begin{equation} \label{eq:Ndephi}
N(\phi) = \frac{\mu^2}{4 \Mpl^2} \left[ \left(\frac{\phi_{\rr i}}{\mu} \right)^2 - \left( \frac{\phi}{\mu}  \right)^2 - 2 \ln \left( \frac{\phi}{\phi_{\rr i}} \right) \right],
\end{equation}
where $\phi_{\rr i}$ is the initial inflaton value.  Therefore, if $\phi \ll \mu$, classical oscillations become dominated by quantum fluctuations in a range of inflaton value 
\begin{equation}
\frac{\Delta \phi }{ \phi} \simeq - \rr e^{- \frac{2 \Mpl^2}{\mu^2} N_{\rr {qu} } }~,
\end{equation} 
where $N_{\rr{qu}} $ is the required number of e-folds for the classical oscillations to be dominated by the transverse field quantum fluctuations.  $\Delta \phi $ is typically very small for a nearly flat valley.   
Thus the oscillations of $\psi$ are expected to be dominated by quantum fluctuations after a very small range of variation for $\phi$.    

To go beyond this approximation and determine, for the full potential, how generic are trajectories whose classical oscillations become smaller than quantum fluctuations,  the classical 2-field dynamics have been integrated numerically.  We have followed the method used in~\cite{Clesse:2009ur} and run an identical Monte-Carlo-Markov-Chains analysis
of the 7D space of initial field values, initial velocities and potential parameters.   The result of this analysis is that around $99.9 \% $ of trajectories trapped inside the valley perform transverse oscillations whose amplitude is below the most restrictive level of transverse quantum fluctuations $\Delta \psi_{\rr{qu}} \sim 10^{-35} \mpl$.  

From these considerations, at instability, the slight transverse displacement essential for a waterfall phase to take place is not supplied by classical oscillations of the auxiliary field but by its quantum fluctuations.  In sections 6.3 and 6.4, the waterfall phase will be studied classically taking initial values $\phi_{\rr i} = \phi_{\rr c}$ and $\psi_{\rr i} \simeq \Delta \psi_{\rr {qu}} $.   We will assume initial field velocities given by the slow-roll approximation.   Actually, due to the slow-roll attractor, different choices of initial velocities only marginally influence the resulting waterfall dynamics.    

\section{Fast waterfall phase} \label{sec:fast_waterfall}

\subsection{Linear perturbation theory}

Since the auxiliary field is well anchored at its minimum $\psi = 0$ before the waterfall (up to quantum fluctuations), it can be regarded as the same as its fluctuation, $\psi = \delta \psi$.  On the other hand, one can assume that the $\phi$ field evolves independently according to Eq.~(\ref{eq:Ndephi}). By Fourier expanding $\delta \psi$ and neglecting non linear terms, one obtains the mode evolution equation
\begin{equation} \label{eq:mode_waterfall}
\delta \ddot \psi_{\bf{k}} + 3 H \delta  \dot \psi_{\bf{k}} + \left[ \frac{k^2}{a^2} - m^2(\phi) \right] \delta \psi_{\bf{k}} = 0 ~,
\end{equation}
where $m (\phi) $ is the time-dependent mass of the tachyonic field $\psi$.   If inflation is driven by the field $\phi$ in the false vacuum, $H$ is nearly constant. In a specific case where the mass would be constant,  long wavelength modes $k / a \ll  m $ would grow like
\begin{equation}
\delta \psi_{\bf{k}} \propto  \rr e^{-\frac 3 2 H t \left( 1 - \sqrt{1  + 4 m ^2 / 9 H^2 }  \right)}
\end{equation}
If the expansion term is neglected ($H\rightarrow 0$), long wavelength modes therefore grow exponentially $\propto \exp (m t) $ such that the linear regime becomes quickly inappropriate.   However, if $4 m^2 \ll 9 H^2 $, the long wavelength modes evolve as 
$\delta \psi_{\mathbf k} \propto \exp[- H t m^2 / (3 H^2) ]$.  Their growth thus only explodes when $N \simeq H t > 3 H^2 / m^2 \gg 1$.  In this case, the number of e-folds created in the linear regime can thus be very large.  In the hybrid model, since the mass is vanishing around the instability point, such a phase can exist, but the amount expansion during it is usually neglected and only the growing phase is considered.  In the next section, it will be shown that inflation can continue for more than 60 e-folds during the waterfall.

\subsection{Tachyonic preheating}

The linear perturbation theory is only adequate to describe the first stage of the waterfall phase.   Since long wavelength quantum fluctuations are exponentially growing, they can be interpreted as classical waves of the scalar field.  The spontaneous symmetry breaking occurs when the amplitude of these fluctuations reaches the scale $\sqrt{\langle \delta \psi^2 \rangle} \sim M$ and leads to the formation of topological defects.  The inhomogeneities of the scalar field absorb some part of the initial energy density such that the expected coherent field oscillations are suppressed in only one or a few oscillations.   This process of rapid energy transfer of the homogeneous scalar field into the energy of inhomogeneous oscillations was called \textit{tachyonic preheating}~\cite{Felder:2000hj,Felder:2001kt}.   

Studying the tachyonic preheating phase requires numerical lattice simulations.  Such simulations have been performed for the original hybrid model~\cite{Felder:2000hj,Felder:2001kt,Copeland:2002ku,Barnaby:2006cq}.  However, these neglect expansion or assume that no more than only a few e-folds are realized during the waterfall~\cite{Barnaby:2006cq}.  These results could not stand if inflation continues efficiently after the field trajectories have reached the instability point.

\section{Hybrid inflation along waterfall trajectories}

Before studying the waterfall phase, it must be verified that the classical dynamics is valid and not spoiled by quantum backreactions of both the adiabatic and the entropic fields.  

\subsection{Quantum backreactions}


The collective evolution of the fields can be described by the adiabatic field $\sigma$, defined in Eq.~(\ref{eq:adiabfield}).  It is very light and its classical evolution is valid if the classical evolution is larger than the quantum fluctuation scale, that is
\begin{equation}
\Delta \sigma _{\rr{cl}} = \frac{\dot \sigma}{H} > \Delta \sigma_{\rr{qu}} \simeq \frac{H}{2 \pi} \ .
\end{equation}
which is the case when 
\begin{equation} 
\epsilon_{\rr 1} (\sigma) > \frac{H^2}{\pi \mpl^2} \ .
\end{equation}
Thus we pay a particular attention to only consider waterfall trajectories along which this condition remains true.  

At the critical point $\phi_{\rr c}$, the classical value of the transverse field is about 0, and the transverse quantum fluctuations will determine on which side the system will evolve towards. The overall dynamics remains classical due to the $\phi$ field evolution.  However, it must be ensured that the quantum backreactions of the transverse field do not push the field evolution far from the valley line $\psi = 0$.  Such effects would modify strongly the dynamics such that the waterfall phase would take place in a low number of e-folds.  In other words, it must be ensured that the spread of the probability distribution of the auxiliary field does not become much larger than its classical value during the waterfall.

 The coarsed-grained auxiliary field can be described by a Klein-Gordon equation in which a random noise field $\xi (t)$ is added~\cite{Vilenkin:1983xp}.  This term acts as a classical stochastic source term.  In the slow-roll approximation, the evolution is given by the first order Langevin equation
\begin{equation}
\dot \psi + \frac{1}{3 H} \frac{\dd V}{\dd \psi}= \frac{H^{3/2}}{2 \pi} \xi(t) ~.
\end{equation}
which can be rewritten
\begin{equation}
\dot \psi = \frac{H^{3/2}}{2 \pi} \xi(t) + H \frac{4 \psi \Mpl^2}{M^2} \left( 1- \frac{\phi^2}{\phi_{\rr c} ^2}\right) ~.
\end{equation}
The two-points correlation function of the noise field obeys 
\begin{equation}   
\langle \xi(t) \rangle = 0,  \hspace{5mm} \langle \xi(t) \xi(t') \rangle = \delta (t-t') ~.
\end{equation}
When the expansion is governed by the evolution of $\phi$ in the false vacuum, one has $H^2 \propto \Lambda^4 \left[ 1 + \mathcal O(\phi^2 / \mu^2) \right]$ with $\phi \ll \mu$.  In the limit of $H$ constant, this equation can be integrated exactly.  Under  a change of variable \cite{GarciaBellido:1996qt}, $x\equiv \exp \left[-2 r (N-N_{\rr{c}}) \right]$, where  
\begin{equation}
r\equiv \frac 3 2 - \sqrt{\frac 9 4 - 6 \frac{\Mpl^2}{ \mu^2}}~,
\end{equation}
and where $N_{\rr c} $ is the number of e-folds at the critical point $\phi_{\rr c}$, one has
\begin{equation}
\frac{\dd \psi}{\dd x} = - \frac{H^{1/2} } { 4 \pi r x} \xi(x) - \frac {4 \psi \Mpl^2 (1-x)}{2 M^2 r x}~.
\end{equation}
This equation has an exact solution
\begin{equation}
\begin{aligned}
\psi(x) & =  C \exp \left( C_2 x - C_2 \ln x  \right) \\ 
 & - C_1 \exp \left( C_2 x - C_2 \ln x \right)  \\ 
 & \times \int_1^x \exp \left( -C_2 x' + C_2 \ln x' \right) \xi(x') \dd x' ~,
\end{aligned}
\end{equation}
where $C_1 \equiv H^{1/2} / ( 4 \pi r)  $, $C_2 \equiv 2/  (M^2 r)   $ and $C$ is a constant of integration.  Taking the two point correlation function and assuming an initial delta distribution for $\psi$ at $\phi \gg \phi_{\rr c}$, one obtains 
\begin{equation} \label{eq:psi_qudist}
\langle \psi^2(x) \rangle = \frac {H^2}{8 \pi^2 r} \left[ \frac{\exp (x)}{ a x}  \right]^a  \Gamma(a,a x) \ ,
\end{equation}
where $a\equiv 4 \Mpl^2 /(M^2 r)$ and  $\Gamma$ is the upper incomplete gamma function.

In the following, we will consider large values of $\mu$ and relatively small values of $M$ compared to the Planck mass, such that $ r \simeq 2 \Mpl^ 2 / \mu^2$ and $a\simeq 2 \mu^2 /M^2 \gg 1$.  
At instability, $x = 1$ and one thus has
\begin{equation}  \label{eq:qufluct}
\langle \psi^2(x=1) \rangle \simeq \frac {H^2 \mu^2}{16 \pi^2 \Mpl^2} \left( \frac{\rr e M^2}{ 2 \mu^2}  \right)^{\frac{2 \mu^2}{ M^2}}  \Gamma \left(\frac{ 2 \mu^2}{M^2} , \frac{ 2 \mu^2}{M^2}  \right).
\end{equation}
By using recurrence relations as well as the asymptotic behavior of the $\Gamma$ function, one can find
\begin{equation}
\left(  \frac{\rr e}{u} \right)^u \times \Gamma(u,u) \sim \sqrt{\frac \pi 2} \frac 1 {\sqrt u} \hspace{2mm} \rr{when} \hspace{2mm} u \rightarrow \infty ~,
\end{equation} 
such that 
\begin{equation} \label{eq:variance_at_inst}
\langle \psi^2(x=1) \rangle \simeq \frac{H^2 \mu M}{32 \pi^{3/2} \Mpl^2}~. 
\end{equation}
For instance, for the parameter values of Fig.~\ref{fig:traj_phipsi2}, $\mu = 636.4 \ \mpl, M=0.03 \ \mpl$, one obtains $ \sqrt{\langle \psi^2 \rangle} \simeq 2 H $.
It will be shown later (see Sec.~\ref{sec:paramspace}) that when the tachyonic preheating is triggered and forces inflation to end,  one has $x  \lesssim 1$ such that 
$\sqrt{\langle \psi^2(x_{\rr{end}}) \rangle} \sim \sqrt{\langle \psi^2(x=1) \rangle} \sim H$.   Therefore, as long as the dynamics is mainly governed by the evolution of the field $\phi  $, the standard deviation of the transverse field distribution around its classical value does not become much larger than $H$.  

Notice that an identical result can be obtained using the linear perturbation formalism developed in \cite{Gong:2010zf,Fonseca:2010nk,Abolhasani:2010kn,Abolhasani:2010kr,Lyth:2010ch}.   As for the stochastic formalism, one can assume that the $\phi$ field evolves independently according to Eq.~(\ref{eq:Ndephi}).   By Fourier expanding $\delta \psi$, neglecting non linear terms and using slow-roll to express the time dependent tachyonic mass as a function of the number of e-folds, one can rewrite the mode evolution equation Eq.~(\ref{eq:mode_waterfall}) using the number of e-folds as a time variable,
\begin{equation} \label{eq:mode_N_waterfall}
\frac {\dd^2 \delta \psi_k}{\dd N^2}  + 3 \frac{\dd \delta \psi_k }{\dd N} + \left\{  \frac{k^2}{a^2 H^2 } - 12 \frac{\Mpl^2}{M^2} \left[1 -  \rr e ^{-2 r (N-N_{\rr c})} \right] \right\} \delta \psi_k = 0 ~.  
\end{equation}
 Following Ref.~\cite{Gong:2010zf}, in the high frequency limit this equation can be solved in terms of the WKB approximation.  In the low frequency limit its exact solution is a combination of the Hankel functions of first and second kind.   Let us introduce $k_{\rr c} \equiv a_{\rr c} H$, the comoving mode leaving the Hubble radius at the critical point of instability $\phi_{\rr c}$, and $n \equiv N-N_{\rr c}$.  Then, under the assumption that $12 \Mpl^2 / M^2 \gg 1$, one finds for the evolution of the small scale modes near the instability [$k \gg k_{\rr c} \exp (n)$, i.e. sub-Hubble modes at the critical instability], 
\begin{equation} \begin{split}
&\left|   \delta \psi_{\rr S} (k,n) \right|= \frac{H}{\sqrt{2 k}  k_{\rr c}} A \\
& \times  \exp \left(  \frac 2 3 \alpha n^{3/2} - \frac 3 2 n - \frac{1}{4} \log n \right) ~,
\end{split} \end{equation}
and for the large scales modes that are already super-Hubble at the critical instability [$k \ll k_{\rr c}  \exp (n)$], 
\begin{equation} \begin{split} \label{eq:smallk}
& \left|  \delta \psi_{\rr L} (k,n) \right| =  \frac{H}{\sqrt{2 \alpha k_{\rr c} ^3}  } \\ 
 & \times \exp \left( \frac 2 3 \alpha n^{3/2} - \frac 3 2  n - \frac{1}{4} \log n \right)~,
\end{split}
\end{equation}
where $A \equiv 3^{2/3} \Gamma (2/3) \alpha^{-1/6}/(2 \sqrt \pi)$ is a typically order unity factor, and 
\begin{equation}
\alpha  \equiv \sqrt{24 r \frac{\Mpl^2 }{M^2 } }~.
\end{equation}
In the regime $2 r n \ll 1 $, from Eq.~(\ref{eq:mode_N_waterfall}) the modes which become tachyonic satisfy 
\begin{equation}
\left(\frac k {k_{\rr c}}\right) ^2 \le \alpha^2 n \  \rr e ^{2n} ~.
\end{equation} 
In Ref.~\cite{Gong:2010zf}, published a few months after our paper~\cite{Clesse:2010iz}, it is then assumed that $\alpha \gg 1$ and $n \sim \mathcal O (1)$ to find that the quantum back-reactions from the small scales entropy perturbations dominate and force inflation to end quickly after waterfall instability.   We are here interested by the opposite case, $\alpha \lesssim 1$.  
As shown later in Sec.\ref{sec:paramspace}, the total number of e-folds that can be realized classically between the instability and the beginning of the tachyonic preheating is larger than 60 and is roughly given by $ n \sim \mu^2 M^2 $.  Therefore the tachyonic modes are super-Hubble during all this phase.   In that case, the variance of $\delta \psi$ is dominated by the large scale mode contribution 
\begin{equation} 
\langle \delta \psi^2 (n)  \rangle  =  \int_0 ^{k_{\rr c} \rr e^n} \frac{\dd ^3 k}{(2 \pi)^3}\delta \psi_L ^2 (k,n) ~.
\end{equation} 

One obtains just after the critical instability 
\begin{equation} \begin{split}
\langle \delta \psi^2 [n\sim \mathcal O(1)]  \rangle &\simeq  \frac{H^2}{12 \pi^2 \alpha} \exp \left( \frac 4 3 \alpha n^{3/2}  - \frac{1}{2} \log n \right) \\
& \simeq   \frac{H^2}{12 \pi^2 \alpha}
\end{split}
\end{equation}
which is identical to Eq.~(\ref{eq:variance_at_inst}) up to an order unity numerical factor\footnote{Notice that a similar result can be obtained for $n=0$, from the Eqs.~(2.27) and (2.30) of Ref. \cite{Gong:2010zf}.  In that case, an additional factor $\alpha^{-1/6}$ is obtained, but it can be due to a problem of matching between the small scale and the large scale evolution of the modes. This problem is discussed in Ref. \cite{Gong:2010zf}}.

In the section \ref{sec:classical}, initial values of $\psi$ at the critical point of instability are taken to follow a gaussian random distribution verifying Eq.~(\ref{eq:qufluct}).   From this point, the classical value of $\psi$ moves away from its initial amplitude and increases such that it becomes quickly much larger than its quantum fluctuations, even if the overall dynamics is still governed mainly by the $\phi$ evolution.  Therefore, the classical dynamics is quickly recovered and from this point it is not spoiled by transverse quantum fluctuations.


\subsection{Transverse field gradient contribution}

Another effect susceptible to spoil the inflationary dynamics is the backreaction due to the transverse field gradient contribution to the energy density\footnote{D. Lyth, private communication}.  

Assuming the statistical homogeneity, the mean-square value of transverse field gradient after smoothing on a length $L = 1/ (a H) $  is given by 
\begin{equation}
\langle |  \nabla \psi |^2 \rangle = \frac{1}{(2 \pi)^3} \int_0 ^{a H} (\dd k)^3  \left( \frac k a \right)^2 | \psi_k | ^2   ~.
\end{equation}
During the waterfall, $\psi_k$  is given by Eq.~(\ref{eq:smallk}).   After integration over the modes, one obtains
\begin{equation}
\langle |  \nabla \psi |^2 \rangle  \sim H^2  \langle | \psi |^2 \rangle \sim H^4~,
\end{equation}
since $\langle | \psi |^2 \rangle \sim H^2$ during the waterfall in the regime of interest.  From the amplitude of the scalar power spectrum, one knows that $H^2 \ll \mpl^2 $.  So the gradient term is negligible compared to the potential term $ V \simeq 3 H^2  \Mpl^2 $ in the energy density.   The background dynamics thus remains mostly homogeneous.


\subsection{Inflation along classical waterfall trajectories} \label{sec:classical}

Once the critical instability point is reached, field trajectories deviates from the valley line and fall through one of the global minima $(\phi=0,\psi=\pm M)$ of the potential.  In a first approximation, we can follow Ref.~\cite{Copeland:2002ku} and assume that the auxiliary field reacts faster than the inflaton field such that trajectories follow the ellipse defined by the minima of the potential in the $\psi$ direction,
\begin{equation} \label{eq:ellipse}
\frac {\dd V(\phi,\psi)}{\dd \psi} = 0 \ \rr{ with } \  -\phi_c \le \phi \le \phi_c \Longrightarrow \frac{\psi^2}{M^2} + \frac{\phi^2}{\phi_c^2} =  1 \ .
\end{equation} 
We will compare thereafter this approximation to the exact numerical integration of the dynamics. 

Near the critical instability point, the effective potential defined by this ellipse is similar to the potential of a small-field inflation model.   It is very flat at its top, where its curvature is negative.  Depending on the potential parameters and the initial value of $\psi$, it is therefore in principle possible for inflation to continue for a certain amount of e-folds along the classical waterfall trajectory.

The  collective evolution of the fields along the classical trajectory is described by the adiabatic field $\sigma$, introduced in section~\ref{sec:multifield}~\cite{Gordon:2000hv}.  It is defined such that
\begin{equation} \label{eq:adiabaticfield}
\dot \sigma = \sqrt{\dot \phi^2 + \dot \psi^2  } \ ,
\end{equation}
and its equation of motion reads
\begin{equation}
\ddot \sigma + 3 H \dot \sigma + V_\sigma = 0 \ ,
\end{equation}
where 
\begin{equation}
V_\sigma = \frac{\dot \phi}{\dot \sigma} \frac{\dd V}{\dd \phi}+ \frac{\dot \psi}{\dot \sigma} \frac{\dd V}{\dd \psi} \ .
\end{equation}
On the ellipse of Eq.~(\ref{eq:ellipse}), one obtains 
\begin{equation} \label{eq:potadiabatic}
V_\sigma = \Lambda^4 \frac{2 \dfrac{\phi}{\mu^2} + 4 \dfrac{\phi}{\phi_{\rr c}^2 } \left( 1 - \dfrac{\phi^2}{\phi_{\rr c}^2 } \right)  }{ \sqrt{1+ \dfrac{M^2 \phi^2}{\phi_{\rr c} ^2 ( \phi_{\rr c}^2 - \phi^2 )  }   }  }\  ,
\end{equation}
where $\phi$ is related to the adiabatic field through the relation 
\begin{equation}
\sigma(\phi) = \int_{\phi_{\rr c}} ^{\phi} \dd \phi'  \sqrt{1+ \dfrac{M^2 \phi'^2}{\phi_{\rr c} ^2 ( \phi_{\rr c}^2 - \phi'^2 )  }   }  \  .
\end{equation}
Then the slow-roll can be assumed and one may think that all the required ingredients for calculating the dynamics are given.  If inflation lasts for more than $60$ e-folds during the waterfall, one may also predict the power spectrum of adiabatic perturbations in the slow-roll approximation.  
To do that, one has to evaluate the field value at Hubble exit of observable modes, that is when
\begin{equation}
N(\phi) = \int^{\phi} _{\phi_{\epsilon_{\rr 1} = 1 } } \dd \phi' \frac{V}{\Mpl^2 V_\sigma}  \sqrt{1+ \frac{M^2 \phi'^2}{\phi_{\rr c} ^2 ( \phi_{\rr c}^2 - \phi'^2 )  }   } \simeq 60 \  .
\end{equation}
Then the spectral index is directly determined with Eq.~(\ref{eq:slowrollparams}).   

But in practice, at the critical instability point, the gradient of the potential is along the $\phi$ direction. The field evolution first follows this direction and thus does not follow exactly the ellipse of Eq.~(\ref{eq:ellipse}).   Therefore the predictions are expected to be modified more or less importantly.  These modifications are studied by solving numerically the exact classical dynamics.  


\begin{figure}[p]
\begin{center}
 \includegraphics[width=8.0cm]{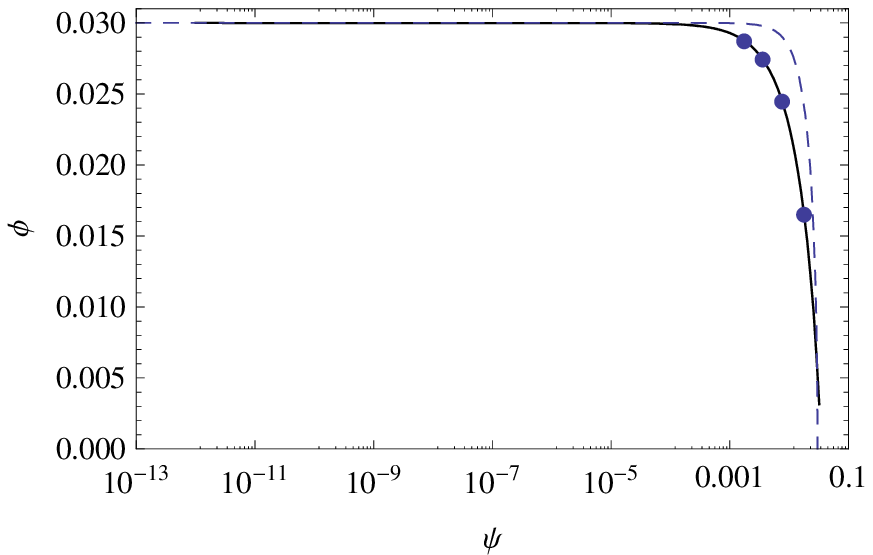}
  \includegraphics[width=8.5cm]{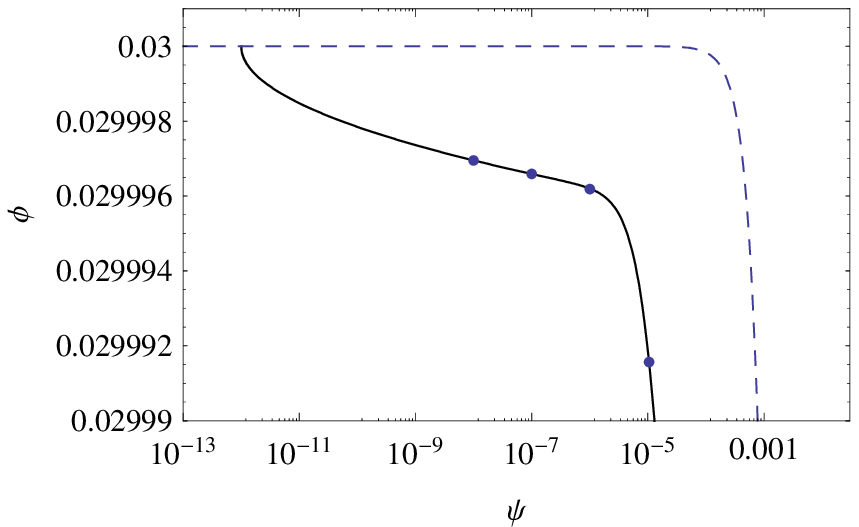}
  \includegraphics[width=8cm]{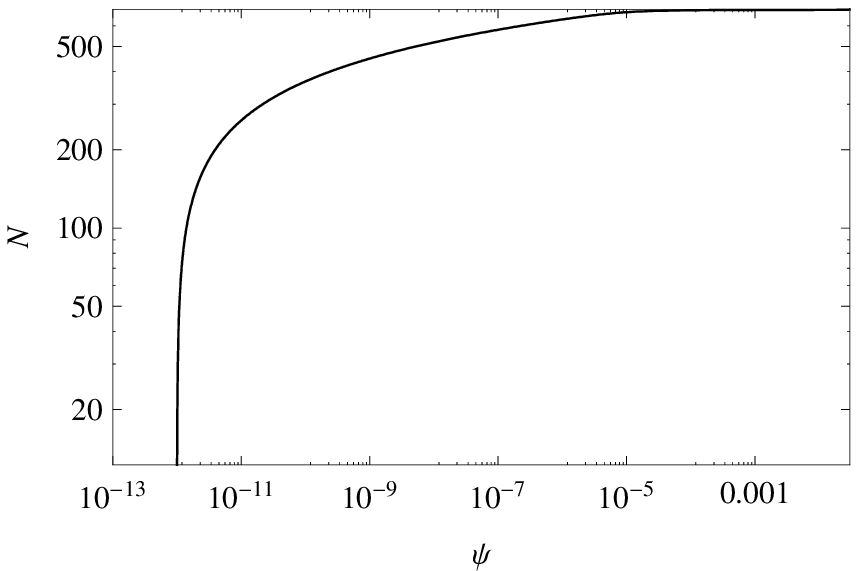}
  \caption{Top: typical trajectory (solid line) in the $(\phi , \psi) $ plane ($\phi$ and $\psi$ are in $\mpl$ units), for initial field values $\phi_i = \phi_c = 0.03 \mpl, \psi_i  = 10^{-12} \mpl $ and potential parameters $M=0.03 \mpl, \mu = 636.4 \mpl $.   The ellipse of minima of Eq.(\ref{eq:ellipse}) is also represented (dashed line).  The points on the trajectory indicate where $\epsone = 10^{-3} / 10^{-2}  / 0.1  / 1 $ respectively from left to right.  Center:  zoom around $\phi_{\rr c}$ for the same trajectory.  The points on the trajectory indicate where $n_{\rr s} = 1. / 0.97 / 0.91 / 0.65$.  Bottom:  Number of e-folds realised along this trajectory, from the critical instability point.  }
  \label{fig:traj_phipsi2}
  \end{center}
\end{figure}

One may discuss the validity of the homogeneous dynamics at the very end of the waterfall phase, when the mass of the auxiliary field becomes larger than $H$, that is when the growth of the long wavelength modes of $\psi$ reach the non-linear regime.   Actually, it is not trivial to determine if inflation ends due to the transverse field gradient terms or due to slow-roll violating field velocities.   Such calculation would require lattice simulations taking account the expansion during the waterfall.  Nevertheless, we have checked that the number of e-folds generated during the waterfall only differs marginally (it is slightly lower) if we assume that inflation ends when the mass of the auxiliary field, calculated along the axis $\psi = 0$, becomes larger than $H$.

\begin{figure}[h!]
\begin{center}
\includegraphics[width=9.cm]{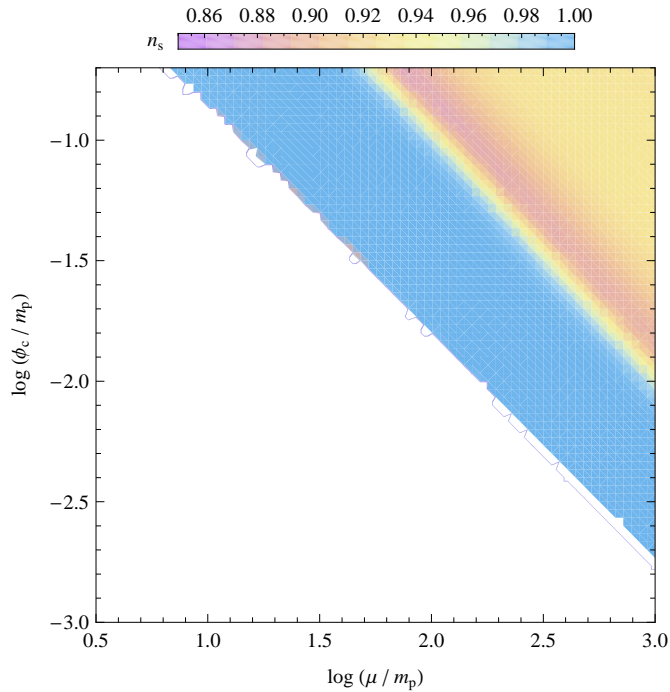}
\caption{$200 \times 200$ grid of spectral index values of the power spectrum of adiabatic perturbations, in the plane ($\mu, M$) for the exact classical dynamics, with $\phi_{\rr c} = 10^{-3} \mpl $. White region correspond to trajectories leading to no more than $60$ e-folds after instability. }
\label{fig:nsexact}
\end{center}
\end{figure}

Fig.~\ref{fig:traj_phipsi2}  shows a typical trajectory,  for typical potential parameters.  
 More than $600$ e-folds are found to be realised before inflation ends, when $\epsilon_{\rr 1} = 1$.   
 Therefore, 60 e-folds before the effective end of inflation, when observable modes exit the Hubble radius, the critical instability point has already been crossed.  The spectral index of the adiabatic power spectrum can also be determined numerically.  
It has been plotted for several values of the fields at Hubble exit of the observable modes, along the trajectory of Fig.~\ref{fig:traj_phipsi2} as well as for a grid in the parameter space ($\mu,M$), in Fig.~\ref{fig:nsexact}.    Because $\epsilon_{1*} \ll 1$ and $\epsilon_{2*} > 0$, it is generically red

The effective potential of Eq.~(\ref{eq:potadiabatic}) is thus an ideal case.  Two regimes during which inflation is possible are identified: 

\begin{enumerate}
\item PHASE I:  Driven by their velocity along $\phi$, the field trajectories first follow the slope of the potential in the $\phi$ direction before turning, following roughly the gradient of the potential until the ellipse defined in Eq.~(\ref{eq:ellipse}) is reached.   As shown in Fig.\ref{fig:traj_phipsi2} a large number of e-folds can be  realized already during this first phase.

\item PHASE II:  Trajectories reach and mostly follow the above defined ellipse.  A large number of e-folds is realized if the effective potential along the ellipse is sufficiently flat.
\end{enumerate}

The tachyonic preheating is not triggered immediately after the critical instability point, but it only takes place at the end of inflation, similarly to what happens for the new inflation model \cite{Desroche:2005yt}.  


In this section, some waterfall trajectories leading to more than 60-folds have been shown to exist.  But before to draw conclusions, it is essential to measure how generic such trajectories are in the parameter space.  This is the point of the following section, in which the full potential parameter space will be explored using a statistical MCMC method.

\section{Exploration of the parameter space} \label{sec:paramspace}

The number of e-folds generated after crossing the instability point depends on the form of the potential through its three parameters $M, \mu, \phi_{\rr c}$.  The classical dynamics depends also on the initial value of the auxiliary field.  At the critical instability point $\phi_{\rr c}$, the distribution of $\psi$ is given by Eq.~(\ref{eq:qufluct}).   So it is related to the potential parameter $\Lambda$ through the Friedmann-Lemaitre equation.    From this point, the auxiliary field is assumed to evolve classically.

To explore this 4D space, we have used a Monte-Carlo-Markov-Chains method.  Flat priors have been chosen on the logarithm of these parameters, in order to not favor any precise scale.     The chosen ranges of parameters are the following:
\begin{eqnarray}
0.3 \mpl & < & \mu < 10^4 \  \mpl \\
10^{-6}\mpl & < & M < \Mpl \\
10^{-6}\mpl & < & \phi_{\rr c} <  \Mpl \\
 10^{-70} \mpl^4 & < & \Lambda^4 < 10^{-12} \mpl^4    
\end{eqnarray}
The lower bound on $\mu$ comes from its posterior probability distribution \cite{Clesse:2009ur} to generate sufficiently long inflation inside the valley from arbitrary subplankian initial conditions.  As discussed in chapters 4 and 5, this bound is induced by the slow-roll violations preventing inflation to take place along the valley at small values of the field $\phi$.  This effect is avoided for trajectories remaining in the regime $\phi \ll \mu$ along the valley, but these are generated in only a fine-tuned region of the space of initial conditions.  Upper bounds on $M$ and $\phi_c$ stand because we only consider the dynamics at field values smaller than the reduced Planck mass.
The lower bounds on $M$ and $\phi_c$ and the upper bound on $\mu$ have been chosen for numerical convenience.  The bounds on the parameter $\Lambda $ are such that the present constraints on the energy scale of inflation, given by the nucleosynthesis and the observations of the CMB, are respected.

While the initial inflaton value is $\phi _{\rr i}= \phi_{\rr c}$, the initial auxiliary field values are assumed to follow a gaussian distribution around $\psi = 0$, with a dispersion given by Eq.~(\ref{eq:qufluct}).
  In order to avoid strong quantum backreactions of the adiabatic field, a hard prior coming from Eq.~(\ref{eq:hardprior}) is enforced,
\begin{equation} \label{eq:hardprior}
\epsilon_{\rr 1} (\phi = \phi_{\rr c}, \psi \simeq 0 ) \simeq  \frac{\phi_{\rr c}^2 \mpl^2}{\mu^4}  > \frac{H^2}{\pi \mpl^2} \sim \Lambda^4,
\end{equation}
 such that each trajectory that do not verify this condition is excluded of the Markov chain.
Integration stops at $\epsilon_{\rr 1} = 1$, the end of inflation.  The acceptance condition for the Markov chain is a realization of at least $60$ e-folds after $\phi_{\rr c}$.

\begin{figure}[h!]
\begin{center}
\includegraphics[width=60mm]{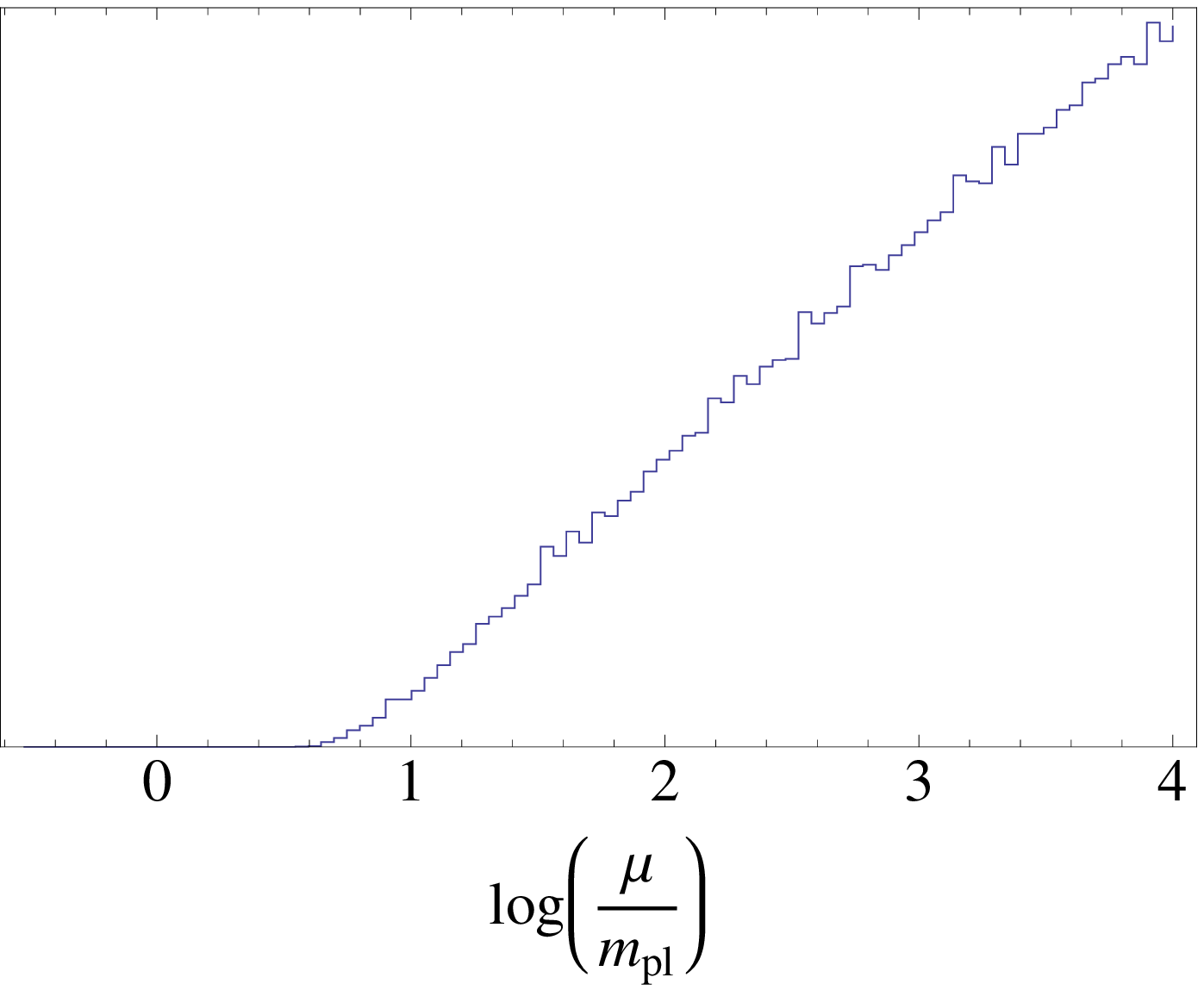}  
\includegraphics[width=60mm]{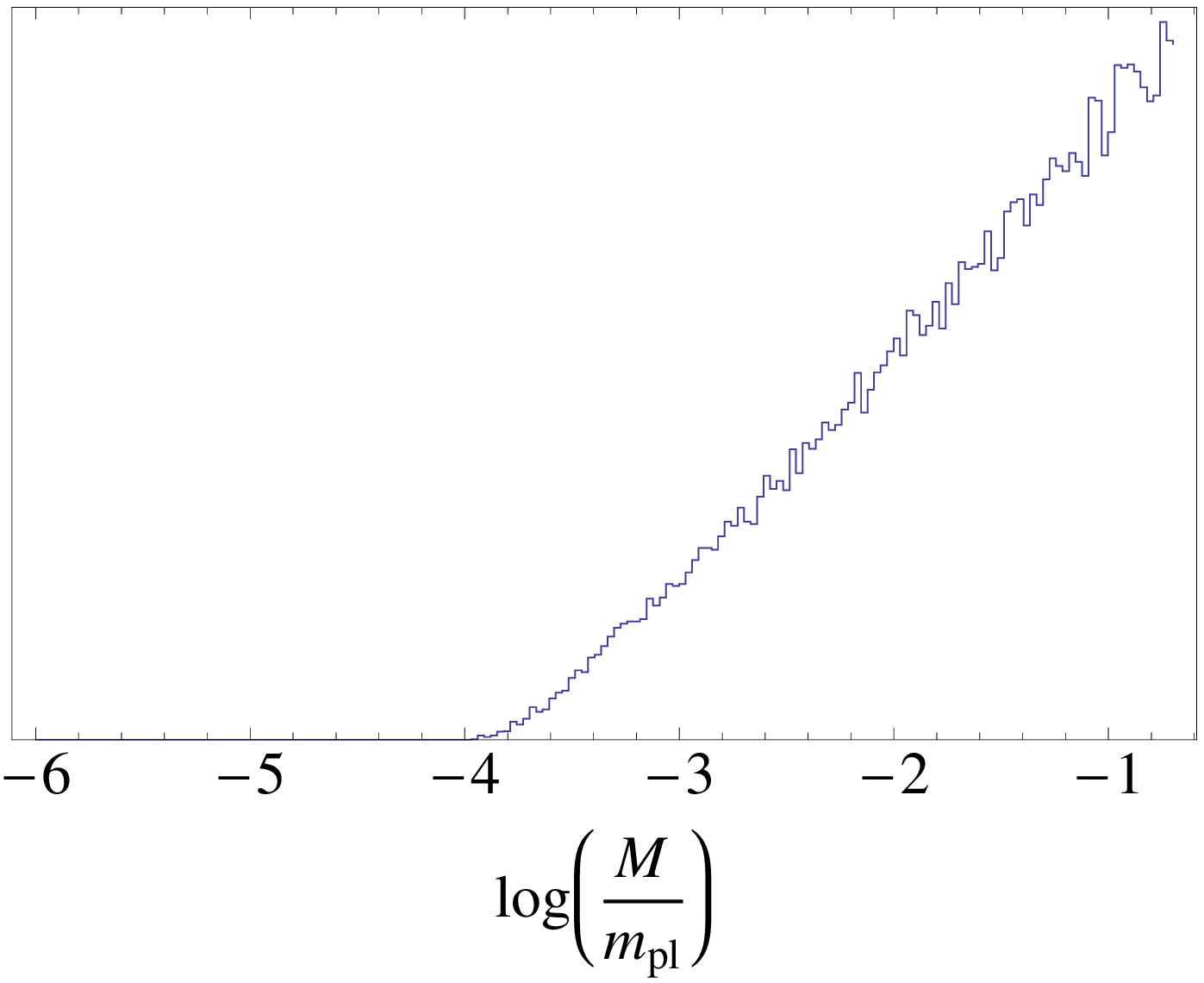} 
\caption{Marginalized posterior probability density distributions of the potential parameters $\mu$ (left) and $M$ (right), for classical field trajectories realizing more than $60$ e-folds of inflation during the waterfall. The vertical axis is normalized such that the total area under the distribution is $1$.   $\mu$ and $M$ are correlated and the posterior distributions depend on the prior ranges.}
\label{fig:probamuM}
\end{center}
\end{figure}

\begin{figure}[h!]
\begin{center}
\includegraphics[width=60mm]{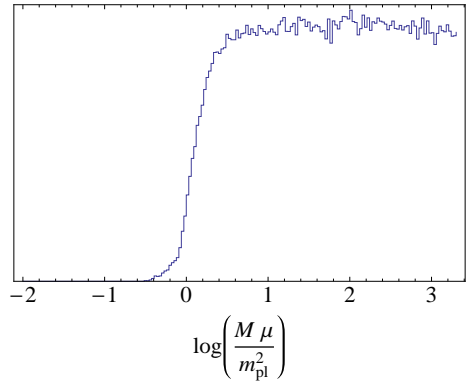}  
\caption{Marginalized posterior density probability distribution of the product  $M \mu$.  Inflation can continue for more than $60$ e-folds along waterfall trajectories when the condition $ \mu M \gtrsim \mpl^2 $ is satisfied. The posterior distribution does not depend on the prior ranges. }
\label{fig:probaMmu}
\end{center}
\end{figure}

\begin{figure}[h!]
\begin{center}
\includegraphics[width=60mm]{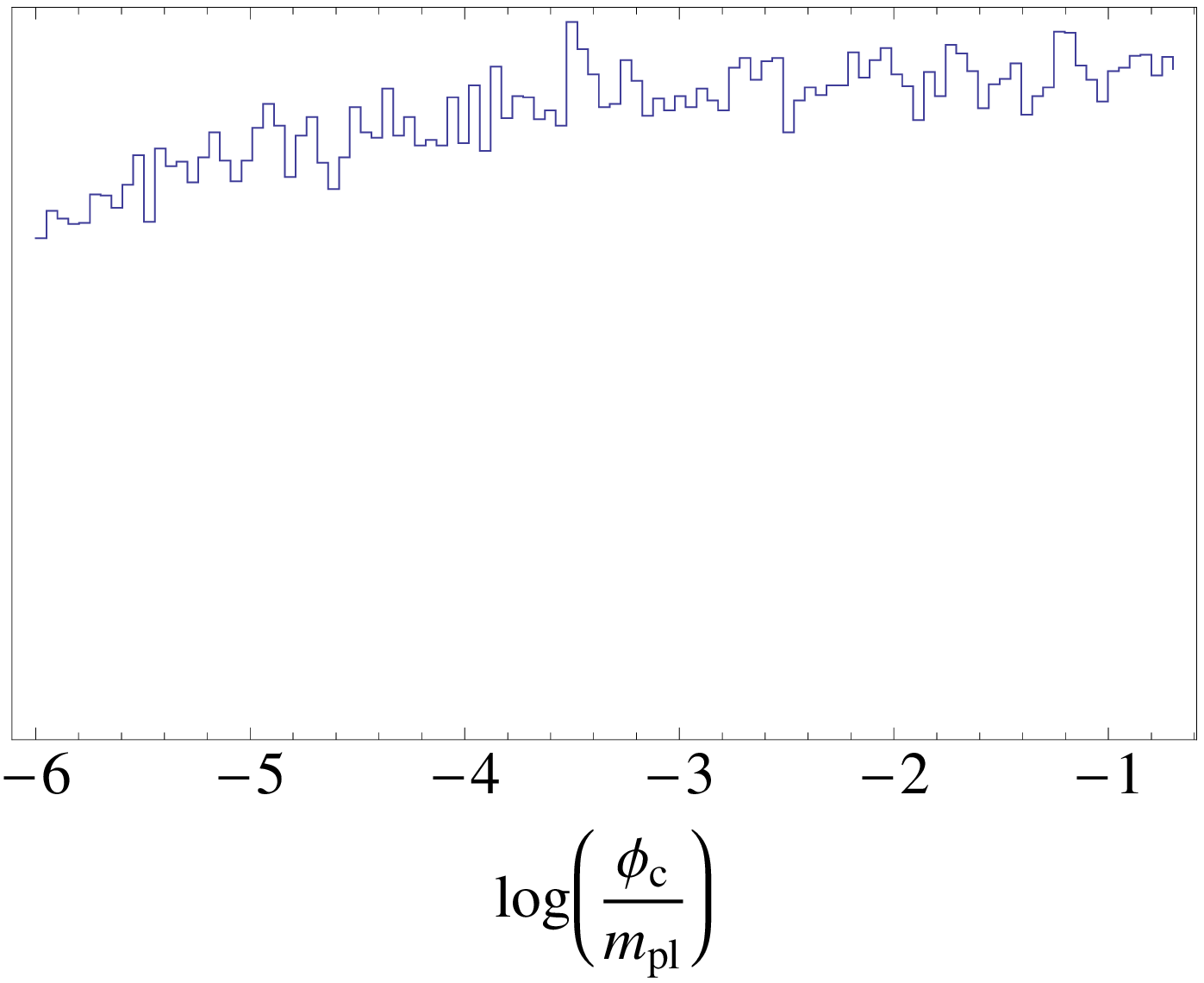}  
\includegraphics[width=60mm]{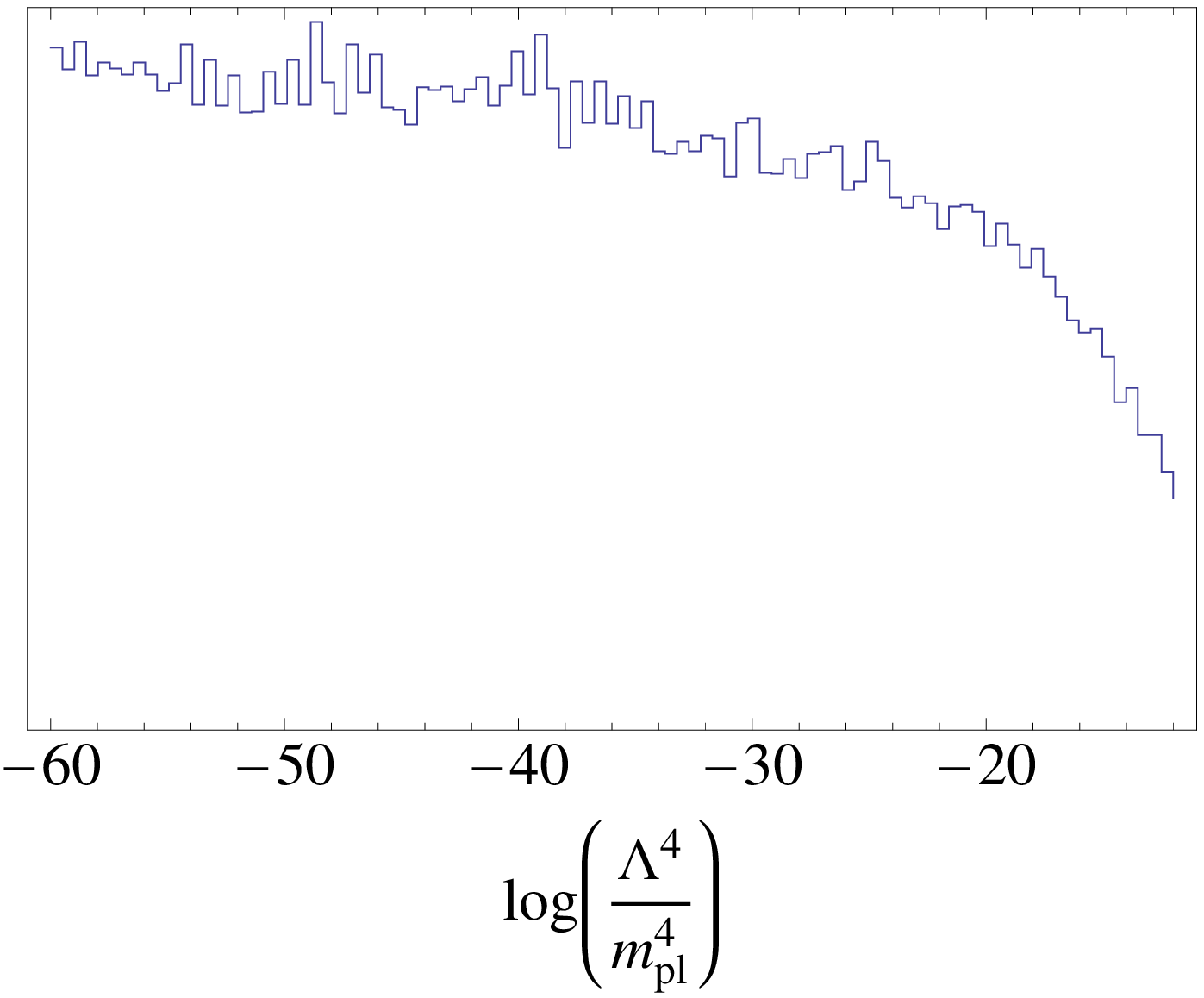} 
\caption{Marginalized posterior probability density distributions of the critical point of instability  $\phi_c$ (left) and the potential parameter $\Lambda^4$ (right).  Distributions are nearly flat and thus $\phi_{\rr c}$ and $\Lambda^4$ do not influence the possibility for inflation to continue for more than $60$ e-folds during the waterfall.   A decrease of the posterior distribution of $\Lambda^4$ is observed at high energy.  This is due to: 1) the fact that more trajectories are removed from the Markov chains at high values of $\Lambda^4$, because of the hard prior of Eq.~(\ref{eq:hardprior}). 2) the initial value of $\psi$, given by the distribution of Eq.~\ref{eq:qufluct}, depending on $\Lambda^4$ through $H$.   For high values of $\Lambda^4$, field trajectories start far from the valley line and the resulting number of e-folds is lower. }
\label{fig:probaphiclambda4}
\end{center}
\end{figure}


The marginalized posterior  probability density distributions, normalized such that the area under each distribution is $1$,  are shown in Figs. \ref{fig:probamuM} and \ref{fig:probaphiclambda4}.    The posteriors on the parameters $M$ and $\mu$ seem to indicate that these parameters are bounded.  However, they are affected by our prior choices.  If one changes the upper limit on $\mu$ or the lower limit on $M$, a modification to the values at which the posteriors fall off is observed.  Such a situation is typical of the existence of correlations between these parameters.   But we have determined that the posterior distribution of their product\footnote{in that case, the parameter $\mu$ is replaced by the product $M \mu$ in the MCMC simulation, using a flat prior on the log of $\mu M$.} $\mu M$ (see Fig.\ref{fig:probaMmu}) does not depend on the prior ranges.   A bound on this combination is obtained:
\begin{equation} \label{eq:boundmuM}
\log \left( \frac{ \mu M}{\mpl^2}  \right) > 0.21 \ \ 95 \% \rr{C.L.} .
\end{equation}
This bound can be rewritten $M/ \mpl \gtrsim  \mpl / \mu   $ and understood intuitively. A large number of e-folds have to be generated after the instability point, and before the magnitude of the effective negative mass of the auxiliary field 
\begin{equation}
m_{\psi} (\phi) = - \sqrt 2 \frac{\Lambda^2}{M} \sqrt{ 1 - \frac{\phi^2}{\phi_{\rr c} ^2} }~, 
\end{equation}
increases and becomes larger than $H$.  From this time the tachyonic modes indeed grow exponentially (see section~\ref{sec:fast_waterfall}).    Following \cite{GarciaBellido:1996qt}, this happens in the range 
\begin{equation} \label{eq:rangephi}
\phi_{\rr c} > \phi > \phi_{\rr c} \sqrt{1- \frac{M^2}{\mpl^2}}.  
\end{equation}
During this period, the slow-roll approximation for $\phi$ is still valid and the number of e-folds generated given by Eq.~(\ref{eq:Ndephi}).  In the limit $\phi \ll \mu$,  using the range of $\phi$ obtained in Eq.~(\ref{eq:rangephi}), straightforward manipulations give $ \Delta N \sim  \mu^2 M^2 / 4 \Mpl^4$ and thus the number of e-folds is roughly fixed by the combination of the parameters $\mu$ and $M$.  From this reasoning comes also the argument that, at the end of the Phase I, $x \simeq \exp[-  M^2 /\Mpl^2 ] \sim 1 $.
Since the analysis is restricted to the sub-planckian field dynamics, and thus to sub-planckian values of $M$, there exist a gap between
\begin{equation}
0.3 \mpl \lesssim \mu \lesssim 6 \mpl ,
\end{equation}
for which the number of e-folds generated is less than $60$.  In this regime, the slope of the potential in the $\phi$ direction is too large to generate a sufficient number of e-folds in Phase I, and the negative mass of the auxiliary field pushes the trajectory away from the $\psi = 0 $ line.

The posterior probability distribution of $\phi_{\rr c}$ is nearly flat and this parameter does not influence significantly the duration of the waterfall inflationary phase.   

The marginalized posterior probability distribution of $\Lambda^4$ is also almost flat and decreases for values corresponding to the highest energy scales of inflation, without becoming negligible.  This suppression is induced by two effects.   One is the hard prior.  Since $H$ is directly related to $\Lambda^4$ through the F.L. equations,  more trajectories are affected by quantum stochastic effects and are removed from the Markov chains for high values of $\Lambda^4$.  
 On the other hand, the distribution of $\psi_{\rr i} $  is also related to $\Lambda^4$.   When the classical trajectories start more far away from the $\psi = 0$ axis, the phase I is less efficient and the number of e-folds generated during the waterfall is reduced.

The MCMC analysis therefore provides an explicit answer to the question of how generic are trajectories realizing a large number of e-folds after the critical instability point $\phi_{\rr c}$.  They are found to occupy a large part of the parameter space, gathered in the region given by Eq.~(\ref{eq:boundmuM}).   Since $\Lambda^4$ is directly linked to the energy scale of inflation, waterfall inflation is also found to be more favorable at low energies.

Finally, if one needs the posterior distributions of the parameters for having more than 60 e-folds along classical waterfall trajectories together with initial field values anywhere is the sub-planckian field space,  the posterior distributions obtained in this chapter should be combined with the posterior distributions of $\phi_{\rr i}, \psi_{\rr i}, \dot \phi_{\rr i}, \dot \psi_{\rr i}, M, \mu, \phi_{\rr c}$~\cite{Clesse:2009ur}, obtained in the chapter 5.  

All these results stand for the original hybrid model, but the general features are expected to be reproducible with more or less efficiency, for all models in which inflation mostly occur in a nearly flat valley and end due to a tachyonic instability, like in many SUSY realizations (e.g. F-term hybrid model \cite{Dvali:1994ms}).
Notice that when Eq.~(\ref{eq:boundmuM}) is not verified, then the standard mechanism does work:  namely inflation stops soon after $\phi_{\rr c}$.

\section{Conclusion and discussion}

In hybrid models, the standard picture assumes slow-roll inflation along a nearly flat valley ending quasi-instantaneously due to a tachyonic instability at $\phi = \phi_{\rr c}$, triggering the tachyonic preheating \cite{Kofman:1997yn,Garcia-Bellido:1997wm, Felder:2000hj, Felder:2001kt, Copeland:2002ku, Senoguz:2004vu, Micha:2004bv,Allahverdi:2007zz}.  
  
In this chapter, the waterfall phase has been studied in a regime during which inflation continues for a long time.   It has been shown that more than $60$ e-folds can be realised classically during the waterfall, after crossing the instability point.   
Particular attention has been given to study regions in the parameter space where the classical dynamics is valid and not spoiled by quantum backreactions of adiabatic and entropic fields. 
Let us also mention that our results have been later confirmed by H. Kodama et al. in Ref.~\cite{Kodama:2011vs} by using analytical approximations\footnote{Similar results have also been obtained recently in Ref.~\cite{Abolhasani:2011yp}.}. 


Observable modes therefore leave the Hubble radius when the effective potential is very flat with a negative curvature.  Instead of blue, the power spectrum of adiabatic perturbations becomes generically red.   
However, the calculation does not include the contribution of potentially observable iso-curvature modes.  This contribution was shown to be negligible in recent developments \cite{Gong:2010zf, Abolhasani:2010kn, Fonseca:2010nk, Abolhasani:2010kr, Lyth:2010ch,Lyth:2010zq}, but in these references a fast waterfall phase is assumed and their conclusion does not apply to the regime of potential parameters studied here.  The full numerical calculation of the primordial power spectrum should be realised soon in \cite{Clesse:2010prepa}.  The effect on the non-gaussianities produced during the tachyonic preheating \cite{Barnaby:2006cq, Barnaby:2006km} could be also important.  

Therefore, we have shown that it is premature to conclude that the original hybrid model is disfavored by CMB experiments.   In addition, a bayesian MCMC analysis demonstrates that such trajectories are generic in a large part of the potential parameters space, and thus cannot be ignored.  


These observations may have also an important impact on questions related to the end of inflation.  In particular, if a large number of e-folds occur after symmetry breaking, the eventually formed topological defects will be diluted by expansion and thus will not affect our observable universe.   Therefore, some works constraining the schemes of symmetry breaking in grand unified theories with topological defects~\cite{Rocher:2004et} may be reviewed.   Our results should have also some impact on tachyonic preheating.  The previous studies of tachyonic preheating in hybrid inflation~\cite{Felder:2000hj, Felder:2001kt, Copeland:2002ku} consider an instantaneous waterfall, or consider the regime in which the waterfall lasts no more than a few e-folds~\cite{Barnaby:2006km}.  Lattice simulations indicates that cosmic strings and domain walls strongly affect the way the preheating phase occurs.  However if such defects are diluted by a phase of inflation after the symmetry breaking, lattice simulations should be updated to include a long period of inflation after the critical instability, or to start after this period of inflation, like in the new inflation models \cite{Desroche:2005yt}.

Finally, let us comment about stochastic effects.  For the 1-field effective potential of hybrid inflation, these were found to not affect the classical trajectories along the valley in~\cite{Martin:2005ir, Martin:2005hb}.  The authors of~\cite{Martin:2005ir, Martin:2005hb} also notice that in small field inflation, a stable solution of eternal inflation should exist at the top of the potential.   At $\phi \simeq 0$, the hybrid potential along $\psi$ reduces to a small field type and thus this observation should apply to the waterfall phase in hybrid inflation.   In particular, it would be interesting  to determine, in a full 2-field approach of stochastic effects, how the field dynamics is affected if initial conditions for the waterfall are taken along the border of the stochastic patch.

%% file: bounce.tex
\chapter{Hybrid inflation in a classical bounce scenario}
\label{chap:bounce}

\begin{center}
\textit{based on}\\
S. Clesse, M. Lilley, L. Lorenz\\
How natural is a classical bounce plus hybrid inflation scenario?\\
\textit{in preparation} \\
\end{center}

\section{Introduction}


Despite undeniable advantages, most models of inflation do not provide a solution to the important question of the initial singularity.    A quantum theory of gravitation is believed to be needed to describe the Universe at the Planck energy scale.  
To avoid the initial singularity, scenarios of closed Universe performing a classical bounce have been proposed~\cite{Starobinsky:1980te}.  But such models alone do not provide a solution to the flatness problem.  To avoid this problem, the classical bounce can be followed by phase of inflation~\cite{Falciano:2008gt,Starobinsky:1980te}.   As an example, a classical bounce could occur when a scalar field evolves at the top of a small-field Higgs-type potential of the form $V(\phi) \propto (\psi^2 - M^2  )^2$~\cite{Falciano:2008gt,Lilley:2011ag}.  For such models, we have shown with M. Lilley and L. Lorenz in Ref.~\cite{Lilley:2011ag} that specific signatures on the scalar power spectrum are expected, on the form of super-imposed oscillations.  Such signatures could be potentially observable in the CMB or in the matter power spectrum.   

However, these scenarios are expected to suffer from an extreme fine-tuning problem of the initial conditions during the contracting phase~\cite{Falciano:2008gt}.   Indeed, for a Higgs-type potential, for slow-roll inflation to take place near the top of the potential, the bounce needs to occur in a field range fine-tuned at the top of the potential and the field velocity at this point needs to be very small.   This makes the scenario highly improbable.  Indeed, if the field was evolving around one of the global minima of the potential (at $\phi = \pm M$)  during the contraction, there is no reason that it reaches the top of the potential (at $\phi = 0$) with an extremely small velocity just after the classical bounce.  

In this chapter, it is investigated whether this fine-tuning remains if the Universe is filled with two homogeneous scalar fields evolving on a hybrid potential.   In chapter 5, we have shown in the flat case that hybrid inflation does not suffer from a problem of fine-tuning of the initial field values.   
Here we examine whether the inflationary attractor along the valley can be reached without extreme fine-tuning by field trajectories initially evolving in a contracting phase and undergoing a classical bounce due to the positive curvature of the Universe.  
Since in hybrid inflation much more than 60 e-folds of accelerated expansion are realized generically, any considered value of the curvature during the contraction will lead today to a Universe whose curvature is generically inside the observable bounds.


For sets of initial conditions and potential parameter values in the contracting phase,  the exact field dynamics have been integrated numerically.   Three behaviors are possible:  
\begin{enumerate}
\item The field trajectories evolve in the contracting phase, the energy density growth and reach the Planck scale before the bounce occurs.  When the Planck energy scale is reached, the classical dynamics is possibly not valid anymore and the integration is stopped.   
\item A classical bounce occurs and hybrid inflation takes place, for more than $60$ e-folds. 
\item  A classical bounce occurs but the field evolution is such that the inflationary era is not triggered.  Since the Universe is still dominated by the curvature, the expansion phase does not last a long time and the Universe contracts again.  The process is therefore iterated until either case 1 or case 2 occurs.   The maximal integration time has been chosen such that most of the field trajectories have reached case 1 or case 2 at the end.  
\end{enumerate}

Since our aim is to extend the results of the chapter 5 to the case of a closed Universe, the approach is similar.  In a first step, we plot grids of initial conditions in field space, for fixed potential parameters and vanishing velocities.  Then, since the space of initial conditions and potential parameter values is 9-dimensional, we perform a Monte--Carlo--Markov--Chains (MCMC) analysis to determine in which regions of the parameter space the realization of a bounce plus inflation is more probable.   Finally, the genericity of the results for the original hybrid model are tested by considering a second model, F-term hybrid inflation in supergravity. 


\section{2-field hybrid dynamics in a closed Universe}

The expansion/contraction and the field dynamics in a closed Universe are governed by the Friedmann-Lema\^itre equations
\begin{equation} \label{eq:FLtc12field_bounce}
H^2 = \frac {8\pi }{3 \mpl^2}  \left[ \frac 1 2 \left(\dot
\phi^2 + \dot \psi^2 \right)  + V(\phi,\psi) \right]  - \frac{K}{a^2}~, 
\end{equation}
\begin{equation}
\frac{\ddot a }{a} = \frac {8\pi}{3 \mpl^2} \left[ - \dot \phi^2
- \dot \psi^2 + V(\phi,\psi ) \right]~,
\end{equation}
in which $K=1$, as well as the Klein-Gordon equations
\begin{equation} \label{eq:KGtc2field_bounce}
\ddot \phi + 3 H \dot \phi + \frac {\partial
V(\phi,\psi)}{\partial \phi} = 0~,
\end{equation}
\begin{equation}
\ddot \psi + 3 H \dot \psi + \frac {\partial 
V(\phi,\psi)}{\partial \psi} = 0~.
\end{equation}

These equations have been integrated numerically from initial conditions in the contracting phase.  Besides initial field values and velocities, one needs an initial condition for the curvature 
\begin{equation}
\Omega_K = \frac{-1}{H^2 a^2}~.
\end{equation} 
We have assumed that the field trajectories evolve initially in the phase of contraction around one of the global minima $(\phi=0,\psi = -M)$ of the potential\footnote{For the F-term SUGRA model, the global minimum is $(\phi = 0, \psi = - 2M)$.}.  

\section{Initial conditions in the contracting phase}

\subsection{Grids of initial conditions}

As a first step, we have calculated grids of initial field values,  for fixed potential parameters and initial curvature, and for vanishing initial velocities.  Such grids give an indication of the required tuning and show how are distributed in the field space the trajectories leading to a classical bounce plus a phase of hybrid inflation.  

For the original hybrid model as well as for the F-term SUGRA model, we found that a small but non-negligible proportion (about 3\%) of the initial field space generates a classical bounce plus a phase of hybrid inflation.  As for the initial conditions in a flat universe (see Chapter 5), the successful patches are arranged in complex structures.  These are observed to be self-similar when zooming over particular regions, an indication for fractal boundaries.  

The amount of successful initial conditions depends on the initial curvature.  As illustrated in Tab.~\ref{tab:curv}, a large initial value of  $- \Omega_{K}$ is needed for producing at least one bounce, typically $- \Omega_K \gtrsim 100 $, and for reaching a value of a few percents of trajectories triggering a phase of hybrid inflation.  
These observations remain roughly valid for various sets of potential parameters.  


We have performed a similar analysis for a Higgs-type potential (i.e. the model of Refs.~\cite{Falciano:2008gt,Lilley:2011ag}), but even by increasing very much the resolution, no successful trajectory has been found.  This result confirms that the scenario of a classical bounce plus a phase of inflation at the top of a Higgs potential suffers from an severe fine-tuned problem of the initial conditions.  
\vspace{3mm}

\begin{table} \label{tab:curv} \begin{center}
\begin{tabular}{|c|c|c|c|c|} 
\hline
$\Omega_K$  & \ $> 1$ CB \   &  CB + HI & \  $> 1$ CB \ & CB + HI     \\
\hline
\multicolumn{3}{|c|}{\textit{Original hybrid model} } & \multicolumn{2}{|c|}{\textit{F-term SUGRA model} } \\
\hline
$- 0.1 $ & 0 \%  & 0 \%& 0 \%  & 0 \% \\
$-1. $ & 0 \%  &  0 \% & 0 \%  &  0 \% \\
$-10. $ & 2.4 \%  & 0.9 \%  &2.2 \%  & 0.8 \%  \\
$-100 $ & 6.9 \%  & 2.8 \%  &9.0 \%  & 2.5 \%  \\
$-10^{3} $ & 8.7 \%  & 3.1 \% & 15 \%  & 4.2 \% \\
$-10^{4} $ & 48 \%  & 3.3 \%  & 33 \%  & 4.0 \% \\
$-10^{5} $ & 95 \%  & 2.8 \%  & 94 \%  & 3.9 \% \\
$-10^{6} $ & 100 \%  & 2.9 \% & 99 \%  & 4.1 \% \\
$-10^{7} $ & 100 \% & 3.1 \% & 100 \% & 4.3 \% \\
$-10^{8} $ & 100 \%  & 2.9 \% & 100 \%  & 3.8 \% \\
$-10^{9} $ & 100 \%  & 2.8 \% & 100 \%  & 3.9 \% \\
$-10^{10} $ & 100 \%  & 2.9 \% & 100 \%  & 4.2 \% \\
\hline 
 \end{tabular}
 \caption{Percentages of field trajectories performing at least one classical bounce (columns 2 and 4), for various initial values of the curvature $\Omega_K$.  The percentages of field trajectories performing at least one classical bounce plus a phase of hybrid inflation are given in columns 3 and 5.   Results are for $100 \times 100$ grids of initial conditions.   The potential parameter values are $M = \phi_{\rr c} = 0.03 m_{\rr p}, \mu  = 636 m_{\rr p}, \Lambda^4 = 2 \times 10^{-7}  m_{\rr p}^4 $ for the original hybrid model, $\kappa = 5 \times 10^{-4}, M = 0.01 m_{\rr p}$ for the F-term SUGRA model. }
\end{center}
\end{table}

\begin{figure}[p] \begin{center}
 \includegraphics[width=12.0cm]{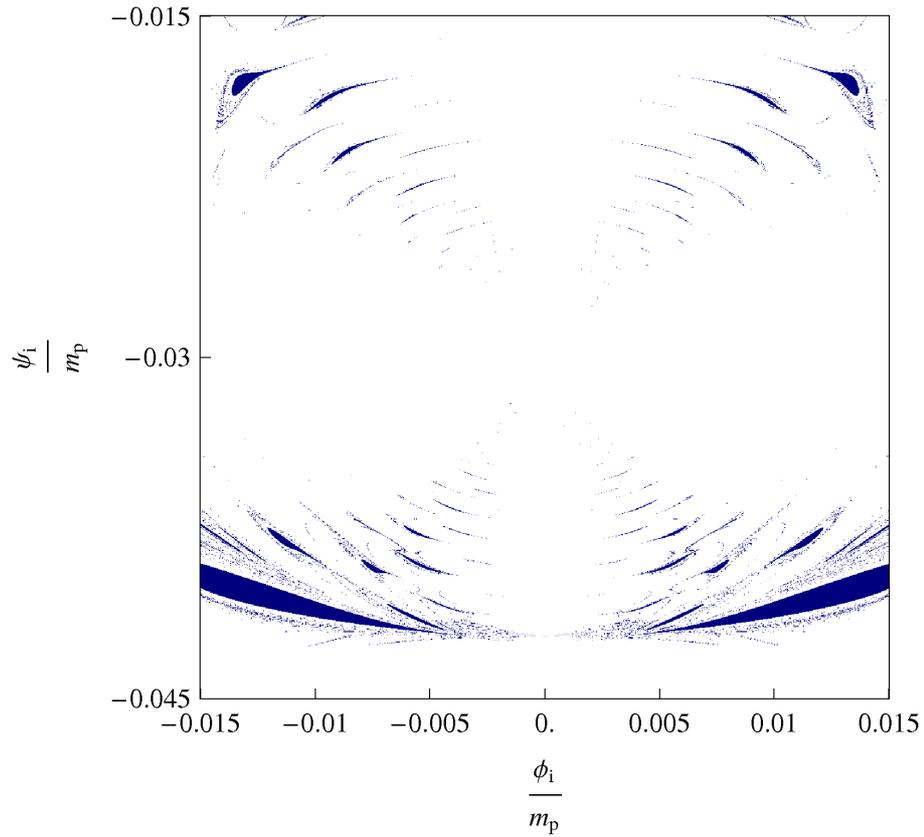}
  \caption{$1000\times 1000$ grid of initial conditions for the original hybrid model, for $M= \phi_{\rr c} = 0.03 \ \mpl$, $\mu = 636.4 \ \mpl $, $\Lambda^4 = 2\times 10^{-7} \Mpl^4$, $\Omega_{\rr k} = -10^{6} $.  White regions correspond to trajectories reaching the Planck energy density.  Dark blue points correspond to trajectories undergoing a classical bounce plus a phase of hybrid inflation. These occupy about 3\% of the field space. }  
  \end{center}
\end{figure}

\begin{figure}[p]   \begin{center}
 \includegraphics[width=12.0cm]{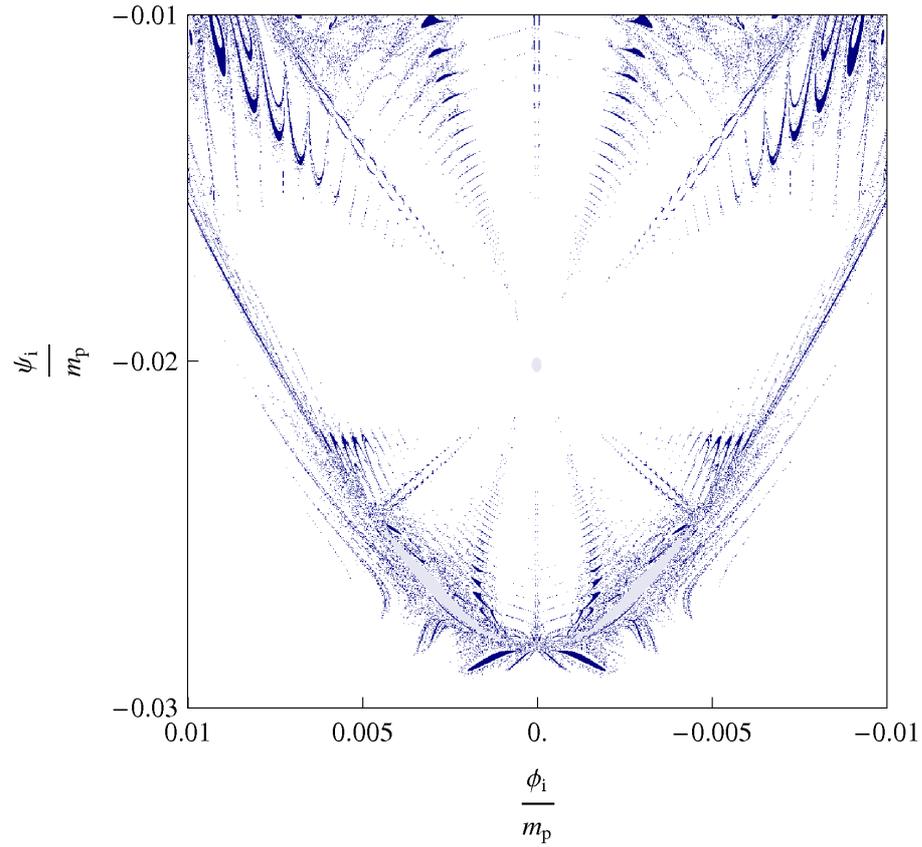}
  \caption{$1000\times 1000$ grid of initial conditions for the F-term SUGRA model, for $M= 0.01 \ \mpl$, $\kappa = 5.\times 10^{-4}$, $\Omega_{\rr k} = -10^{6} $.  White regions correspond to trajectories reaching the Planck energy density. Dark blue points (about 4 \%) correspond to trajectories undergoing a classical bounce plus a phase of hybrid inflation.  Subdominant light blue points are the initial conditions of trajectories that do not reach neither the Planck scale, neither the inflationary valley, at the end of the numerical integration.  }
   \end{center}
\end{figure}

\subsection{MCMC exploration of the parameter space}

In order to probe the whole parameter space of the models, we use a MCMC method identical to the one used in chapter 5, except that there are two additional parameters:  $\Lambda^4$ and the initial curvature $\Omega_K $.  

\subsubsection{Original hybrid model}

Compared to the chapter 5, the priors are identical except for
\begin{eqnarray}
10^{-10} M_{\rr{pl}}^4 & < \Lambda^4 & < M_{\rr{pl}}^4~,  \\
10^{-1}   & < - \Omega_{K} & < 10^{11}~,  \\
0  & <  \dfrac{v}{C \sqrt{1 - \Omega_K}}  & < \sqrt 6   M_{\rr{pl}}~.
\end{eqnarray}
For these parameters, we took flat priors on the logarithm.  The ranges of $- \Omega_{\rr K} $ and $\Lambda^4$ are adjusted for numerical convenience.  It has nevertheless been verified that the behavior of their posterior distributions can be extended to respectively lower and higher values.   In order to focus on the regime of small initial field velocities (so that the field trajectories are not thrown directly at the Planck scale but initially evolve around one of the global minima of the potential),  we consider $v / C \sqrt{1- \Omega_K} $, with $C = 100 $.   The case $C=1$ will be only briefly discussed.  

Furthermore, for the initial field values $\phi_{\rr i} $ and $\psi_{\rr i}$, we define $\phi_{\rr{rel}}  \equiv  \phi_{\rr i}    / \phi_{\rr c} $ and $\psi_{\rr{rel}} \equiv  - (  M + \psi_{\rr i} )/M $ and we take a flat prior on these parameters within the range
\begin{eqnarray}
- 1 &< \phi_{\rr{rel}}  < 1 \\
-1 &< \psi_{\rr{rel}} < 1  ~.
\end{eqnarray}
In this way, the initial field space we probe is around one of the global minima of the potential and is scaled with the parameters $M$ and $\phi_{\rr c} $. 

\begin{figure}[p]  \begin{center}
\begin{minipage}[t]{5.2cm}
\vspace{0pt}   \includegraphics[width=5.2cm]{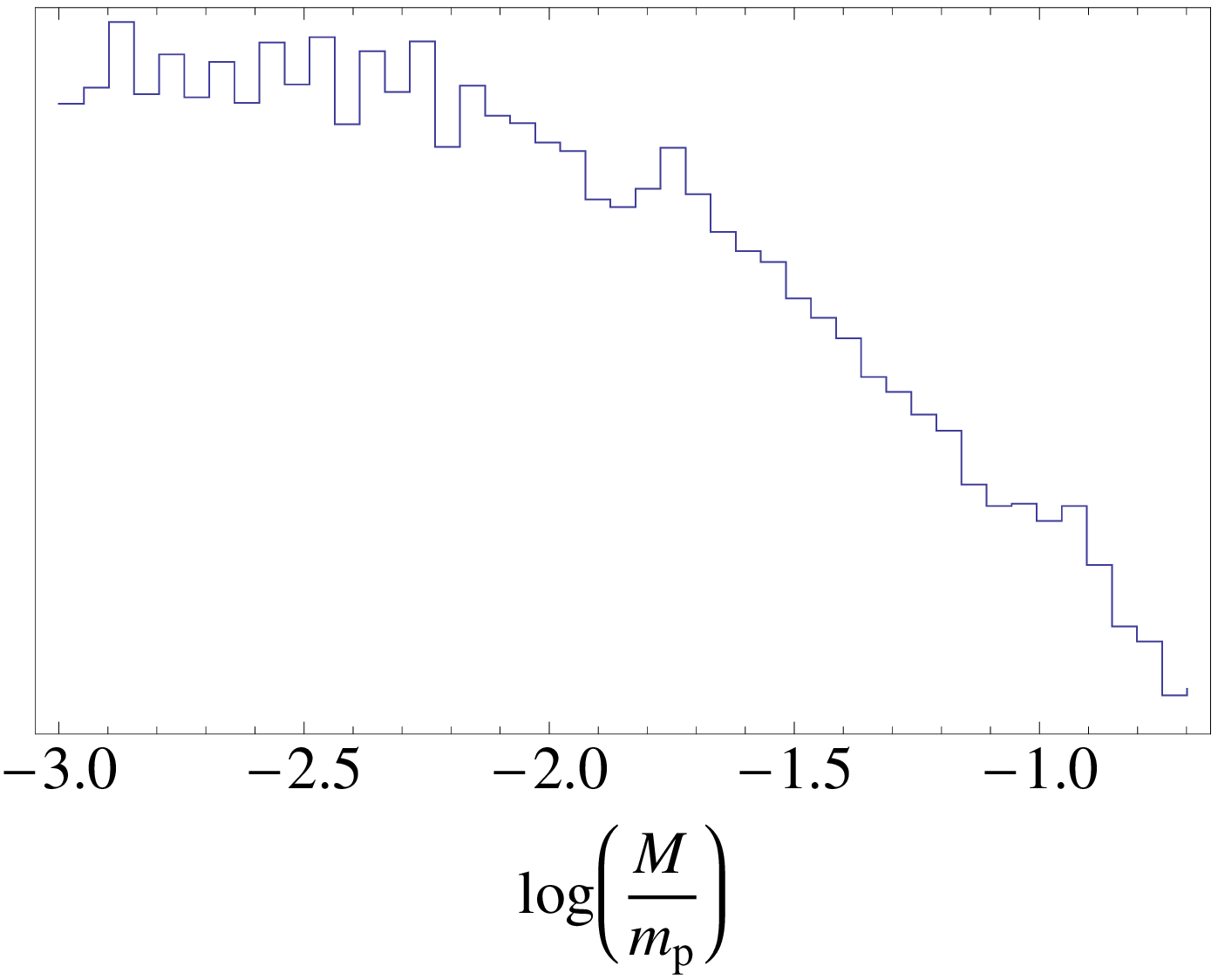}
\end{minipage}
\begin{minipage}[t]{5.2cm}
\vspace{0pt}   \includegraphics[width=5.2cm]{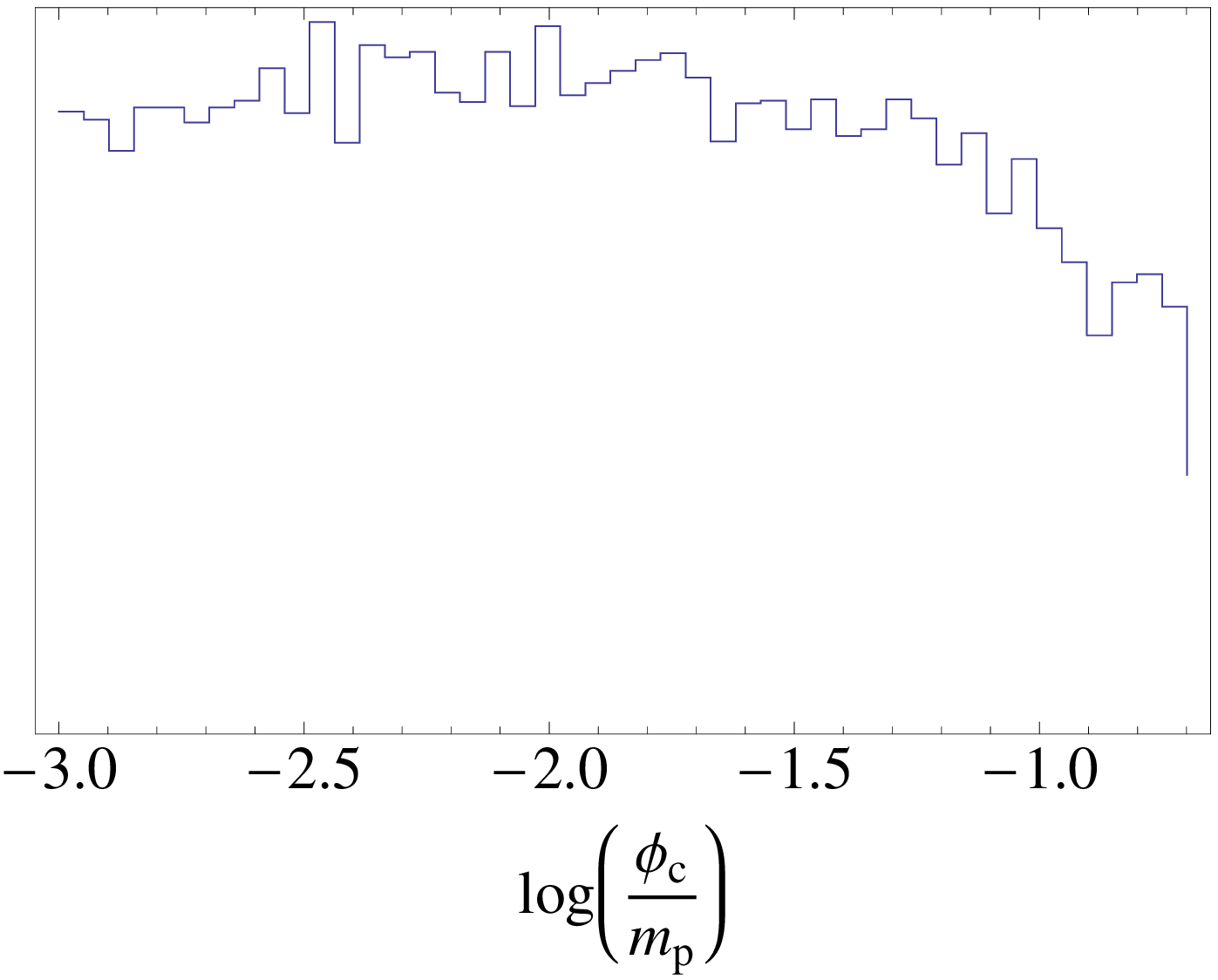}
\end{minipage}

\begin{minipage}[t]{5.2cm}
\vspace{0pt}  \includegraphics[width=5.2cm]{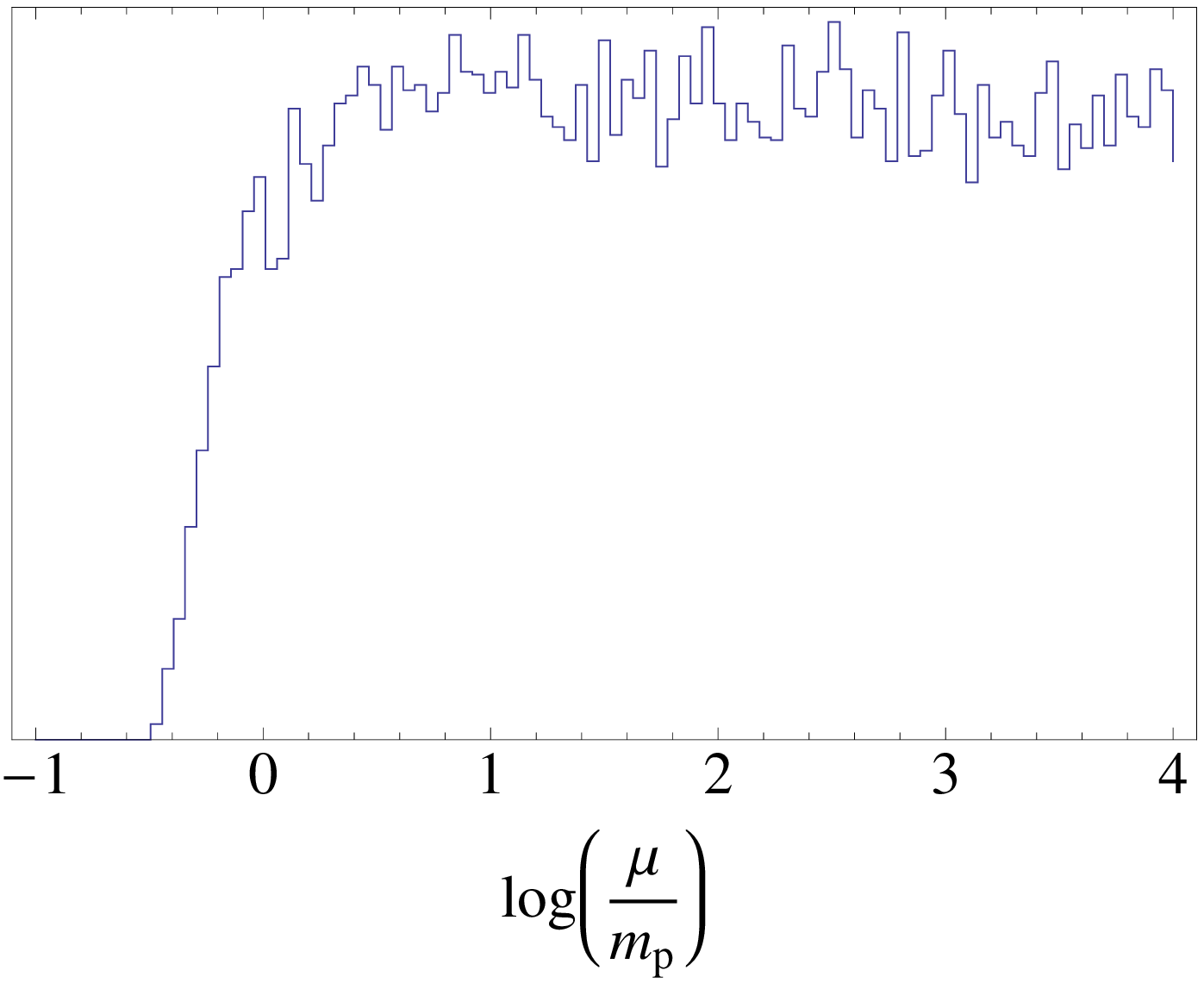}
\end{minipage}
\begin{minipage}[t]{5.2cm}
\vspace{0pt}   \includegraphics[width=5.2cm]{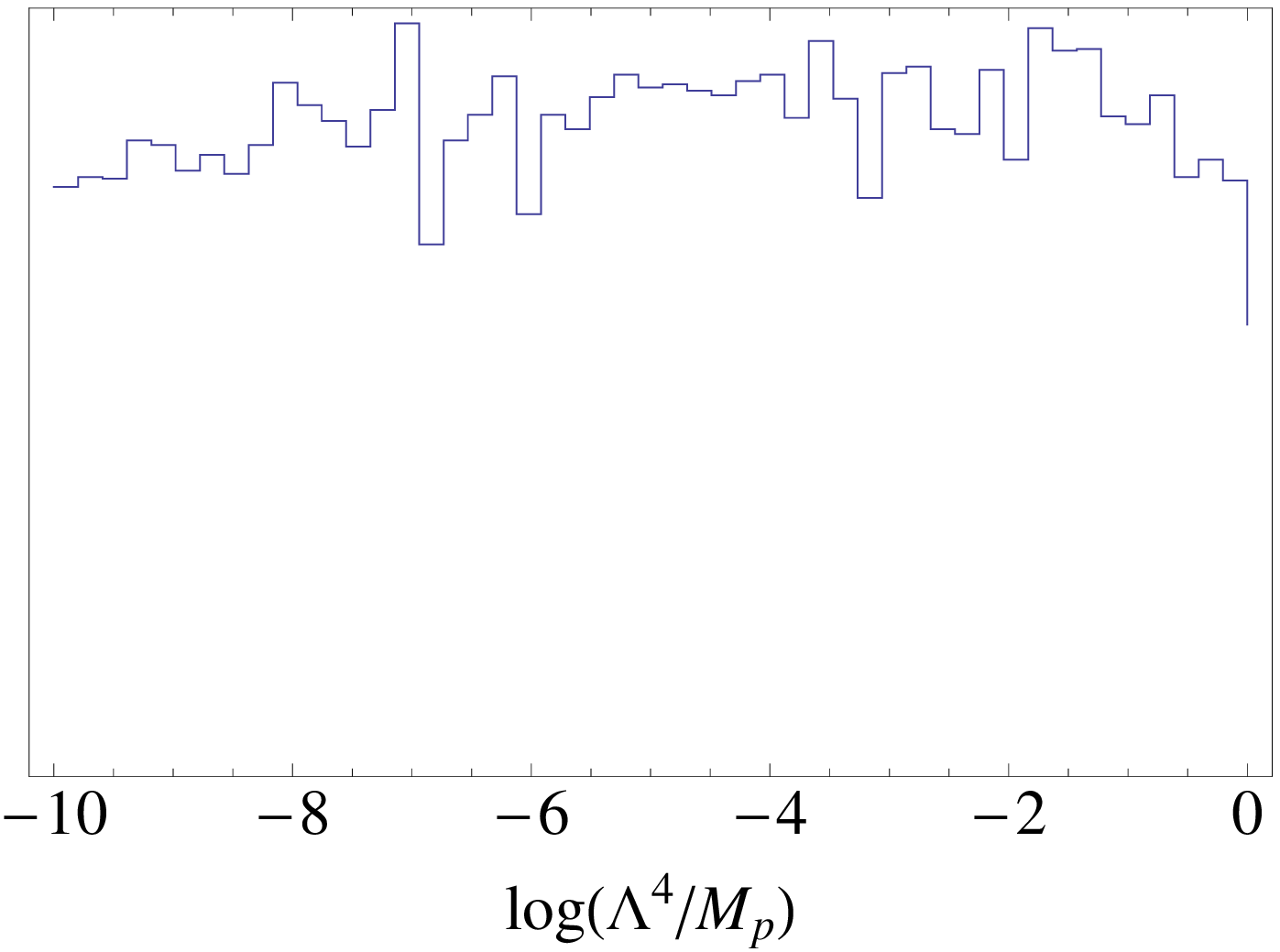}
\end{minipage}

\begin{minipage}[t]{5.2cm}
\vspace{0pt}  \includegraphics[width=5.2cm]{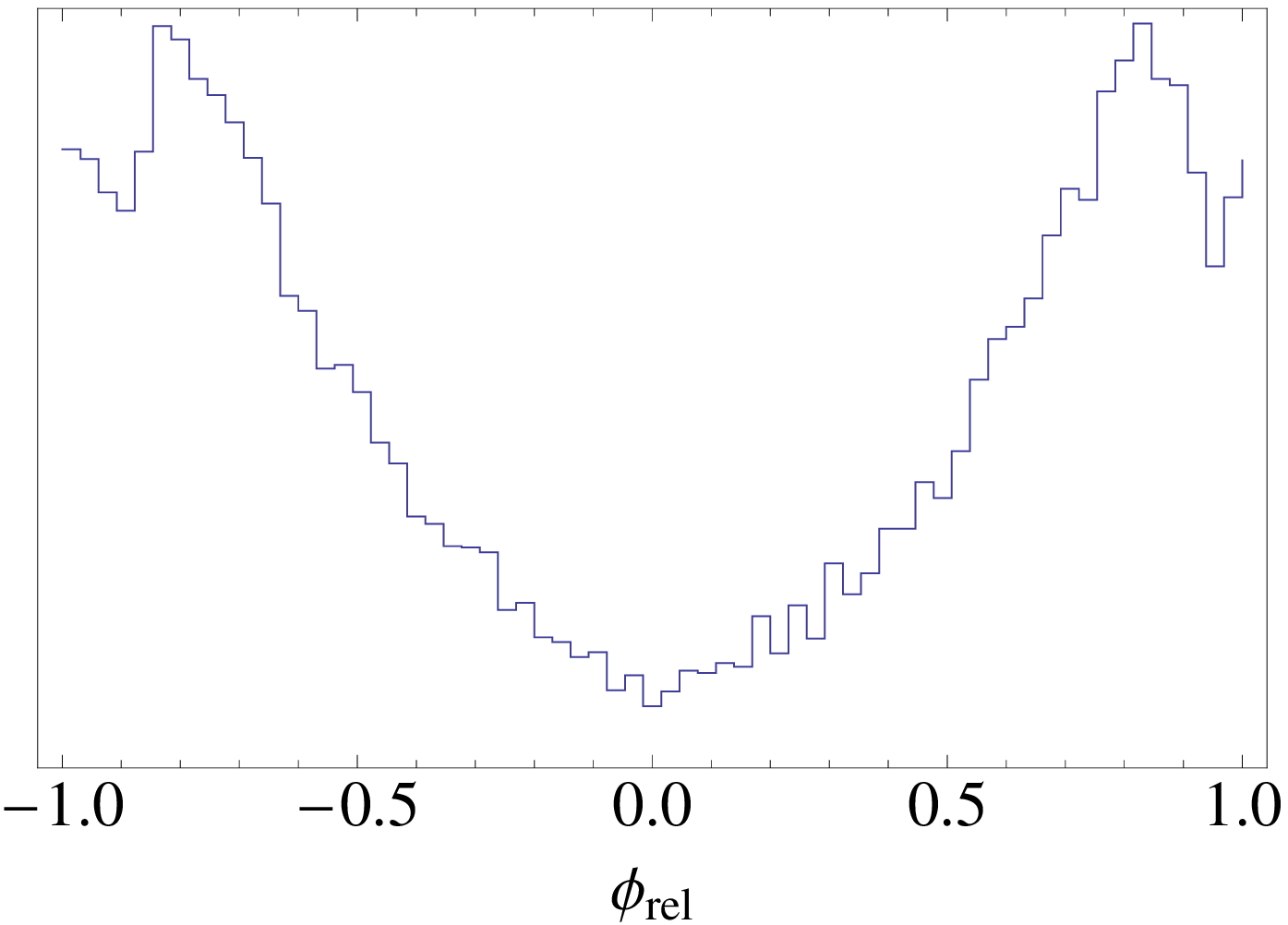}
\end{minipage}
\begin{minipage}[t]{5.2cm}
\vspace{0pt}  \includegraphics[width=5.2cm]{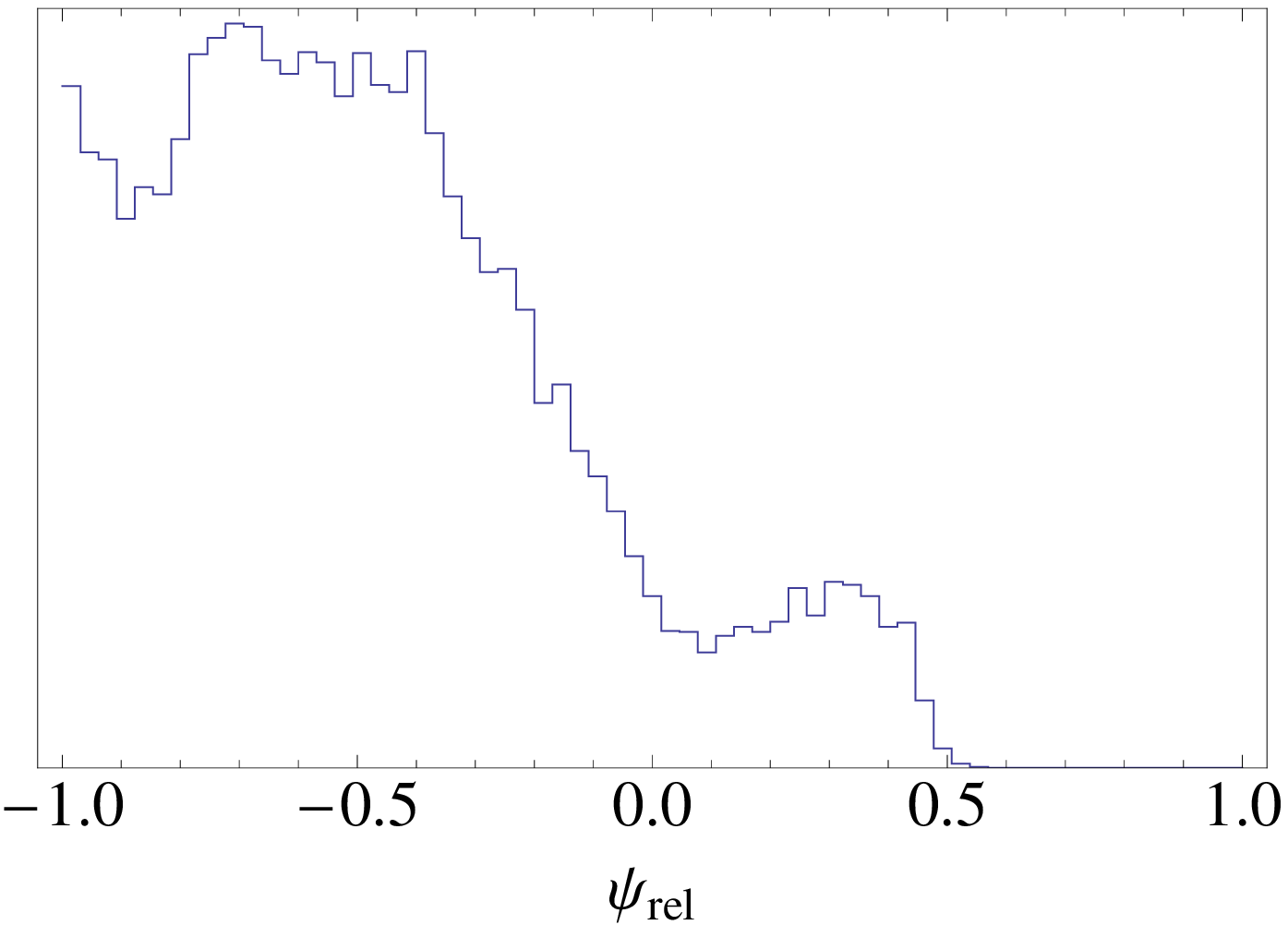}
\end{minipage}

\begin{minipage}[t]{5.2cm}
\vspace{0pt}   \includegraphics[width=5.2cm]{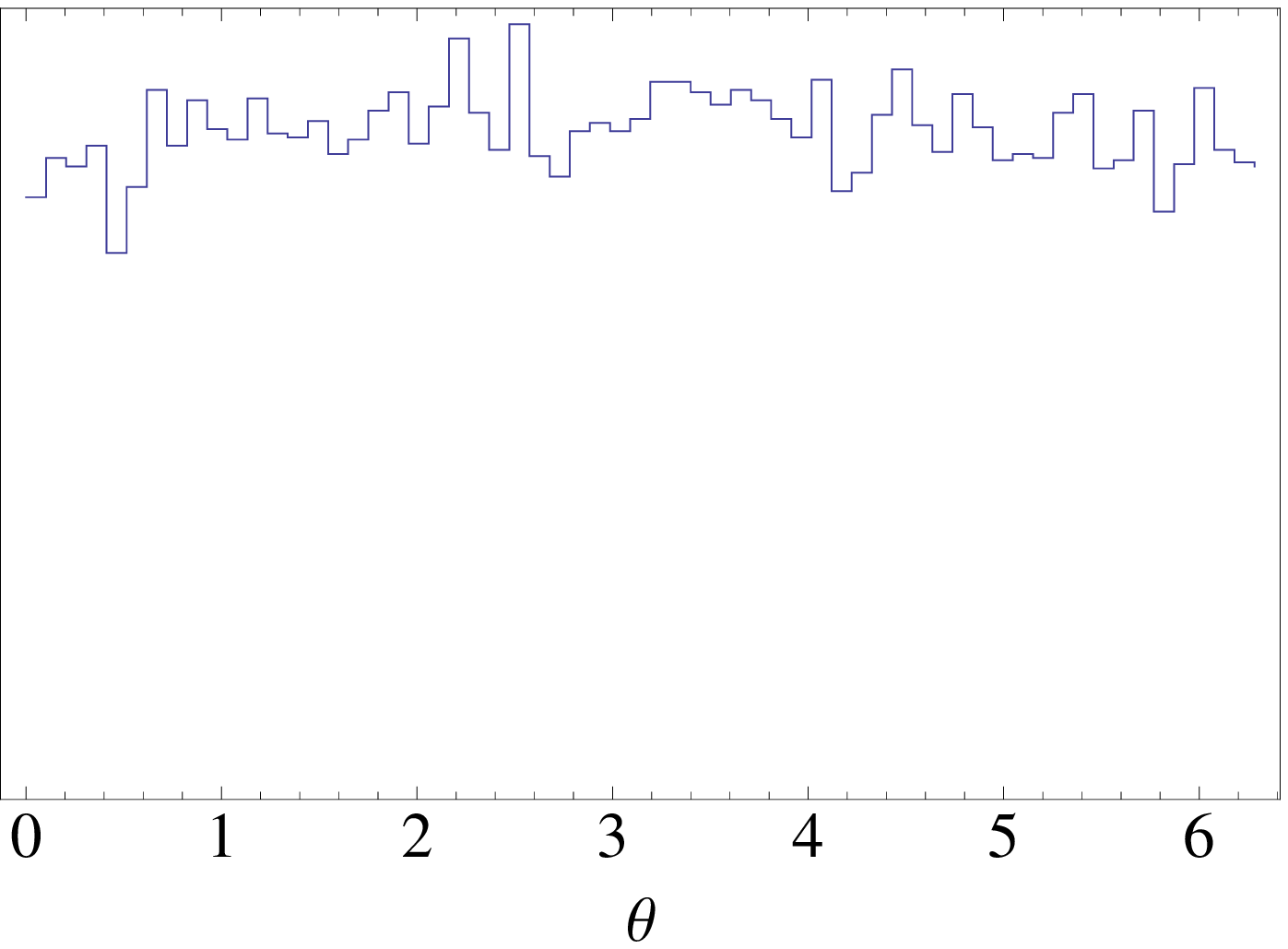}
\end{minipage}
\begin{minipage}[t]{5.2cm}
\vspace{0pt}   \includegraphics[width=5.2cm]{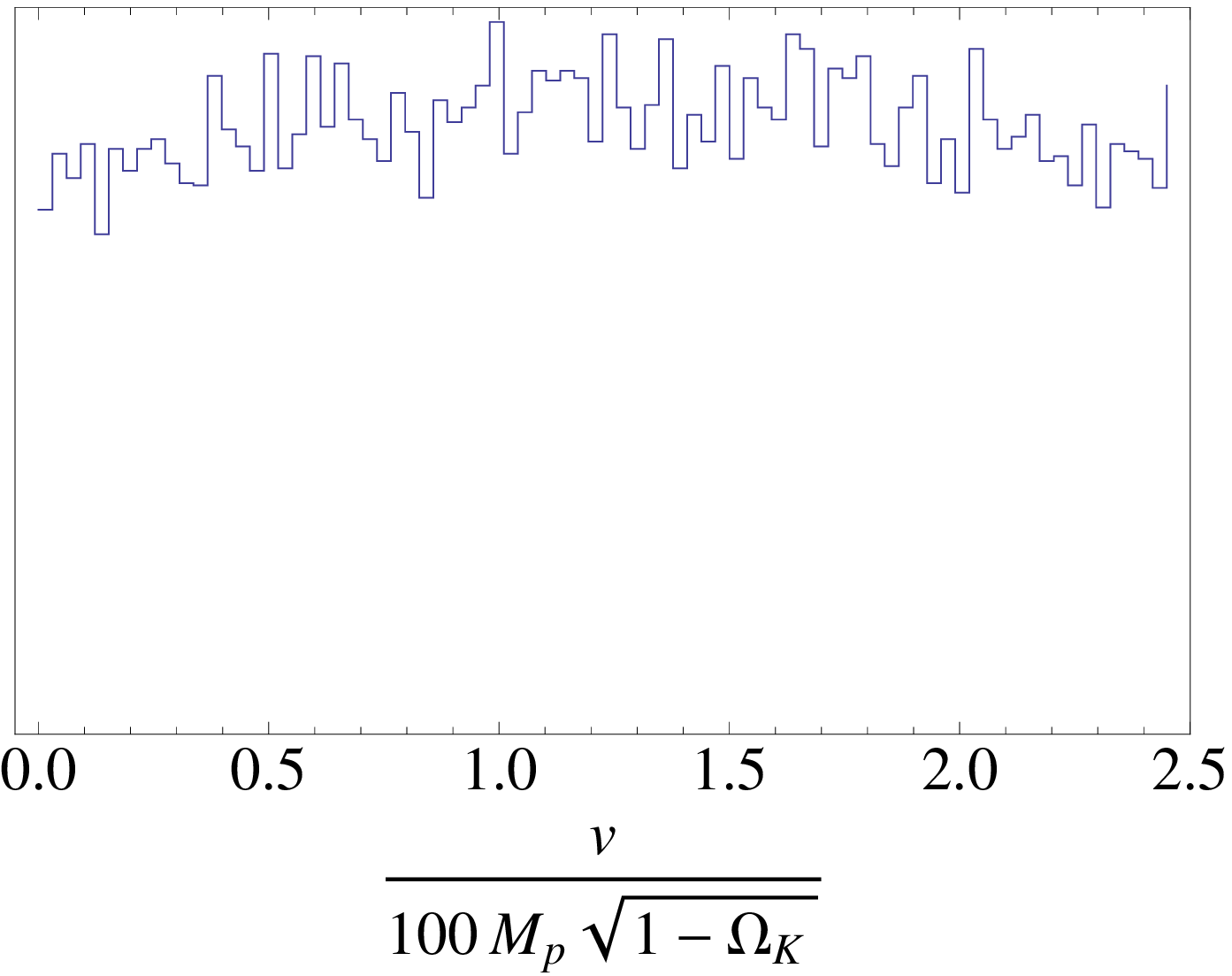}
\end{minipage}

\begin{minipage}[t]{5.2cm}
\vspace{0pt}  \includegraphics[width=5.2cm]{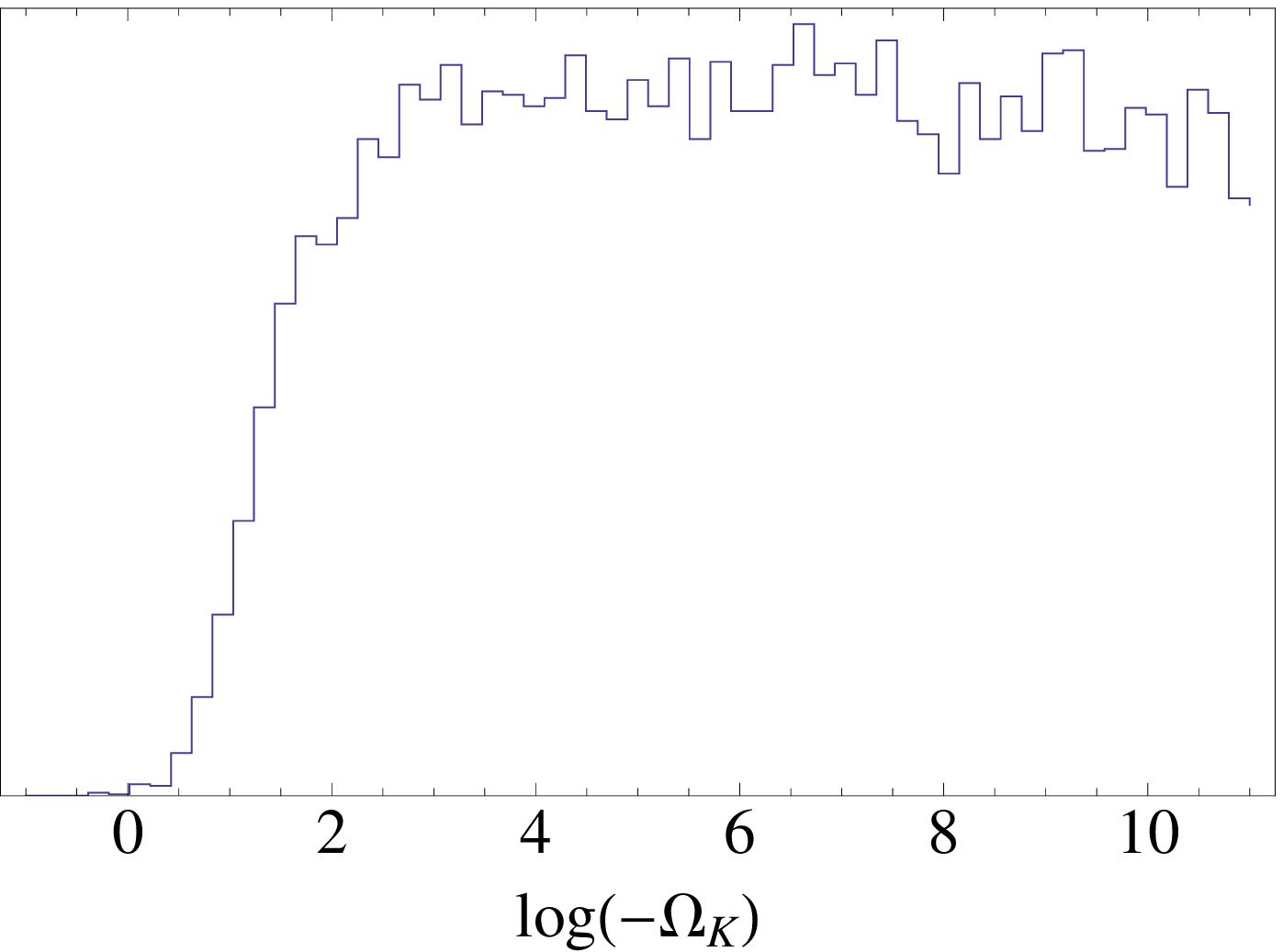}
\end{minipage}
\begin{minipage}[t]{5.2cm}
\vspace{0pt}   \includegraphics[width=5.2cm]{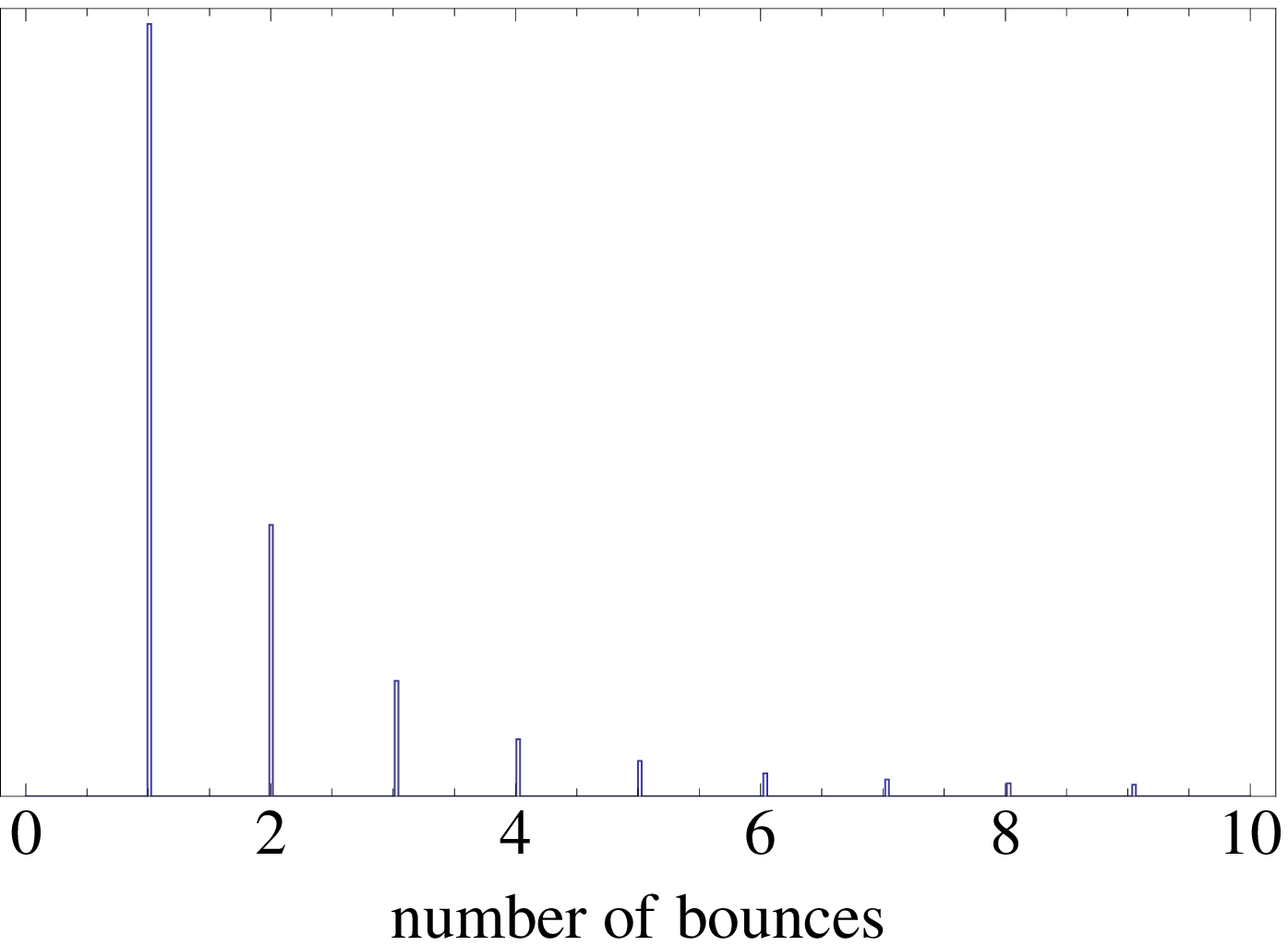}
\end{minipage}

  \caption{Marginalized posterior probability density distributions (normalized to unity) to generate a classical bounce followed by a phase of inflation, for the parameters of the original hybrid model.  Planck-like values of $M$ are slightly disfavored. A bound on the initial curvature is observed.  The distribution of the number of bounces realized before reaching the phase of hybrid inflation decreases exponentially.  }
  \label{fig:mcmcbouncehybrid}  \end{center}
\end{figure}

The marginalized posterior probability density distributions, normalized to unity, are given in Fig.~\ref{fig:mcmcbouncehybrid}.
As in Chapter 5, the lower bound $\mu \simeq m_{\rr p} $ is observed.  For smaller values of $\mu$, field trajectories reach the valley mostly at large field phase ($\phi > \mu$) and slow-roll violations prevent inflation to be triggered at small field values, as discussed in chapter 4.   Otherwise, the posterior on $\mu$ is almost flat, like for the posteriors of the potential parameters $\phi_{\rr c} $ and $\Lambda^4$ (but which exhibits a slight decreasing for the posterior $\phi_{\rr c} $ 
distribution near $m_{\rr p} $), and for the initial velocity parameters $v/ (100 \sqrt{1 - \Omega_K})$ and $\theta$.  These parameters therefore only influence marginally the probability to generate a classical bounce followed by a phase of hybrid inflation.   On the contrary, planck-like values of the potential parameter $M$ are observed to be disfavored.  But below $M \simeq 10^{-2} \mpl $, the posterior distribution remains almost flat.  

As expected given the grids of initial conditions, the posterior distributions of $\phi_{\rr{rel}} $ and $\psi_{\rr{rel}} $ peak near $0.7 m_{\rr p} $.  They are nevertheless smoothed by the marginalization over the other parameters.   The posterior distributions fall off near $\phi_{\rr{rel}} = 0$ and  
$\psi_{\rr{rel}} = 0$, but this observation is sensitive to the prior.  For a flat prior taken on the logarithm of 
$\phi_{\rr{rel}} $ and  $\psi_{\rr{rel}} $, the posterior distributions are actually almost flat.  This indicates that the structures of successful initial field values observed with the gridding method are reproduced at smaller scales around the global minima.    However, for $\phi_{\rr{rel} } \gtrsim 0.5 $, i.e. for initial field values far from the valley, all the trajectories reach the Planck scale and the posterior distribution vanish.  

The posterior for the initial velocity parameters $v / (100 \sqrt{\Omega_K})$ and $\theta $ are almost flat. Thus one can conclude, as in the chapter 5, that these parameters do not influence the probability to generate a classical bounce plus hybrid inflation scenario.  This is valid for initial velocities relatively small compared to their maximal allowed value.  In the case $C=1$, posterior distributions are not flat and high velocities are disfavorized.  Indeed, contrary to chapter 5, in the contracting phase the Hubble term in the K.G. equation is not a friction term and thus the initial velocity can push the field trajectories directly to the Planck scale.  

The main result concerns the posterior probability density distribution of the initial curvature $-\Omega_K$.   The classical bounce plus hybrid inflation scenario is found to require high values of the curvature in the contracting phase and lower bound can be determined, 
\begin{equation}
- \Omega_K \gtrsim 10~. 
\end{equation}
However, for $ - \Omega_K \gtrsim 10^3 $ the posterior distribution is observed to be nearly flat.   This confirms and generalizes the results of Tab.~\ref{tab:curv} to the entire parameter space.   If this bound is satisfied, any precise scale of the spatial curvature is therefore favored. 

Finally, it is interesting to notice that the distribution of the number of classical bounces realized before the trajectories reach the slow-roll attractor along the inflationary valley, is exponentially decreasing.  Nevertheless, the integrated probability for more than one bounce is higher than the probability to realize inflation directly after the first classical bounce.   As an example, several trajectories performing more than 100 bounces have been observed in the Markov chains.  Thus a multi-bounces scenario is possible.

\begin{figure}[p]  \begin{center}
\begin{minipage}[t]{6.cm}
\vspace{0pt}  \includegraphics[width=6.0cm]{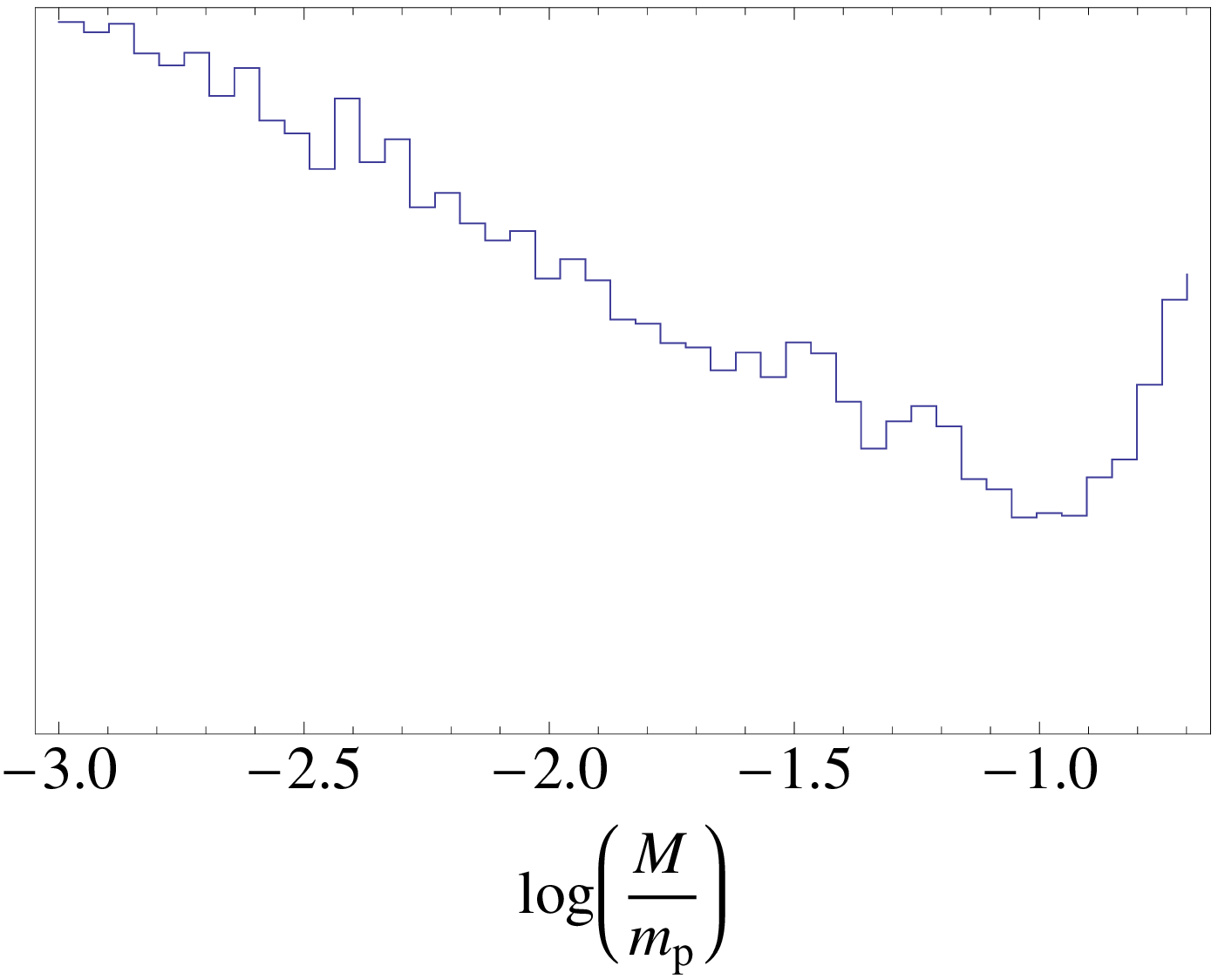}
\end{minipage}
\begin{minipage}[t]{6.cm}
\vspace{0pt}  \includegraphics[width=6.0cm]{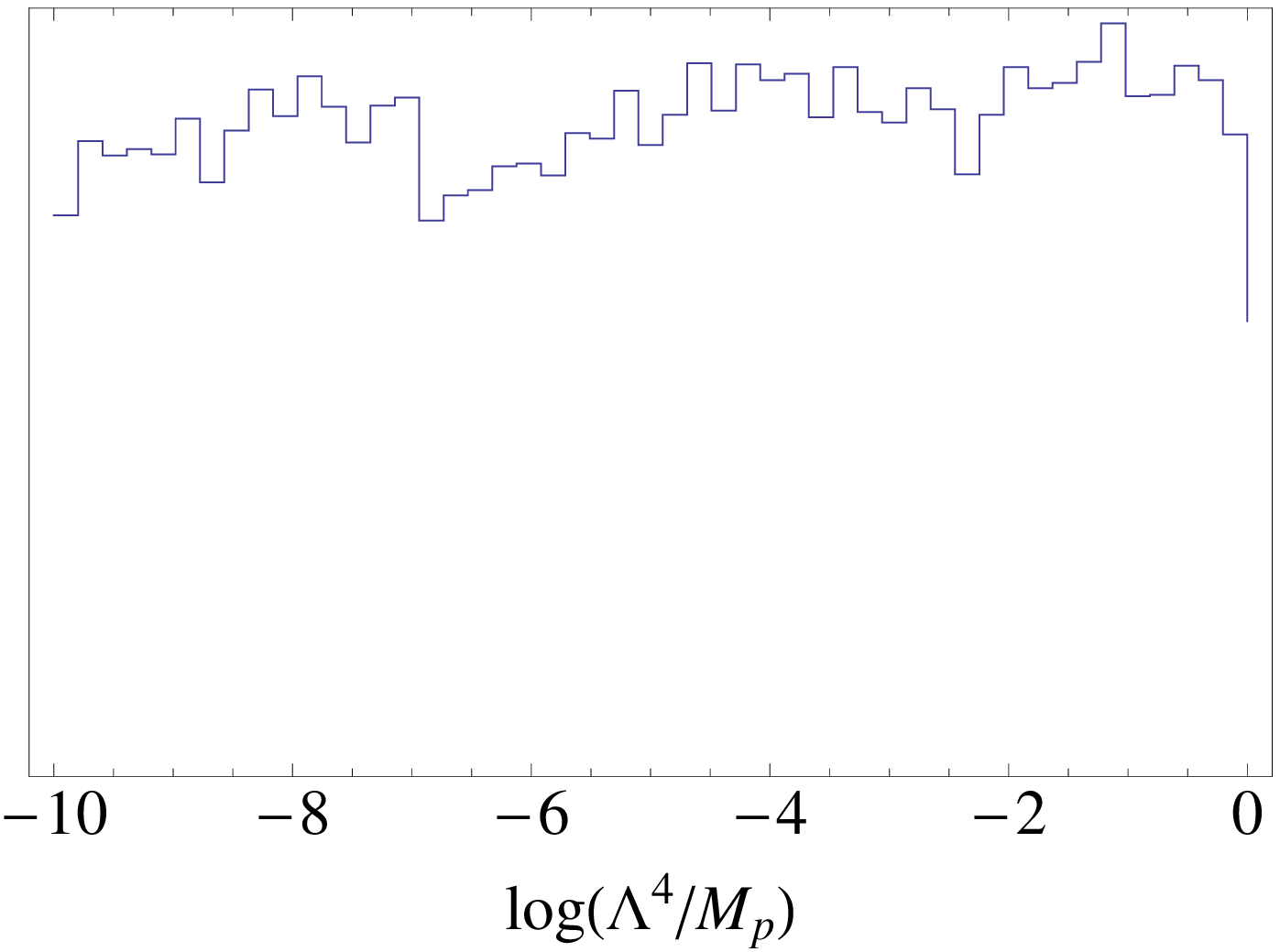}
\end{minipage}

\begin{minipage}[t]{6.cm}
\vspace{0pt}  \includegraphics[width=6.0cm]{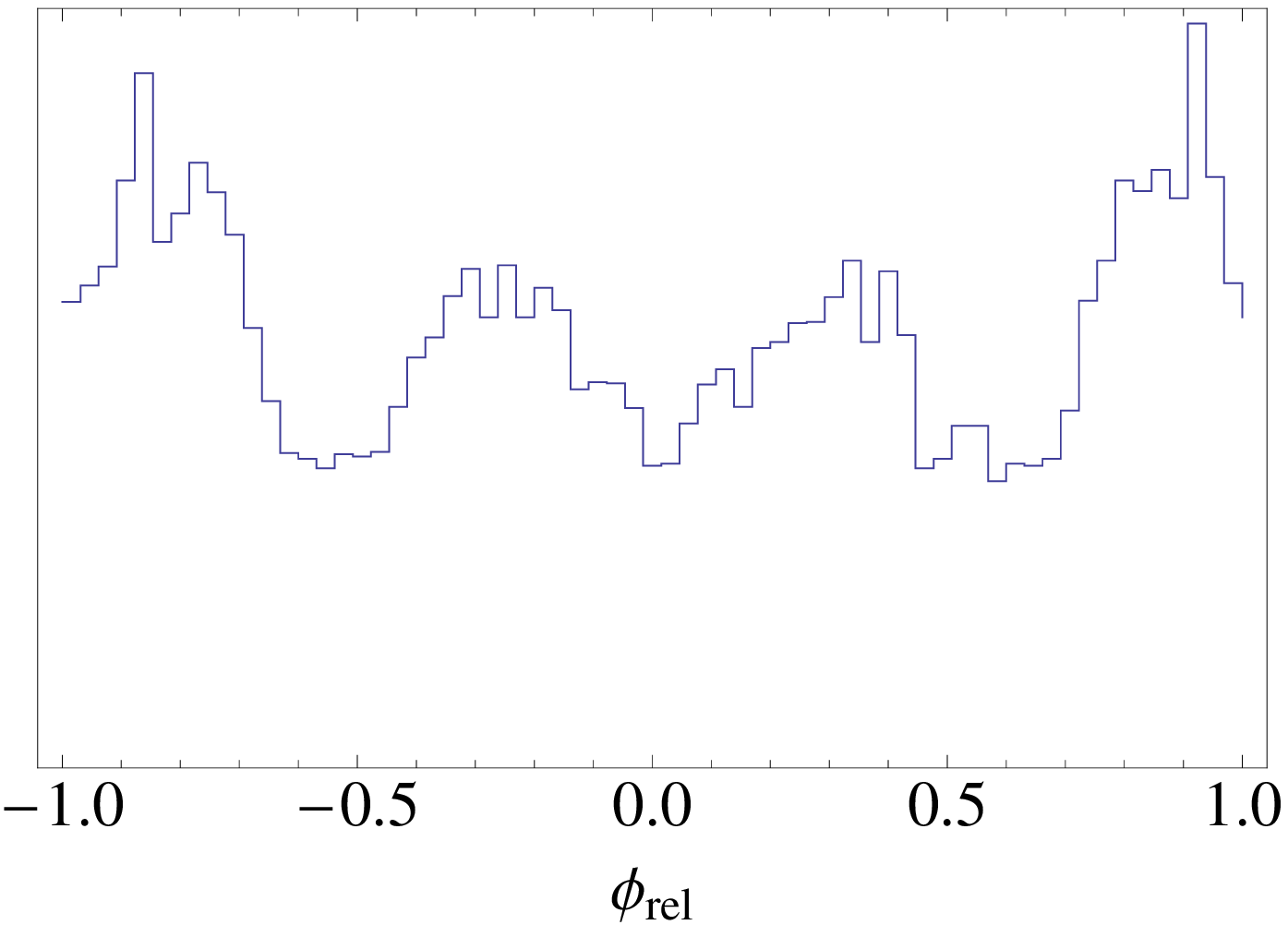}
\end{minipage}
\begin{minipage}[t]{6.cm}
\vspace{0pt}  \includegraphics[width=6.0cm]{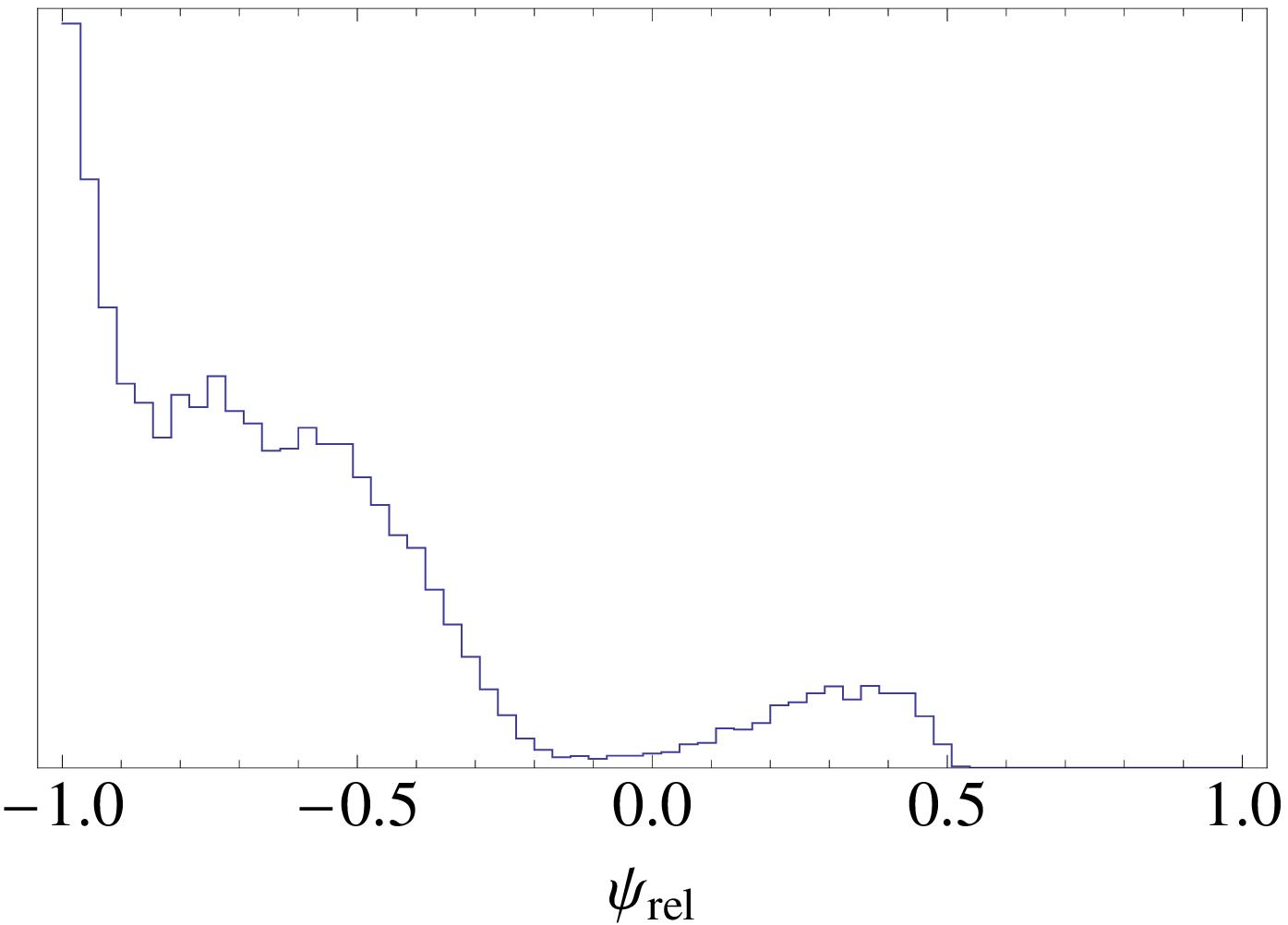}
\end{minipage}

\begin{minipage}[t]{6.cm}
\vspace{0pt}  \includegraphics[width=6.0cm]{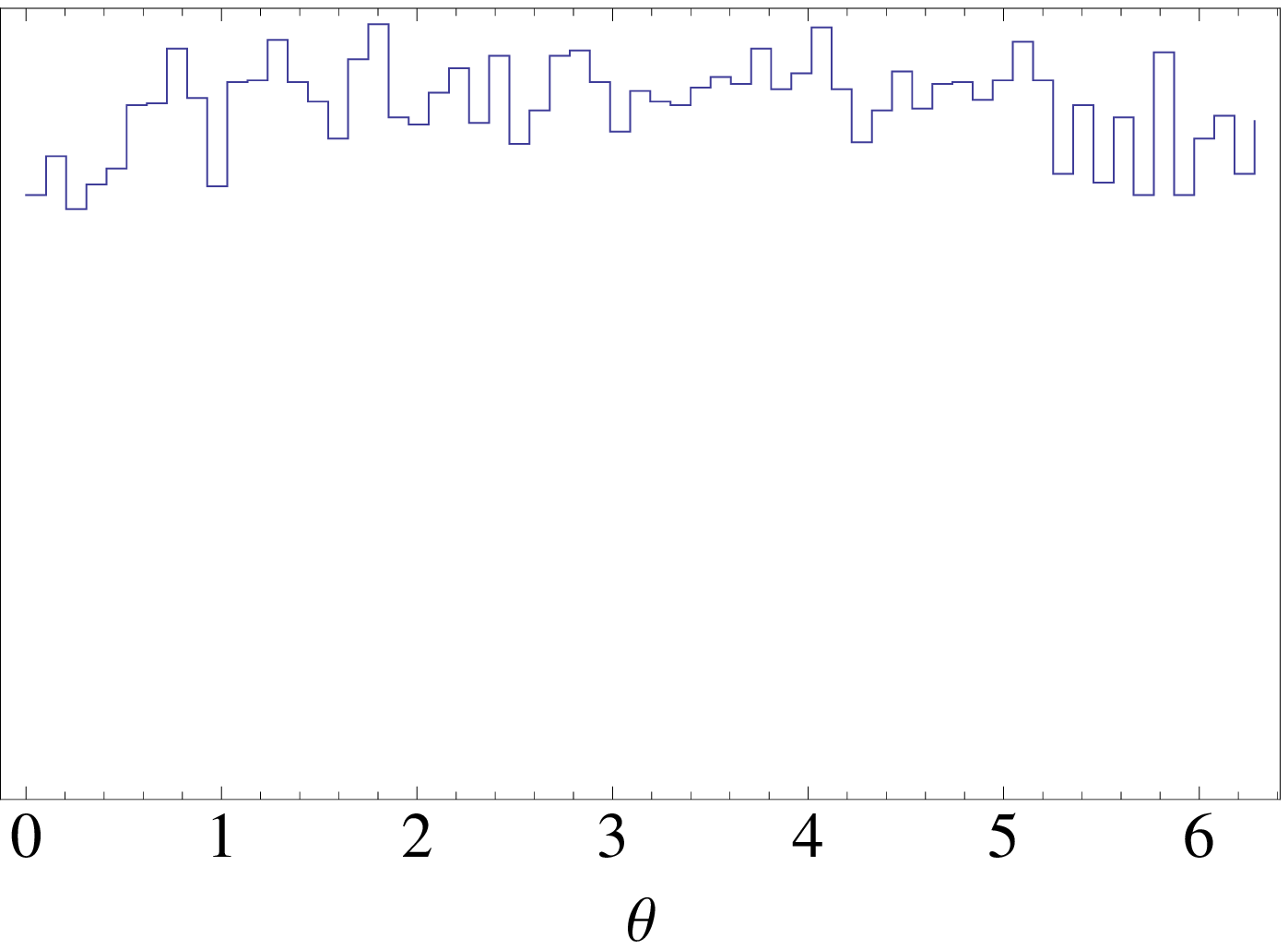}
\end{minipage}
\begin{minipage}[t]{6.cm}
\vspace{0pt}  \includegraphics[width=6.0cm]{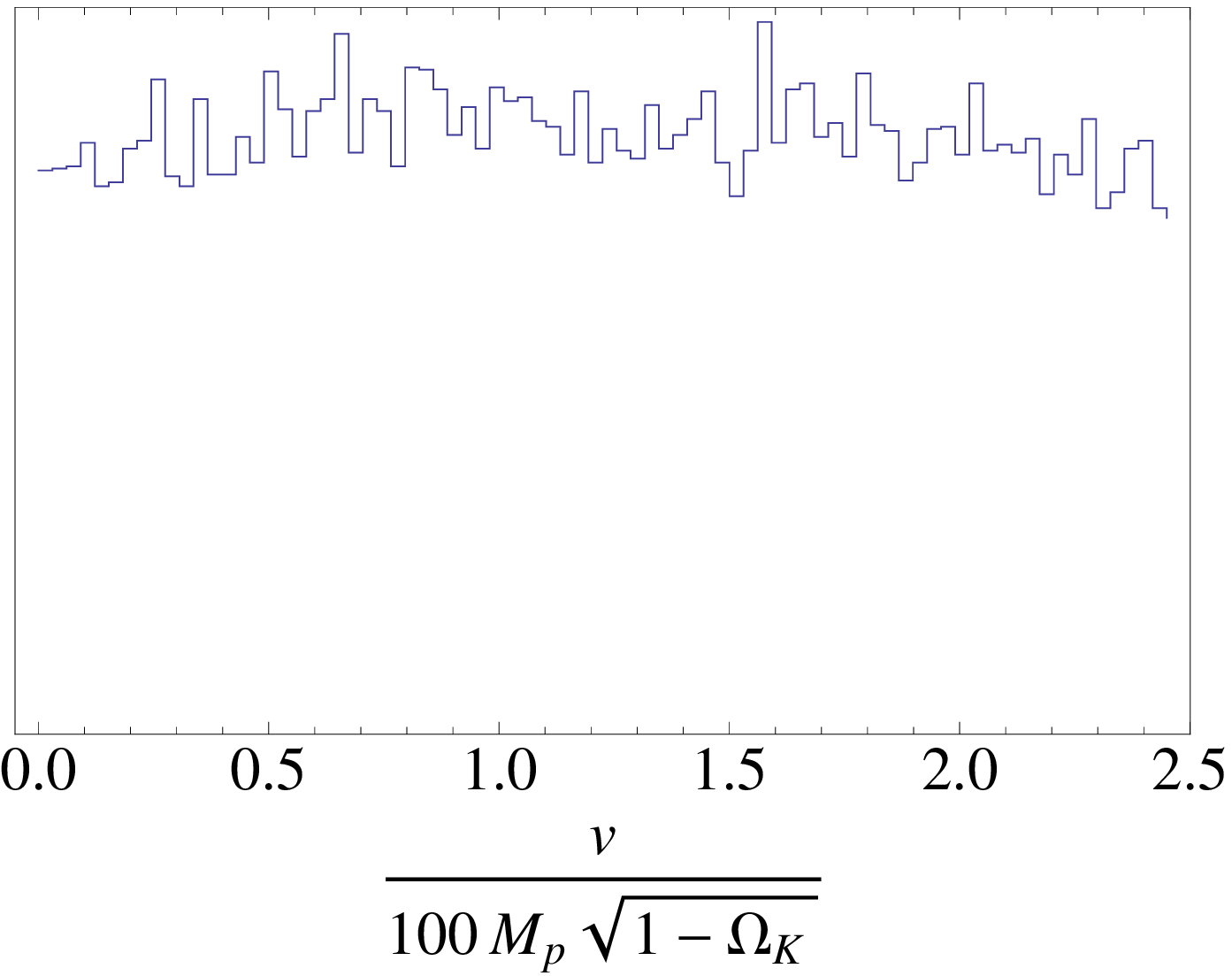}
\end{minipage}

\begin{minipage}[t]{6.cm}
\vspace{0pt}  \includegraphics[width=6.0cm]{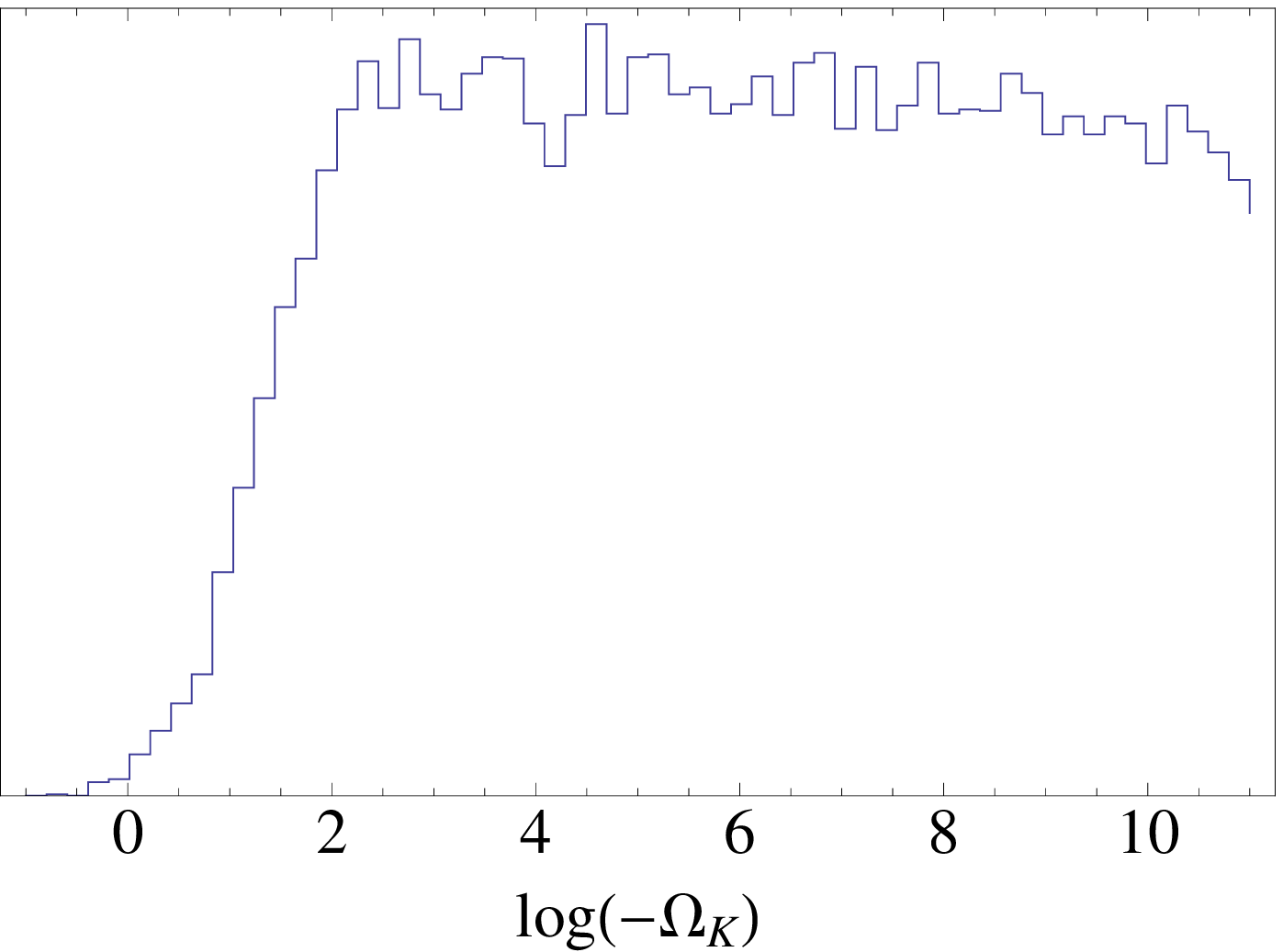}
\end{minipage}
\begin{minipage}[t]{6.cm}
\vspace{0pt}  \includegraphics[width=6.0cm]{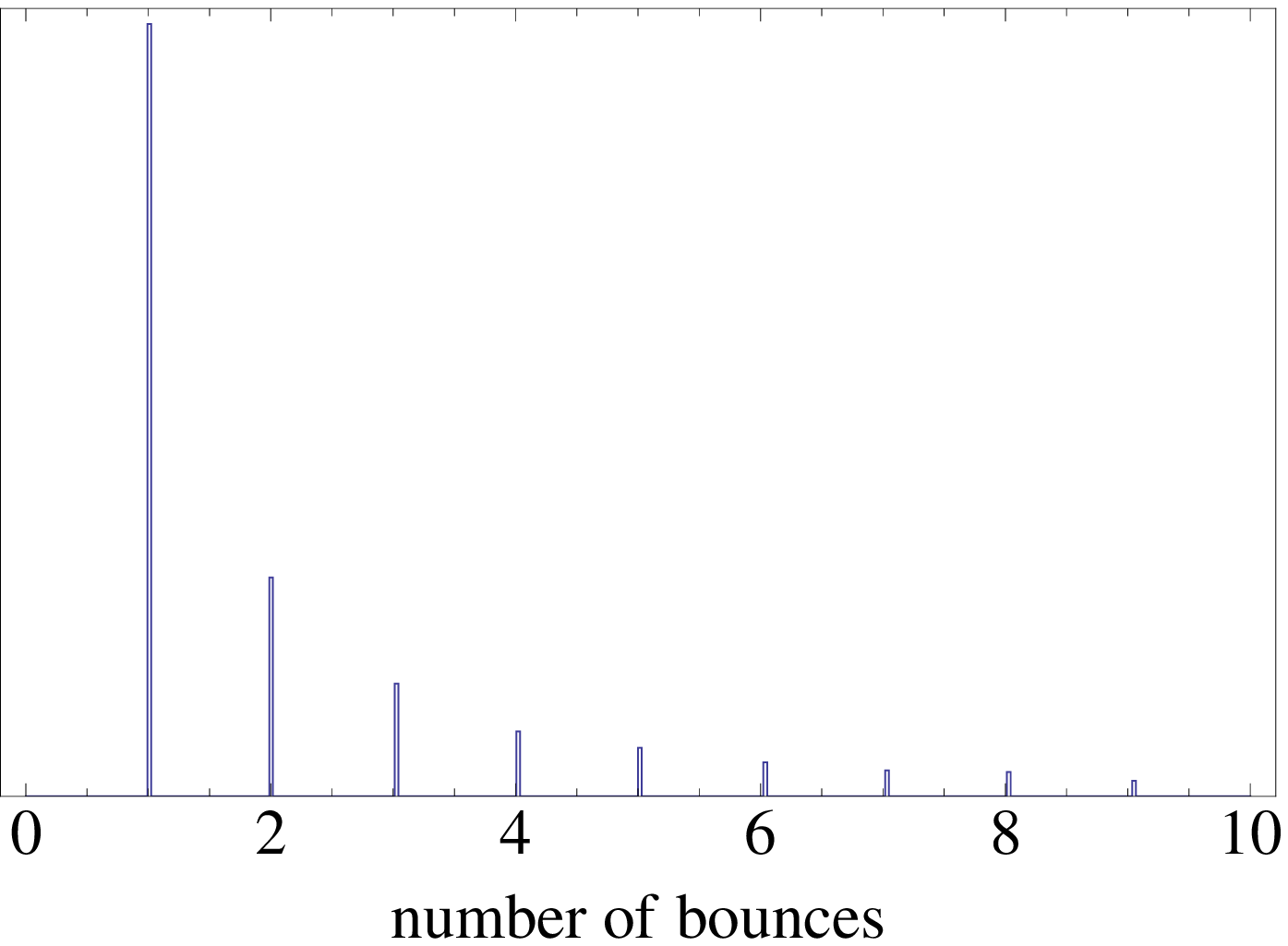}
\end{minipage}

  \caption{Marginalized posterior probability density distributions (normalized to unity) for the parameters of theF-term SUGRA model.  As for the original model, a bound on the initial curvature is observed and the distribution of the number of bounces decreases exponentially.  }
  \label{fig:mcmcSUGRAhybrid}  \end{center}
\end{figure}

\subsubsection{F-term SUGRA hybrid model}

As mentioned in the Chapter 3, an interest of the F-term SUGRA potential is that it contains only two potential parameters, $M$ and $\Lambda^4 \equiv \kappa^2 M^4 $.  Thus the dimensionality of the parameter space is reduced.  We took for this model identical priors than for the original hybrid one, except that we have defined $\phi_{\rr{rel}}  \equiv  \phi_{\rr i}    /  2 M $ and $\psi_{\rr{rel}} \equiv (- 2 M - \psi_{\rr i} )/ 2M $ in order to probe the initial field space around the global minima of the potential, localized at $(0,\pm 2M)$.   

The posterior probability density distributions for all the parameters are given in 
Fig.~\ref{fig:mcmcSUGRAhybrid}.   These are almost identical to the posteriors of the original hybrid model parameters.  It can be noticed that the posterior distribution on $\phi_{\rr {rel}}$ is slightly modified and the posterior on the parameter $M$ is redressed at about the Planck mass, but the general behaviors remain very similar.  Thus our previous observations are generic and apply to both of the models.  


\section{Conclusion}

In this chapter, we have extended the results of the chapter 5 to the case of a closed Universe, in which the initial singularity is replaced by a classical bounce, fully described within GR.  We have studied how natural are field trajectories reaching the slow-roll attractor along the valley after performing a bounce, for the original hybrid and the F-term SUGRA models.





The initial field values leading to this scenario are found to occupy a sub-dominant, but non-negligible (about a few percents), proportion of the field space, provided that the spatial curvature in the contracting phase was initially $-\Omega_K \gtrsim 10$.  The rest of the space corresponds to trajectories reaching the Planck energy scale before the bounce occurs.   The MCMC analysis of the whole parameter space reveals that only the potential parameter $M$ affects the probability of such a scenario.  Planck-like values of $M$ are slightly disfavored in this context.   


Compared to the scenario of a classical bounce plus a phase of inflation at the top of a Higgs-type potential,  the fine-tuning of initial conditions is much less severe for hybrid models, due to the attractor nature of the inflationary valley.  
It is also interesting to remark that several bounces can occur before the slow-roll attractor is reached.  Even if the probability distribution of the number of bounces decreases exponentially, this is nevertheless a plausible scenario.  Several trajectories performing more than 100 bounces before inflation have even been observed. 

These results have an interest in the context of the self-reproducing Universe.  If our observable Universe was initially only a small patch of a much more larger Universe, and  if in this patch the spatial curvature was positive and sufficiently large, it could have been locally bouncing and some spatial regions can have reached the inflationary attractor.  

Finally, we would like to remind that presently we have no observational evidence for the sign of the spatial curvature of the Universe prior to inflation.  The classical bounce scenario should be therefore considered with no less interest than scenarios with an initial singularity.   


%% file: 21cmb.tex
\chapter{21cm cosmic background from dark ages and reionization }

%

\section{Introduction}

The theory and observations of the CMB temperature anisotropies have proved to be a formidable tool to probe the physics of the early Universe and to measure the cosmological parameters.   Their ability to constrain cosmology relies on the relatively simple physics of the acoustic wave propagation in the primordial photon-matter plasma.  
On the other hand, astrophysical observations are a very valuable tool to determine the statistical properties of the large scale structures today, as well as the Universe's expansion history, up to a redshift $z \sim 6$.   

But there is a large gap between $z \sim 1100 $ and $z \sim 6$ from which almost no signal has been detected yet.  Indeed, during this period, the Universe is almost transparent to CMB photons.  Nevertheless, this period contains a potentially large amount of information about the formation of structures in the linear and the non-linear regimes, about the first luminous objects and about the resulting reionization of the Universe.

\begin{figure}[h!]
\begin{center}
\includegraphics[height=100mm]{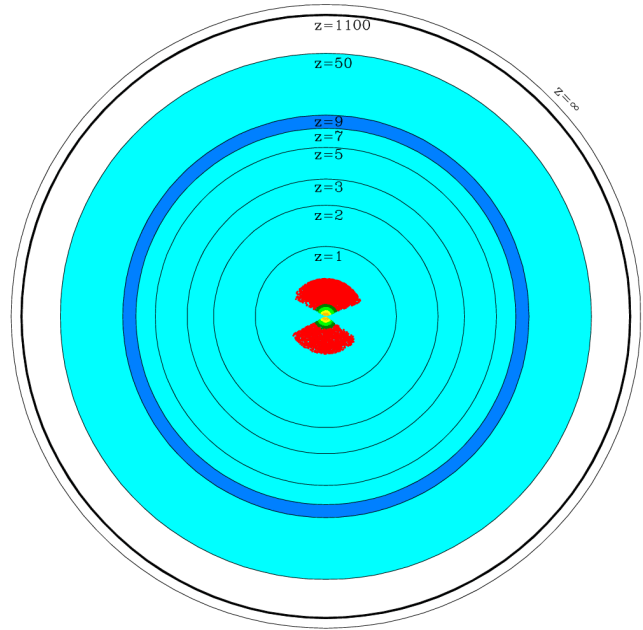}
\caption{ Whereas the CMB probes only a thin redshift slice at $z \sim 1100$ and the large scale structure surveys a volume up to $z \sim 0.5$ (red region), the 21cm tomography (blue regions) can potentially map most of the observable universe, from $z \sim 1$ to $z \sim 200 $.  The figure is from Ref.~\cite{Mao:2008ug}.}  \label{fig:21cm_redshifts}
\end{center}
\end{figure}

After recombination, the remaining small fraction of free electrons interact with CMB photons through Compton scattering and with the atoms via Coulomb interactions.  As a result, the baryon gas temperature is coupled to the photon temperature, until $z \sim 200$ (see Section~\ref{sec:darkages}).  From this time, the free electron fraction is insufficient to maintain the thermal equilibrium and the baryon gas starts to cool adiabatically.    

\begin{figure}[h!]
\begin{center}
\includegraphics[height=90mm]{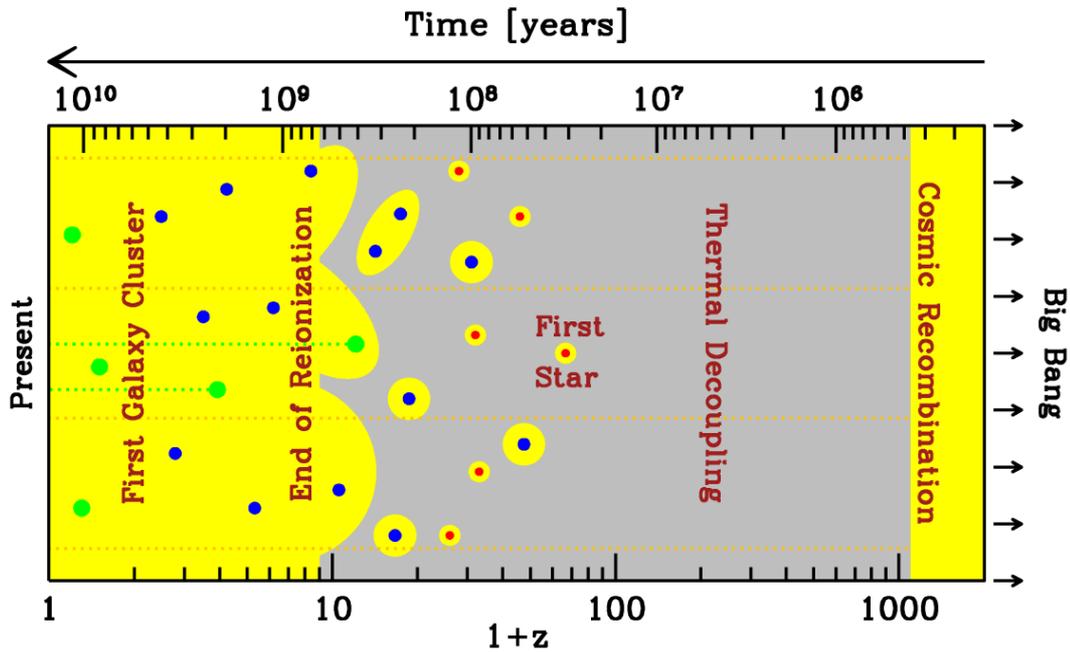}
\caption{Overview of the cosmic history~\cite{Barkana:2006ep}.  After recombination, down to a redshift $z \sim 200$, photon and baryon gas temperatures are coupled.  Gray regions represent the neutral regions, the yellow ones represent the regions in which hydrogen is ionized.  The first stars appear at $z \sim 20$ in galaxies whose typical mass is $\sim 10^5 M_{\odot}$ (red circles).  Galaxy masses raise to $10^7 - 10^9 M_{\odot}$ for the sources of reionization (blue circles) and reach $ \sim 10^{12} M_{\odot} $ for present-day galaxies (green circles).    The Universe is totally reionized at $z \sim 10$.  
The top axis is the age of the Universe.  The corresponding redshifts are given on the bottom axis. }  \label{fig:history_reion}
\end{center}
\end{figure}

During dark ages ($1100 \gsim z \gsim 20$), the growth of the inhomogeneities is governed by only two mechanisms:  the Universe's expansion and the gravitational attraction.   As long as the growth of density perturbations is linear, the physics is therefore rather simple, so that the computation of the linear matter power spectrum during the dark ages is relatively straightforward, provided that the baryon and cold dark matter power spectra are known at last scattering.   After hundred billions years of contraction, the initially thin small scale inhomogeneities can enter in a phase of non-linear growth.  They lead at $z \sim 20 $ to the formation of the first stars and galaxies (see Fig.~\ref{fig:history_reion}).   

Around $z \sim 10 $ the first luminous objects inject a large amount of radiation in the intergalactic medium (IGM).  The IGM temperature is heated up to typically thousands of Kelvins and all the Universe is reionized.   Our current knowledge of the reionization period relies on one hand on the measurement of the optical depth of the CMB photons, affected by the diffusion with free electrons.  The reionization redshift can be constrained in this way, but it depends on the considered reionization model.  In the case of instantaneous reionization, one has $z \sim 11$~\cite{Komatsu:2010fb}.  On the other hand, the high redshift quasar spectra show that the Universe is fully ionized up to a redshift $z \sim 6$.  At higher redshift, a Gunn-Peterson trough in the quasar spectra is observed~\cite{Becker:2001ee}, due to the absorption of Lyman-$\alpha$ photons by neutral hydrogen in the IGM (see Section~\ref{sec:darkages}).   These observations are used to fix a lower bound ($z \gsim 6$) on the reionization redshift.   

Because no signal has been observed directly from the reionization epoch, the details of the reionization process are only investigated theoretically, using complex numerical and semi-numerical methods.  These are based on N-body simulations of the growth of the structures in the non-linear regime and on complex model dependent algorithms for the calculation of the radiative transfer to the IGM (see e.g.~\cite{Santos:2009zk,Thomas:2010mz}).   
  
A promising observational technique to probe directly the dark ages and the reionization is the 21cm spectral line of the neutral hydrogen (HI) atoms.  This technique has been used intensively in radio-astronomy at low redshifts.  Due to the interaction between the spins of the proton and the electron, the ground state of neutral hydrogen atoms exhibits a hyperfine splitting.  The state corresponding to parallel spins (triplet state) has a slightly higher energy than the state associated to anti-parallel spins.  
The 21cm hyperfine spin-flip transitions between these singlet and triplet states can be used to detect the presence of neutral hydrogen in the Universe.   This is therefore of particular interest for probing the reionization process and the dark ages.  

Hyperfine transitions can occur via the spontaneous emission of 21cm photons, via the stimulated emission due to the CMB photon background, or via the absorption of 21cm CMB photons.    
As long as no additional physical process can modify the hyperfine level populations, the temperature associated to the hyperfine atomic levels (the so-called \textit{spin temperature}) is in equilibrium with the photon temperature.   


But during the dark ages, after the thermal decoupling of the baryon gas at $z \sim 200$, the spin changing collisions between neutral hydrogen (HI) atoms, and between HI and free electrons, shift the hyperfine level populations away from thermal equilibrium with photons.   They drive the spin temperature to the gas temperature.  Because collisional de-excitation of the triplet state are favored, one expect an absorption signal of 21cm CMB photons that could to be in principle observable in the CMB spectrum.   However, this process becomes inefficient when the collision rates are strongly reduced, at $z \sim 20$.   

\begin{figure}[h!]
\begin{center} 
\scalebox{0.4}{\includegraphics{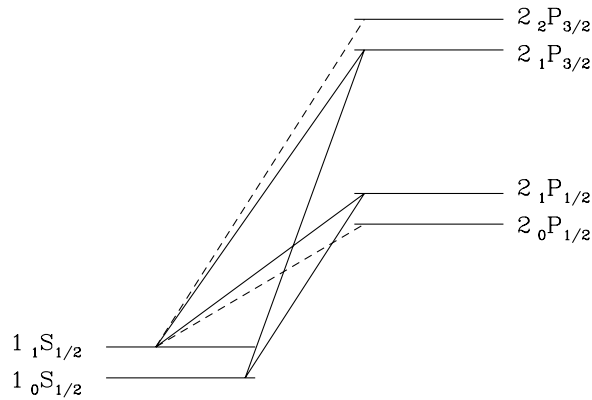}}
\caption{Illustration of the Wouthuysen-Field effect~\cite{Furlanetto:2006jb}.  Hyperfine transitions are possible during the reionization via the absorption-reemission of a Ly$\alpha$ photons.  These transitions obey to the electric dipole selection rules.  The solid lines represent the transitions contributing to the hyperfine level mixing, via one of the  $|2p, F=1 \rangle$ states.  Dashed lines represent the transitions that do not contribute to the mixing.   }
\label{fig:WFeffect}  \end{center}
\end{figure}

A 21cm signal in emission/absorption against CMB from the reionization is also potentially observable.  In this case, the hyperfine transitions are driven by the so-called \textit{Wouthuysen-Field effect}.  Suppose that a hydrogen atom in the hyperfine singlet state absorbs a Ly$\alpha$ photon emitted by the first stars.  The electric dipole selection rules allow $\Delta F = 0, 1$, except $F= 0 \rightarrow 0$, where $F$ is the total atomic angular momentum.  The atom thus jumps in one of the $|2p, F=1 \rangle$ states (see Fig.~\ref{fig:WFeffect}).   Then it can decay to the $|1s, F=1 \rangle$  triplet hyperfine state.  Atoms can therefore change their hyperfine state by absorbing and re-emitting Ly$\alpha$ photons.  These come from the first luminous objects that reionize the Universe.
Because the neutral IGM is highly opaque to resonant scattering, and because Lyman-$\alpha$ photons receive Doppler kicks in each scattering, the radiation spectrum near the resonance is well approximated by a black-body at the baryon gas temperature~\cite{Furlanetto:2006jb}.  It results that the spin temperature is driven to the gas temperature, and since the IGM is heated to temperature much higher than the photon temperature, the 21cm signal during reionization is mostly due to the stimulated emission of 21cm photons against CMB.  


21cm observations could be use to learn more about reionization by determining how the averaged ionized fraction evolve during the reionization.   For cosmology, the angular power spectrum of the 21cm signal is much more interesting.  The observation of the 21cm power spectrum could be used to realize a tomography in redshift of the Universe by mapping the matter distribution during the dark ages.  This would help to constrain the cosmological parameters as well as the primordial power spectra of density perturbations.  By extension, inflation and reheating models could also be constrained.  
21cm cosmology have two key advantages compared to the CMB:  
\begin{itemize}   
\item The potential amount of data is much larger than for the CMB.  Whereas the CMB is an unique quasi-instantaneous image of the Universe, limited by the cosmic variance, observing the dark ages would help to realize a 3D-tomography of the matter distribution over a wide range of redshifts.  The cosmic variance would therefore be limited. 
\item Because their density perturbations are not affected by the photon diffusion after recombination, the baryon fluctuations can be probed on length scales order of magnitudes smaller than before recombination. 
\end{itemize}

During the reionization ($z \sim 10$), the astrophysical contribution becomes important and complicates the extraction of the cosmological parameters from an eventual 21 signal.
The reionization process is interesting in itself,  and observations from this period would provide information about the nature and the formation of the first luminous objects. 
And if the cosmological signal can be de-correlated from the astrophysics, they will provide also information about cosmology.  

Due to technology limitations and a low signal to noise ratio, the 21cm cosmic background and its power spectrum from the dark ages or the reionization should only be observed by the next generations of radio-telescopes.  Some instruments like the LOw Frequency ARray (LOFAR)~\cite{website:lofar}, the Murchison Widefield Array (MWA)~\cite{website:mwa} and the Square Kilometre Array (SKA)
 ~\cite{website:ska} should detect in the next few years the 21cm signal from the reionization.  However, their sensitivity should not be sufficient for putting significant constraints on cosmology~\cite{Mao:2008ug}.  Some concepts of 21-cm dedicated radio-telescopes have been proposed, like the Fast Fourier Transform Telescope (FFTT)~\cite{Tegmark:2008au}.   The ability for a FFTT-type experiment to put strong constraints on the cosmological parameters have been demonstrated recently~\cite{Mao:2008ug}.  But the 21cm cosmology is clearly in its infancy and a lot of (experimental and theoretical) work has still to be realized.  
 
In this chapter, we describe the theory of the 21cm signal from the dark ages and the reionization.  We detail and explain both the homogeneous background and the linear perturbation theory.  The next chapter will be dedicated to the determination of 21cm forecasts, for a FFTT-type experiment, on the cosmological parameters, especially on the scalar spectral index since it is related to the inflationary and the reheating history of the Universe.   
 
This chapter is organized as follows.   In the next section, the spin temperature and the brightness temperature are defined.  In section 3, we focus on the evolution with redshift of the homogeneous 21cm signal from the dark ages.   We discuss also the calculation of the power spectrum of the brightness temperature fluctuations.   Section 4 is dedicated to the 21cm signal from the reionization.  The homogeneous evolution and the power spectrum of the brightness temperature fluctuations are determined.   

\section{Spin and brightness temperatures}

\subsection{Spin temperature $T_{\rr s} $}

Let us denote respectively by the subscripts $0$ and $1$ the singlet and triplet hyperfine levels of the  $|1s\rangle$ state of neutral hydrogen.  To quantify the relative number densities of these two levels, it is convenient to define the \textit{spin temperature} $T_{\rr s} $ as
\begin{equation}  \label{eq:Ts}
\frac {n_1}{n_0} = 3 \rr e^{-\frac{T_{\rr{21}} } { T_{\rr s} }}~,
\end{equation}
where $T_{\rr{21}} \equiv E_{\rr{21} } / k_{\rr B} \simeq 0.068 \rr K $, and $E_{\rr{21} } \simeq 5.9 \times 10^{-6} \rr{eV} $ is the energy splitting between the two hyperfine levels, corresponding to a wavelength $\lambda = 21 \ \rr{cm}$.  The factor $3$ accounts for the triplet state degeneracy.   The physical processes changing the relative number densities, and thus the spin temperature, will be described in the next section for the particular cases of the dark ages and the epoch of reionization.  

\subsection{Brightness temperature $T_{\rr B} $}

The energy flux of photons traveling along a given direction, per unit of area, frequency, solid angle and time is called the brightness $I_\nu$.   It is convenient to define the equivalent \textit{brightness temperature} $T_{\rr B}(\nu) $ corresponding to the temperature of the blackbody with a Planck spectrum $B_\nu$ that would lead to $B_\nu (T_{\rr B}) = I_\nu$.  For the frequency range of interest, and at the temperatures relevant for the dark ages and the reionization, the Rayleigh-Jeans formula is an excellent approximation of the blackbody spectrum, so that 
\begin{equation} \label{eq:Tbdef}
T_{\rr B} (\nu) \simeq \frac{I_\nu c^2} { 2 \nu^2 k_{\rr B}}~.
\end{equation}
As for the photon temperature and frequency, the brightness temperature is redshifted by the Universe's expansion.

One can measure the photon flux density for a frequency $\nu$, through a solid angle $\Delta \Omega $, 
\begin{equation}
S_\nu = I_\nu \Delta \Omega = \frac {2 k_{\rr B} T_{\rr B} \nu^2 \Delta \Omega }{c^2}~.
\end{equation}
This quantity is expressed in Jansky ($1 \rr{Jy} = 10^{-26} \ \rr{W m^{-2} Hz^{-1}} $).  Therefore, measurements of the redshifted 21cm photons directly probe the brightness temperature during the reionization or during the dark ages.   

Because the 21cm signal is seen in emission or absorption against the CMB, it is usual to refer to the 21cm brightness temperature as the difference between the observed brightness temperature and the expected brightness temperature of CMB photons only.  It is thus negative for an absorption and positive for an emission against CMB.   

\section{21cm tomography from dark ages}

In this section, the 21cm homogeneous brightness temperature from the dark ages is derived.  Following~\cite{Lewis:2007kz}, we assume that there are no sources of Ly$\alpha$ photons, that the CMB spectrum is exactly a blackbody, we do not account for the full distribution of spin and velocity states~\cite{Hirata:2006bn}, neither for the 21cm line profile.  The collision rates are assumed to be given by Ref.~\cite{Furlanetto:2006jb}.  The background cosmology is given by the standard $\Lambda$-CDM model, described in chapter 1.    

\subsection{Homogeneous brightness temperature}

For obtaining the brightness temperature of the 21cm signal, we need to solve the Boltzmann equation for the distribution function of the emitted or absorbed 21cm photons.  
In the rest frame of the hydrogen gas, the number of 21cm photons $\dd n_{\rr{21}}$ emitted in a time interval $\dd t$ within a solid angle $\dd \Omega$ and per unit volume is obtained from detailed balance equilibrium (in which there is no net production of photons), 
\begin{equation} \label{Eq:dn21}
 \dd n _{\rr{21}} = \frac 1 {4 \pi} \left[ \left( n_{\rr 1} - 3 n_{\rr 0}  \right) \mathcal N_\nu + n_{\rr 1}  \right]
 A_{\rr{10}} \delta(E- E_{21} ) \dd t \dd \Omega~,
 \end{equation}
where $A_{\rr{10} } = 2 \pi \alpha \nu_{\rr{21}} ^3 h_{\rr p}^2 / (3 c^4 m_{\rr e} ^2 )  \simeq 2.869 \times 10^{-15} \rr{s^{-1}}$  is the Einstein coefficient of spontaneous emission, and where we have assumed monochromatic emission of 21cm photons.
The first term on the right hand side corresponds to the stimulated emission and the absorption of 21cm photons while the second term accounts for spontaneous emission. $\mathcal N_\nu $ is the incident number of photons at a frequency $\nu$, due to the isotropic CMB blackbody spectrum plus a term due to the previous emission or absorption of 21cm photons $ \mathcal N_{\rr{21}}$.  Because during the dark ages the CMB temperature $T_{\rr CMB} \simeq 2.7 (1+z) \rr K$ is much larger than the temperature of 21cm photons ($T_{21} = 0.068 \rr K$), the Rayleigh-Jeans approximation is valid and one has
\begin{equation} \label{eq:Nnu}
\mathcal N_{\nu}  = \frac{1}{\rr e^{E / k_{\rr B} T_\gamma  } - 1  } + \mathcal N_{\rr{21}} \simeq \frac{T_\gamma}{E  } +  \mathcal N_{\rr{21}}~.
\end{equation}
Let notice that the spontaneous emission rate ($\sim 10^7$ years) is much lower than the stimulated emission rate ($\sim A_{\rr{10}}T_\gamma / T_{\rr{21}} \sim 10^4 $ years at $z\sim 30 $). 

Let us consider the distribution function $f$ of the 21cm photons emitted or absorbed during the dark ages. The number density of these photons, in an energy interval $(E, E+ \dd E)$, through a solid angle $\dd \Omega $ during a time interval $\dd t $, reads
\begin{equation} \label{eq:ntof}
\dd n_{21} = \frac{\dd f}{c^3} \dd \Omega E^2 \dd E  = \frac{2 \dd \mathcal N_{\rr{21} }}{c^3}  \dd \Omega \nu^2 \dd \nu~.
\end{equation}
An interval $\dd \lambda $ along the photon path corresponds to a proper time $\dd t = E \dd \lambda$.  
On one hand, by using Eq.~(\ref{eq:Ts})  and the relation $n_{\rr {H}} = n_0 + n_1$ ($n_{\rr {H}}$ is the total number density of neutral hydrogen), one can express $n_1 - 3 n_0 $  in Eq.~(\ref{Eq:dn21}) as a function of the spin temperature.  One has
\begin{eqnarray}
n_1 - 3 n_0 & = & - 3 n_0 \left(1- \rr e^{- \frac{T_{21}}{T_{\rr s}}   } \right) \\
 & = & - \frac{3 n_{\rr {H}}}{1+ 3  \rr e^{- \frac{T_{21}}{T_{\rr s}}   } }  \left(1- \rr e^{- \frac{T_{21}}{T_{\rr s}}   } \right)~.
\end{eqnarray}
On the other hand, $\mathcal N_\nu $ is given by Eq.~(\ref{eq:Nnu}).   Then the Boltzmann equation for the distribution $f$, by using Eq.~(\ref{eq:ntof}), reads
\begin{equation}
 \frac{\dd f}{\dd \lambda} = \frac{c^3 E_{\rr 21} 3 n_{\rr{H} }  A_{\rr{10}}  }{ 4 \pi E_{\rr {21} }^2  \left( 3+ \rr e^{T_{\rr{21} } / T_{\rr s}  } \right) } \left[  \left( 1 - \rr e ^{T_{\rr {21} } / T_{\rr s} } \right)  \left( \frac{T_\gamma}{T_{\rr{21} } } + \frac 1 2 h_{\rr p} ^3 f \right) + 1  \right]  \delta \left( E - E_{\rr{21}}   \right) ~.
 \end{equation}
In the approximation that $ T_{\rr{21}} \ll T_{\rr s} $ (this approximation will be shown to be very good during the dark ages below), the Boltzmann equation reduces to 
 \begin{equation} \label{eq:bolt21}
 \frac{\dd f}{\dd \lambda} = \frac{ 3 c^3  n_{\rr{H} }  A_{\rr{10}}  }{ 16 \pi E_{\rr {21} }  } \left[  1 - \frac{T_\gamma}{T_{\rr s}  } - \frac 1 2 h_{\rr p}^3 f \frac{T_{\rr{21} } } {T_{\rr s} }  \right]  \delta \left( E - E_{\rr{21}}   \right) ~.
 \end{equation}
In conformal time $ \eta $, the Boltzmann equation can be rewritten,
 \begin{equation} \label{eq:dfdlambda}
 \frac{\partial f}{\partial \eta} =  a \rho_{\rr s} \delta (E- E_{21} ) -  \tau'  f ~,
 \end{equation}
in which we have defined
\begin{equation}
 \rho_{\rr s}  \equiv   \frac{ 3 c^3  n_{\rr{H} }  A_{\rr{10}}  }{ 16 \pi E_{\rr {21} } ^2 } \left( \frac{T_{\rr s} - T_\gamma } {T_{\rr s}} \right)~,
 \end{equation}
and from Eq.~(\ref{eq:bolt21}),
 \begin{equation}
 \tau' = \frac{3 a c^3 n_{\rr{H}} A_{10} h_{\rr p} ^3 T_{21} }{32 \pi E_{21} T_s} \delta (E-E_{21} ) ~.
 \end{equation}
 The optical depth $\tau$ of photons with energy $E$ at  time $\eta$ is obtained by integrating this relation,
 \begin{eqnarray}
 \tau(\eta,E) & = & \int_0 ^\eta  \dd \eta' \frac{3 a c^3 n_{\rr{H}} A_{10} h_{\rr p} ^3 T_{21} }{32 \pi E_{21} T_s} \delta \left[\frac{E}{a(\eta') } -E_{21} \right]~,\\
 & = & \frac{3 c^3 n_{\rr{H}}(a_{E}) A_{10} h_{\rr p} ^3 T_{21} }{32 \pi E_{21} T_s(a_{ E}) H(a_{ E})} \Theta [a(\eta) - a_{E} ]~,
 \end{eqnarray}
 in which $a_{ E} $ is defined so that $E  a(\eta) / a_{E}= E_{\rr{21}}$, and where $h_{\rr p}$ is the Planck constant.   So it corresponds to the scale factor at the time of emission or absorption of a 21cm photon whose redshifted energy is $E$ at the time $\eta$.  $ \Theta$ is the Heaviside step function.  
  
 The solution to the Boltzmann equation is obtained by integrating Eq.~(\ref{eq:dfdlambda}).  If we define 
 $ \tau_{E}$ such that $\tau(\eta, E) = \tau_E \Theta  [a(\eta) - a_{ E} ] $, one obtains
 \begin{equation}  \label{eq:f_eta}
  f(\eta,E) = \frac{1 - \rr e^{- \tau} }{ \tau_{E}  } \left[ \frac{\rho_{\rr s} } {E_{\rr 21} a H}  \right]_{\eta(a_E)} ~, 
  \end{equation}
where the last factor is evaluated at the time corresponding to the scale factor $a_E$.  
From Eq.~(\ref{eq:Tbdef}), the brightness temperature today due to redshifted 21cm photons emitted or absorbed during the dark ages is given by $T_{\rr B} = E_{\rr{obs} } h_{\rr p}^3 f / 2 k_{\rr B} $.    From Eq.~(\ref{eq:f_eta}), it reads
\begin{equation}  \label{eq:brightness}  \boxed{
T_{\rr B} (E_{\rr{obs} } ) = \left( 1 - \rr e ^{- \tau_E} \right) \left. \left( \frac{T_{\rr s} - T_\gamma }{1+z} \right) \right| _{\eta(a_E)}~.}
\end{equation}
If the spin temperature is below the CMB temperature (it is shown thereafter that this is the case during the dark ages), the brightness temperature is negative and signal corresponds to a net absorption of 21cm CMB photons.   

The spin temperature evolution still needs to be determined.  To do so, let us consider a number of atoms $N_{\rr{H}} = N_0 + N_1 $.  If recombinations are to the singlet and triplet state in the 1-3 ratio, then one can read
\begin{equation} \label{eq:dNdt}
\frac{\dd N_1}{\dd t} = - N_0 \left( C_{\rr{01}} + 3 A_{\rr 10 } \mathcal N_\nu  \right) + N_1 \left( C_{10} + A_{10} + A_{10} \mathcal N_\nu  \right) - \frac{\dd x_{\rr i}}{ \dd t} \frac{N_{\rr{H}  } + N_{\rr i} }{4}~,
\end{equation}
where $x_{\rr i} $
 is the ionized fraction, and where 
\begin{equation}
C_{\rr{01}} = 3 C_{\rr{10}} \rr e ^{-T_{21} / T_{\rr g}}
\end{equation}
are the collision terms.  The first term of Eq.~(\ref{eq:dNdt}) accounts for the transitions from the singlet to the triplet state, induced by  a spin changing collision or by the absorption of a 21cm CMB photon.  The second term accounts for the transitions from the triplet to the singlet state, induced by collisions or by the stimulated or spontaneous emission of a 21cm photon.   The last term takes into accounts the possibility for a proton and a free electron to recombine in the singlet state.

The collision term comprises the spin-changing collisions between neutral hydrogen (HH), between electrons and hydrogen (eH) and between protons and hydrogen (pH), so that
\begin{equation}
C_{10} =  n_{\rr{H}} \kappa ^{\rr{HH}} + n_{\rr e} \kappa^{\rr{eH}} + n_{\rr i } \kappa ^{\rr{pH}}~,
\end{equation}
where the $\kappa^{ii} $ are the corresponding collision rates, whose values are taken from Ref.~\cite{Furlanetto:2006jb}.  Because the ionized fraction is small during the dark ages ($ x_{\rr i} \sim 2 \times 10^{-4} $, see section~\ref{sec:recomb}), collisions are mainly between neutral hydrogen atoms.  At high redshift ($z \gsim 70$ ), the collision rate $C_{01}$ ($\sim 10^{-3}$years$^{-1}$) is larger than the rate of spontaneous emission and absorption of 21cm photons $A_{10} \mathcal N_\nu$ ($\sim 10^{-4}$years$^{-1}$). At low redshifts ($z \lsim 70 $), the number density of hydrogen atoms is sufficiently reduced by the expansion for the collision rate to be negligible compared to the rate of 21cm photon emission or absorption.  

\begin{figure}[h!]
\begin{center} 
\scalebox{1.}{\includegraphics{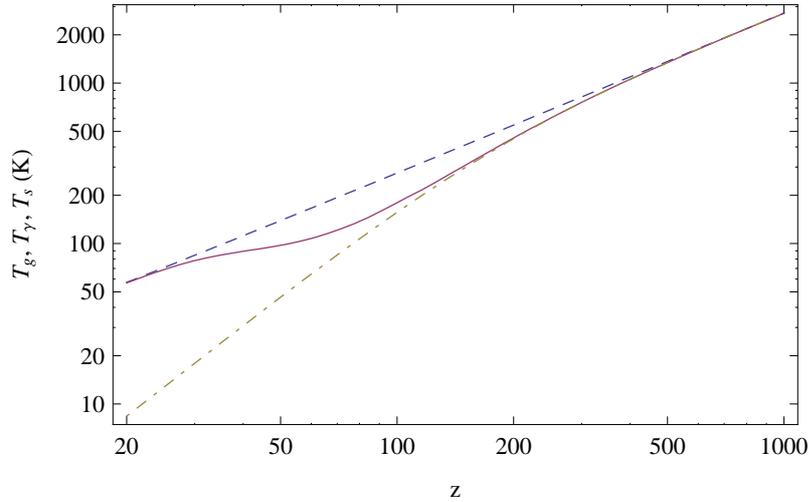}}
\caption{Evolution of the spin temperature $T_{\rr s}$ (plain line), from Eq.~(\ref{eq:Tspinevol}), of the gas temperature $T_{\rr g}$ (dashed-dotted line) and the photon temperature $ T_\gamma$ (dashed line), in the standard $\Lambda$-CDM model.  The collisions drive $T_{\rr s}$ to $T_{\rr g}$ at high redshift ($z \gtrsim 100 $).  At low redshifts ($z \sim 30$), collisions are rarefied and the photon interactions drives $T_{\rr s}$ to $T_\gamma $.  From Eq.~(\ref{eq:brightness}), the 21cm signal is seen in absorption against CMB photons in the redshift range $30 \lesssim z \lesssim 200$.  } \label{fig:TsTgTgamma}
\end{center}
\end{figure}

The evolution of the spin temperature can be derived from Eq.~(\ref{eq:dNdt}).   But it is convenient to define $\beta_{\rr s} \equiv 1/T_{\rr s}, \beta_{\rr g} \equiv 1/ T_{\rr g} $ and $\beta_{\rr{21}} \equiv 1 / T_{\rr{21}} $.  After inserting the spin temperature in Eq.~(\ref{eq:dNdt}), and by using Eq.~(\ref{eq:Nnu}), the equation governing the evolution $\beta_{\rr s} $, at first order in $T_{21}/ T_{\rr s} $, can be obtained: 
\begin{equation}  \label{eq:Tspinevol}
\frac{\dd \beta_{\rr s }}{ \dd t} + \frac{\beta_{\rr s}}{1- x_{\rr i}  } \frac{\dd x_{\rr i}}{\dd t}= 4 \left[ \left( \beta_{\rr g} - \beta_{\rr s}   \right) C_{\rr {10}}   + A_{10} \beta_{\rr{21}}  \left( 1 - \beta_{\rr s} T_\gamma - \beta_{\rr s} T_{\rr{21}} \mathcal N_{\rr{21}}  \right) \right]~.
\end{equation}   
If we neglect the recombination between protons and free electrons, two regimes can be identified.  When the collision term on the right hand side is dominant, the spin temperature is driven to the gas temperature.   Since it is smaller than the photon temperature, the 21 signal is seen in absorption.  When the photon interaction dominates over the collisions, because $T_{\rr 21} \mathcal N_{\rr 21} \ll T_\gamma$, the spin temperature is driven back to the photon temperature, and the 21cm brightness temperature becomes negligible.    The overall evolution of the spin temperature during the dark ages is represented in Fig.~\ref{fig:TsTgTgamma}. 
 
\subsection{Perturbations}

As for the CMB, the statistical properties of the 21cm brightness temperature anisotropies could be a powerful tool to constrain cosmology.  
The angular power spectrum of the 21cm brightness temperature anisotropies today is obtained by integrating the first-order Boltzmann equation for the perturbed distribution function $ \delta f (\mathbf{x},\eta,E,  \mathbf{e} )  $, depending on the position $\mathbf x$ and the direction of observation $\mathbf e$.  This is obtained by considering the perturbations of all the relevant quantities, like the baryon number density, the free electron fraction, the gas temperature and the CMB photon field, and by considering the peculiar velocities of the gas and the Thomson optical depth of the CMB photons.  The complete calculation has been realized in Ref.~\cite{Lewis:2007kz}, and the results have been included in the CAMB numerical code~\cite{Lewis:1999bs}.   

For the purpose of this thesis, the CAMB code has been used to calculate of the angular power spectrum of the 21cm brightness temperature from the dark ages, for the best fit values of the cosmological parameters.

\section{21cm signal from the Reionization}

In this section, the 21cm brightness temperature from the reionization epoch is given and its power spectrum is determined in the limit $T_{\rr s} \gg T_\gamma$.  In particular, we discuss the cosmological and the astrophysical contribution to the signal and explain how they can be de-correlated.



\subsection{Homogeneous brightness temperature}

As for dark ages, the homogeneous evolution of the 21cm brightness temperature during reionization is given by Eq.~(\ref{eq:brightness}).
Because typically $\tau_E \ll 1$, the 21cm brightness temperature today is well approximated by 
\begin{equation} \label{eq:Tbrionexact}
T_{\rr B} (E ) =  \frac{3 c^3 n_{\rr{H}}(a_{E}) A_{10} h_{\rr p} ^3 T_{21} }{32 \pi E_{21}  H(a_{ E})}  \left. \frac{T_{\rr s} - T_\gamma }{ T_s
(1+z)} \right|_{\eta(a_E)}~.
 \end{equation}
As explained in the introduction, the spin temperature evolution is not dictated by the collisions anymore but by the Wouthuysen-Field effect, corresponding to hyperfine transitions via the absorption and re-emission of Lyman-$\alpha$ photons.  
The spin temperature depends also on the baryon gas temperature in the IGM, that is heated by X-ray photons of the first stars to typically thousands of Kelvin during reionization~\cite{Mao:2008ug}.  
This process is until now relatively unknown.   In absence of direct observations, our knowledge of the reionization relies on complex semi-analytical calculations and numerical simulations of the structure formation and radiative transfer to the IGM (see e.g. \cite{Santos:2009zk,McQuinn:2007dy})   
As a consequence, it is difficult to predict the spin temperature evolution during the reionization.

Let us follow Ref.~\cite{Mao:2008ug} and assume that there exists a redshift range during which the following two conditions are satisfied:
\begin{itemize}
\item  Due to the X-ray heating, the gas temperature $T_{\rr g} $ is much larger that the CMB photon temperature $T_\gamma $.
\item  The spin temperature $T_{\rr s} $ is driven to the gas temperature $T_{\rr g} $ via the Wouthuysen-Field effect, so that the approximation $ (T_{\rr s} - T_\gamma ) / T_{\rr s} \simeq 1 $ is valid.  More precisely, during reionization, the collision terms $C_{01}$ and $C_{10}$ in Eq.~(\ref{eq:dNdt}) are not dominant and can be replaced by the rates of transition due to the Wouthuysen-Field effect, denoted $P_{01}$ and $P_{10}$.  Then one can introduce a \textit{color temperature} $T_{\rr c} $, defined as 
\begin{equation} 
\frac{P_{01}}{ P_{10}} \equiv 3 \left(1 - \frac{T_{21}} { T_{\rr c} }) {\dd \nu} \right)~.
\end{equation}
In a good approximation, one has $T_{\rr c} \approx T_{\rr g}$: indeed, because the IGM is extremely optically thick, the large number of Lyman-$\alpha$ scatterings brings the Lyman-$\alpha$ profile to a blackbody of temperature $T_{\rr g} $ near the line center~\cite{Furlanetto:2006jb}.  
\end{itemize}
In this limit,  the 21cm homogeneous brightness temperature takes the simpler form
\begin{equation} \label{TblargeTs}
T_{\rr B} (E  ) \simeq  \frac{3 c^3 A_{10} h_{\rr p} ^3 T_{21} a_E n_{\rr{b}}(a_{E}) \left[1 - f_{\rr{He}} - x_{\rr i} (a_{E}) \right]  }{32 \pi E_{21}  H(a_{ E})  }~,
\end{equation}
where we have used the relation $n_{\rr{H}} = n_{\rr b} (1 - f_{\rr{He}} - x_i ) $ taking account for the Helium fraction\footnote{Let notice that the Helium fraction was not included in Ref.~\cite{Mao:2008ug}, whereas its effect on the brightness temperature is not negligible since $f_{\rr{He}} \simeq 0.24 $.}.  The brightness temperature does not depend on the spin temperature anymore.   It depends on the cosmology through the variables $n_{\rr b} $ and $H$.  However, it still depends also on the reionization model through the ionized fraction $x_{\rr i} = 1 - x_{\rr{H}}$ at the considered redshift. 

For instance, if we follow Ref.~\cite{Mao:2008ug} and take $x_{\rr i} = 0.1$ at $z = 9.2$, together with the $\Lambda$-CDM best fit values of the cosmological parameters, the 21cm homogeneous brightness temperature is $T_{\rr B} \simeq 28 \rr{mK} $.  Let notice that the spin temperature evolution during the reionization period has been evaluated in Ref.~\cite{Thomas:2010mz}, and is found to be no more than a few hundreds of Kelvins at the very beginning of the reionization.  So the approximation $ (T_{\rr s} - T_\gamma ) / T_{\rr s} = 1 $ can introduce errors of a few percents.  In the next chapter, we will assume a toy model coherent with Ref.~\cite{Thomas:2010mz} for the evolution of $T_{\rr s} $ during the reionization.

\subsection{Perturbations}

The  brightness temperature fluctuation in the limit $T_{\rr s} \gg T_\gamma $ is obtained by perturbing the baryon number density $n_{\rr b}$ as well as the ionized fraction $x_{\rr i} $ in Eq.~(\ref{TblargeTs}).  One has also to consider the gradient of the peculiar velocity along the line of sight, $\partial v_{\rr r} / \partial \eta $.  Let us denote  $\bar T_{\rr B} $ and $\bar x_{\rr H}$, respectively the homogeneous brightness temperature given by Eq.~(\ref{TblargeTs}) and the mean neutral hydrogen fraction.  In a given direction $\mathbf e$, the brightness temperature now reads
\begin{equation}   \label{eq:T_B_reion}
T_{\rr B} (\mathbf{e} ) = \frac{ \bar T_{\rr B} }{\bar x_{\rr H}} \left[  1 - \bar x_{\rr i} ( 1 + \delta_{x_{\rr i}} ) \right] ( 1 + \delta _{\rr b}  ) \left( 1 - \frac{1}{aH}  \frac{\partial v}{ \partial \eta} \right)~,
\end{equation}
Then let  us define 
\begin{equation}
\delta_{\rr v} \equiv \frac{1}{aH}  \frac{\partial v}{ \partial \eta} ~.
\end{equation}
After a Fourier expansion, and as long as the linear perturbation is valid, one can show that $\delta_v (\mathbf k) = - \mu^2 \delta_{\rr b} $ where $\mu \equiv \mathbf{k \cdot n} $ is the cosine of the angle between the Fourier mode $\mathbf{k} $ and the line of sight.   In the linear regime, the brightness temperature perturbation $\Delta T_{\rr B} (\mathbf k) $ is obtained from Eq.~(\ref{eq:T_B_reion}),
\begin{eqnarray}
 \Delta T_{\rr B} (\mathbf k) & = & \frac{ \bar T_{\rr B} }{\bar x_{\rr H}} \left[ \delta_{\rr b} (\mathbf k) - \delta_{\rr v}  (\mathbf k)- 
 \bar x_{\rr i} \delta_{x_{\rr i}}  (\mathbf k) - \bar x_{\rr i} \delta_{\rr b}  (\mathbf k) + \bar x_{\rr i} \delta_{\rr v}   (\mathbf k) \right]\\
 & = &  \frac{ \bar T_{\rr B} }{\bar x_{\rr H}} \left[ \delta_{\rr b} (\mathbf k) + \mu^2  \delta_{\rr b}  (\mathbf k)- 
 \bar x_{\rr i} \delta_{x_{\rr i}}  (\mathbf k) - \bar x_{\rr i} \delta_{\rr b}  (\mathbf k) - \bar x_{\rr i} \mu^2 \delta_{\rr b}   (\mathbf k) \right]~,
\end{eqnarray}
and power spectrum of the brightness temperature fluctuations reads~\cite{Mao:2008ug}, 
\begin{equation} \begin{split} \label{eq:PTb}
P_{\Delta T_{\rr B}} (\mathbf k )  = \left( \frac { \bar T_{\rr B} }{\bar x_{\rr H}} \right)^2 \left\{  \left[  \bar x_{\rr H}^2 P_{\rr{bb}}(\mathbf{k} ) - 2 \bar x_{\rr H} \bar x_{\rr i} P_{\rr{ib}}( \mathbf k) +  \bar x_{\rr i}^2 P_{\rr{ii}} (\mathbf(k))\right] \right. \\
\left. + 2 \mu^2 \left[ \bar x_{\rr H}^2 P_{bb}(\mathbf{k} ) -   \bar x_{\rr H} \bar x_{\rr i} P_{\rr{ib}}( \mathbf k)  \right] + 
\mu^4 \bar x_{\rr H}^2 P_{\rr{bb}}(\mathbf{k} ) \right\} ~.  \end{split}
\end{equation}
The key point is that the $\mu^4$ component only depends on the baryonic matter power spectrum, and thus on the cosmology.  This property is essential to extract the cosmological signal from the astrophysical contaminants involved in the $ P_{\rr{ib}}( \mathbf k) $ and $ P_{\rr{ii}}( \mathbf k) $ power spectra.  

Let us now discuss the evolution of the perturbations in baryon and ionized fraction.  After recombination, the linear growth of dark matter and baryon perturbations is only due to gravity.  Let us define  $f_{\rr b} \equiv \Omega_{\rr b} / \Omega_{\rr m} $ and $f_{\rr c} \equiv \Omega_{\rr c} / \Omega_{\rr m}$, respectively the baryon and dark matter fraction.   On sub-Hubble scales, dark matter and baryon perturbations ($\delta_{\rr c}$ and $\delta_{\rr b}$) are coupled and evolve at the linear level according to~\cite{Barkana:2005xu}
\begin{equation} \label{eq:deltab}
\ddot \delta_{\rr b} + 2 H \dot \delta_{\rr b} = 4 \pi G (\bar \rho_{\rr b} + \bar \rho_{\rr c} ) ( f_{\rr b} \delta_{\rr b} + f_{\rr c} \delta_{\rr c}  )~,
\end{equation}
\begin{equation} \label{eq:deltac}
\ddot \delta_{\rr c} + 2 H \dot \delta_{\rr c} = 4 \pi G (\bar \rho_{\rr b} + \bar \rho_{\rr c} ) ( f_{\rr b} \delta_{\rr b} + f_{\rr c} \delta_{\rr c}  )~,
\end{equation}
where $\bar \rho_{\rr b} $ and $\bar \rho_{\rr c} $ are the mean densities of baryon and dark matter.   The second term on the left hand side accounts for the Universe's expansion and the right hand side describes in the Newtonian limit the gravitational collapse due to both the baryon and dark matter over-densities.   Since they do not involve spatial gradient, the same equations stand for perturbations in the real and in the Fourier space. 

Before recombination, as mentioned in chapter 1, the baryonic matter is tightly coupled to the radiation, and the determination of the matter density perturbations require the integration of the first order Boltzmann equations for all the fluid and metric fluctuations.   
But if initial conditions on $\delta_{\rr b} $ and $\delta_{\rr c} $ can be fixed just after the recombination (e.g. by using a numerical code for the cosmic evolution, like CAMB~\cite{Lewis:1999bs}), the calculation of the power spectrum of baryons at reionization consists simply in integrating Eqs.~(\ref{eq:deltab}) and (\ref{eq:deltac}).   

For the purpose of this thesis, the baryon perturbations at reionization have been calculated with three different methods:
\begin{itemize}

\item By integrating numerically the first order Boltzmann equations for all the fluids and for scalar metric perturbations, from before recombination, and by using and integrating Eqs.~(\ref{eq:saha}) and (\ref{eq:recomb}) for the evolution of the free electron fraction through recombination.  Our code does not include the effects of high multipoles photon perturbations and the details of the recombination process.  We obtain nevertheless relative errors less than 10\%.  

\item By using the CAMB code to provide initial conditions on $\delta_{\rr b} $ and $\delta_{\rr c} $ after recombination ($z \sim 900 $), and then by integrating numerically Eqs.~(\ref{eq:deltab}) and (\ref{eq:deltac}).

\item By using the CAMB code to obtain the transfer functions to apply to the primordial power spectrum of density perturbations at the considered redshift.  This method is the most accurate since several non trivial effects (e.g. more detailed recombination) are taken into account.   

\end{itemize}

The ionized gas fluctuations are however strongly dependent on the reionization model.   Indeed, ionized bubbles grow and can eventually merge with one another.   For simplicity, we have followed
~\cite{Mao:2008ug} and have considered two reionization models, one simple optimistic and one more realistic:
\begin{enumerate}
\item In the first scenario, we assume that the hydrogen gas has been heated sufficiently before the reionization proceeds.  In this particular case, one thus has 
\begin{equation} \label{eq:reion_opt_case}
P_{\rr{ii}} = P_{\rr{iH}} = 0~,
\end{equation}
 for all the observable perturbation wavelengths.   
\item In the second scenario, we assume that $P_{\rr{ii}} $ and $ P_{\rr{iH}}  $  are smoothed function that can be parametrized in the following way:
\begin{equation} \label{eq:reion_realistic1}
 \bar x_{\rr i}^2 P_{\rr{ii}} (k) = b_{\rr {ii}}^2 \left[ 1 + \alpha_{\rr{ii}} (k R_{\rr{ii}} ) + (k R_{\rr{ii}})^2 \right]^{-\frac{\gamma_{\rr{ii}}}{ 2} } P_{\rr{bb}}~,
 \end{equation} 
\begin{equation} \label{eq:reion_realistic2}
 \bar x_{\rr b} \bar x_{\rr i} P_{\rr{ib}} (k)= b_{\rr{ib}}^2 \exp \left[ - \alpha_{\rr{ib}} (i R _{\rr{ib}}) - (k R_{\rr{ib}} )^2 \right]  P_{\rr{bb}}~.
 \end{equation}
 The fiducial parameters, according to Refs.~\cite{Mao:2008ug,McQuinn:2007dy}, are given in Tab~\ref{tab:Reion} for $z=9.2$.
\end{enumerate}

\begin{table} \begin{center}
\begin{tabular}{|c|c|} 
\hline
$b_{\rr{ii}}^2 $ & 0.208  \\
$  \alpha_{\rr{ii}} $  & -1.63  \\
$ R_{\rr{ii}} $  & 1.24  \\
$\gamma_{\rr{ii}} $ &  0.38 \\
$b_{\rr{ib}}^2 $ & 0.45  \\
$  \alpha_{\rr{ib}} $   & -0.4   \\
$ R_{\rr{ib}} $  & 0.56  \\
\hline
\end{tabular}
\caption{Fiducial values of the of the reionization parameters for the realistic case of Eqs.~(\ref{eq:reion_realistic1}) and (\ref{eq:reion_realistic2}), given in Refs.~\cite{Mao:2008ug,McQuinn:2007dy}, for $z=9.2$. } \label{tab:Reion} 
\end{center}
\end{table}
 
 \begin{figure}[h!]
\begin{center} 
\includegraphics[height=80mm]{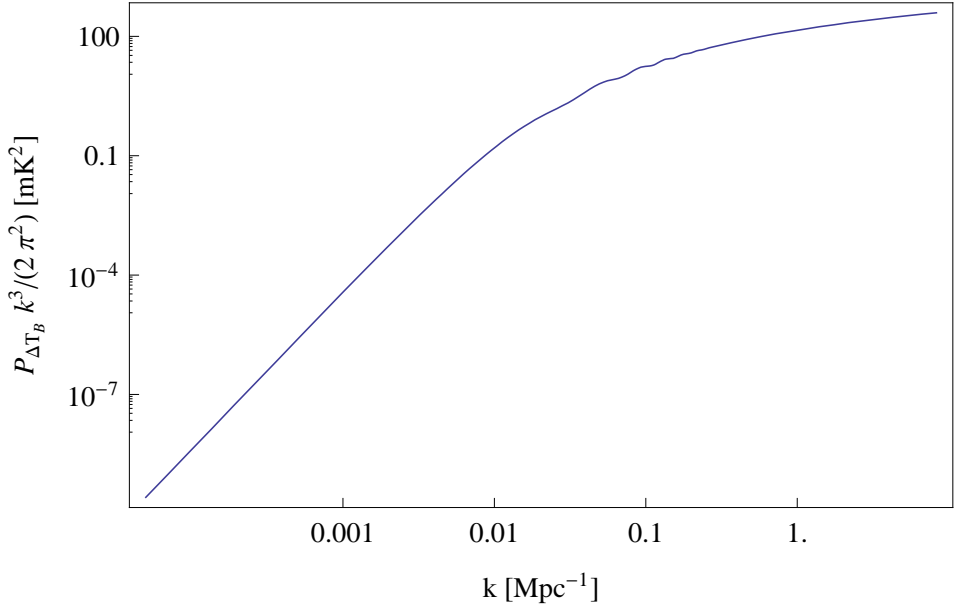}
\caption{ 21-cm brightness temperature power spectrum at $z=11$, for $\mu = 1$, for the optimistic reionization model of Eq.~(\ref{eq:reion_opt_case}) and best fit values of the cosmological parameters.  BAO are seen at about $k=0.1$ Mpc$^{-1}$. } \label{fig:PTBz11}
\end{center}
\end{figure}

\chapter{21cm forecasts}

\section{Introduction}

The purpose of this chapter is to calculate the forecasts on the cosmological parameters, for typical experiments dedicated to the observation of the 21cm power spectrum from the dark ages and the reionization.  In recent works~\cite{Mao:2008ug,Barkana:2005xu,Jelic:2010qt,Mao:2007ti,Wyithe:2007rq}, such forecasts have already been determined for the 21cm signal from the reionization, by using Fisher matrix methods.  In these studies, the forecasts are determined for the next generation of radio-telescopes (LOFAR~\cite{website:lofar}, MWA~\cite{website:mwa}, SKA~\cite{website:ska}) and for the concept of Fast Fourier Transform Telescope (FFTT)~\cite{Tegmark:2008au}.  The influences of various theoretical and experimental free parameters are also discussed.  They found that the precision on the cosmological parameter measurements could be improved significantly principally for the FFTT.   However, to our knowledge, little has been done for the 21cm signal from the dark ages (see nevertheless~\cite{Adshead:2010mc}).  Here, we are more specific and focus on:
\begin{enumerate}
\item The comparison between the forecasts obtained for a realistic (1 km$^2$) and an idealistic (10 km$^2$) FFT radio-Telescope.   
\item The comparison between the  forecasts obtained for the same experiment, but for two different redshifts:  the first one in the dark ages ($z = 40$), the second one at the beginning of the reionization ($z=11 $).     
\end{enumerate}
For the first time, we use a Monte-Carlo-Markov-Chain (MCMC) bayesian method to determine the forecasts.  But we have also developed a code based on the Fisher matrix method for the 21cm reionization signal.  In this way, the two methods shall be compared directly and the consistency of the results can be checked by comparison with previous studies.  
Moreover, the bayesian method is of particular interest for identifying the degeneracies between parameters and for probing non gaussian posterior likelihood functions of the model parameters.  

Furthermore, we improve the calculation of Ref.~\cite{Mao:2008ug} for the 21cm brightness temperature from the reionization, by considering the Helium fraction in Eq.~(\ref{eq:Tbrionexact}) and by including explicit values for $(T_{\rr s} - T_{\gamma}) / T_{\rr s} $ instead of considering the limit where it is equal to unity.  
This affects the homogeneous brightness temperature typically at first order and modifies it  at a few percent level.  We  nevertheless  neglect the spin temperature fluctuations, that are second order effects.   We model the ionized fraction and the spin temperature evolutions by hyperbolic tangent functions, for which the central redshift $z_{\rr{re}} $, the width $\Delta z $ and the maximal value of the spin temperature $T_{\rr s}^{\rr {max}} $ are the three free parameters.   In a way consistent with the numerical simulations of the reionization of Ref.~\cite{Santos:2009zk} and the CMB constraints on the optical depth, we choose as fiducial values $z_{\rr{re}} = 10$, $\Delta z = 0.5$, $T_{\rr s}^{\rr {max}} = 1000 K $.  At $z = 11$, the corresponding neutral fraction and spin temperature are respectively $x_{\rr H} = 0.87$ and $T_{\rr s} = 133 \ \rr K$.

For the sake of simplicity, we consider for the reionization the optimistic case in which the perturbations in the ionized fraction are negligible at the redshift of interest, so that 
Eq.~(\ref{eq:reion_opt_case})  is valid.   Nevertheless, for the FFTT, it is shown in Ref.~\cite{Mao:2008ug} that this assumption leads to weaker constraints on the cosmological parameters than for the realistic case described by Eqs~(\ref{eq:reion_realistic1}) and (\ref{eq:reion_realistic2}).   This is due to the fact that the power spectra of the ionized fraction are added to the power spectrum of baryons in Eq.~(\ref{eq:PTb}).  For the FFTT, the astrophysical uncertainties are compensated by the higher amplitude of the 21cm power spectrum $P_{\Delta T_{\rr B}}$.   
For this reason, the case we consider here is not so optimistic, since in the realistic case the forecasts are expected to be better.  

The chapter is organized as follows:  in the next section, the two considered experiments are introduced.  Then we describe how the forecasts can be obtained by using the Fisher matrix and the bayesian MCMC methods.  The last two sections concern the forecasts themselves, for the 21cm signal from the dark ages and from the period of the reionization.   We discuss the ability to put significant constraints on cosmology, and especially on inflation and reheating models, in the conclusion.    

\section{Two typical experiments} \label{sec:exper}

We base our analysis on a hypothetic Fast Fourier Transform Telescope experiment dedicated to 21cm cosmology.  In the FFTT concept, a large number of dipole antennas are distributed on a rectangular grid.   Compared to standard radio-interferometry, the summation over all the baselines is replaced by a Fast-Fourier-Transform (see Appendix A for further details).  As a result, the computational cost scales as $N_{\rr A} \log N_{\rr A}$, where $N_{\rr A}$ is the number of antennas, instead of $N_{\rr A}^2 $ for standard interferometers.  Since the total cost of the present and projected giant radio-telescopes is dominated by the computational costs, this principle could be used advantageously to increase the number of antennas, and thus the sensitivity to the signal~\cite{Tegmark:2008au}.   
We chose to consider a FFTT telescope because it is an intermediate case between the standard radio-telescopes like LOFAR and MWA,
whose sensitivity is not expected to be sufficient to improve cosmological constraints, and non realistic  concepts like building a radio-interferometer on the Moon~\cite{Carilli:2007eb}.  We consider two configurations of the FFTT, one realistic and one idealistic:
\vspace{5mm}
\begin{enumerate}
\item Experiment 1:  a $1 \rr{km} \times 1 \rr{km}$ FFTT telescope, with a minimum distance of $1\rr m$ between dipole antennas (realistic).  This configuration is used in Refs.~\cite{Tegmark:2008au,Mao:2008ug}.
\item Experiment 2:  a  $10 \rr{km} \times 10 \rr{km}$  FFTT telescope,  with a minimum distance of $1\rr m$ between dipole antennas (idealistic).
\end{enumerate}
For the other characteristics of the experiments, like the bandwidth, the observation time and the noise spectrum, we refer to Refs.\cite{Tegmark:2008au,Mao:2008ug}.  These specifications are given in tab~\ref{tab:FFTT}.  
For simplicity, we assume ideal foreground removals.  However, it is important to notice that the extraction of the cosmological and astrophysical signals is very challenging due to the high level of atmospheric and galactic foregrounds (fore further details on the foregrounds and the removal techniques, see e.g.~\cite{Morales:2005qk}).  
\vspace{3mm}
\begin{table}  \begin{center}
\begin{tabular}{|c|c|} 
\hline
Total size & $D=1$ km (realistic) / $10$ km (idealistic) \\
Min. baseline & d= 1 m \\
Number of antennas & $N_{\rr A} = 10^6$ (realistic) / $10^{8}$ (idealistic) \\
Bandwidth & $\Delta \nu = 1$ Mhz \\
System temperature & $T_{\rr{sys} } = 300 \ \rr K  $ \\
Observation time & $t_{\rr o} = 1$ year  \\
Angular resolution & $\theta_{\rr{res}} =  \lambda / D $ (Beam FWHM) \\
Field of view &$ \Omega = 2 \pi $ \\
\hline
\end{tabular}
\caption{Characteristics of the two considered FFTT experiments.  $\lambda$ is the redshifted wavelength of the 21cm signal.  
We assume a gaussian beam and use the FWHM convention for the beam width, as well as ideal foreground removal. The FFTT covers half of the sky sphere.  }
\label{tab:FFTT}
\end{center}
\end{table}

\section{Fisher Matrix and MCMC bayesian methods}

Assuming that the true Universe is described by a known set of parameters, the Fisher matrix formalism and the bayesian MCMC method are two techniques that can be used to determine, for a typical experiment, the expected errors on the parameter measurements.  
The Fisher matrix and the MCMC methods are described respectively in Appendix B and Appendix C.  In this section, we give the guidelines for the calculation of the forecasts, given the theoretical 21cm signal and the specifications of the experiment.  

\subsection{Fisher matrix analysis}

In the previous chapter, the 3D power spectrum of the 21cm brightness temperature fluctuations from the reionization has been calculated [see Eq.~(\ref{eq:PTb})].   But the power spectrum $P_{\Delta T_{\rr B}} (\mathbf k)$ is not directly observed by 21cm experiments.    What is observed are angular positions in the sky plane and frequency differences $\Delta f$ from the central redshift of a $z$-bin.  The FFTT is designed to map the full sky on a hemisphere ($\Omega = 2 \pi $), but the angular scales relevant for cosmological information are essentially much smaller than a radian.  Therefore, we follow~\cite{Mao:2008ug} and consider the flat sky approximation in which the angular distances are approximatively proportional to comoving distances.   Let us define $\mathbf \Theta_{\bot} $ the angular distance in the sky.  If there are no peculiar velocities, $\mathbf \Theta_{\bot} $ and $\Delta f $ are related to comoving distances at reionization decomposed in components perpendicular $\mathbf r_\bot$ and  parallel $r_{\parallel} $ to the line of sight, through the relation
\begin{equation} \label{eq:thetabot}
\mathbf \Theta_\bot = \frac{\mathbf r _\bot}{ D_{\rr A} (z_{\rr{re}} )}~,
\end{equation}
\begin{equation} \label{eq:thetaparallel}
\Delta f = \frac{r_{\parallel} }{y(z_{\rr{re}}  )}~,
\end{equation}
where $D_{\rr A} $ is the comoving angular distance, given in a flat Universe by
\begin{equation}
D_{\rr A} (z) = c  \int_0 ^z \frac{1}{H(z')} \dd z'~,
\end{equation}
and where $y(z)$ is the conversion factor between comoving distances and frequency intervals,
\begin{equation}
y(z) = \frac{\lambda_{21} (1+z)^2 }{H(z)}~.
\end{equation}
Since the 21-cm brightness temperature fluctuations are given in the Fourier space, for the comoving modes $\mathbf k$ that are the Fourier duals of  comoving space-like vectors $\mathbf r$, it is convenient to define the Fourier dual of $\mathbf \Theta$ as $\mathbf u$.  From Eqs.~(\ref{eq:thetabot}) and (\ref{eq:thetaparallel}), the components $\mathbf u_\bot$ and $ u_{\parallel} $  are related to the comoving mode components  $\mathbf k_\bot $ and $k_{\parallel} $ through the relations
\begin{equation}
\mathbf u_\bot = D_{\rr A} \mathbf k_\bot~,
\end{equation}
\begin{equation}
u_{\parallel} = y k_{\parallel}~.
\end{equation}
It results that the power spectrum of the 21-cm brightness temperature in the $\mathbf u$ space, that is directly measurable by observations, is given by
\begin{equation}
P_{\Delta T_{\rr B} } ( \mathbf u ) = \frac{P_{\Delta T_{\rr B} }  (\mathbf k) }{D_{\rr A} ^2 y}~.
\end{equation}

If for computational convenience we subdivide the $u$-space in cells so small that the power spectrum remains almost constant in each one, the Fisher matrix is given by~\cite{Mao:2008ug} (see Appendix B)
\begin{equation}
\mathbf F_{ab} = \sum_{\rr{pixels}} \frac{N_{\rr c}}{\left[P_{\Delta T_{\rr B} } (\mathbf u) + P^{\rr n}\right]^2 }
\left( \frac{\partial P_{\Delta T_{\rr B} } (\mathbf u) }{\partial \lambda_a}   \right) 
 \left(  \frac{\partial P_{\Delta T_{\rr B} } (\mathbf u)}{ \partial \lambda_b } \right)~,
\end{equation}
where $N_{\rr c} = V_{\rr c}  ( \Omega \times B) / (2 \pi)^3  $ is the number of cells, $V_{\rr c} $ is the volume of a cell in the $u$-space, $B $ is the frequency size of a $z$-bin and the $\lambda_a $ are the model parameters (i.e. the cosmological plus eventual reionization parameters).  $P^{\rr n} $ is the noise power spectrum.  For the FFTT radio-interferometer, it is given by~\cite{Tegmark:2008au}
\begin{equation}
P^{\rr n} = \frac{4 \pi \lambda^2 T_{\rr{sys}} ^2 }{D^2 \Omega t_{\rr o} }~.
\end{equation}
The diagonal elements of the Fisher matrix determine the errors on the parameters $\lambda_a$,
\begin{equation}
\Delta \lambda_a = \sqrt{(\mathbf F^{-1} )_{aa}}.
\end{equation}

\subsection{MCMC bayesian method}

We calculate for the first time the forecasts on the cosmological parameters, for the two considered 21cm experiments, by using a MCMC technique.  

To do so, we consider the 21cm angular power spectrum at a given redshift, instead of the 3D power spectrum.   Let us assume that the real Universe is correctly described by a set of fiducial cosmological (and astrophysical) parameters $\bar \lambda_a$.  They lead to  21cm brightness temperature fluctuations in the sky, characterized by a set of $\hat C_l ^{21} (z) $.  These so-called \textit{mock} $\hat C_l^{21}(z) $  can be estimated with 21cm dedicated experiments like the FFTT.  

Our objective is to estimate the likelihood function of the parameters $\lambda_a $, for measuring these mock $\hat C_l ^{21} (z) $, given the uncertainty $\delta C_l^{21} (z)$ due to the cosmic variance and the noise of the experiment,  
\begin{equation}
\delta C_l^{21} (z) = \frac{1}{\sqrt{2 l +1} } \left( C_l^{21} (z)  + C_l^{\rr n}   \right)~.
\end{equation}
The noise term for the FFT Telescope is given in Ref.~\cite{Tegmark:2008au} (see also Appendix 1),
\begin{equation} \label{eq:noise21}
C_l^{\rr n}  =  C_ 0^{\rr n} B_l^{-2}~,
\end{equation}
where $B_l $ is the beam function of the experiment.  It is well approximated by a Gaussian function of width $\lambda / D$.  $C_0^{\rr n}$ is a normalization constant that reads
\begin{equation} 
C_0^{\rr n} = \frac{4 \pi \lambda^2 T_{\rr{sys} }  ^2}{D^2 \Omega t_{\rr 0} \Delta \nu}~.
\end{equation}
The likelihood for measuring the parameters $\lambda_a$ is given by Eq.~(\ref{annex:mcmclik}) of Appendix C
\begin{equation}
- 2 \ln \mathcal L(\lambda_a | \hat C_l)  = \sum _l (2 l +1 ) \times \left( \frac{\hat C_l^{\rr{tot}}}{C_l^{\rr{tot}} } + \ln \frac{C_l^{\rr{tot}} }{\hat C_l^{\rr{tot}} }  - 1\right)~.
\end{equation}
In the context of Bayesian analysis, the MCMC method is used to probe this likelihood function multiplied by the prior of the model parameters
\footnote{We took flat priors on cosmological parameters, in the range $0.005 < \Omega_{\rr b} h^2 < 0.1$, $0.01 < \Omega_{\rr c} h^2 < 0.99$, $0.001 < \tau < 0.8 $, $0.9 < n_{\rr s} < 1.1$, $ 2.8 < \ln (10^{10} A_{\rr s} ) < 3.6$, $0 < r < 0.5 $.  We fixed $\Omega_{\rr K} = 0$. }.  
Then the marginalized posterior probability density distributions of the parameters $\lambda_a $ can be calculated and the 1-$\sigma$ or 2-$\sigma$ forecasts can be deducted.  

For given sets of cosmological parameters, the $C_l^{21} (z = 40)$ are calculated with the numerical CAMB code.  For the 21cm signal from the reionization, the CAMB code has been modified to calculate the 21cm angular power spectrum with the brightness temperature given by Eq.~(\ref{eq:PTb}).  

\section{Forecasts for the dark ages}

\begin{figure}[h!]
\begin{center} 
\includegraphics[height=80mm]{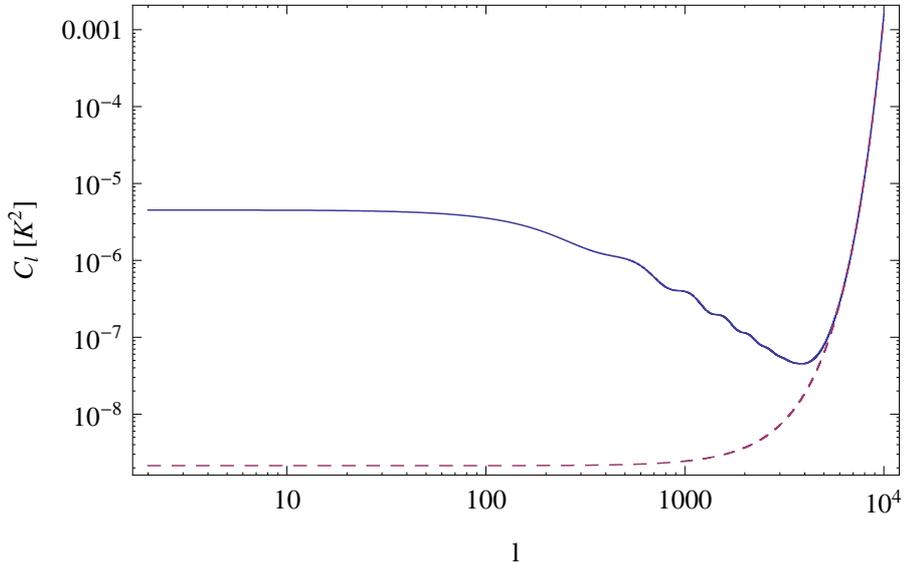}
\caption{ Total 21cm brightness temperature angular power spectrum $C_l ^{\rr{tot} } = C_l^{21} + C_l ^{\rr n}$, at $z= 40$ (plain curve).  The noise term $ C_l ^{\rr n}$ (dashed curve) is given by Eq.~(\ref{eq:noise21}) for the Experiment 2 - FFTT radio-telescope.  The noise dominates over the cosmological signal at $l > 5000 $.  The theoretical $C_l^{21}$ are calculated for the best fits of the $\Lambda$-CDM model.} \label{fig:Cls21DA}
\end{center}
\end{figure}

\begin{figure}[p]
\begin{center}
\includegraphics[width=160mm]{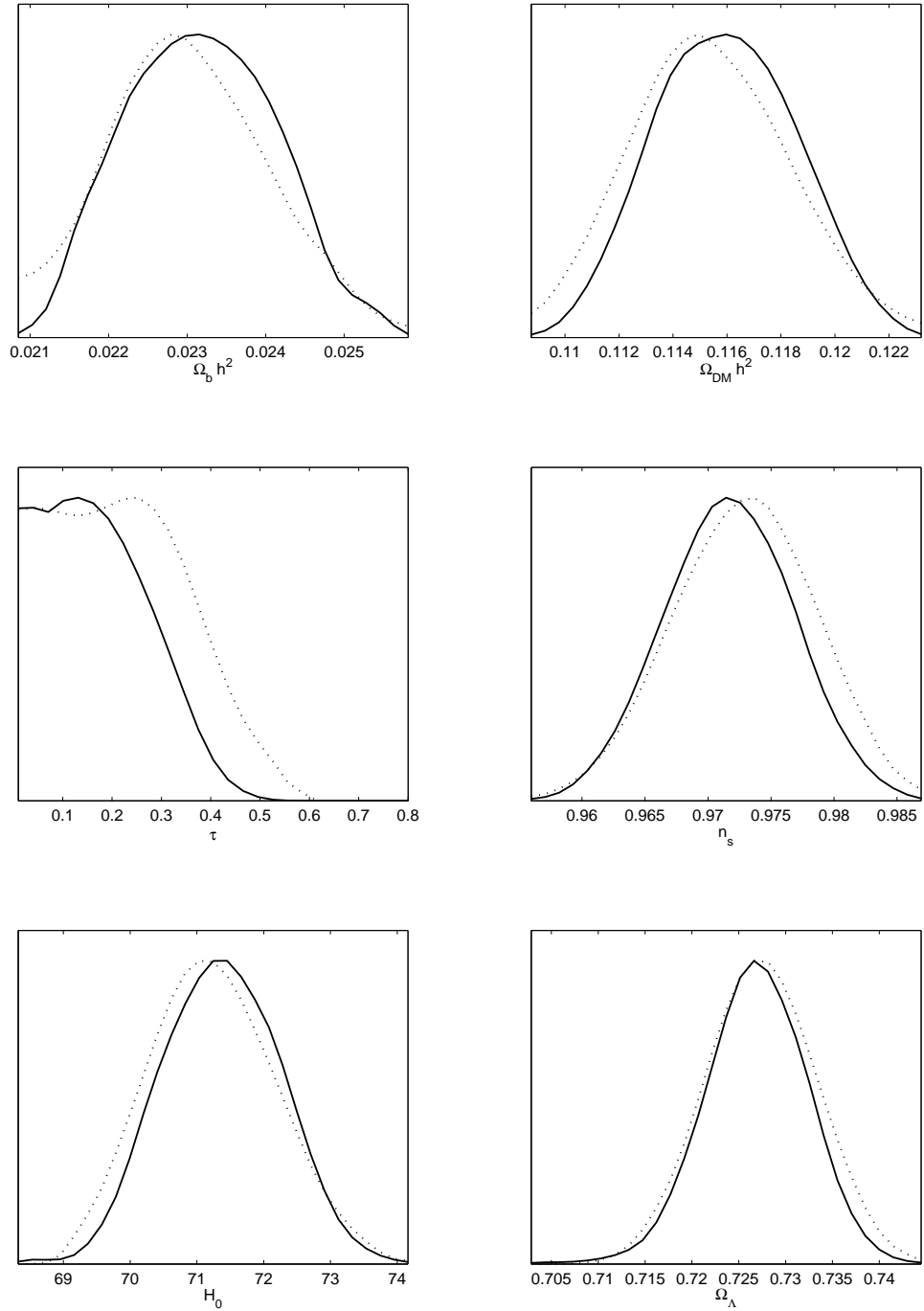}
\caption{ Marginalized posterior probability distributions of the $\Lambda$-CDM cosmological parameters, for the same experiment of Fig.~\ref{fig:Cls21DA}.  Forecasts are at the level or better than the present CMB constraints.  Dotted curves correspond to the average over the parameter space.  
} \label{fig:forecastDA1}
\end{center}  
\end{figure}

\begin{figure}[p]
\begin{center} 
\includegraphics[height=160mm]{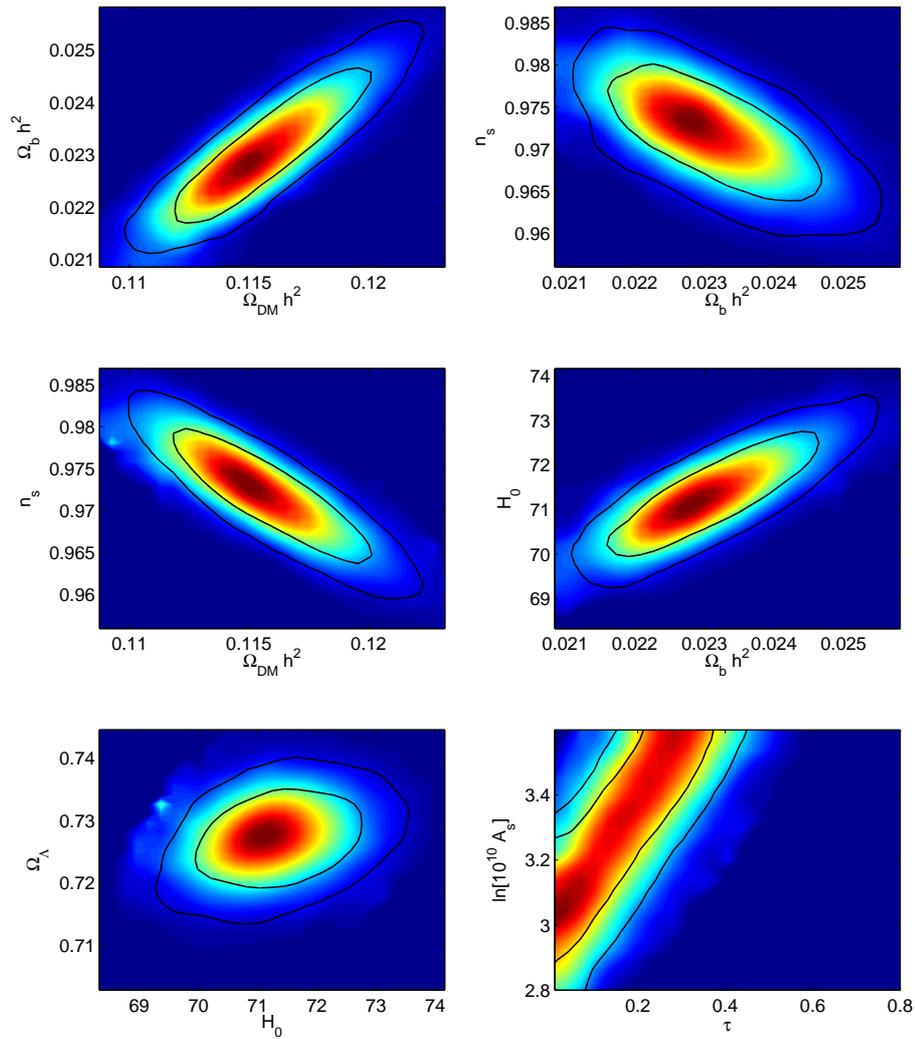}
\caption{ 2D-marginalized posterior probability distributions of the $\Lambda$-CDM cosmological parameters, for the same experiment of Fig.~\ref{fig:Cls21DA}.  The optical depth $\tau $ and the 
amplitude of the primordial scalar power spectrum $A_{\rr s} $ are observed to be degenerate.   }
\label{fig:forecastDA2}
\end{center}
\end{figure}

\begin{figure}[p]
\begin{center} 
\includegraphics[height=130mm]{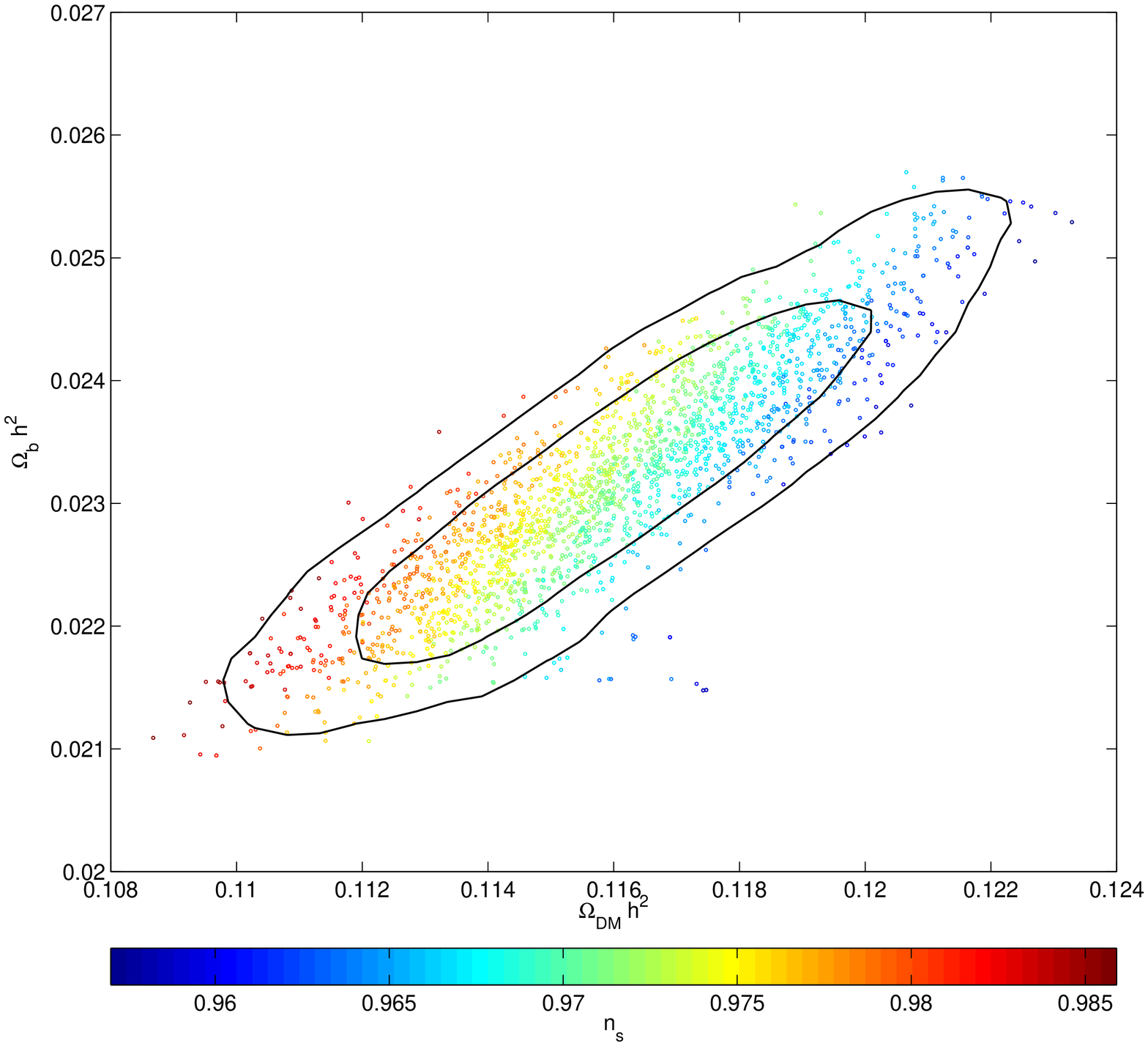}
\caption{ 3D distribution of points from the Markov chains, in the space ($\Omega_{\rr c} h^2, \Omega_{\rr b} h^2, n_{\rr s}$).  Marginalized 1-$\sigma$ and 2-$\sigma$ contours, for the same experiment of Fig.~\ref{fig:Cls21DA} are also given.   } \label{fig:forecastDA3}

\end{center}
\end{figure}

The total 21cm brightness temperature angular power spectrum $C_l ^{\rr{tot} } = C_l^{21} + C_l ^{\rr n}$, at $z = 40$, for the Experiment 2 described in Sec.~\ref{sec:exper}, is given in Fig.~\ref{fig:Cls21DA}.   The noise term dominates over the cosmological signal for $l > 5000 $ and the relic of the acoustic oscillations are visible.   
The forecasts on the cosmological parameters, obtained with the MCMC bayesian method, are given in Figs.~\ref{fig:forecastDA1}, \ref{fig:forecastDA2} and \ref{fig:forecastDA3}.   One sees that the optical depth $\tau $ and the scalar power spectrum amplitude $A_{\rr s} $ are degenerated.   Because the 21cm signal is only sensitive to the scalar perturbations, the tensor to scalar ratio $r$ is not constrained. The forecast for the scalar spectral index is much better than present CMB constraints.  For other parameters, forecasts are at the level or slightly better than the present CMB constraints. 

In the case of the Experiment 1,  the noise term dominates at $l > 500$, resulting in no improvement of the parameter estimations.   

\section{Forecasts for the reionization}

The total 21cm brightness temperature angular power spectrum $C_l ^{\rr{tot} } = C_l^{21} + C_l ^{\rr n}$, at $z = 11$, for the Experiment 2 described in Sec.~\ref{sec:exper}, is given in Fig.~\ref{fig:Cls21DA}.   The noise term dominates over the cosmological signal only at very small scales, for $l > 20000 $.

\begin{figure}[h!]
\begin{center}
\includegraphics[height=80mm]{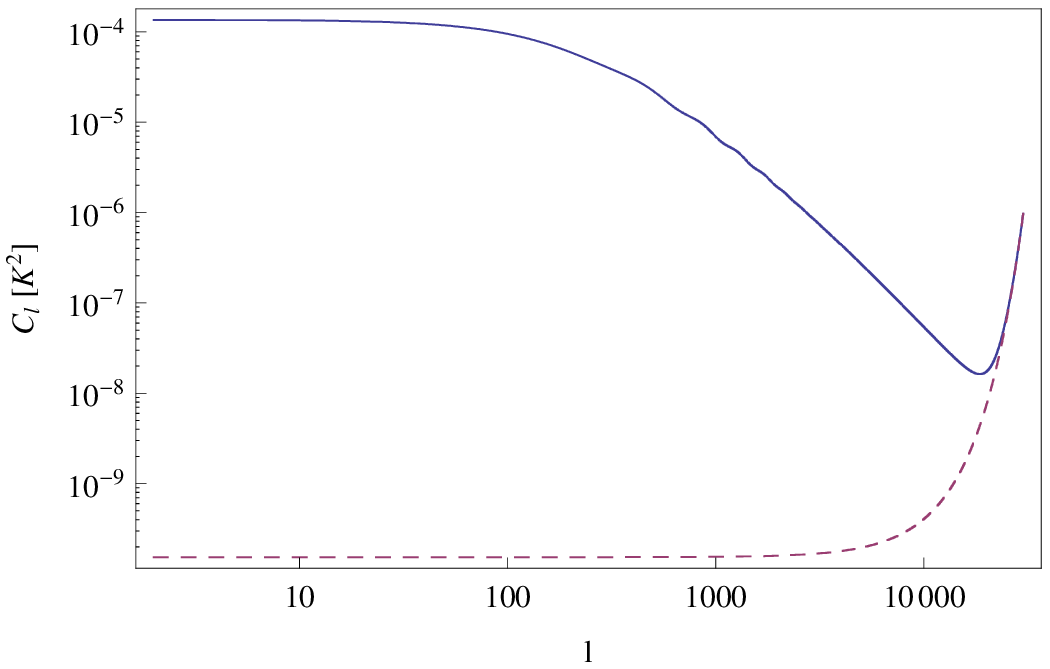}
\caption{ Total 21cm brightness temperature angular power spectrum $C_l ^{\rr{tot} } = C_l^{21} + C_l ^{\rr n}$, at $z= 11$ (plain curve).  The noise term $ C_l ^{\rr n}$ (dashed curve) is given by Eq.~(\ref{eq:noise21}) for the Experiment 2  FFTT radio-telescope.  The noise dominates over the cosmological signal at $l > 20000 $.  The theoretical $C_l^{21}$ are calculated for the best fits of the $\Lambda$-CDM model, for our toy model of the reionization process [$x_{\rr H} (z=11) = 0.87 , T_{\rr s} (z=11) = 133 \ \rr K$].} \label{fig:Cls21RE}
\end{center}
\end{figure}

The forecasts on the cosmological parameters for the Experiment 2, obtained with the Fisher matrix formalism, for a redshift bin $\Delta z = 0.5$, are given in Tab.~\ref{tab:fft1sigma}.  For comparison the 1-$\sigma$ marginalized forecast for the dark ages ($z =40$) are also given.   Since $\tau $ and $A_{\rr s} $ are degenerated, these parameters are not included in the Fisher matrix analysis.  Significant improvements on the parameter measurements are expected.   
In the context of this thesis, the expected improvements of the spectral index measurements must be emphasized.  The precision reached by the idealistic experiment could be sufficient to rule out or to put strong constraints on many inflation models, like the large field models, or the hybrid ones.  We have added to the Fisher analysis the running scalar index.  The running (linked to the slow-roll parameter $\epsilon_3$), could be measured with a good accuracy and therefore could contribute to distinguish between inflation models and to put new constraints on the reheating temperature.  

The forecasts for the Experiment 1 are given in the last column of Tab~\ref{tab:fft1sigma}.  Our results have been checked to  be coherent with Ref.~\cite{Mao:2008ug}.  Significant improvements of the present constraints are expected, for all the parameters.  Finally, let remind that the study is performed for an unique redshift bin, at the beginning of the reionization.  The combination of the 21cm signal at various redshifts are expected to improve the expected precision of the parameter measurements~\cite{Mao:2008ug}.  

 
\begin{table} 
 \begin{center}
 \begin{tabular}{|c|c|c|c|} 
 \hline
  & Exp 2. $z=40$ & Exp. 2, $z=11$ & Exp.1, $z=11$ \\ \hline
  $\Delta \Omega_{\rr b} h^2$ & 0.0009   & 0.00004  &  0.00006 \\
 $\Delta  \Omega_{\rr c} h^2$ & 0.0026  & 0.0003   & 0.0006  \\
  $\Delta  n_{\rr s}   $         & 0.005  & 0.0007  & 0.002 \\
  $ \Delta \alpha_{\rr s} $ &  & 0.0002 & 0.001 \\ \hline
  \end{tabular}
  \caption{1-$\sigma$ forecasts on cosmological parameters, for the Experiment 2 at $z=40$ (column 2) obtained with the MCMC method, and at $z=11$ (column 3) with the Fisher matrix method, for a redshift bin $\Delta z = 0.5$, as in Ref.~\cite{Mao:2008ug}.  In column 3, forecasts for the Experiment 1 calculated with the Fisher matrix method are given.  For the two experiments at $z=11$, constraints on the spectral index and its running are improved.  
  } 
  \label{tab:fft1sigma}
 \end{center}
 
 \end{table}

\section{Conclusion}

The 21cm signal from the dark ages and the reionization period is a promising signal that should play in the future a major role in the game of improving the present constraints on the cosmological parameters.  Contrary to the CMB that is a nearly instantaneous image, 21cm observations could be used realize a 3D tomography in redshift of the Universe, by mapping the hydrogen gas (and thus the matter) distribution over the range $ 3 \lesssim z \lesssim 200 $.   Since it is not affected by the photon diffusion, the perturbation length scales in principle accessible are much lower than for the CMB. The lever-arm for measuring the primordial scalar power spectrum is therefore larger.  The spectral index (and the running) could be measured with a high accuracy.   The 21cm signal is therefore a good laboratory to learn more about the hypothetic phase of inflation and the reheating era that follows.  

In this chapter, we have contributed to study the 21-cm forecasts on the cosmological parameters.  We have focus on the 21cm signal from two arbitrary redshifts, $z=40$ and $z=11$, i.e. respectively during the dark ages and at the onset of the reionization era.  Our analysis relies on two typical configurations of the FFTT concept.   In order to be sensitive to deviations from gaussianity of the likelihood function of the parameters, we have extended previous Fisher matrix analysis by considering also a bayesian MCMC technique.   

Forecasts for the 21cm signal from the dark ages are only competitive with CMB observations in the idealistic experimental configuration, i.e. a $10$ km $\times 10$ km radio-interferometer.    Indeed, because the signal is more far than the 21cm from reionization, a higher angular resolution is required to probe the first baryon acoustic oscillations in the matter power spectrum.  Thus very large configurations of the experiments are necessary.  Moreover, the bandwidth of the experiment corresponds to a larger redshift bin.  Because the signal is convoluted with the window function of the experiment,  the ability to put significant cosmological constraints is reduced.   
 
Forecasts for the 21cm signal from the reionization are better, even in the case of a realistic configuration ($1$ km $\times 1$ km) of the FFTT radio-telescope.  
However, our analysis is based on the strong assumption that the ionized fraction follow a simple evolution and that the contribution to the 21cm brightness temperature power spectrum of the ionized matter density perturbations can be neglected at the beginning of the reionization.  
Let nevertheless remark that for a specific parametrization of the power spectrum of the ionized matter density perturbations, the forecasts are not expected to be degraded~\cite{Mao:2008ug}.     

Let us comment and discuss with more details the differences between the 21 signal from the dark ages and the reionization period.  First, it must be noticed that our results for the dark ages are for only one redshift slice.  They should be ameliorated when enlarging the study to a wide range of redshifts over the dark ages period.   Because the growth of the observable perturbations is still in the linear regime, the physics of the signal is rather simple, and it is not affected by complex astrophysical processes like it is during the reionization.   However, galactic foregrounds are expected to be order of magnitudes higher than for the reionization.  But even if it is very challenging, their frequency dependance could be used to provide good foreground removal techniques~\cite{Morales:2005qk}.  Atmospheric opacity to low frequencies ($\sim 20$ Mhz) should nevertheless prevent the detection of the 21cm signal at very high redshifts ($z> 70$) with Earth-based radio-telescopes~\cite{Carilli:2007eb}.  Our ability to constrain cosmology with 21cm observations from the dark ages therefore principally depends on the foreground removal techniques, and on our technology limitations for building sufficiently large large radio-interferometers with low bandwidths.   

On the contrary, the 21cm power spectrum from the dark ages could be observed by more realistic experiments, and the galactic foregrounds should be removed with less difficulties.  But the physics during the reionization is much more complex, possibly non-gaussian~\cite{Cooray:2004kt}, and the range of accessible redshifts is lower.  Non-linear growth of perturbations also limits the range of perturbation length scales interesting for cosmology.  Therefore the ability to improve cosmological parameter estimations with observations of the 21cm signal from the reionization mainly depends on how confident we are to the calculations and simulations of the physics of the reionization process.  Especially we need to know if the proposed parametrizations~\cite{Mao:2008ug} for the  ionized fraction evolution and the density fluctuations of the ionized matter are sufficiently accurate.   

In order to avoid the complications of the reionization process, it has been recently proposed to observe the 21cm signal at lower redshifts (at about $z \sim 2$)~\cite{Visbal:2008rg,Wyithe:2008mv,Loeb:2008hg,Wyithe:2007gz}, i.e. when the reionization has been nearly totally completed, and when the foreground level is lower.  Even if the 21cm brightness temperature is reduced due to the lower neutral fraction, the 21cm power spectrum could be detectable and useful for cosmology.  But a major limitation comes from the fact that the non-linear growth affect perturbations in the observable range, and thus the range of length scales interesting for cosmology is reduced.

Our study should be extended to determine directly forecasts on the parameters of inflation and reheating models.  As an example, one can see in Fig.~\ref{fig:nsrplane}, for a large field model of inflation, with $V(\phi) \propto \phi^2$, that in absence of improvements of the tensor to scalar ratio, an accuracy of $\Delta n_{\rr s} \approx 0.001 $ is required for determining the reheating temperature scale.  One sees in Tab.~\ref{tab:fft1sigma} that such a precision could be nearly reached by a $1$km$^2$ 21cm FFTT radio-telescope, provided ideal foreground removals and assuming the optimistic reionization case.  

Let us mention also that a detection of the slow-roll parameter $\epsilon_3$ could be used to put strong constraints on inflation models (and possibly to rule out several of them) and on the energy scale of the reheating.   

But the interest of the 21cm signal is not limited to the cosmological parameter estimation and to the inflation and reheating models.   It is expected to be a valuable tool in several domains, from dark matter~\cite{Natarajan:2009bm} and dark energy models~\cite{Wyithe:2007rq} to the study of the cosmic strings~\cite{Hernandez:2011ym, Brandenberger:2010hn}, the non-gaussianities~\cite{Joudaki:2011sv},  the variations of the fundamental constants~\cite{Khatri:2009aw} and the formation of the first stars.

%% file: conclu.tex
\chapternonum{Conclusion}
\label{cha:conclu}

In this thesis, the exact multi-field classical dynamics of hybrid inflation models has been investigated.  
Several models from various high energy frameworks have been considered:  the original non-supersymmetric model, the F-term, shifted and smooth hybrid models both in their SUSY and SUGRA versions, as well as the radion assisted gauge inflation model in which the inflaton is the phase of a Wilson loop wrapped in an extra dimension.   

We have focus on three specific issues:  the set of the initial field values, the effects of slow-roll violations  during the field evolution along the valley, and the final waterfall phase.  Our contributions are summarized below:

\begin{itemize}
\item Instead of fined-tuned along the inflationary valley, the set of initial field values leading to more than 60 e-folds of inflation has been found to occupy a considerable part of the field space exterior to the valley.  These form a complex structure with fractal boundaries that is the basin of attraction of the inflationary valley.  Moreover, by using bayesian MCMC methods, it has been shown that inflation is realized without fine-tuning of initial field values in a large part of the parameter space, independently of the initial field velocities.  Natural bounds on the potential parameters have been determined.

\item This analysis has been extended to the case of a closed Universe, for which the initial singularity is replaced by a classical bounce.  In the contracting phase, the initial conditions of the field trajectories performing at least one classical bounce followed by a phase of hybrid inflation are sub-dominant but non negligible, provided that the curvature was initially sufficiently large.   Compared to some other scenarios of classical bounce plus inflation, initial conditions are not extremely fine-tuned and the model is found to be viable on this point of view.  
\item For the original hybrid model, when the inflationary valley is reached by the field trajectories, the slow-roll conditions can be violated at the transition between the large field and the small field phase of inflation.  By integrating the exact dynamics, we have shown that these slow-roll violations induce the non-existence of the phase of inflation at small field values.  In this case, the hybrid model is similar to a large field model and super-planckian initial conditions are required for inflation to last more than 60 e-folds.  We have determined numerically a condition on the potential parameter $\mu $ (see Eq.~\ref{cond_redspectrum_2}) for which this mechanism is triggered.  
\item For the original hybrid model, the integration of the exact 2-field dynamics has revealed that inflation can continue for more than $60$ e-folds along the classical waterfall trajectories, provided that a condition on the potential parameters is satisfied (see Eq.~\ref{eq:boundmuM}).   
\end{itemize}

As a result, for hybrid inflation to be realized in a natural way, without fine-tuning of the initial conditions, natural constraints on the parameter space need to be imposed.   Moreover, compared to the effective 1-field slow-roll approximation, the observable predictions can be modified. 

For instance, for the hybrid model, the primordial scalar power spectrum can be red, in accord with CMB observations, in several cases.  The first two of them are trivial and well-known, but two new mechanisms have been identified:
\begin{enumerate}
\item When the tachyonic instability is developed in the large field phase of inflation.
\item When inflation is generated along radial field trajectories instead of along the inflationary valley.  In this case, the model is similar to the double inflation model.  
\item When the small field phase of inflation is avoided due to slow-roll violations.  
However, these first three ways require super-planckian initial field values.  

\item When inflation continues for more than 60 e-folds after the critical instability point, during the first stages of the waterfall.   Observable modes leave the Hubble radius during the waterfall and the power spectrum of adiabatic perturbations is found to be generically red.   
\end{enumerate}

To this list one may add the effects of cosmic strings, produced at the end of hybrid inflation when a $U(1)$ symmetry is fully broken.   In this case, it has been shown~\cite{Bevis:2007gh} that a value of the scalar spectral index $n_{\rr s} = 1 $ is not disfavored.  The original hybrid model can therefore be in agreement with CMB observations if it is modified (e.g. by considering a complex auxiliary field $\psi$) to lead to the formation of cosmic strings.   


Several perspectives have risen from this work.  Some of them are briefly described below and should be the subject of future work and collaborations:  
\begin{itemize}
\item For waterfall trajectories performing inflation, the power spectrum of curvature perturbations can be affected more or less significantly by iso-curvature perturbations.  Their contribution should be calculated. 
\item The tachyonic preheating process can be also affected by the phase of inflation between the instability point and its triggering.   However, this phase can only be investigated by lattice numerical methods.  These should be extended to include inflation during the waterfall.
\item Due to inflation along the waterfall trajectories, the eventually formed topological defects are strongly diluted by the expansion and can be pushed exterior to the observable Universe.  This result may be of interest for the determination of the allowed schemes of symmetry breaking in GUT.  
\item Quantum stochastic effects are susceptible to play a significant role in some parts of the parameter space and could modify the observable predictions.   
\item For the classical bounce plus hybrid inflation scenario, the signatures on the primordial scalar and tensor power spectra should be determined and could be eventually in the range of observations. 
\item Finally, our results could be extended by studying other hybrid models.  
\end{itemize}

Today, our ability to constrain inflation and reheating models relies mainly on CMB observations.  
Parallel to our work on hybrid inflation, we have studied the possibility to improve these constraints with the observation of the 21cm signal form the dark ages and the reionization.   Our analyze is based on two hypothetic (one realistic and one optimal) radio-telescope experiments, based on the concept of Fast-Fourier-Transform radio-Telescope.    

We have determined the forecasts by using a Fisher Matrix method and a full MCMC method.  Our results confirm the forecasts of Ref.~\cite{Mao:2008ug} for the 21cm signal from the reionization:  observing the 21cm power spectrum could improve significantly the  measurements of the primordial scalar power spectrum.  However, these forecasts rely on the strong assumption that the reionization process is well described by simple parametrization of the mean ionized fraction and spin temperature evolutions, as well as of the power spectrum of the ionized matter density perturbations.  

We obtain that observations of the 21cm signal from the dark ages would only improve the measurements of the cosmological parameters for giant idealistic experimental configurations.  This is without taking account foregrounds that are several order of magnitude higher for the 21cm signal from the dark ages.  However, the physics during the dark ages is fairly simple and thus the 21cm signal can be used directly to probe cosmology.

At short term, we expect to extend this work to determine forecasts directly on the parameters of some inflation models, including the hybrid ones, in a way consistent with the reheating history.  Forecasts on the reheating temperature will thus be included and compared to the present bounds from CMB observations.  



%% file: annexes/FFTT.tex
\chapter{Fast Fourier Transform Telescope}
\label{A:FFTT}

From a mathematical point of view, what does a telescope is a Fourier transform.  Indeed, the aim of a telescope consists in extracting from the received total electromagnetic field, the individual Fourier modes $\mathbf k$, giving the source position in the sky and the wavelength of the signal.   But the electromagnetic field received by the telescope at a given position and time $(\mathbf r, t)$ is only the sum of these Fourier modes weighted by phase factors  $\exp [ i (\mathbf{k \cdot r} + \omega ) t ]$.  

This Fourier transform is performed by using different techniques, depending on the type of the telescope.  Let us distinguish
\begin{itemize}
\item \textbf{Single-dish telescopes}:  The spatial and temporal Fourier transforms are performed by using analog techniques like lenses or mirrors as well as slits, gratings or band-pass filters. 
\item \textbf{Interferometers:}   The frequency separation and the correlation between different receivers are done by using analog techniques, but then the Fourier transform to the $\mathbf r$ space is realized digitally.   This method can be used to increase the resolution without need of extremely large single-dish telescopes, since instead the signals from several receivers separated by long baselines are combined.  Correlations between each pair of receivers need to be calculated, so that the computational cost scales like $N^2$, where $N$ is the number of antennas.   

\item \textbf{Fast Fourier Transform Telescopes (FFTT):}  The Fourier transforms are realized fully digitally and antennas do not need to be pointable anymore.  The receivers are equidistant and distributed on a plane such that each baseline corresponds to a large number of pair receivers.   The computational cost scales as $N \log_2 N $, instead of $N^2$ for standard interferometers.    
\end{itemize}

\section{The Fast Fourier Transform Telescope concept}

Any telescope composed of some antennas is characterized at fixed frequency by the sky response $\mathbf B _n(\hat {\mathbf k}   )  $ of each antenna ($\mathbf{\hat k} $ denotes a unit vector in the direction of $\mathbf k$).  The data measured by the antenna $n$ (imaging half of the sky for the FFTT) is
\begin{equation} \label{eq:ffttsignal}
\mathbf d_n = \int \mathbf B _n(\hat {\mathbf k} )  \mathbf s (\hat {\mathbf k} ) \rr e ^{- i (\mathbf{k \cdot r}_n + \omega t) }  \dd \Omega_k~,
\end{equation}
where $  \mathbf s (\hat {\mathbf k} )  $ is the sky signal\footnote{$\mathbf s$ is a 2-component complex vector giving the electric field in two orthogonal directions.}, $ \Omega_k $ is the solid angle in the $k$ space, and $\mathbf{r_n} $ is the location of the $n$-th antenna.  The last factor accounts for the extra path length to reach the $n$-th antenna.    This stands for any telescope array.  

In the FFTT concept, all the antennas are distributed in a plane $(x,y)$ such that $\mathbf r_n =  (x_n,y_n)$, and have an identical beam pattern $\mathbf B$.  It is therefore convenient to decompose any wavevector into orthogonal and parallel components to the $k_z$ axis, $(k_\bot \equiv \sqrt{k_x^2 + k_y^2} ,k_\parallel \equiv k_z )$.  Then Eq.~(\ref{eq:ffttsignal}) can be rewritten on the form of a two-dimensional Fourier transform,
\begin{equation} \label{eq:ffttsignalb}
\mathbf d_n = \int  \frac{ \mathbf{B(\hat k) s(\hat k)} \rr e^{-i (k_x x_n + k_y y_n + \omega t )} }{ k  \sqrt{k^2 - k_\bot ^2}}        \dd k_x \dd k_y~.
\end{equation}
It is the Fourier transform of the function
\begin{equation} 
\mathbf {s_B} (\mathbf k) \equiv \frac{\mathbf B (\mathbf{\hat k}) \mathbf {s (\hat k)}}{k \sqrt{k^2 - k_\bot ^2} }~,
\end{equation}
and one can read
\begin{equation}  \label{eq:fftdn}
\mathbf{d}_n = \mathbf{ \hat  s _B } (x_n,y_n)  \rr e^{- i \omega t}~.
\end{equation}
The sky signal coming from different directions of the sky is usually considered to be uncorrelated, 
\begin{equation}
\langle \mathbf{s(\hat k), s(\hat k ') ^\dagger} \rangle = \delta(\mathbf{\hat k, \hat k'}) \mathbf{S(\hat k)}~.
\end{equation}
This relation defines $\mathbf {S(\hat k )} $, the $2 \times 2 $ complex \textit{Stokes matrix}.  It can be applied to Eq.~(\ref{eq:ffttsignalb}) to obtain the so-called \textit{visibility}, that is the correlation between two measurements,
\begin{eqnarray} \label{eq:ffttvisibility}
\langle \mathbf d_m,  \mathbf d_n ^\dagger  \rangle & = & \int  \dfrac{ \mathbf{B(\hat k)^\dagger S(\hat k) B(\hat k) } \rr e^{-i [ k_x (x_m - x_n) + k_y (y_m - y_n)  ] } }{ k  \sqrt{k^2 - k_\bot ^2}}        \dd k_x \dd k_y~\\
 & =  & \hat {\mathbf S}_B (\mathbf{r_m - r_n} )~,
\end{eqnarray}
where the last line is obtained after defining the $2 \times 2$ complex matrix 
\begin{equation} \label{eq:S_B}
\mathbf {S_B} (k) \equiv \frac{ \mathbf{B(\hat k)^\dagger S(\hat k) B(\hat k) }}{ k  \sqrt{k^2 - k_\bot ^2}}~.
\end{equation}

Therefore, the statistical properties of the sky map can be recovered from data measurements by implementing the following procedure:
\begin{enumerate}
\item Evaluation of  the data correlations for a large number of baselines, $\mathbf{\hat S_B (\Delta r)}$.
\item Fourier transform in the $x$ and in the $y$ directions to obtain $ \mathbf {S_B} (k)  $.
\item Evaluation of the Stokes matrix by inverting Eq.~(\ref{eq:S_B})
\end{enumerate}
Finally, it is important to remark that the sky has not been assumed to be flat and thus this result is valid for any size of the field of view.  

In standard interferometry, this process is performed for each pair of antennas, and thus the number of computations scales like $N_{\rr a} ( N_{\rr a} - 1 ) / 2 \sim N_{\rr a} ^2 $.   For the FFTT concept, proposed in Ref.~\cite{Tegmark:2008au}, all the antennas are placed on a rectangular grid and have an  identical separation distance.  Therefore each baseline corresponds to a large number of pairs of antennas.  It results that the number of computations is of the order of $ N_{\rr a} \log_2 N_{\rr a} $.   More precisely, the convolution of the 2D grid by itself is realized by a FFT in the $x$ and $y$ directions,  a squaring and an inverse FFT.  Actually, the inverse Fourier transform is not needed, since after FFT-ing the 2D antenna grid, one already has the electric field components via $\mathbf {s_B} $ [see Eq.~(\ref{eq:fftdn})].

Practically, the above procedure needs to be repeated for each time sample and each frequency.  

Finally, it must be noticed that a first Fourier Transform in the time domain is required to separate out the different frequencies from the total signal.  This can be done using the standard digital filtering methods.  

\section{Beam function and sensibility}

Let us consider the distribution function of baselines $W(\Delta x,\Delta y)$.  Ignoring polarization and assuming that the radiation wavevector is along the zenith $k_z$ direction, the response of the interferometer to the radiation (in the flat sky approximation) is given by its Fourier transform $ \hat W (k_x, k_y ) $~\cite{Tegmark:2008au}. 

The FFTT is a square of length $D$, the distribution function of baselines is obtained after convolving the square with itself,
\begin{equation}
 W(\Delta x, \Delta y )  \propto ( D - \Delta x) (D - \Delta y)~,
\end{equation} 
and the synthesized beam is its Fourier transform
\begin{equation}
 \hat W( k_x, k_y )  =  j_0 \left( \frac D 2  k_x \right)  j_0 \left( \frac D 2 k_y \right)~.
\end{equation} 
Following~\cite{Tegmark:2008au}, we have been interested in Chapter 9 to the azimuthally averaged beam, that only depends on $\theta$, the angle to the zenith.   In~\cite{Tegmark:2008au}, the beam function $B(\theta) \simeq W(l/k) $ is shown to be well approximated by a Gaussian function, characterized only by its FWHM (i.e. twice the $\theta$ value when $B(\theta)$ is reduced by a factor 2).   


The \textit{noise power spectrum} $C_l^{\rr n} $ quantifies the sensibility of the telescope to the signal on various angular scales.  It is usual to factorize the noise power spectrum as 
\begin{equation}
C_l^{\rr n} = C_0^{\rr n} B_l^{-2}~,
\end{equation}
where $B_l$ is the beam function, normalized such that it is unity at its maximum.  For a FFTT telescope with total collecting area $A$, a total observation time $t_{\rr o}$ and the bandwidth $\Delta \nu$ around the observation frequency $\nu = c / \lambda $ (corresponding to the redshifted 21cm wavelength for 21cm mapping), the normalization factor $C_0^{\rr n} $ is given by~\cite{Tegmark:2008au}
\begin{equation} 
C_0^{\rr n} = \frac{4 \pi \lambda^2 T_{\rr{sys}}^2 }{A \Omega t_{\rr 0} \Delta \nu }~,
\end{equation}
where $\Omega $ is the field of view ($2 \pi$ for the FFTT) and $T_{\rr{sys}} $ is the so-called \textit{system temperature}.  These parameters are given by the specifications of the experiment.

%% file: annexes/fisher.tex
\chapter{Fisher matrix formalism}
\label{A:fisher}

Current experiments provide a huge amount of data that need to be analyzed to obtain at the end the maximum of the likelihood function for the cosmological model parameters as well as the 95\% confidence regions.   If one uses brutal force, that is estimating the likelihood everywhere in the parameter space (that has usually more than 10 dimensions),   an extremely huge computation time is required because for each estimation one needs to inverse a non-diagonal covariance matrix~\cite{Dodelson} whose size is extremely high (of the order $n_{\rr{pixels}} \times n_{\rr{pixels}}$, where $n_{\rr {pixels} }$ is the number of pixels on the sky).   As a consequence, some more expeditious statistical techniques have been developped.   The Fisher matrix formalism is one of them and is described in this appendix. 

\section{Optimal quadratic estimator}

Let us follow Ref.~\cite{Dodelson} and consider for simplicity a 1-dimensional parameter space.  The following results will be generalized easily in the multi-dimensional case.   Our objective is to find the value of the parameter $\lambda $ for which the likelihood function $\lik (\lambda) $ is maximal.  Let us denote this value $\bar \lambda $, such that one has
\begin{equation}
\left. \frac{\partial \lik }{\partial \lambda} \right|_{\bar \lambda} = 0~. 
\end{equation}
If $ \lik$ was a quadratic function of $\lambda$, the root would be found easily by evaluating the likelihood function and its derivatives at an a arbitrary point $\lambda^{(0)} $.   After a Taylor expansion, one has
\begin{equation}
\lik_{,\lambda} (\bar \lambda ) = \lik _{,\lambda} (\lambda^{(0)}) + \lik_{,\lambda \lambda} (\lambda^{(0)}) (\bar \lambda - \lambda^{(0)} )~,
\end{equation}
where the subscripts $_{,\lambda} $ and $_{, \lambda \lambda}$ denote respectively the first and second derivatives of the likelihood function with respect to the parameter $\lambda$. The left hand side is zero and thus one finds 
\begin{equation}
\bar \lambda = \lambda^{(0)} - \frac{\lik_{,\lambda} (\lambda^{(0)})  }{\lik_{,\lambda \lambda} (\lambda^{(0)}  )}~.
\end{equation}
If the likelihood is only nearly quadratic, one can use the Newton-Raphton method and iterate the process.  The parameter value will converge through the true $\bar \lambda $ in typically a few number of iterations.  

In practice, the Likelihood function is not at all a quadratic function of $\lambda$.  Indeed, it is expected to become exponentially small at parameter values far from $\bar \lambda$.   It is much better to assume that $\lik $ is a nearly gaussian function.    In this case, what is quadratic in $\lambda$ is the function $\ln (\lik) $.   The method described above can therefore be applied on this function and one can estimate
\begin{eqnarray}  
\bar \lambda & \simeq & \lambda^{(0)} - \frac{(\ln \lik)_{,\lambda} (\lambda^{(0)})  }{(\ln \lik)_{,\lambda \lambda} (\lambda^{(0)}) }~\\  \label{eq:fisher1}
& \simeq &   \lambda^{(0)} +  \mathcal F^{-1} (\lambda^{(0)})  (\ln \lik)_{,\lambda} (\lambda^{(0)})~,
\end{eqnarray}
where the curvature of the likelihood function, 
\begin{equation}
\mathcal F \equiv - \frac{\partial ^2 \ln \lik }{ \partial \lambda^2}~,
\end{equation} 
has been introduced.  When the curvature is evaluated at the maximum of $\lik$, it measures how rapidly the likelihood falls away from the maximum.  For a high curvature, the uncertainty on the model parameter $\lambda$ will be small, for a low curvature it will be more important.    

It is therefore possible to evaluate the parameter value for which the likelihood function is maximum only by evaluating it and its derivatives at one or at a small number of points in the parameter space.  Let us remind that the validity of the method relies on the assumption that the likelihood is a Gaussian fonction.  It is therefore inefficient if the likelihood behavior is strongly non-Gaussian, e.g. if it has several local maxima.  For this reason, the interest of the Fisher matrix formalism is limited and for accurate results, it is better to use bayesian methods like Monte-Carlo-Markov-Chains.  

In practice, the likelihood function and its curvature are related to the covariance matrix of the experiment (see e.g.~\cite{Dodelson} for details), and it is convenient to replace $\mathcal F$ by 
\begin{equation}
F \equiv \langle \mathcal F \rangle~,
\end{equation}
corresponding to the average of the curvature over many realizations of signal and noise.  

For a $n$-dimensional parameter space $\{\lambda_{i = 1,2,\cdots n} \} $, Eq.~(\ref{eq:fisher1}) can be generalized directly.  One obtains
\begin{equation} \label{eq:estimator}
\hat \lambda_i =  \lambda_i^{(0)} + F^{-1}_{ij} (\lambda_i^{(0)})  (\ln \lik)_{,\lambda_j} (\lambda_i^{(0)})~,
\end{equation}
where 
\begin{equation}
F_{ij} \equiv \langle - \frac{\partial^2 ( \ln \lik )}{\partial \lambda_i \partial \lambda_j}  \rangle
\end{equation}
is the so-called \textit{Fisher matrix}, and were the set $\hat \lambda_i $ estimates the true parameters $\bar \lambda_i$.  Here again there is no need to cover the whole parameter space to estimate the parameter values for which the likelihood is maximal.  Since Eq.~(\ref{eq:estimator}) is an estimator for the best fit values of the parameters, it is possible to study its distribution, its expectation value and variance.  In the case the signal and noise are gaussian, one has for the expectation value
\begin{equation}
\langle \hat \lambda_i \rangle = \bar \lambda_i~,
\end{equation}
and for the variance
\begin{equation}
\langle (\hat \lambda_i - \bar \lambda_i  ) (\hat \lambda_j - \bar \lambda_j  ) \rangle = F^{-1}_{ij}~.  
\end{equation}
The expected errors are thus simply the diagonal elements of the inverse Fisher matrix.  Actually, there exists a theorem\footnote{The \textit{Cramer-Rao inequality theorem} } proving that no method can measure the parameter best-fits with lower errors.  

\section{Forecasting}

An interesting property of the Fisher matrix formalism is the possibility to use it in absence of data, in order to forecast the uncertainties on the (cosmological) parameters, given the specifications of the projected experiment. 

Let us consider a future experiment that will map the sky and thus will measure a set of $C_l^{\rr{obs}}$.   In absence of data, let us consider that the true Universe leads to a set of mock $\hat C_l$'s, and that the given experiment will measure them with uncertainties $\delta C_l$.   Let us consider the function
\begin{equation} 
\chi^2 (\{ \lambda_i \}) \equiv \sum_l \frac{\left[ C_l ( \{ \lambda_i \} )-  \hat C_l   \right]^2}{\delta C_l^2}~,
\end{equation}
expected to reach a minimum at the point in the parameter space corresponding to the "true" cosmological parameters $\{ \bar \lambda_i\}$.   It is important to remind that these parameter values are \textit{fiducial}, i.e. we \textit{assume} that these describe the real Universe but of course we are not sure about that.  If the errors on the $C_l$'s are Gaussian, then the likelihood function is given by $\exp (- \chi^2 / 2) $.  

For simplicity we can first consider only one parameter $\lambda$ and then generalize to the multidimensional case.   The function $\chi^2$ can be expanded about its minimum
\begin{equation}
  \chi^2 (\lambda ) = \chi^2 (\bar \lambda) + \mathcal F (\lambda - \bar \lambda)^2~,
  \end{equation}
  where 
  \begin{equation}
  \mathcal F \equiv \frac 1 2 \left. \frac{\partial^2 \chi^2}{\partial \lambda^2}  \right|_{\bar \lambda} = 
  \sum_ l \frac{1}{(\delta C_l)^2} \left[ \left( \frac{\partial C_l}{\partial \lambda} \right)^2  + \left( C_l - \hat C_l \right)  \frac {\partial ^2 C_l}{\partial \lambda^2}\right]~.
  \end{equation}
Over many realizations, the second term can be neglected, because on average the difference between the $C_l$'s and the $\hat C_l$ will cancel.  That corresponds to replace the curvature by the Fisher matrix.   One therefore can read
\begin{equation}
F = \sum_l \frac{1}{(\delta C_l)^2} \frac{\partial C_l}{\partial \lambda} \frac{\partial C_l}{\partial \lambda}~.
\end{equation}
This result can be generalized easily to the case of a multi-dimensional parameter space,
\begin{equation}
F_{ij} = \sum _l \frac{1}{(\delta C_l)^2} \frac{\partial C_l}{\partial \lambda_i} \frac{\partial C_l}{\partial \lambda_j}~.
\end{equation}
In a realistic case, the errors are not exactly gaussian so that $F_{ij} $ is not exactly the true Fisher matrix.  But the method remains very efficient if the error distributions do not depart to much from the gaussianity.   Once given the experiment specifications (necessary to determine $\delta C_l$)  and the derivatives of the $C_l$ with respect to the model parameters, evaluated at the fiducial values, it is straightforward to determine the forecasts for the 1-$\sigma$ errors on these parameters, given by $\sqrt{F^{-1}_{ii}} $.  
The uncertainties $\delta C_l$ are both due to the specifications of the experiments (beam function, observation time,...) and the cosmic variance.  One usually rewrites 
\begin{equation}
\delta C_l = \sqrt{\frac{2}{(2l + 1) f_{\rr{sky}}   }} \left(C_l + C_l ^{\rr n} \right),
 \end{equation}
where $C_l ^{\rr n}$ is the noise of the experiment and where $f_{\rr{sky}}$ is the covered fraction of the sky.  

 In Chapter 9, the Fisher matrix formalism is used for the 3D power spectrum of the 21cm brightness temperature $P_{\Delta T_{\rr B}  }$, in the $u$ space.    The previous calculation can be directly generalized to this case.  The Fisher matrix reads
 \begin{equation}
F_{ij} = \sum \frac{1}{\left[  \delta P_{\Delta T_{\rr B}  }(u) \right]^2} \frac{\partial P_{\Delta T_{\rr B}  }(\mathbf u)}{\partial \lambda_i} \frac{\partial P_{\Delta T_{\rr B}  }(\mathbf u) }{\partial \lambda_j}~,
\end{equation}
where sum goes on all the cells in the $u$ space, and where
\begin{equation} 
\delta P_{\Delta T_{\rr B}  } (\mathbf u) = P_{\Delta T_{\rr B}  } (\mathbf u) + P^{\rr n} (\mathbf u) ~,   
 \end{equation}
 with a noise spectrum  $ P^{\rr n} (\mathbf u)$ given by the specifications of the experiment.

%% file: annexes/mcmc.tex
\chapter{MCMC Bayesian methods}

\label{A:MCMC}

\section{Bayes' Theorem}

Bayesian methods are developed in the presence of observations, denoted $d$.  Initially, these observations are uncertain and only described with a probability density function $f(d|\theta)$, where $\theta$ is an index of the possible distribution for the observations.  As an exemple, if one wants to measure a physical quantity $\mu $, and if the measurements are controlled by an error characterized by a normal distribution of variance $\sigma^2$, the probability of measuring $d$ is 
\begin{equation}
f(d|\mu,\sigma^2 ) = \frac{1}{\sqrt{2 \pi \sigma^2}} \exp\left[  - \frac {(d- \mu)^2}{\sigma^2} \right]~.
\end{equation}
But in general, to obtain a complete description of the process, the quantity of interest is $\theta$ (in the previous exemple, one would need to estimate the physical quantity $\mu$ from a series of observations $d_i$).  It is likely that the researcher has some knowledge about its value.  Let us denote it $p(\theta)$, the so-called \textit{prior distribution}.  This prior knowledge should be incorporated in the analysis treatment.   This is in contrast with statistical frequencist methods in which the prior information is not included.    

We thus need to asses the probability density distribution of $\theta$ after observing $d$.  This so-called \textit{posterior distribution} is denoted $p(\theta  | d) $.   This is obtained with the \textit{Bayes' theorem},
\begin{equation} \label{eq:bayes}
p(\theta  | d) = \frac{f(d| \theta) p(\theta) }{\int f(d| \theta') p(\theta') \dd \theta'}~.
\end{equation}
The Bayes' theorem is derived directly from conditional probabilities.  Let us consider the probability of an event $A$ given the event $B$,
\begin{equation} 
P( A | B) = \frac{P(A \cap B)}{P(B)}~,
\end{equation}
and equivalently the probability of the event B given the event A,
\begin{equation} 
P( B | A) = \frac{P(A \cap B)}{P(A)}~.
\end{equation}
Combining these two equations, one has $P(A|B) P(B) = P(B|A) P(A)$, and therefore
\begin{equation}
P( A | B)  = \frac{P(B|A) P(A) }{P(B)}~,
\end{equation}
that has the form of Eq.~(\ref{eq:bayes}).

In the next section, we describe the Metropolis-Hastings algorithm that can be used to estimate the probability density function $p(\theta  | d) $.  This algorithm is used in Chapters 5, 6, 7 and 9 of the thesis.  

\section{Metropolis-Hastings algorithm}

\section{Overview}

The Monte--Carlo--Markov--Chains (MCMC) method is a widespread
technique in Bayesian analysis.  It is used for estimating the posterior probability density distribution $p(\theta  | d)$.   Its main power is that
it numerically scales linearly with the number of dimensions of the space to probe, instead 
of exponentially for standard Monte-Carlo techniques.  The principle is to construct Markov chains, that are chains of points whose the $n$-th point only depends on the $(n-1)$-th point.   After a relaxation period, the density of chain
elements in the probed space directly samples the posterior probability distribution
$p(\theta|d)$ of the model, given the data.   

The Metropolis-Hastings algorithm is probably the
simplest~\cite{Metropolis53,Hastings70} to implement in the context of MCMC bayesian analysis.
Its characteristic is that each point $x_{i+1}$ is
obtained from a (usually Gaussian) random distribution $q(x)$, the so-called \textit{proposal
density function}, around the previous point $x_i$ of the chain.  This point is accepted to be the next
element of the Markov chain with the probability
\begin{equation}
\label{eq:mcmcaccept}
P(x_{i+1}) = \min \left[1,\frac{\pi(x_{i+1})}{\pi(x_i)}  \right],
\end{equation}
where $\pi(x)$ is the function that has to be sampled via the Markov
chain [e.g. $p(\theta|d$].  In this way, the Markov chain will move more probably to a region where the function $\pi(x)$ is higher, together with maintaining a low probability to probe regions in the space where it is lower.  
After a relaxation period, one can show that
Eq.~(\ref{eq:mcmcaccept}) ensures that $\pi$ is the asymptotic
stationary distribution of the chain~\cite{MacKayBook}. 

\section{Step by step implementation}

In practical terms, MCMC simulations using the Metropolis-Hastings algorithm for probing a probability density distribution $\pi(x)$ can be set up as follows:
\begin{enumerate}
\item Initialization of the iteration counter to i=1.  Set of an arbitrary initial value $x_1$ and calculation of $\pi (x_1)$.
\item Move to a new value $x_{i+1}$ generated from the proposal density $q(x_i)$. \\
\item Evaluation of the acceptance probability with Eq.~(\ref{eq:mcmcaccept}), and generation of a random number between $0$ and $1$ to determine if the point $x_{i+1}$ is accepted of not.  If it is rejected the process is reiterated for a new value of  $x_{i+1}$.
\item Change the counter from $i$ to $i+1$ and return to step 2 until convergence is reached.
\end{enumerate}
In order to determine if the Markov chain has converged after a large number of points, one can implement several chains and check if the errors on their variances is lower than the desired precision. 

\section{Applications}

\subsection{CMB data analysis}

MCMC methods have been intensively used in the context of CMB
data analysis~\cite{Christensen:2001gj, Lewis:2002ah, Martin:2006rs,
  Lorenz:2007ze, Dunkley:2008ie}.  In this case, the function to probe is 
 \begin{equation}
 \pi(\theta|d) \propto \calL(d|\theta)P(\theta)~,
 \end{equation} 
 the posterior probability density
distribution in the $n$-th dimensional space of the model parameters $\theta=\{\theta_{i=1 \cdots n}\}$ (e.g. cosmological parameters), given the data (CMB maps).  $\calL(d|\theta)$ is the likelihood of the experiment.   The probability density distribution for a specific parameter $\theta_i $ is obtained by marginalizing the function $\pi(\theta|d) $ over the parameter space.  
.
\subsection{Forecasts}

In chapter 9, we are interested in forecasting the posterior likelihood of the cosmological parameters, for hypothetic 21cm experiments, assuming that the real universe is described by fiducial values of these parameters.  These lead to a theoretical set of mock $\hat C_l$.   Assuming gaussian brightness temperature fluctuations, as well as an experimental noise $C_l^{n}$, the likelihood function for measuring $C_{l} $ is given by~\cite{Percival:2006ss},
\begin{equation}
\mathcal L (C_l | \hat C_l ) = \left( \frac{C_l^{\rr {tot} }  }{\hat C_l ^{\rr {tot} } }\right)^{\frac{2l+1}{2}} \exp \left( -\frac 1 2 (2l +1) \frac{C_l^{\rr {tot} }  - \hat C_l^{\rr {tot} }  }{C_l^{\rr {tot} } }  \right)~,
\end{equation}
where 
\begin{equation}
C_l^{\rr {tot} } = C_l + C_l^{\rr n}  = C_l + C_0 B_l^{-2}~,
\end{equation}
is the noise and $B_l$ is the characteristic beam function of the experiment.  The likelihood for measuring the parameters $\theta$ is then obtained by multiplying the likelihood function at each $l$.  One obtains
\begin{equation} \label{annex:mcmclik}
- 2 \ln \mathcal L(\theta | \hat C_l)  = \sum _l (2 l +1 ) \times \left( \frac{\hat C_l^{\rr{tot}}}{C_l^{\rr{tot}} } + \ln \frac{C_l^{\rr{tot}} }{\hat C_l^{\rr{tot}} }  - 1\right)~,
\end{equation}
The function probed with the MCMC method is this likelihood function factorized by the prior on the model parameters.  

\subsection{Realization of more than 60 e-folds of inflation}

In Chapters 5, 6 and 7, MCMC methods are used to assess the posterior probability density distributions 
of the model parameters, that are initial field values, velocities and potential parameters of the inflation model, for realizing more than 60 e-folds of inflation. 

In this case, the Metropolis-Hastings algorithm is simplified.  The likelihood $\calL$ is simply a
binary function:   either the field trajectory ends up on the slow-roll attractor and produces more 
than $60$ e-folds of inflation ($\calL=1$), either it does not ($\calL=0$).

%% file: These.bbl
\begin{thebibliography}{100}

\bibitem{Halyo:1996pp}
Edi Halyo.
\newblock {Hybrid inflation from supergravity D-terms}.
\newblock {\em Phys. Lett.}, B387:43--47, 1996, hep-ph/9606423.

\bibitem{Binetruy:1996xj}
P.~Bin\'etruy and G.~R. Dvali.
\newblock {D-term inflation}.
\newblock {\em Phys. Lett.}, B388:241--246, 1996, hep-ph/9606342.

\bibitem{Dvali:1994ms}
G.~R. Dvali, Q.~Shafi, and Robert~K. Schaefer.
\newblock {Large scale structure and supersymmetric inflation without fine
  tuning}.
\newblock {\em Phys. Rev. Lett.}, 73:1886--1889, 1994, hep-ph/9406319.

\bibitem{Kallosh:2003ux}
Renata Kallosh and Andrei Linde.
\newblock {P-term, D-term and F-term inflation}.
\newblock {\em JCAP}, 0310:008, 2003, hep-th/0306058.

\bibitem{Jeannerot:1997is}
R.~Jeannerot.
\newblock {Inflation in supersymmetric unified theories}.
\newblock {\em Phys. Rev.}, D56:6205--6216, 1997, hep-ph/9706391.

\bibitem{Jeannerot:2003qv}
Rachel Jeannerot, Jonathan Rocher, and Mairi Sakellariadou.
\newblock {How generic is cosmic string formation in SUSY GUTs}.
\newblock {\em Phys. Rev.}, D68:103514, 2003, hep-ph/0308134.

\bibitem{Dvali:1998pa}
G.~R. Dvali and S.~H.~Henry Tye.
\newblock {Brane inflation}.
\newblock {\em Phys. Lett.}, B450:72--82, 1999, hep-ph/9812483.

\bibitem{Brax:2007xq}
Philippe Brax, Anne-Christine Davis, Stephen~C. Davis, Rachel Jeannerot, and
  Marieke Postma.
\newblock {Warping and F-term uplifting}.
\newblock {\em JHEP}, 09:125, 2007, 0707.4583.

\bibitem{Koyama:2003yc}
Fumikazu Koyama, Yuji Tachikawa, and Taizan Watari.
\newblock {Supergravity analysis of hybrid inflation model from D3-D7 system}.
\newblock {\em Phys. Rev.}, D69:106001, 2004, hep-th/0311191.

\bibitem{Fukuyama:2008dv}
Takeshi Fukuyama, Nobuchika Okada, and Toshiyuki Osaka.
\newblock {Realistic Hybrid Inflation in 5D Orbifold SO(10) GUT}.
\newblock {\em JCAP}, 0809:024, 2008, 0806.4626.

\bibitem{Linde:2005dd}
Andrei Linde.
\newblock {Inflation and string cosmology}.
\newblock {\em ECONF}, C040802:L024, 2004, hep-th/0503195.

\bibitem{ArkaniHamed:2003wu}
Nima Arkani-Hamed, Hsin-Chia Cheng, Paolo Creminelli, and Lisa Randall.
\newblock {Extranatural inflation}.
\newblock {\em Phys. Rev. Lett.}, 90:221302, 2003, hep-th/0301218.

\bibitem{Felder:2000hj}
Gary~N. Felder et~al.
\newblock Dynamics of symmetry breaking and tachyonic preheating.
\newblock {\em Phys. Rev. Lett.}, 87:011601, 2001, hep-ph/0012142.

\bibitem{Felder:2001kt}
Gary~N. Felder, Lev Kofman, and Andrei~D. Linde.
\newblock {Tachyonic instability and dynamics of spontaneous symmetry
  breaking}.
\newblock {\em Phys. Rev.}, D64:123517, 2001, hep-th/0106179.

\bibitem{Linde:1993cn}
Andrei~D. Linde.
\newblock {Hybrid inflation}.
\newblock {\em Phys. Rev.}, D49:748--754, 1994, astro-ph/9307002.

\bibitem{Copeland:1994vg}
Edmund~J. Copeland, Andrew~R. Liddle, David~H. Lyth, Ewan~D. Stewart, and David
  Wands.
\newblock {False vacuum inflation with Einstein gravity}.
\newblock {\em Phys. Rev.}, D49:6410--6433, 1994, astro-ph/9401011.

\bibitem{Coleman:1973jx}
Sidney~R. Coleman and Erick~J. Weinberg.
\newblock {Radiative Corrections as the Origin of Spontaneous Symmetry
  Breaking}.
\newblock {\em Phys. Rev.}, D7:1888--1910, 1973.

\bibitem{Senoguz:2003zw}
Vedat~Nefer Senoguz and Q.~Shafi.
\newblock {Testing Supersymmetric Grand Unified Models of Inflation}.
\newblock {\em Phys. Lett.}, B567:79, 2003, hep-ph/0305089.

\bibitem{Jeannerot:2005mc}
Rachel Jeannerot and Marieke Postma.
\newblock {Confronting hybrid inflation in supergravity with CMB data}.
\newblock {\em JHEP}, 05:071, 2005, hep-ph/0503146.

\bibitem{Jeannerot:2006jj}
Rachel Jeannerot and Marieke Postma.
\newblock {Enlarging the parameter space of standard hybrid inflation}.
\newblock {\em JCAP}, 0607:012, 2006, hep-th/0604216.

\bibitem{Battye:2010hg}
Richard Battye, Bjorn Garbrecht, and Adam Moss.
\newblock {Tight constraints on F- and D-term hybrid inflation scenarios}.
\newblock {\em Phys. Rev.}, D81:123512, 2010, 1001.0769.

\bibitem{Clesse:2008pf}
Sebastien Clesse and Jonathan Rocher.
\newblock {Avoiding the blue spectrum and the fine-tuning of initial conditions
  in hybrid inflation}.
\newblock {\em Phys. Rev.}, D79:103507, 2009, 0809.4355.

\bibitem{Clesse:2009ur}
Sebastien Clesse, Christophe Ringeval, and Jonathan Rocher.
\newblock {Fractal initial conditions and natural parameter values in hybrid
  inflation}.
\newblock 2009, 0909.0402.

\bibitem{Clesse:2009zd}
Sebastien Clesse.
\newblock {Anamorphosis in hybrid inflation: How to avoid fine-tuning of
  initial conditions?}
\newblock {\em AIP Conf. Proc.}, 1241:543--550, 2010, 0910.3819.

\bibitem{Clesse:2010ht}
Sebastien Clesse.
\newblock {Initial conditions in hybrid inflation: exploration by MCMC
  technique}.
\newblock 2010, 1006.4435.

\bibitem{Clesse:2010iz}
Sebastien Clesse.
\newblock {Hybrid inflation along waterfall trajectories}.
\newblock {\em Phys. Rev.}, D83:063518, 2011, 1006.4522.

\bibitem{Mao:2008ug}
Yi~Mao, Max Tegmark, Matthew McQuinn, Matias Zaldarriaga, and Oliver Zahn.
\newblock {How accurately can 21 cm tomography constrain cosmology?}
\newblock {\em Phys. Rev.}, D78:023529, 2008, 0802.1710.

\bibitem{Friedmann1922}
Aleksandr Friedmann.
\newblock {\"Uber die Kr\"ummung des Raumes}.
\newblock {\em Zeitschrift f}.

\bibitem{Friedmann1924}
Aleksandr Friedmann.
\newblock {\"Uber die M\"oglichkeit einer Welt mit konstanter negativer
  Kr\"ummung des Ramues}.
\newblock {\em Zeitschrift f}.

\bibitem{Lemaitre1927}
Georges Lema\^itre.
\newblock {Un Univers homogene de masse constante et de rayon croissant rendant
  compte de la vitesse radiale des n\'ebuleuses galactiques}.
\newblock {\em Annales de la Soci\'et\'e Scientifique de Bruxelles},
  A47:49--59, 1927.

\bibitem{Hubble1929}
E.~Hubble.
\newblock {A relation between distance and radial velocity among extragalactic
  nebulae}.
\newblock {\em Proceedings of the National Academy of Science (U.S.A.)},
  15:168--173, 1929.

\bibitem{Slipher1915}
V.M. Slipher.
\newblock {Spectrographic Observations of Nebulae}.
\newblock {\em Popular Astronomy}, 23:21--24.

\bibitem{PeterUzan}
Patrick Peter and Jean-Philippe Uzan.
\newblock {\em {Primordial Cosmology}}.
\newblock {Oxford University Press}, Oxford, UK, 2005.

\bibitem{Starobinsky:1980}
A.~A. Starobinsky.
\newblock {A new type of isotropic cosmological models without singularity}.
\newblock {\em Physics Letters B}, 91:99--102, 1980.

\bibitem{Mukhanov:1990me}
Viatcheslav~F. Mukhanov, H.~A. Feldman, and Robert~H. Brandenberger.
\newblock {Theory of cosmological perturbations. Part 1. Classical
  perturbations. Part 2. Quantum theory of perturbations. Part 3. Extensions}.
\newblock {\em Phys. Rept.}, 215:203--333, 1992.

\bibitem{Alpher:1948}
R.A. Alpher, H.~Bethe, and G~Gamow.
\newblock {The origin of chemical elements}.
\newblock {\em Phys. Rev. Lett.}, 73:803, 1948.

\bibitem{Gamow:1948}
G.~Gamow.
\newblock {The evolution of the universe}.
\newblock {\em Nature}, 162:680, 1948.

\bibitem{Alpher:1948b}
R.A. Alpher and R.~Herman.
\newblock {Evolution of the Universe}.
\newblock {\em Nature}, 162:774, 1948.

\bibitem{Peebles:1966}
P.J.E. Peebles.
\newblock {Primordial Helium abundance and the primeval fireball}.
\newblock {\em Astrophys. J.}, 146:542, 1966.

\bibitem{Mukhanov:2003xs}
Viatcheslav~F. Mukhanov.
\newblock {Nucleosynthesis Without a Computer}.
\newblock {\em Int. J. Theor. Phys.}, 43:669--693, 2004, astro-ph/0303073.

\bibitem{Komatsu:2010fb}
E.~Komatsu et~al.
\newblock {Seven-Year Wilkinson Microwave Anisotropy Probe (WMAP) Observations:
  Cosmological Interpretation}.
\newblock {\em Astrophys. J. Suppl.}, 192:18, 2011, 1001.4538.

\bibitem{Penzias:1965}
A.~Penzias and R.~Wilson.
\newblock {A measurement of excess antenna temperature at 4080 Mc/s}.
\newblock {\em Astrophys. J.}, 142:L419, 1965.

\bibitem{Smoot:1998iq}
George~F. Smoot and D.~Scott.
\newblock {Cosmic Background Radiation: in Review of Particle Physics (RPP
  1998)}.
\newblock {\em Eur. Phys. J.}, C3:127--131, 1998.

\bibitem{website:recfast}
Recfast website.
\newblock \url{http://www.astro.ubc.ca/people/scott/recfast.html}.

\bibitem{Seager:1999bc}
Sara Seager, Dimitar~D. Sasselov, and Douglas Scott.
\newblock {A New Calculation of the Recombination Epoch}.
\newblock {\em Astrophys. J.}, 523:L1--L5, 1999, astro-ph/9909275.

\bibitem{Seager:1999km}
Sara Seager, Dimitar~D. Sasselov, and Douglas Scott.
\newblock {How exactly did the Universe become neutral?}
\newblock {\em Astrophys. J. Suppl.}, 128:407--430, 2000, astro-ph/9912182.

\bibitem{Dodelson}
Scott Dodelson.
\newblock {\em {Modern Cosmology}}.
\newblock {Academic Press}, Elsevier, 2003.

\bibitem{1965ApJ...142.1633G}
J.~E. {Gunn} and B.~A. {Peterson}.
\newblock {On the Density of Neutral Hydrogen in Intergalactic Space.}
\newblock {\em Astrophys. J.}, 142:1633--1641, November 1965.

\bibitem{Becker:2001ee}
Robert~H. Becker et~al.
\newblock {Evidence for Reionization at z ~ 6: Detection of a Gunn- Peterson
  Trough in a z=6.28 Quasar}.
\newblock {\em Astron. J.}, 122:2850, 2001, astro-ph/0108097.

\bibitem{Santos:2009zk}
M.~G. Santos, L.~Ferramacho, M.~B. Silva, A.~Amblard, and A.~Cooray.
\newblock {Fast Large Volume Simulations of the 21 cm Signal from the
  Reionization and pre-Reionization Epochs}.
\newblock {\em Mon. Not. Roy. Astron. Soc.}, 406:2421--2432, 2010, 0911.2219.

\bibitem{Thomas:2010mz}
Rajat~M. Thomas and Saleem Zaroubi.
\newblock {On the spin-temperature evolution during the epoch of reionization}.
\newblock 2010, 1009.5441.

\bibitem{Webb:1999}
S.K. Webb.
\newblock {\em {Measuring the Universe, the cosmological distance ladder}}.
\newblock Springer-Verlag, 1999.

\bibitem{Udalski:1999pc}
A.~Udalski et~al.
\newblock {The Optical Gravitational Lensing Experiment. Cepheids in the
  Magellanic Clouds. IV. Catalog of Cepheids from the Large Magellanic Cloud}.
\newblock {\em Acta Astron.}, 49:223--317, 1999, astro-ph/9908317.

\bibitem{Tully:1977fu}
R.~B. Tully and J.~R. Fisher.
\newblock {A New method of determining distances to galaxies}.
\newblock {\em Astron. Astrophys.}, 54:661--673, 1977.

\bibitem{1979ApJ...232..404C}
S.~A. {Colgate}.
\newblock {Supernovae as a standard candle for cosmology}.
\newblock {\em Astrophys. J.}, 232:404--408, September 1979.

\bibitem{Riess:1998cb}
Adam~G. Riess et~al.
\newblock {Observational Evidence from Supernovae for an Accelerating Universe
  and a Cosmological Constant}.
\newblock {\em Astron. J.}, 116:1009--1038, 1998, astro-ph/9805201.

\bibitem{Freedman:2000cf}
W.~L. Freedman et~al.
\newblock {Final Results from the Hubble Space Telescope Key Project to Measure
  the Hubble Constant}.
\newblock {\em Astrophys. J.}, 553:47--72, 2001, astro-ph/0012376.

\bibitem{Izotov:1999wa}
Y.~I. Izotov et~al.
\newblock {Helium abundance in the most metal-deficient blue compact galaxies:
  I Zw 18 and SBS 0335-052}.
\newblock {\em Astrophys. J.}, 527:757--777, 1999, astro-ph/9907228.

\bibitem{Luridiana:2003jy}
V.~Luridiana, A.~Peimbert, M.~Peimbert, and M.~Cervino.
\newblock {The effect of collisional enhancement of Balmer lines on the
  determination of the primordial helium abundance}.
\newblock {\em Astrophys. J.}, 592:846--865, 2003, astro-ph/0304152.

\bibitem{Izotov:2003xn}
Y.~I. Izotov and T.~X. Thuan.
\newblock {Systematic effects and a new determination of the primordial
  abundance of 4He and dY/dZ from observations of blue compact galaxies}.
\newblock {\em Astrophys. J.}, 602:200--230, 2004, astro-ph/0310421.

\bibitem{Kirkman:2003uv}
David Kirkman, David Tytler, Nao Suzuki, John~M. O'Meara, and Dan Lubin.
\newblock {The cosmological baryon density from the deuterium to hydrogen ratio
  towards QSO absorption systems: D/H towards Q1243+3047}.
\newblock {\em Astrophys. J. Suppl.}, 149:1, 2003, astro-ph/0302006.

\bibitem{Ryan:2000zz}
Sean~G. Ryan, Timothy~C. Beers, Keith~A. Olive, Brian~D. Fields, and John~E.
  Norris.
\newblock {Primordial Lithium and Big Bang Nucleosynthesis}.
\newblock {\em Astrophys. J.}, 530:L57--L60, 2000.

\bibitem{Shimon:2010ug}
M.~Shimon et~al.
\newblock {Using Big Bang Nucleosynthesis to Extend CMB Probes of Neutrino
  Physics}.
\newblock {\em JCAP}, 1005:037, 2010, 1001.5088.

\bibitem{Dent:2007zz}
Thomas Dent, Steffen Stern, and Christof Wetterich.
\newblock {Big Bang nucleosynthesis as a probe of varying fundamental
  'constants'}.
\newblock {\em AIP Conf. Proc.}, 957:383--386, 2007.

\bibitem{Peiris:2005dt}
H.~Peiris.
\newblock {First year Wilkinson Microwave Anisotropy Probe results:
  Implications for cosmology and inflation}.
\newblock In *Thompson, J.M.T. (ed.): Advances in astronomy* 99-122.

\bibitem{Larson:2010gs}
D.~Larson et~al.
\newblock {Seven-Year Wilkinson Microwave Anisotropy Probe (WMAP) Observations:
  Power Spectra and WMAP-Derived Parameters}.
\newblock {\em Astrophys. J. Suppl.}, 192:16, 2011, 1001.4635.

\bibitem{Coc:2003ce}
Alain Coc, Elisabeth Vangioni-Flam, Pierre Descouvemont, Abderrahim Adahchour,
  and Carmen Angulo.
\newblock {Updated Big Bang Nucleosynthesis confronted to WMAP observations and
  to the Abundance of Light Elements}.
\newblock {\em Astrophys. J.}, 600:544--552, 2004, astro-ph/0309480.

\bibitem{1992ApJ...396L...1S}
G.~F. {Smoot}, C.~L. {Bennett}, A.~{Kogut}, E.~L. {Wright}, J.~{Aymon}, N.~W.
  {Boggess}, E.~S. {Cheng}, G.~{de Amici}, S.~{Gulkis}, M.~G. {Hauser},
  G.~{Hinshaw}, P.~D. {Jackson}, M.~{Janssen}, E.~{Kaita}, T.~{Kelsall},
  P.~{Keegstra}, C.~{Lineweaver}, K.~{Loewenstein}, P.~{Lubin}, J.~{Mather},
  S.~S. {Meyer}, S.~H. {Moseley}, T.~{Murdock}, L.~{Rokke}, R.~F. {Silverberg},
  L.~{Tenorio}, R.~{Weiss}, and D.~T. {Wilkinson}.
\newblock {Structure in the COBE differential microwave radiometer first-year
  maps}.
\newblock {\em Astrophys. J. Letters}, 396:L1--L5, September 1992.

\bibitem{Grishchuk:1997pk}
L.~P. Grishchuk and Jerome Martin.
\newblock {Best Unbiased Estimates for the Microwave Background Anisotropies}.
\newblock {\em Phys. Rev.}, D56:1924--1938, 1997, gr-qc/9702018.

\bibitem{1968ApJ...151..459S}
J.~{Silk}.
\newblock {Cosmic Black-Body Radiation and Galaxy Formation}.
\newblock {\em Astrophys. J.}, 151:459--+, February 1968.

\bibitem{Reichardt:2008ay}
C.~L. Reichardt et~al.
\newblock {High resolution CMB power spectrum from the complete ACBAR data
  set}.
\newblock {\em Astrophys. J.}, 694:1200--1219, 2009, 0801.1491.

\bibitem{Pryke:2008xp}
:~C. Pryke et~al.
\newblock {Second and third season QUaD CMB temperature and polarization power
  spectra}.
\newblock {\em Astrophys. J.}, 692:1247--1270, 2009, 0805.1944.

\bibitem{riazuello}
A.~Riazuello.
\newblock {\em {Signature de divers mod\`eles d'Univers primordial dans les
  anisotropies du rayonnement fossile (PhD thesis)}}.

\bibitem{Percival:2001hw}
Will~J. Percival et~al.
\newblock {The 2dF Galaxy Redshift Survey: The power spectrum and the matter
  content of the universe}.
\newblock {\em Mon. Not. Roy. Astron. Soc.}, 327:1297, 2001, astro-ph/0105252.

\bibitem{website:2dF}
2df website.
\newblock \url{http://www.mso.anu.edu.au/2dFGRS/}.

\bibitem{website:SDSS}
Sloan digital survey (sdss) website.
\newblock \url{http://www.sdss.org/}.

\bibitem{Eisenstein:2005su}
Daniel~J. Eisenstein et~al.
\newblock {Detection of the Baryon Acoustic Peak in the Large-Scale Correlation
  Function of SDSS Luminous Red Galaxies}.
\newblock {\em Astrophys. J.}, 633:560--574, 2005, astro-ph/0501171.

\bibitem{Colless:1998yu}
Matthew Colless.
\newblock {First results from the 2dF galaxy redshift survey}.
\newblock 1998, astro-ph/9804079.

\bibitem{Tegmark:2001jh}
Max Tegmark, Andrew J.~S. Hamilton, and Yong-Zhong Xu.
\newblock {The power spectrum of galaxies in the 2dF 100k redshift survey}.
\newblock {\em Mon. Not. Roy. Astron. Soc.}, 335:887--908, 2002,
  astro-ph/0111575.

\bibitem{Bartelmann:1999yn}
Matthias Bartelmann and Peter Schneider.
\newblock {Weak Gravitational Lensing}.
\newblock {\em Phys. Rept.}, 340:291--472, 2001, astro-ph/9912508.

\bibitem{Voit:2004ah}
G.~Mark Voit.
\newblock {Tracing cosmic evolution with clusters of galaxies}.
\newblock {\em Rev. Mod. Phys.}, 77:207--258, 2005, astro-ph/0410173.

\bibitem{Viel:2006yh}
Matteo Viel, Martin~G. Haehnelt, and Antony Lewis.
\newblock {The Lyman-alpha forest and WMAP year three}.
\newblock {\em Mon. Not. Roy. Astron. Soc.}, 370:L51--L55, 2006,
  astro-ph/0604310.

\bibitem{Battaner:2000ef}
Eduardo Battaner and Estrella Florido.
\newblock {The rotation curve of spiral galaxies and its cosmological
  implications}.
\newblock {\em Fund. Cosmic Phys.}, 21:1--154, 2000, astro-ph/0010475.

\bibitem{Bertone:2004pz}
Gianfranco Bertone, Dan Hooper, and Joseph Silk.
\newblock {Particle dark matter: Evidence, candidates and constraints}.
\newblock {\em Phys. Rept.}, 405:279--390, 2005, hep-ph/0404175.

\bibitem{DiCiaccio:2011zz}
Anna Di~Ciaccio.
\newblock {LHC and the dark matter search: A review}.
\newblock {\em Nucl. Instrum. Meth.}, A630:273--278, 2011.

\bibitem{website:CDMS}
Cryogenic dark matter search (cdms) website.
\newblock \url{http://cdms.berkeley.edu/}.

\bibitem{Angloher:2008jj}
G.~Angloher et~al.
\newblock {Commissioning Run of the CRESST-II Dark Matter Search}.
\newblock 2008, 0809.1829.

\bibitem{website:Edelweiss}
Edelweiss experiment website.
\newblock \url{http://edelweiss.in2p3.fr/}.

\bibitem{Adriani:2008zr}
Oscar Adriani et~al.
\newblock {An anomalous positron abundance in cosmic rays with energies 1.5-100
  GeV}.
\newblock {\em Nature}, 458:607--609, 2009, 0810.4995.

\bibitem{Delahaye:2008ua}
T.~Delahaye et~al.
\newblock {Galactic secondary positron flux at the Earth}.
\newblock {\em Astron. Astrophys.}, 501:821--833, 2009, 0809.5268.

\bibitem{Bernabei:2000qi}
R.~Bernabei et~al.
\newblock {Search for WIMP annual modulation signature: Results from DAMA /
  NaI-3 and DAMA / NaI-4 and the global combined analysis}.
\newblock {\em Phys. Lett.}, B480:23--31, 2000.

\bibitem{Angle:2007uj}
J.~Angle et~al.
\newblock {First Results from the XENON10 Dark Matter Experiment at the Gran
  Sasso National Laboratory}.
\newblock {\em Phys. Rev. Lett.}, 100:021303, 2008, 0706.0039.

\bibitem{Sapone:2010iz}
Domenico Sapone.
\newblock {Dark Energy in Practice}.
\newblock {\em Int. J. Mod. Phys.}, A25:5253--5331, 2010, 1006.5694.

\bibitem{Ringeval:2010hf}
Christophe Ringeval, Teruaki Suyama, Tomo Takahashi, Masahide Yamaguchi, and
  Shuichiro Yokoyama.
\newblock {Dark energy from primordial inflationary quantum fluctuations}.
\newblock {\em Phys. Rev. Lett.}, 105:121301, 2010, 1006.0368.

\bibitem{Weinberg:2000yb}
Steven Weinberg.
\newblock {The cosmological constant problems}.
\newblock 2000, astro-ph/0005265.

\bibitem{Nakahara:1990th}
M.~Nakahara.
\newblock {Geometry, topology and physics}.
\newblock Bristol, UK: Hilger (1990) 505 p. (Graduate student series in
  physics).

\bibitem{Langacker:1980kd}
Paul Langacker and So-Young Pi.
\newblock {Magnetic Monopoles in Grand Unified Theories}.
\newblock {\em Phys. Rev. Lett.}, 45:1, 1980.

\bibitem{Zeldovich:1978wj}
Ya.~B. Zeldovich and M.~Yu. Khlopov.
\newblock {On the Concentration of Relic Magnetic Monopoles in the Universe}.
\newblock {\em Phys. Lett.}, B79:239--241, 1978.

\bibitem{PhysRevLett.43.1365}
John~P. Preskill.
\newblock Cosmological production of superheavy magnetic monopoles.
\newblock {\em Phys. Rev. Lett.}, 43(19):1365--1368, Nov 1979.

\bibitem{Kaiser:1984iv}
Nick Kaiser and A.~Stebbins.
\newblock {Microwave Anisotropy Due to Cosmic Strings}.
\newblock {\em Nature}, 310:391--393, 1984.

\bibitem{Battye:2010xz}
Richard Battye and Adam Moss.
\newblock {Updated constraints on the cosmic string tension}.
\newblock {\em Phys. Rev.}, D82:023521, 2010, 1005.0479.

\bibitem{Hindmarsh:1994re}
M.~B. Hindmarsh and T.~W.~B. Kibble.
\newblock {Cosmic strings}.
\newblock {\em Rept. Prog. Phys.}, 58:477--562, 1995, hep-ph/9411342.

\bibitem{Rovelli:1995ac}
Carlo Rovelli and Lee Smolin.
\newblock {Spin networks and quantum gravity}.
\newblock {\em Phys. Rev.}, D52:5743--5759, 1995, gr-qc/9505006.

\bibitem{PhysRevD.79.084008}
Petr Ho\ifmmode~\check{r}\else \v{r}\fi{}ava.
\newblock Quantum gravity at a lifshitz point.
\newblock {\em Phys. Rev. D}, 79(8):084008, Apr 2009.

\bibitem{Langlois:1999dw}
David Langlois.
\newblock {Correlated adiabatic and isocurvature perturbations from double
  inflation}.
\newblock {\em Phys. Rev.}, D59:123512, 1999, astro-ph/9906080.

\bibitem{Martin:2010kz}
Jerome Martin and Christophe Ringeval.
\newblock {First CMB Constraints on the Inflationary Reheating Temperature}.
\newblock 2010, 1004.5525.

\bibitem{Gangui:2002qc}
Alejandro Gangui, Jerome Martin, and Mairi Sakellariadou.
\newblock {Single field inflation and non-Gaussianity}.
\newblock {\em Phys. Rev.}, D66:083502, 2002, astro-ph/0205202.

\bibitem{Creminelli:2004yq}
Paolo Creminelli and Matias Zaldarriaga.
\newblock {Single field consistency relation for the 3-point function}.
\newblock {\em JCAP}, 0410:006, 2004, astro-ph/0407059.

\bibitem{Komatsu:2010hc}
Eiichiro Komatsu.
\newblock {Hunting for Primordial Non-Gaussianity in the Cosmic Microwave
  Background}.
\newblock {\em Class. Quant. Grav.}, 27:124010, 2010, 1003.6097.

\bibitem{Collins:2009pf}
Hael Collins and R.~Holman.
\newblock {Trans-Planckian enhancements of the primordial non- Gaussianities}.
\newblock {\em Phys. Rev.}, D80:043524, 2009, 0905.4925.

\bibitem{PhysRevD.83.103511}
Johannes Noller and Jo\~ao Magueijo.
\newblock Non-gaussianity in single field models without slow-roll conditions.
\newblock {\em Phys. Rev. D}, 83(10):103511, May 2011.

\bibitem{Sugiyama:2011jt}
Naonori~Suma Sugiyama, Eiichiro Komatsu, and Toshifumi Futamase.
\newblock {Non-Gaussianity Consistency Relation for Multi-field Inflation}.
\newblock 2011, 1101.3636.

\bibitem{Ringeval:2005yn}
Christophe Ringeval, Philippe Brax, Carsten van~de Bruck, and Anne-Christine
  Davis.
\newblock {Boundary inflation and the WMAP data}.
\newblock {\em Phys. Rev.}, D73:064035, 2006, astro-ph/0509727.

\bibitem{Leach:2002ar}
Samuel~M. Leach, Andrew~R. Liddle, Jerome Martin, and Dominik~J Schwarz.
\newblock {Cosmological parameter estimation and the inflationary cosmology}.
\newblock {\em Phys. Rev.}, D66:023515, 2002, astro-ph/0202094.

\bibitem{Liddle:1994dx}
Andrew~R. Liddle, Paul Parsons, and John~D. Barrow.
\newblock {Formalizing the slow roll approximation in inflation}.
\newblock {\em Phys. Rev.}, D50:7222--7232, 1994, astro-ph/9408015.

\bibitem{PhysRevD.22.1882}
James~M. Bardeen.
\newblock Gauge-invariant cosmological perturbations.
\newblock {\em Phys. Rev. D}, 22(8):1882--1905, Oct 1980.

\bibitem{Jedamzik:2010dq}
Karsten Jedamzik, Martin Lemoine, and Jerome Martin.
\newblock {Collapse of Small-Scale Density Perturbations during Preheating in
  Single Field Inflation}.
\newblock {\em JCAP}, 1009:034, 2010, 1002.3039.

\bibitem{Jedamzik:2010hq}
Karsten Jedamzik, Martin Lemoine, and Jerome Martin.
\newblock {Generation of gravitational waves during early structure formation
  between cosmic inflation and reheating}.
\newblock {\em JCAP}, 1004:021, 2010, 1002.3278.

\bibitem{Ringeval:2007am}
Christophe Ringeval.
\newblock {The exact numerical treatment of inflationary models}.
\newblock {\em Lect. Notes Phys.}, 738:243--273, 2008, astro-ph/0703486.

\bibitem{Liddle:2003as}
Andrew~R Liddle and Samuel~M Leach.
\newblock {How long before the end of inflation were observable perturbations
  produced?}
\newblock {\em Phys. Rev.}, D68:103503, 2003, astro-ph/0305263.

\bibitem{PhysRevD.35.419}
Joseph Silk and Michael~S. Turner.
\newblock Double inflation.
\newblock {\em Phys. Rev. D}, 35(2):419--428, Jan 1987.

\bibitem{Gordon:2000hv}
Christopher Gordon, David Wands, Bruce~A. Bassett, and Roy Maartens.
\newblock {Adiabatic and entropy perturbations from inflation}.
\newblock {\em Phys. Rev.}, D63:023506, 2001, astro-ph/0009131.

\bibitem{Martin:2004um}
Jerome Martin.
\newblock {Inflationary cosmological perturbations of quantum- mechanical
  origin}.
\newblock {\em Lect. Notes Phys.}, 669:199--244, 2005, hep-th/0406011.

\bibitem{Martin:2006rs}
Jerome Martin and Christophe Ringeval.
\newblock {Inflation after WMAP3: Confronting the slow-roll and exact power
  spectra to CMB data}.
\newblock {\em JCAP}, 0608:009, 2006, astro-ph/0605367.

\bibitem{Davis:2008dj}
Anne-Christine Davis, Philippe Brax, and Carsten van~de Bruck.
\newblock {Brane Inflation and Defect Formation}.
\newblock {\em Phil. Trans. Roy. Soc. Lond.}, A366:2833--2842, 2008, 0803.0424.

\bibitem{Brax:2007fe}
Philippe Brax, Anne-Christine Davis, Stephen~C. Davis, Rachel Jeannerot, and
  Marieke Postma.
\newblock {D-term Uplifted Racetrack Inflation}.
\newblock {\em JCAP}, 0801:008, 2008, 0710.4876.

\bibitem{Fairbairn:2003yx}
M.~Fairbairn, Laura Lopez~Honorez, and M.~H.~G. Tytgat.
\newblock {Radion assisted gauge inflation}.
\newblock {\em Phys. Rev.}, D67:101302, 2003, hep-ph/0302160.

\bibitem{Brax:2006ay}
Philippe Brax, Carsten van~de Bruck, Anne-Christine Davis, and Stephen~C.
  Davis.
\newblock {Coupling hybrid inflation to moduli}.
\newblock {\em JCAP}, 0609:012, 2006, hep-th/0606140.

\bibitem{Kofman:1997yn}
Lev Kofman, Andrei~D. Linde, and Alexei~A. Starobinsky.
\newblock Towards the theory of reheating after inflation.
\newblock {\em Phys. Rev.}, D56:3258--3295, 1997, hep-ph/9704452.

\bibitem{Garcia-Bellido:1997wm}
Juan Garcia-Bellido and Andrei~D. Linde.
\newblock Preheating in hybrid inflation.
\newblock {\em Phys. Rev.}, D57:6075--6088, 1998, hep-ph/9711360.

\bibitem{Copeland:2002ku}
Edmund~J. Copeland, S.~Pascoli, and A.~Rajantie.
\newblock {Dynamics of tachyonic preheating after hybrid inflation}.
\newblock {\em Phys. Rev.}, D65:103517, 2002, hep-ph/0202031.

\bibitem{Senoguz:2004vu}
V.~N. Senoguz and Q.~Shafi.
\newblock Reheat temperature in supersymmetric hybrid inflation models.
\newblock {\em Phys. Rev.}, D71:043514, 2005, hep-ph/0412102.

\bibitem{Micha:2004bv}
Raphael Micha and Igor~I. Tkachev.
\newblock Turbulent thermalization.
\newblock {\em Phys. Rev.}, D70:043538, 2004, hep-ph/0403101.

\bibitem{Allahverdi:2007zz}
Rouzbeh Allahverdi and Anupam Mazumdar.
\newblock {Reheating in supersymmetric high scale inflation}.
\newblock {\em Phys. Rev.}, D76:103526, 2007, hep-ph/0603244.

\bibitem{Rocher:2004et}
Jonathan Rocher and Mairi Sakellariadou.
\newblock {Supersymmetric grand unified theories and cosmology}.
\newblock {\em JCAP}, 0503:004, 2005, hep-ph/0406120.

\bibitem{Mazumdar:2010sa}
Anupam Mazumdar and Jonathan Rocher.
\newblock {Particle physics models of inflation and curvaton scenarios}.
\newblock 2010, 1001.0993.

\bibitem{Fraisse:2006xc}
Aurelien~A. Fraisse.
\newblock {Limits on Defects Formation and Hybrid Inflationary Models with
  Three-Year WMAP Observations}.
\newblock {\em JCAP}, 0703:008, 2007, astro-ph/0603589.

\bibitem{Bevis:2007gh}
Neil Bevis, Mark Hindmarsh, Martin Kunz, and Jon Urrestilla.
\newblock {Fitting CMB data with cosmic strings and inflation}.
\newblock {\em Phys. Rev. Lett.}, 100:021301, 2008, astro-ph/0702223.

\bibitem{Vilenkin:1994}
A.~Vilenkin and P.~Shellard.
\newblock {Cosmic strings and other topological defects}.
\newblock 1994.
\newblock Cambridge monographs on mathematical physics, Cambridge University
  Press, (1994).

\bibitem{Lazarides:1995vr}
George Lazarides and C.~Panagiotakopoulos.
\newblock {Smooth hybrid inflation}.
\newblock {\em Phys. Rev.}, D52:559--563, 1995, hep-ph/9506325.

\bibitem{Jeannerot:2000sv}
R.~Jeannerot, S.~Khalil, George Lazarides, and Q.~Shafi.
\newblock {Inflation and monopoles in supersymmetric SU(4)c x SU(2)L x SU(2)R}.
\newblock {\em JHEP}, 10:012, 2000, hep-ph/0002151.

\bibitem{Yamaguchi:2004tn}
Masahide Yamaguchi and Jun'ichi Yokoyama.
\newblock {Smooth hybrid inflation in supergravity with a running spectral
  index and early star formation}.
\newblock {\em Phys. Rev.}, D70:023513, 2004, hep-ph/0402282.

\bibitem{ArkaniHamed:2003mz}
Nima Arkani-Hamed, Hsin-Chia Cheng, Paolo Creminelli, and Lisa Randall.
\newblock {Pseudonatural inflation}.
\newblock {\em JCAP}, 0307:003, 2003, hep-th/0302034.

\bibitem{Kaplan:2003aj}
David~E. Kaplan and Neal~J. Weiner.
\newblock {Little inflatons and gauge inflation}.
\newblock {\em JCAP}, 0402:005, 2004, hep-ph/0302014.

\bibitem{Freese:1990rb}
Katherine Freese, Joshua~A. Frieman, and Angela~V. Olinto.
\newblock {Natural inflation with pseudo - Nambu-Goldstone bosons}.
\newblock {\em Phys. Rev. Lett.}, 65:3233--3236, 1990.

\bibitem{Kallosh:1995hi}
Renata Kallosh, Andrei~D. Linde, Dmitri~A. Linde, and Leonard Susskind.
\newblock {Gravity and global symmetries}.
\newblock {\em Phys. Rev.}, D52:912--935, 1995, hep-th/9502069.

\bibitem{Goldwirth:1991rj}
Dalia~S. Goldwirth and Tsvi Piran.
\newblock {Initial conditions for inflation}.
\newblock {\em Phys. Rept.}, 214:223--291, 1992.

\bibitem{Vachaspati:1998dy}
Tanmay Vachaspati and Mark Trodden.
\newblock {Causality and cosmic inflation}.
\newblock {\em Phys. Rev.}, D61:023502, 2000, gr-qc/9811037.

\bibitem{Kaloper:2002cs}
Nemanja Kaloper, Matthew Kleban, Albion Lawrence, Stephen Shenker, and Leonard
  Susskind.
\newblock {Initial conditions for inflation}.
\newblock {\em JHEP}, 11:037, 2002, hep-th/0209231.

\bibitem{Linde:1986fd}
Andrei~D. Linde.
\newblock {Eternally Existing Selfreproducing Chaotic Inflationary Universe}.
\newblock {\em Phys. Lett.}, B175:395--400, 1986.

\bibitem{Tetradis:1997kp}
N.~Tetradis.
\newblock {Fine tuning of the initial conditions for hybrid inflation}.
\newblock {\em Phys. Rev.}, D57:5997--6002, 1998, astro-ph/9707214.

\bibitem{Lazarides:1997vv}
George Lazarides and N.~D. Vlachos.
\newblock {Initial conditions for supersymmetric inflation}.
\newblock {\em Phys. Rev.}, D56:4562--4567, 1997, hep-ph/9707296.

\bibitem{Mendes:2000sq}
Luis~E. Mendes and Andrew~R. Liddle.
\newblock {Initial conditions for hybrid inflation}.
\newblock {\em Phys. Rev.}, D62:103511, 2000, astro-ph/0006020.

\bibitem{Lazarides:1996rk}
George Lazarides, C.~Panagiotakopoulos, and N.~D. Vlachos.
\newblock {Initial conditions for smooth hybrid inflation}.
\newblock {\em Phys. Rev.}, D54:1369--1373, 1996, hep-ph/9606297.

\bibitem{Calzetta:1992bp}
Esteban Calzetta and Maria Sakellariadou.
\newblock {Semiclassical effects and the onset of inflation}.
\newblock {\em Phys. Rev.}, D47:3184--3193, 1993, gr-qc/9209007.

\bibitem{Dimopoulos:2005ac}
S.~Dimopoulos, S.~Kachru, J.~McGreevy, and Jay~G. Wacker.
\newblock {N-flation}.
\newblock 2005, hep-th/0507205.

\bibitem{Easther:2005zr}
Richard Easther and Liam McAllister.
\newblock {Random matrices and the spectrum of N-flation}.
\newblock {\em JCAP}, 0605:018, 2006, hep-th/0512102.

\bibitem{Underwood:2008dh}
Bret Underwood.
\newblock {Brane Inflation is Attractive}.
\newblock {\em Phys. Rev.}, D78:023509, 2008, 0802.2117.

\bibitem{Ramos:2001zw}
Rudnei~O. Ramos.
\newblock {Fine-tuning solution for hybrid inflation in dissipative chaotic
  dynamics}.
\newblock {\em Phys. Rev.}, D64:123510, 2001, astro-ph/0104379.

\bibitem{PhysRevLett.75.3218}
Arjun Berera.
\newblock Warm inflation.
\newblock {\em Phys. Rev. Lett.}, 75(18):3218--3221, Oct 1995.

\bibitem{Panagiotakopoulos:1997if}
C.~Panagiotakopoulos and N.~Tetradis.
\newblock {Two-stage inflation as a solution to the initial condition problem
  of hybrid inflation}.
\newblock {\em Phys. Rev.}, D59:083502, 1999, hep-ph/9710526.

\bibitem{Lazarides:2007fh}
George Lazarides and Achilleas Vamvasakis.
\newblock {New smooth hybrid inflation}.
\newblock {\em Phys. Rev.}, D76:083507, 2007, arXiv:0705.3786 [hep-ph].

\bibitem{Jeannerot:2002wt}
R.~Jeannerot, S.~Khalil, and George Lazarides.
\newblock {New shifted hybrid inflation}.
\newblock {\em JHEP}, 07:069, 2002, hep-ph/0207244.

\bibitem{Linde:1993xx}
Andrei~D. Linde, Dmitri~A. Linde, and Arthur Mezhlumian.
\newblock {From the Big Bang theory to the theory of a stationary universe}.
\newblock {\em Phys. Rev.}, D49:1783--1826, 1994, gr-qc/9306035.

\bibitem{Martin:2005ir}
Jerome Martin and Marcello Musso.
\newblock {Solving stochastic inflation for arbitrary potentials}.
\newblock {\em Phys. Rev.}, D73:043516, 2006, hep-th/0511214.

\bibitem{Schwarz:2001vv}
Dominik~J. Schwarz, Cesar~A. Terrero-Escalante, and Alberto~A. Garcia.
\newblock {Higher order corrections to primordial spectra from cosmological
  inflation}.
\newblock {\em Phys. Lett.}, B517:243--249, 2001, astro-ph/0106020.

\bibitem{Ott:fractals}
{{Ott}, Edward}.
\newblock {\em {Chaos in Dynamical Systems}}.
\newblock {Cambridge University Press}, Cambridge, UK, 2002.

\bibitem{Falconer:fracgeo}
Kenneth Falconer.
\newblock {\em {Fractal Geometry}}.
\newblock {Wiley}, Chichester, UK, 2006.

\bibitem{1997:Dieci}
L.~Dieci, R.~D. Russel, and E.~S.~V. Vleck.
\newblock {\em SIAM Journal on Numerical Analysis}, 34:402, 1997.

\bibitem{2003:dieci}
L.~Dieci and E.~S.~V. Vleck.
\newblock {\em SIAM Journal on Numerical Analysis}, 40:516, 2002.

\bibitem{Mandelbrot:1980}
Beno\^it Mandelbrot.
\newblock {Fractal aspects of the iteration of $z\rightarrow\lambda z(1-z)$ for
  complex $\lambda,z$}.
\newblock {\em Annals NY Acad. Sci.}, 357:249, 1980.

\bibitem{Metropolis53}
Nicholas Metropolis, Arianna~W. Rosenbluth, Marshall~N. Rosenbluth, Augusta~H.
  Teller, and Edward Teller.
\newblock Equation of state calculations by fast computing machines.
\newblock {\em Journal of Chemical Physics}, 21:1087--1092, 1953.

\bibitem{Hastings70}
W.K. Hastings.
\newblock Monte carlo samping methods using markov chains and their
  applications.
\newblock {\em Biometrika}, pages 97--109, 1970.

\bibitem{Christensen:2001gj}
Nelson Christensen, Renate Meyer, Lloyd Knox, and Ben Luey.
\newblock {II: Bayesian Methods for Cosmological Parameter Estimation from
  Cosmic Microwave Background Measurements}.
\newblock {\em Class. Quant. Grav.}, 18:2677, 2001, astro-ph/0103134.

\bibitem{Lewis:2002ah}
Antony Lewis and Sarah Bridle.
\newblock Cosmological parameters from cmb and other data: a monte- carlo
  approach.
\newblock {\em Phys. Rev.}, D66:103511, 2002, astro-ph/0205436.

\bibitem{Lorenz:2007ze}
Larissa Lorenz, Jerome Martin, and Christophe Ringeval.
\newblock {Brane inflation and the WMAP data: a Bayesian analysis}.
\newblock {\em JCAP}, 0804:001, 2008, 0709.3758.

\bibitem{Dunkley:2008ie}
J.~Dunkley et~al.
\newblock {Five-Year Wilkinson Microwave Anisotropy Probe (WMAP) Observations:
  Likelihoods and Parameters from the WMAP data}.
\newblock {\em Astrophys. J. Suppl.}, 180:306--329, 2009, 0803.0586.

\bibitem{MacKayBook}
D.~J.~C. MacKay.
\newblock {\em Information Theory, Inference and Learning Algorithms}.
\newblock Cambrdige University Press, 2003.
\newblock \url{http://www.inference.phy.cam.ac.uk/mackay/itprnn/book.html}.

\bibitem{Guth:2000ka}
Alan~H. Guth.
\newblock {Inflation and eternal inflation}.
\newblock {\em Phys. Rept.}, 333:555--574, 2000, astro-ph/0002156.

\bibitem{Linde:2008xf}
Andrei Linde, Vitaly Vanchurin, and Sergei Winitzki.
\newblock {Stationary Measure in the Multiverse}.
\newblock {\em JCAP}, 0901:031, 2009, 0812.0005.

\bibitem{GarciaBellido:1996qt}
Juan Garcia-Bellido, Andrei~D. Linde, and David Wands.
\newblock {Density perturbations and black hole formation in hybrid inflation}.
\newblock {\em Phys. Rev.}, D54:6040--6058, 1996, astro-ph/9605094.

\bibitem{Lyth:2011kj}
David~H. Lyth.
\newblock {Primordial black hole formation and hybrid inflation}.
\newblock 2011, 1107.1681.

\bibitem{Martin:2010hh}
Jerome Martin, Christophe Ringeval, and Roberto Trotta.
\newblock {Hunting Down the Best Model of Inflation with Bayesian Evidence}.
\newblock 2010, 1009.4157.

\bibitem{Barnaby:2006cq}
Neil Barnaby and James~M. Cline.
\newblock {Nongaussian and nonscale-invariant perturbations from tachyonic
  preheating in hybrid inflation}.
\newblock {\em Phys. Rev.}, D73:106012, 2006, astro-ph/0601481.

\bibitem{Vilenkin:1983xp}
Alexander Vilenkin.
\newblock {QUANTUM FLUCTUATIONS IN THE NEW INFLATIONARY UNIVERSE}.
\newblock {\em Nucl. Phys.}, B226:527, 1983.

\bibitem{Gong:2010zf}
Jinn-Ouk Gong and Misao Sasaki.
\newblock {Waterfall field in hybrid inflation and curvature perturbation}.
\newblock 2010, 1010.3405.

\bibitem{Fonseca:2010nk}
Jose Fonseca, Misao Sasaki, and David Wands.
\newblock {Large-scale Perturbations from the Waterfall Field in Hybrid
  Inflation}.
\newblock 2010, 1005.4053.

\bibitem{Abolhasani:2010kn}
Ali~Akbar Abolhasani, Hassan Firouzjahi, and Mohammad~Hossein Namjoo.
\newblock {Curvature Perturbations and non-Gaussianities from Waterfall Phase
  Transition during Inflation}.
\newblock 2010, 1010.6292.

\bibitem{Abolhasani:2010kr}
Ali~Akbar Abolhasani and Hassan Firouzjahi.
\newblock {No Large Scale Curvature Perturbations during Waterfall of Hybrid
  Inflation}.
\newblock 2010, 1005.2934.

\bibitem{Lyth:2010ch}
David~H. Lyth.
\newblock {Issues concerning the waterfall of hybrid inflation}.
\newblock 2010, 1005.2461.

\bibitem{Desroche:2005yt}
Mariel Desroche, Gary~N. Felder, Jan~M. Kratochvil, and Andrei~D. Linde.
\newblock {Preheating in new inflation}.
\newblock {\em Phys. Rev.}, D71:103516, 2005, hep-th/0501080.

\bibitem{Kodama:2011vs}
Hideo Kodama, Kazunori Kohri, and Kazunori Nakayama.
\newblock {On the waterfall behavior in hybrid inflation}.
\newblock 2011, 1102.5612.

\bibitem{Abolhasani:2011yp}
Ali~Akbar Abolhasani, Hassan Firouzjahi, and Misao Sasaki.
\newblock {Curvature perturbation and waterfall dynamics in hybrid inflation}.
\newblock 2011, 1106.6315.

\bibitem{Lyth:2010zq}
David~H. Lyth.
\newblock {Contribution of the hybrid inflation waterfall to the primordial
  curvature perturbation}.
\newblock 2010, 1012.4617.

\bibitem{Clesse:2010prepa}
Sebastien Clesse and Christophe Ringeval.
\newblock {in preparation}.
\newblock 2010.

\bibitem{Barnaby:2006km}
Neil Barnaby and James~M. Cline.
\newblock {Nongaussianity from Tachyonic Preheating in Hybrid Inflation}.
\newblock {\em Phys. Rev.}, D75:086004, 2007, astro-ph/0611750.

\bibitem{Martin:2005hb}
Jerome Martin and Marcello Musso.
\newblock {On the reliability of the Langevin pertubative solution in
  stochastic inflation}.
\newblock {\em Phys. Rev.}, D73:043517, 2006, hep-th/0511292.

\bibitem{Starobinsky:1980te}
Alexei~A. Starobinsky.
\newblock {A new type of isotropic cosmological models without singularity}.
\newblock {\em Phys. Lett.}, B91:99--102, 1980.

\bibitem{Falciano:2008gt}
Felipe~T. Falciano, Marc Lilley, and Patrick Peter.
\newblock {A classical bounce: constraints and consequences}.
\newblock {\em Phys. Rev.}, D77:083513, 2008, 0802.1196.

\bibitem{Lilley:2011ag}
Marc Lilley, Larissa Lorenz, and Sebastien Clesse.
\newblock {Observational signatures of a non-singular bouncing cosmology}.
\newblock 2011, 1104.3494.

\bibitem{Barkana:2006ep}
Rennan Barkana and Abraham Loeb.
\newblock {The Physics and Early History of the Intergalactic Medium}.
\newblock {\em Rept. Prog. Phys.}, 70:627, 2007, astro-ph/0611541.

\bibitem{Furlanetto:2006jb}
Steven Furlanetto, S.~Peng Oh, and Frank Briggs.
\newblock {Cosmology at Low Frequencies: The 21 cm Transition and the
  High-Redshift Universe}.
\newblock {\em Phys. Rept.}, 433:181--301, 2006, astro-ph/0608032.

\bibitem{website:lofar}
Lofar radio-telescope website.
\newblock \url{http://www.lofar.org/}.

\bibitem{website:mwa}
Mwa radio-telescope website.
\newblock \url{http://www.mwatelescope.org/}.

\bibitem{website:ska}
Ska project website.
\newblock \url{http://www.skatelescope.org/}.

\bibitem{Tegmark:2008au}
Max Tegmark and Matias Zaldarriaga.
\newblock {The Fast Fourier Transform Telescope}.
\newblock {\em Phys. Rev.}, D79:083530, 2009, 0805.4414.

\bibitem{Lewis:2007kz}
Antony Lewis and Anthony Challinor.
\newblock {The 21cm angular-power spectrum from the dark ages}.
\newblock {\em Phys. Rev.}, D76:083005, 2007, astro-ph/0702600.

\bibitem{Hirata:2006bn}
Christopher~M. Hirata and Kris Sigurdson.
\newblock {The Spin-Resolved Atomic Velocity Distribution and 21-cm Line
  Profile of Dark-Age Gas}.
\newblock {\em Mon. Not. Roy. Astron. Soc.}, 375:1241--1264, 2007,
  astro-ph/0605071.

\bibitem{Lewis:1999bs}
Antony Lewis, Anthony Challinor, and Anthony Lasenby.
\newblock Efficient computation of {CMB} anisotropies in closed {FRW} models.
\newblock {\em Astrophys. J.}, 538:473--476, 2000, astro-ph/9911177.

\bibitem{McQuinn:2007dy}
Matthew McQuinn, Lars Hernquist, Matias Zaldarriaga, and Suvendra Dutta.
\newblock {Studying Reionization with Ly-alpha Emitters}.
\newblock {\em Mon. Not. Roy. Astron. Soc.}, 381:75--96, 2007, 0704.2239.

\bibitem{Barkana:2005xu}
Rennan Barkana and Abraham Loeb.
\newblock {Probing the Epoch of Early Baryonic Infall Through 21cm
  Fluctuations}.
\newblock {\em Mon. Not. Roy. Astron. Soc. Lett.}, 363:L36--L40, 2005,
  astro-ph/0502083.

\bibitem{Jelic:2010qt}
Vibor Jelic.
\newblock {Cosmological 21cm experiments: Searching for a needle in a
  haystack}.
\newblock {\em PoS}, ISKAF2010:028, 2010, 1008.4356.

\bibitem{Mao:2007ti}
Xiao-Chun Mao and Xiang-Ping Wu.
\newblock {Signatures of the Baryon Acoustic Oscillations on 21cm Emission
  Background}.
\newblock 2007, 0709.3871.

\bibitem{Wyithe:2007rq}
Stuart Wyithe, Abraham Loeb, and Paul Geil.
\newblock {Baryonic Acoustic Oscillations in 21cm Emission: A Probe of Dark
  Energy out to High Redshifts}.
\newblock 2007, 0709.2955.

\bibitem{Adshead:2010mc}
Peter Adshead, Richard Easther, Jonathan Pritchard, and Abraham Loeb.
\newblock {Inflation and the Scale Dependent Spectral Index: Prospects and
  Strategies}.
\newblock {\em JCAP}, 1102:021, 2011, 1007.3748.

\bibitem{Carilli:2007eb}
C.~L. Carilli, J.~N. Hewitt, and Abraham Loeb.
\newblock {Low frequency radio astronomy from the moon: cosmic reionization and
  more}.
\newblock 2007, astro-ph/0702070.

\bibitem{Morales:2005qk}
Miguel~F. Morales, Judd~D. Bowman, and Jacqueline~N. Hewitt.
\newblock {Improving Foreground Subtraction in Statistical Observations of 21
  cm Emission from the Epoch of Reionization}.
\newblock {\em Astrophys. J.}, 648:767--773, 2006, astro-ph/0510027.

\bibitem{Cooray:2004kt}
Asantha Cooray.
\newblock {Large Scale Non Gaussianities in the 21 cm Background}.
\newblock {\em Mon. Not. Roy. Astron. Soc.}, 363:1049, 2005, astro-ph/0411430.

\bibitem{Visbal:2008rg}
Eli Visbal, Abraham Loeb, and J.~Stuart~B. Wyithe.
\newblock {Cosmological Constraints from 21cm Surveys After Reionization}.
\newblock {\em JCAP}, 0910:030, 2009, 0812.0419.

\bibitem{Wyithe:2008mv}
Stuart Wyithe and Abraham Loeb.
\newblock {The 21cm Power Spectrum After Reionization}.
\newblock 2008, 0808.2323.

\bibitem{Loeb:2008hg}
Abraham Loeb and Stuart Wyithe.
\newblock {Precise Measurement of the Cosmological Power Spectrum With a
  Dedicated 21cm Survey After Reionization}.
\newblock {\em Phys. Rev. Lett.}, 100:161301, 2008, 0801.1677.

\bibitem{Wyithe:2007gz}
Stuart Wyithe and Abraham Loeb.
\newblock {Fluctuations in 21cm Emission After Reionization}.
\newblock 2007, 0708.3392.

\bibitem{Natarajan:2009bm}
Aravind Natarajan and Dominik~J. Schwarz.
\newblock {Dark matter annihilation and its effect on CMB and Hydrogen 21 cm
  observations}.
\newblock {\em Phys. Rev.}, D80:043529, 2009, 0903.4485.

\bibitem{Hernandez:2011ym}
Oscar~F. Hernandez, Yi~Wang, Robert Brandenberger, and Jose Fong.
\newblock {Angular 21 cm Power Spectrum of a Scaling Distribution of Cosmic
  String Wakes}.
\newblock 2011, 1104.3337.

\bibitem{Brandenberger:2010hn}
Robert~H. Brandenberger, Rebecca~J. Danos, Oscar~F. Hernandez, and Gilbert~P.
  Holder.
\newblock {The 21 cm Signature of Cosmic String Wakes}.
\newblock {\em JCAP}, 1012:028, 2010, 1006.2514.

\bibitem{Joudaki:2011sv}
Shahab Joudaki, Olivier Dore, Luis Ferramacho, Manoj Kaplinghat, and Mario~G.
  Santos.
\newblock {Primordial non-Gaussianity from the 21 cm Power Spectrum during the
  Epoch of Reionization}.
\newblock 2011, 1105.1773.

\bibitem{Khatri:2009aw}
Rishi Khatri and Benjamin~D. Wandelt.
\newblock {21 cm radiation: A new probe of fundamental physics}.
\newblock 2009, 0910.2710.

\bibitem{Percival:2006ss}
Will~J. Percival and Michael~L. Brown.
\newblock {Likelihood methods for the combined analysis of CMB temperature and
  polarisation power spectra}.
\newblock {\em Mon. Not. Roy. Astron. Soc.}, 372:1104--1116, 2006,
  astro-ph/0604547.

\end{thebibliography}
